\documentstyle[prd,aps,amstex,preprint]{revtex}

\input epsf
\begin{document}
\begin{titlepage}
\rightline{
\vbox{\halign{&#\hfil\cr
&Fermilab-Pub-97/328-E \cr}}}
\vspace{1in}

\begin{center}

{\Large\bf
Determination of the Mass of the $W$ Boson Using the D\O\ Detector at
        the Tevatron}

\vskip 2.0cm
\normalsize

{\large The D\O\ Collaboration}
\vskip 0.3cm
{\it Fermi National Accelerator Laboratory \\
Batavia, IL 60510 \\}
(October 8, 1997)

\end{center}

\begin{abstract}
A measurement of the mass of the $W$ boson is presented which is based
on a sample of 5982 $W \rightarrow e \nu$ decays observed
in $p\overline{p}$ collisions at $\sqrt{s}$ = 1.8~TeV with the D\O\ detector
during the 1992--1993 run. 
From a fit to the transverse mass spectrum,
combined with measurements of the $Z$ boson mass,
the $W$ boson mass is measured to be
$M_W =  80.350
              \pm 0.140 ~\rm{\left (stat. \right )}
              \pm 0.165 ~\rm{\left (syst. \right )}
              \pm 0.160 ~\rm{\left (scale \right )}
                  ~\rm {GeV/}c^2$.
Detailed discussions of the determination of the absolute energy scale,
the measured efficiencies, and all systematic uncertainties are presented.
\end{abstract}

\end{titlepage}



\vskip 2.0cm

%
\author{                                                                      
B.~Abbott,$^{30}$                                                             
M.~Abolins,$^{27}$                                                            
B.S.~Acharya,$^{45}$                                                          
I.~Adam,$^{12}$                                                               
D.L.~Adams,$^{39}$                                                            
M.~Adams,$^{17}$                                                              
S.~Ahn,$^{14}$                                                                
H.~Aihara,$^{23}$                                                             
G.A.~Alves,$^{10}$                                                            
E.~Amidi,$^{31}$
% ...
\\                                                                            
\vskip 0.50cm                                                                 
\centerline{(D\O\ Collaboration)}                                             
\vskip 0.50cm                                                                 
}                                                                             
\address{                                                                     
\centerline{$^{1}$Universidad de los Andes, Bogot\'{a}, Colombia}             
\centerline{$^{2}$University of Arizona, Tucson, Arizona 85721}               
\centerline{$^{3}$Boston University, Boston, Massachusetts 02215}             
\centerline{$^{4}$Brookhaven National Laboratory, Upton, New York 11973}      
\centerline{$^{5}$Brown University, Providence, Rhode Island 02912}           
\centerline{$^{6}$Universidad de Buenos Aires, Buenos Aires, Argentina}       
\centerline{$^{7}$University of California, Davis, California 95616}          
\centerline{$^{8}$University of California, Irvine, California 92697}         
\centerline{$^{9}$University of California, Riverside, California 92521}      
\centerline{$^{10}$LAFEX, Centro Brasileiro de Pesquisas F{\'\i}sicas,        
                  Rio de Janeiro, Brazil}                                     
\centerline{$^{11}$CINVESTAV, Mexico City, Mexico}                            
\centerline{$^{12}$Columbia University, New York, New York 10027}             
\centerline{$^{13}$Delhi University, Delhi, India 110007}                     
\centerline{$^{14}$Fermi National Accelerator Laboratory, Batavia,            
                   Illinois 60510}                                            
\centerline{$^{15}$Florida State University, Tallahassee, Florida 32306}      
\centerline{$^{16}$University of Hawaii, Honolulu, Hawaii 96822}              
\centerline{$^{17}$University of Illinois at Chicago, Chicago,                
                   Illinois 60607}                                            
\centerline{$^{18}$Indiana University, Bloomington, Indiana 47405}            
\centerline{$^{19}$Iowa State University, Ames, Iowa 50011}                   
\centerline{$^{20}$Korea University, Seoul, Korea}                            
\centerline{$^{21}$Kyungsung University, Pusan, Korea}                        
\centerline{$^{22}$Institute of Nuclear Physics, Krak\'ow, Poland}            
\centerline{$^{23}$Lawrence Berkeley National Laboratory and University of    
                   California, Berkeley, California 94720}                    
\centerline{$^{24}$Louisiana Tech University, Ruston, Louisiana 71272}        
\centerline{$^{25}$University of Maryland, College Park, Maryland 20742}      
\centerline{$^{26}$University of Michigan, Ann Arbor, Michigan 48109}         
\centerline{$^{27}$Michigan State University, East Lansing, Michigan 48824}   
\centerline{$^{28}$Moscow State University, Moscow, Russia}                   
\centerline{$^{29}$University of Nebraska, Lincoln, Nebraska 68588}           
\centerline{$^{30}$New York University, New York, New York 10003}             
\centerline{$^{31}$Northeastern University, Boston, Massachusetts 02115}      
\centerline{$^{32}$Northern Illinois University, DeKalb, Illinois 60115}      
\centerline{$^{33}$Northwestern University, Evanston, Illinois 60208}         
\centerline{$^{34}$University of Notre Dame, Notre Dame, Indiana 46556}       
\centerline{$^{35}$University of Oklahoma, Norman, Oklahoma 73019}            
\centerline{$^{36}$University of Panjab, Chandigarh 16-00-14, India}          
\centerline{$^{37}$Institute for High Energy Physics, 142-284 Protvino,       
                   Russia}                                                    
\centerline{$^{38}$Purdue University, West Lafayette, Indiana 47907}          
\centerline{$^{39}$Rice University, Houston, Texas 77005}                     
\centerline{$^{40}$Universidade do Estado do Rio de Janeiro, Brazil}          
\centerline{$^{41}$University of Rochester, Rochester, New York 14627}        
\centerline{$^{42}$CEA, DAPNIA/Service de Physique des Particules,            
                   CE-SACLAY, Gif-sur-Yvette, France}                         
\centerline{$^{43}$Seoul National University, Seoul, Korea}                   
\centerline{$^{44}$State University of New York, Stony Brook,                 
                   New York 11794}                                            
\centerline{$^{45}$Tata Institute of Fundamental Research,                    
                   Colaba, Mumbai 400005, India}                              
\centerline{$^{46}$University of Texas, Arlington, Texas 76019}               
\centerline{$^{47}$Texas A\&M University, College Station, Texas 77843}       
}                                                   

\maketitle

\newpage

\section{Introduction}

Among electroweak measurables,  the mass of the $W$ boson  $M_W$ is of
crucial importance.  Along with the determination of the mass of the top 
quark\cite{d0_top,cdf_top} and in conjunction  with other precisely
determined quantities, including the mass of the $Z$ boson $M_Z$ the
electroweak Standard Model \cite{sm}   is constrained.  This paper
discusses the details of the first measurement of $M_W$ by the
D\O\ ~collaboration using data from the 1992--1993 running of the Fermilab
Tevatron Collider. It includes essential calibrations which will be used in
future D\O\ measurements.  A first report of this measurement was
published in Ref.~\cite{D0-96}. 

An early success of the CERN  $p\bar p$ collider was the discovery
and      measurement  of  the    masses  of  both   the  $W$  and  $Z$
bosons\cite{UA1-83,UA2-83,zfirst}. Table~\ref{whistory} gives a history  of
the  published values of the  direct  measurements of $M_W$.  The approach
taken  in this  analysis  is  similar to  those  of the UA2
\cite{UA2-92} and CDF \cite{CDF-90,CDF-95} experiments.

For the physics of $W$ bosons,  the electroweak measurables of interest 
are $M_W$ and
$\sin^{2}\theta _W$, where $\theta_{W}$ is the weak mixing angle. Both can be
measured precisely and  can be predicted from the lowest order relations of the
model\cite{hbarc}:
\begin{eqnarray}
    M_W & = & M_Z \cos \theta_{W} 
        \label{wzc}  \\
    \alpha_{\mbox{\tiny EM}}  & = & \frac{g^{2} \sin ^2
\theta_{W}}{4\pi} 
        \label{egs}   \\
    \frac{G_{\mu}}{\sqrt{2}} & = & \frac{g^{2}}{8M_W^{2}}.
        \label{Ggw}
\end{eqnarray}
Here, $\alpha_{\mbox{\tiny EM}}$  is the 
fine structure constant,
$g$ is the gauge coupling associated with the $SU(2)_{L}$
gauge group, and $G_{\mu}$ is the Fermi coupling constant.
The weak coupling, the electric charge, and the weak mixing 
angle are related by $\tan \theta_{W}=g' /g$, where $g'$ 
is the  coupling of the $U(1)$ gauge group. 

The standard set of measurable input parameters is the following:
  \begin{eqnarray}
  \alpha_{\mbox{\tiny EM}} & = & 1/(137.0359895 \pm 0.0000061) \label{alpha}\\
  G_{\mu} & = & 1.16639\, ( \pm 0.00002)\times 10^{-5}\,  \mbox{\rm GeV}^{-2} 
  	\;  \\
  	M_Z & = & 91.1884 \pm 0.0022\,  \mbox{GeV/}c^{2}. \label{mz}
  \end{eqnarray}
The fine
structure constant is measured from  the  
quantum Hall effect\cite{pdg}; the Fermi
coupling constant is measured from the muon 
lifetime \cite{pdg}, and $M_Z$ is measured
directly by the combined LEP experiments \cite{zmass}.

In order  to confront  the model  beyond the  lowest  order, a self--consistent
theoretical  scheme for dealing with the effects of higher orders of 
perturbation theory is required.   Different theoretical prescriptions 
motivate  particular definitions of the  weak mixing  angle. The
determination of
$M_W$ and the ratio of neutral to charged current cross sections 
in deep inelastic neutrino 
scattering are most naturally interpreted in terms of the ``on shell'' scheme
\cite{Sirlin} in which the weak mixing angle is defined by
Eq.~(\ref{wzc})  with both  masses as  measurable quantities.   

Within a given scheme, radiative  corrections are most easily  included  through
the use  of a  single  measurable parameter,
$\Delta r$,   which is  analogous to  $(g-2)$ in quantum  
electrodynamic radiative
corrections.  Like $(g-2)$, $\Delta r$ can be determined experimentally and its
 measurement can be directly compared with its  theoretical prediction. 
In the leading log approximation,  it can be written
in terms of $M_W$ as
\begin{equation}
  \Delta r = 
  1 - \frac{\pi \alpha_{\mbox{\tiny EM}} /
    \sqrt{2} }{G_\mu M_Z^{2}(M_W/M_Z)^{2}[1-(M_W/M_Z)^{2}]}.
\label{delr}
\end{equation}

Roughly   90\% of  the  value of   $\Delta r$  is due  to  light quark loop
corrections to $\alpha_{\mbox{\tiny EM}}$,  while the balance is due to the
embedded  physics of heavy  quarks and  the Higgs  boson.  Any physics
beyond the Higgs boson and known heavy quarks would also contribute to
$\Delta  r$.   Prior to  the   measurement described here,  
the  world  average of
$M_W$ ($M_W=80.33 \pm 0.170$~GeV/$c^2$),  
$M_Z$\cite{zmass},  and  Eq.~(\ref{delr}),  results in
$\Delta r = 0.0384 \pm 0.0100$, which is $3.8 \sigma$ from 
the tree level prediction. 

Because $\Delta r$ is dominated by QED corrections, it is interesting to
separate out those ``residual'' effects which are  distinguishable from
electrodynamic effects alone.  Such a separation isolates possible new
physics as well as physics directly associated with the top quark and
the Higgs boson. A prescription for doing this has been suggested by
defining 
$(\Delta r)_{\rm res}$~\cite{sirlin-gambino}, 
\begin{eqnarray}
    { \alpha_{\mbox{\tiny EM}} \over 1 \,-\, \Delta r }  & = & 
    { \alpha_{\mbox{\tiny EM}}(M_Z^2) \over 1 \,-\, (\Delta r)_{\rm res} }.
    \label{delr-res}
\end{eqnarray}
With a determination of $\Delta r$ plus a separate evaluation of 
$\alpha_{\mbox{\tiny EM}} (M_Z^2)$\cite{alpha},
$(\Delta r)_{\rm res}$ can be extracted. Evaluation of this quantity makes a
particularly economical probe of the Standard Model possible.

\subsection{Plan of the D\O\ Measurement}

The D\O\ detector is  a calorimetric detector with nearly
full kinematical coverage for electrons, hadrons, and muons. The inner
tracking    region does not include a magnetic field. Calibration  of the
electromagnetic,    and by  extension  the  hadronic,   calorimeter 
was
accomplished by exposing calorimeter test modules to charged 
particle beams of known
energies and compositions, as well as {\it in situ}  decays of known
particles. The
D\O\ determination of
$M_W$  relies on  the   determination of the mass ratio
$M_W/M_Z$ and the
subsequent   scaling of this ratio  by the  precisely determined $M_Z$
from  LEP\cite{zmass}.  This approach is similar  to 
that of the UA2 experiment\cite{UA2-92}.  The  significant 
advantage of determining
$M_W/M_Z$  is  that a  number  of   systematic  uncertainties cancel
in the ratio. This paper addresses those in detail.

The production and decay characteristics of $W$ bosons in a $p\bar p$ 
collider present a variety of challenges which drive the analysis 
strategy.  Because the statistical power of this measurement is at the 
level of less than 150 MeV/$c^2$ ($<0.2\%$), it is
necessary to understand both the experimental and theoretical 
systematic uncertainties to a precision comparable to this level.  The 
$Z$ boson data are used for studying many of the experimental 
uncertainties, so the total uncertainty of an $M_W$ determination is 
strongly coupled with the size of the $Z$ boson data set.  Hence, 
future determinations will gain in statistical and systematic 
precision with the sizes of both the $W$ and $Z$ boson data sets.

Uncertainties in modeling the production of $W$ and $Z$ bosons present 
a different set of challenges which do not necessarily scale with the 
number of events.  For example, at Tevatron energies roughly 80\% of 
the annihilations which
produce  $W$  bosons   involve  sea  quarks. Additionally,  the  substantial
probability of gluon radiation from the initial state quarks results in 
significant transverse momentum of  the $W$ boson. Both of these
theoretical issues involve uncertainties which  complicate  the simulation.

For the $Z$ boson the observables come
from the reaction chain:
\begin{equation}
	p\bar p \rightarrow Z(\rightarrow e^+ + e^-)+{{\cal H}_Z}(\rightarrow 
	\mbox{\rm hadrons}),
	\label{zprod}
\end{equation}
where ${\cal H}_Z$ is the hadronic recoil against the 
transverse motion of the $Z$
boson. Both electrons are fully measured and the 
dielectron mass is determined from
\begin{equation}
  M_Z  = \sqrt{ 2 E^{e_1}  \, E^{e_2}   \,-\,
                  2 {\vec p}^{\,e_1} \cdot {\vec p}^{\,e_2} }.
                  \label{meedef}
\end{equation}

With the D\O\ detector, the electron decay  mode of the $W$ boson leads to the
most  precise mass determination. This is  due to  the   cleaner 
signal  and  better resolution, as compared to the muon or tau decay modes.  
In  this  experiment,   the  single  relevant channel is
$W\rightarrow e \nu _e $. The
electrons are emitted with  transverse momenta $p_T^e$ of order 40
$\rm {GeV/}c$ and the neutrino is emitted with a comparable momentum, escaping
without  detection.  This leaves a large component of missing energy  
\mbox{${\hbox{$E$\kern-0.6em\lower-.1ex\hbox{/}}}$ }
in  the  event   of which only  the transverse component,
$\mbox{${\hbox{$E$\kern-0.6em\lower-.1ex\hbox{/}}}_T$ }$
is determined. Therefore, the defining characteristics of $W$ bosons are a
high $p_T$ electron accompanied by 
significant $\mbox{${\hbox{$E$\kern-0.6em\lower-.1ex\hbox{/}}}_T$}$. 

The  observable quantities for this measurement come from the 
$W$ boson production and decay chain:
\begin{equation}
	p\bar p \rightarrow W(\rightarrow e+\nu)+{{\cal H}_W}(\rightarrow 
	\mbox{\rm hadrons}),
	\label{wprod}
\end{equation}
where ${\cal H}_W$ is the hadronic recoil against the 
transverse motion of the $W$
boson.  Since a complete characterization of the neutrino 4-momentum is
impossible,  the only quantities directly  measured 
are the electron momentum and
the transverse momentum of the recoil  $\vec p_T^{\,  {\text rec}}$. 

Using these measurables, the two body kinematics of this decay 
provide at least two methods for measuring $M_{W}$.  The transverse 
energy spectrum of the electron will exhibit a kinematical edge (the 
``Jacobian edge'') at $M_{W}/2$ for $W$ bosons with transverse 
momentum $p_T^W$ equal to zero.  However, resolution effects and 
nonzero values of $p_T^W$ smear the 
$p_T^e$ and $\mbox{${\hbox{$E$\kern-0.6em\lower-.1ex\hbox{/}}}_T$ }$ spectra 
and therefore affect the use of the sharp edge as a measure of 
$M_W$.

To  control  the   systematic  effects  while   retaining the  highest
statistical  precision,  the  ``transverse mass''  is used to determine
$M_W$. It is defined as\cite{wmt}
\begin{eqnarray}
M_{T}^{2} & = & 
2E_T^{e}E_T^{\nu}-2\vec{p}_{T}^{\; e}{\scriptscriptstyle \, 
\stackrel{\bullet}\, {{}}} 
          \vec{p}_{T}^{\; \nu} \nonumber \\
          & = & 2E^{e}_{T}E^{\nu}_{T}(1-\cos{\phi_{e\nu}}),
          \label{mtdef}
\end{eqnarray}
where  $p_T^{\nu}$  is the  transverse  momentum  of the  neutrino and
$\phi_{e\nu}$  is  the  azimuthal angle  between  the   electron and   
neutrino\cite{light}.  The
transverse mass  also exhibits a  Jacobian edge,   but at the value of
$M_{W}$ and  with much  less sensitivity  to $p_T^W$.  Hence, precise
determination of the  location of this edge  determines $M_W$. The effect  of
both the  finite width   $\Gamma(W)$  and $p_T^W$ does distort the shape of the
$M_T$ spectrum\cite{reduce}.  
While the  transverse  momentum  of the  $W$ boson is  relatively low,
peaking at approximately $p_T^W\sim 5 ~\rm{GeV/}c$ in this analysis, 
even this small amount can be significant.  

Equation   \ref{mtdef}   shows  that the   necessary  ingredients for
determining $M_T$ are
$\vec{p}_{T}^{\;e}$,    $\vec{p}_{T}^{\;\nu}$,   and the angle between them.  
Among these, only $\vec p_T^{\;e}$ is determined directly. Since momentum
transverse to the beam  is  conserved, a measured imbalance can be
attributed to the neutrino.  Therefore, in the absence of detector effects,  
the 
neutrino transverse energy is equal to the missing transverse  energy and
calculated from the measured 
${\vec p}_{T}^{\,{\text rec}}$ and ${\vec p}_{T}^{\,e}$,
\begin{equation}
\vec p_T^{\;\nu} \equiv  
\mbox{${\hbox{${\vec E}$\kern-0.6em\lower-.1ex\hbox{/}}}_T$ } =
-{\vec p}_{T}^{\,{\text rec}} - {\vec p}_{T}^{\,e}.
\end{equation}

\begin{figure}[h]
\begin{center}
\begin{tabular}{c}
   \epsfxsize=10.0cm
   \leavevmode
   \epsffile[55 190 520 670]{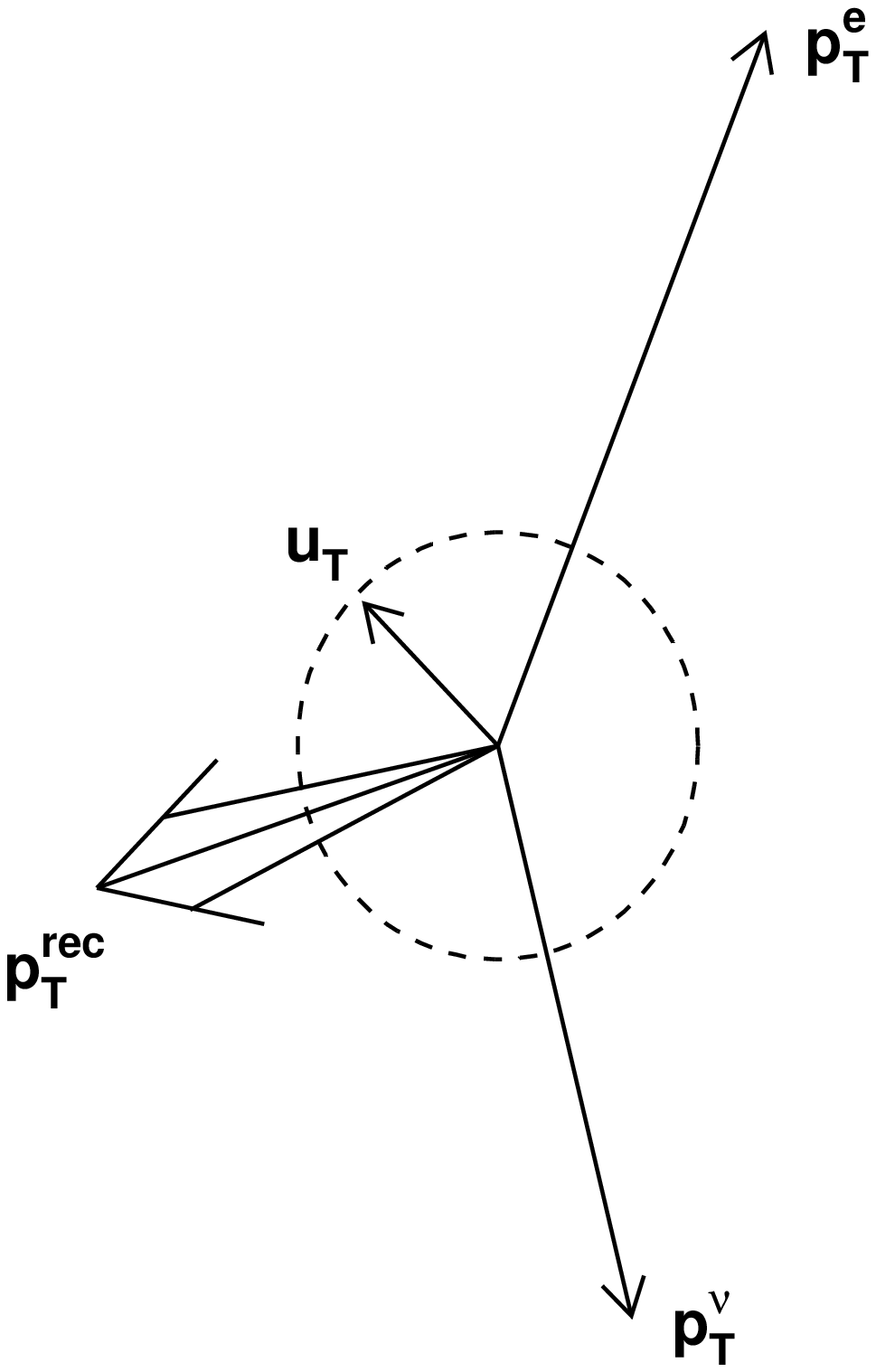}
\end{tabular}
\vspace{8.0cm}
\caption{Kinematic quantities for $W$ events.}
\label{fig:wevent}
\end{center}
\end{figure}

However,   the  reaction  given in   Eq.~(\ref{wprod})   does not fully 
describe the situation since  energy measurement in a  
calorimeter includes other
effects.  Energy  lost in detector  cracks and
inefficiencies can introduce biases in the magnitude 
and direction of the total energy.  The 
interactions  of the  remaining  spectator quarks in 
the proton and antiproton  will add
energy,  as  will noise and ``pileup'' due  to the   
residual  energy  from   multiple
interactions.    Designating these
additional,    non--recoil,   luminosity  dependent  contributions the
``underlying event'',   $\vec{u}_T({\cal  L})$,  the measured neutrino
transverse momentum is given by
\begin{eqnarray}
\mbox{${\hbox{${\vec E}$\kern-0.6em\lower-.1ex\hbox{/}}}_T^\prime$ } \,&=&\,
-{\vec p}_{T}^{\,{\text rec}}-
{\vec u}_T({\cal L})-{\vec p}_{T}^{\,e} \nonumber  \\
                     \,&=&\,
    - [ {\vec p}_{T}^{\,{\text rec}} + {\vec u}_T^{\,{\text rec}}({\cal L})]
    - [ {\vec u}_T^{\,e}({\cal L}) + {\vec p}_{T}^{\,e} ]   \ .
    \label{mtmom}
\end{eqnarray}
The two quantities within the brackets are not distinguished from one another in
the measurement but must be dealt with in the analysis. Figure
\ref{fig:wevent} shows the kinematics of the $W$ boson events.

Since  there  is no   analytic   description of  the   transverse mass
distribution,   determination of  $M_W$  relied on modeling the
transverse mass  spectrum through  a Monte Carlo  simulation.  The $W$ boson
mass was extracted by comparing the measured distribution in transverse
mass to the Monte Carlo distribution generated for different $W$ boson
masses.   The simulation relied
on experimental data as much  as possible  and used  $Z$ boson events,  not 
only to set the energy scale,  but also to  understand the electron
energy resolution, the energy underlying the $W$ boson,  and the scale in
$p_{T}^{W}$.   In the simulation $W$ bosons were
generated with a  relativistic Breit-Wigner  line shape that was skewed
by the mass dependence of the parton luminosities.  The longitudinal and
transverse   momentum   spectrum were  given by  a  double  differential
distribution calculated to next--to--leading order.  

The decay products and  the $W$ boson  recoil  system  
were  traced  through  the 
simulated detector with  resolution smearing.  Minimum 
bias events (collisions which
are recorded with little or no trigger bias) mimic the debris in the event 
produced by the  spectator quarks and  pileup associated with  multiple 
interactions.   The minimum  bias events  also properly 
included  residual  energy 
from  previous  crossings.   The generated spectra in  
transverse mass for  different
values  of
$M_W$ were compared to the measured spectra  by a maximum 
likelihood method,  and the
best fit value of the mass obtained. 

The measurements reported in this paper are: $M_W$, as determined from fits to
the  $M_T$ distribution and fits to the $p_T^e$ and $p_T^\nu$
distributions and $M_W/M_Z$. In addition $\Delta r$ and $(\Delta
r)_{{\text res}}$ are determined. 

The paper is organized as follows: 
\begin{itemize}
\item Section II: a brief description   of  the   detector;   \item Section
III: data   collection, reconstruction,  the corrections applied to the data
and the selection of the final sample;  \item Section IV: determination of 
the parameters  used in the Monte Carlo simulation; \item Section  V: Monte
Carlo simulation; \item Section VI: results of the  fits; \item Section VII:
effects of systematic errors due to the parameters determined in Section IV
and the assumptions described in Section V; \item Section VIII: consistency
checks; \item Section IX: conclusions; \item Appendix A: $W$ Boson and $Z$ Boson
Production Model; \item Appendix B: Bremsstrahlung; and \item Appendix C: Mean
Number of Interactions.
\end{itemize}

\section{D\O\ Detector and Trigger System}

The data collected for this measurement were taken during the exposure of
the  D\O\ detector to collisions of protons and antiprotons at a center of 
mass energy of $\sqrt{s}=1800 ~{\rm GeV}$ in the 1992--1993 running period
of the Fermilab Tevatron collider. This was the first beam exposure of this
experiment and the total luminosity accumulated was $12.8~{\rm pb}^{-1}$. 
The average instantaneous luminosity was 
${\cal L}= 3.4\times 10^{30} ~{\rm cm}^{-2} {\rm s}^{-1}$, which 
corresponded to less than 1 collision per crossing which occurs for
${\cal L}\sim 6\times 10^{30} ~{\rm cm}^{-2} {\rm s}^{-1}$

The D\O\ detector was designed to study a variety of high transverse
momentum physics topics and has been described in   detail
elsewhere~\cite{D0nim}. The detector has nearly full acceptance for
electrons, photons, and muons and measures jets, electromagnetic (EM)
showers,  and $\mbox{${\hbox{$E$\kern-0.6em\lower-.1ex\hbox{/}}}_T$ }$ 
with  good resolution~\cite{topprd}. 
The detector consists of
three major subsystems: a  tracking system, uranium-liquid argon
calorimeters, and a muon toroidal spectrometer. The components of the
detector which are most relevant to this analysis are briefly described 
below.

\subsection{Tracking System}

The tracking system was used to reconstruct charged particle tracks over the
region $|\eta|<3.2$\cite{def_eta}  and to reconstruct the interaction vertex
of the event. It consists of four subsystems: a drift chamber surrounding
the vertex region (VTX), a  transition radiation detector (TRD), a central
drift chamber (CDC) and two forward drift chambers (FDC). The VTX, TRD, and
CDC cover the large angle region and are oriented parallel to the beam axis.
The FDC's  cover the small angle region and are oriented perpendicular to
the beam axis. In addition to the $r\phi$ measurement of hits in the CDC,
delay lines  were used to measure track hit locations in the  $z$ direction.
The TRD provides an independent identification of electrons, in addition to
that provided by the calorimeters, allowing enhanced hadron rejection.

\subsection{Calorimeters}

The calorimeter system consists of one central (CC) and two end (EC)
calorimeters which measure the energy flow in the event over a
pseudorapidity range $|\eta| \leq 4.2$. The calorimeters are enclosed in
three separate cryostats which surround the tracking system. They each have
an electromagnetic, a fine hadronic (FH), and a coarse hadronic (CH)
section.  Liquid argon is employed as the active medium and uranium is the
absorber material in the EM and FH sections and copper (steel) is the
absorber in the CH section for the CC (EC). The inter-cryostat region (ICR)
is instrumented with  scintillator tile detectors 
which are located in the space  between the
EC and CC cryostats. These detectors were used to improve the energy
measurement of jets that  straddle two calorimeters.

The  calorimeters are arranged in a cylindrical geometry with  each EM
section being divided into four  longitudinal readout layers, for a total
depth of 21 radiation lengths. A projective tower  arrangement for 
readout points
toward the interaction region. The hadronic sections are 7--9 nuclear
interaction lengths deep and are divided into four (CC) or five (EC)
longitudinal readout layers. The transverse segmentation of the calorimeters
is 
$0.1\times 0.1$ in $\Delta \eta \times \Delta \phi$, except in the third
layer of the EM calorimeter which is at shower maximum, where it is 
$0.05\times 0.05$ in $\Delta \eta \times \Delta \phi$. 
Measured resolutions will be
discussed below.

\subsection{Muon Spectrometer}

The muon spectrometer provides identification and momentum determination for
muons. It surrounds the calorimeters and consists of planes of proportional
drift tubes  which surround  magnetized iron toroids and covers a region
$|\eta|<3.3$. There is one layer of proportional tubes on the inner face of
the magnet and two layers, separated by $\approx$ 1m, outside  the magnet.
The material in the calorimeter and iron toroids is about 12 interaction
lengths thick, making hadronic punch-through to the outer two layers
negligible. The muon momentum resolution is $\sigma(1/p) = 0.18(p-2)/p^2
\oplus 0.008$ (momentum $p$ in GeV/c).

\subsection{Triggers}

The Tevatron beam crossings occurred every $3.5~ \mu {\rm s}$. For a $p\bar p$ 
total cross section   at 
$\sqrt s = 1.8 ~{\rm TeV}$ of approximately  70 mb\cite{totxsec}, there is 
an interaction rate of $\approx$ 200 kHz at a typical instantaneous
luminosity of $3\cdot 10^{30} ~{\rm cm^{-2}s^{-1}}$. In order to record
events at
$\approx 2$ Hz,  
three stages of hardware and
software triggers were used.

To indicate the presence of a collision within the detector 
and to calculate a fast
approximation to the vertex position, radial scintillation hodoscope arrays are
positioned in the forward directions subtending angles 
of $2.3< |\eta| < 3.9$. To pass
the level 0  (L0)  hardware trigger, coincident hits in these counters  
were
required on both sides of the interaction region, 
signaling an inelastic   collision within the detector volume and also
providing an estimate of the $z$ position of the interaction 
vertex. This trigger
provided the minimum  bias data set used in this analysis. The trigger rate 
depended on the luminosity and for the data analyzed here, was typically 90 kHz.

Events passing the L0 trigger were then passed to the  level~1 (L1)
hardware trigger. Here a decision was made based on the fast analog sums of
all the EM layer calorimeter signals which represent the  energies in
trigger towers. These towers were segmented as 
$0.2 \times 0.2$ in $\Delta \eta
\times\Delta\phi$ with coverage extending to $|\eta| < 3.2$. The L1
electronics  restricted the maximum trigger rate to $\approx
200$ Hz and decisions  made by it and by the L0 trigger were made between beam
crossings.

The final stage of triggering was the level~2 (L2) software trigger which
ran on a farm of 48 VAXstation  4000 M60 processors.  The typical
processing time for an event in L2 was 350 $\mu \rm s$, resulting in an
average deadtime of $\approx 2\%$. The full segmentation of the calorimeter
was available at this trigger level and a full event reconstruction was done,
albeit  with simplified algorithms and coarser segmentation. There were 32
different L1 components and 128 different L2 components which could be
constructed and prescaled at different rates. Each logical combination 
targeted
specific physics for given accelerator conditions. 
These data sets were written out to
corresponding output data streams.

\subsubsection{Trigger requirements for the $W$ boson and $Z$ boson data sets}

For the determination of the $W$ boson mass, the electron decay modes were
required for the selection of both $W$ and $Z$ bosons. While the
characteristics of both are similar---the presence of high transverse
momentum  electrons---the different rates  and backgrounds require distinct
selection criteria.  Because of the presence of a neutrino in $W$ boson decay
events, 
a minimal $\mbox{${\hbox{$E$\kern-0.6em\lower-.1ex\hbox{/}}}_T$ }$ 
requirement was used in the selection of $W$ boson
candidates.

To select electrons, the L1 trigger required the transverse energy in the EM
layers of a trigger tower to be above a preselected threshold. For the
selection of $W$ boson events, the L1 trigger  
required  at least one EM trigger tower
above 10 GeV. For $Z$ boson events, at least two EM 
trigger towers with $E_T > 7$
GeV were required.

The L2 electron algorithm used the full segmentation of the EM calorimeter
to measure the energy deposited by the EM shower and is described in  detail
in Ref.~\cite{topprd}. Trigger towers above threshold were used as seeds to form
energy clusters which included all calorimeter  cells in the four EM layers
and the first FH layer in a window
$\Delta \eta \times \Delta \phi = 0.3 \times 0.3$. For the selection of $W$
events, an energy cluster with $E_T > 20 $ GeV was required by the L2  
filter. For $Z$ boson event candidates, two energy clusters,
each with 
$E_T > 10 $ GeV, were required by the L2  
filter. Transverse and  longitudinal shape
requirements as well as isolation requirements were also placed on the energy cluster
for the selection of
$W$ boson events.

The $\mbox{${\hbox{$E$\kern-0.6em\lower-.1ex\hbox{/}}}_T$ }$ 
in the event was calculated in the L2 trigger and was
required to be above  20 GeV for the $W$ boson event selection. It was computed
using the vector sum of the $E_T$ of all the cells in the calorimeter and
the ICD  with respect to the $z$ position of the interaction vertex, 
determined by L0. Prescaled triggers of $W$ boson events were also recorded 
without the  $\mbox{${\hbox{$E$\kern-0.6em\lower-.1ex\hbox{/}}}_T$ }$ 
trigger requirement in order to study 
efficiencies and biases. Table~\ref{trigger} lists a summary of the trigger
requirements used in the selection of the $W$ boson and $Z$ boson data samples.

\subsubsection{Main Ring veto}

The 150 GeV/$c$ conventional accelerator (main ring) passed through
the coarse hadronic part of the D\O\ calorimeters. Halo particles 
accompanying the circulating
beam can deposit  energy in the calorimeter and corrupt measurements both at
the trigger and offline reconstruction levels.  Such unrelated energy 
depositions in a localized part of the detector will affect the 
$\mbox{${\hbox{$E$\kern-0.6em\lower-.1ex\hbox{/}}}_T$ }$ 
determination and therefore considerable care was required  in
the utilization of triggers taken while the Main Ring beam passed through
the detector. A veto gated on the injection period of the main ring cycle
(the first 0.4s of the 2.4s cycle) was used in some of the L2 filters to
avoid any adverse effects. 
For the mass analysis, all $Z\rightarrow ee$ events were used; 
$Z\rightarrow ee$ candidates recorded during the veto of the main-ring period
were excluded from  resolution studies. No $W$ boson events were taken from
triggers occuring during the main ring cycle.



\section{$W$ and $Z$ Boson Data Sample: Event Selection and Data
Processing}

\subsection{Offline Data Processing and Candidate Event Selection}

\label{sec:evtreco}

Once data were written to tape, the digitized information 
was converted using an initial
calibration.  The initial calibration of the CDC was based 
on  measurements using cosmic ray
data. The calibration of the delay lines of the CDC, 
measuring the $z$ coordinate of
the hits, was derived  from pulser measurements on the bench combined with
cosmic ray data.  The initial calibration of the calorimeter was obtained
from test beam  measurements~\cite{D0nim}.  Corrections to the calorimeter
calibration were required, including corrections for an adjustment to
the operating voltage of  the calorimeter, corrections to the sampling
weights, and to  the gains of individual calorimeter cells.   In addition, a
correction due to a difference in liquid argon temperature  at the test beam
and D\O~was applied~\cite{E-scale}. It should be noted that none of the
central EM calorimeter modules  that were tested were installed in the final
calorimeter. 

The azimuthal uniformity of the central electromagnetic calorimeter  was
determined using approximately 3.5~million triggers from an  inclusive
electron data sample\cite{zhu}.   By equalizing the event rate above a 13~GeV
threshold for  each calorimeter module, relative calibration constants were
determined  to an accuracy of 0.5\%, assuming that 
the observed $\phi$ variations
were  instrumental in origin.   These relative calibration constants showed a
variation in the response between  different modules with a maximum
difference of 5\%. The variations were dependent on which of the 32~EM
modules was  struck by the electron, and not by other features of the
calorimeter  such as a variation in the amount of material in the 
tracking  detector. All of the above corrections to the energy are
propagated into the 
$\mbox{${\hbox{$E$\kern-0.6em\lower-.1ex\hbox{/}}}_T$ }$ calculation.

\subsubsection{EM Clustering}

Electrons and photons were reconstructed as  energy clusters in the EM and
first FH section of the calorimeter.   Towers were defined by adding the
energy measured by the calorimeter in all  four EM layers plus the first FH
layer for cells within $0.1\times  0.1$ in $\Delta \eta \times \Delta
\phi$.   Towers were grouped together with their adjacent neighbors, provided 
their energies are above 50 MeV.  Clusters of adjacent calorimeter  towers
with significant energy depositions were then formed using a nearest 
neighbor clustering algorithm~\cite{youssef}.  

The observed  energy of the EM cluster is given by 
\begin{equation}
E^{\rm meas}  =  \delta^{TB} + C\sum_{i}\beta_{i} \, S_{i} 
\label{SW}
\end{equation}
where $\delta^{TB}$ is an offset in the energy response due to 
energy loss in the material upstream of the test calorimeter. 
$C$ is the conversion constant from the digital signal to energy 
and $\beta_{i}$ are the sampling weights for 
the $i^{\rm th}$ layer with energy deposition (in ADC counts) 
$S_{i}$. 
The sum runs over all five layers in the EM calorimeter which contribute
to the EM cluster.  Both $\beta_i$ and
$\delta^{TB}$ were determined from test beam measurements using electron beams
over a broad range of energy and rapidity. From these test beam
measurements, the offset was determined to be 
$\delta^{TB} = 347$~MeV.  This level of energy scale determination, based on
the test beam  measurements and taking into account corrections due to the
transfer  of the calibration from the test beam to D\O, resulted in an
energy  scale approximately 5\% lower than the nominal energy scale for the
central calorimeter,  as observed from the measured $Z$ boson mass. 
Both the final
overall energy scale and the offset were re--computed {\it in situ}, as will
be discussed in the next section.

\subsubsection{Electron Identification}

Electrons were  identified by a combination of  topological and kinematic
identifiers as described in~\cite{topprd}.   The main electron identification
requirements are
below .

The cluster shower shape can  be characterized by both longitudinal and
lateral energy depositions.  The fraction of the cluster energy which is
deposited in  the EM calorimeter is defined as $f_{EM}$.   Since charged
hadrons  deposit less than $\approx 10\%$ of their energy in the EM
calorimeter, 
$f_{EM}$ provides a powerful discriminant. 
$f_{EM}$ was required to be greater than $90\%$. 

The electron candidate cluster was required to have a topology, both
longitudinal and lateral,  which was consistent with that of electrons
from a detailed {\footnotesize GEANT} Monte Carlo simulation~\cite{geant} 
which was
extensively  compared to test beam measurements.  A covariance matrix of
41~observables was defined to characterize an  electron shower~\cite{topprd}. 
A $\chi^2$ parameter was defined to measure the consistency of the shower 
with that expected for electrons. For central electrons a $\chi^2<100$ was
required; for electrons in the  end calorimeters $\chi^2<200$ was imposed. 
These requirements were   $\approx 94\%$ efficient.   A rejection factor of
about 4 against EM clusters that are not  due to single electrons was
achieved.

Electrons from $W$  and $Z$ boson decays tend to be isolated 
from other particles in the
electromagnetic calorimeter.  
An offline
isolation requirement was used which is defined as 
\begin{equation}
f_{\rm iso} = \frac{E_{tot}(R_{4}) - E_{EM}(R_{2})}
                          {E_{EM}(R_{2})}
\label{fiso}
\end{equation}
where $E_{tot}$ is the total energy in 
cone of radius $R_4=0.4$ in $\eta\times\phi$-space.
$E_{EM}$ is the EM energy in a cone of radius $R_2=0.2$.
An isolation requirement of $f_{\rm iso} < 0.15$ was placed on
the cluster  energies for electrons from both $W$ and $Z$ boson event
candidates.

An important source of background for electrons is photons from 
$\pi^0$ or $\eta$ meson decays which are adjacent to unrelated tracks.   
This background was
reduced by  requiring that a track from a charged particle in the  
tracking  detector be
consistent with the position of the cluster in the  calorimeter.   To
qualify as a match, the shower centroid was required to link with a 
reconstructed  track
with significance $\sigma_{trk}$, 
\begin{equation}
\sigma_{trk} 
\equiv \sqrt{\left(\frac{R\Delta\varphi} {\sigma_{R\Delta\varphi}} \right)^2 
+ \left(\frac{\Delta z}{\sigma_{z}}\right)^2}
\end{equation}
where $R\Delta\varphi$ and $\Delta z$ are the spatial mismatches  between
the track projection and cluster position in the  $\varphi$ and $z$
directions, respectively, and 
$\sigma_{R\Delta\varphi}$ and $\sigma_{z}$ are the associated  experimental
resolutions.   For the data set used in this analysis $\sigma_{trk} < 10$
was  imposed for CC electron candidates. This cut was used to 
minimize bias and results
in an efficiency of $>98$\%\cite{track-top}.

\subsubsection{Electron Direction Determination}

The optimum resolution in the electron polar angle is obtained using 
the $z$ position of the electron cluster as obtained from the 
calorimeter information and the $z$ position of the center-of-gravity 
($cog$) of the CDC track.  These two points thus define the polar 
angle of the electron.  The position of an electron in the calorimeter 
was determined from the energy depositions in the third EM layer of the 
shower using a $\log(E)$ energy weighting algorithm~\cite{pos_loge}.  
The parameters of the algorithm were determined using both test beam 
and Monte Carlo data.  Further study with collider data demonstrated 
the need to remove a residual bias in the $z$ position of the $cog$ of 
the track.  This was accomplished {\it in situ} using an inclusive 
muon data set which demonstrated that the 
$z$ position of the $cog$ of the CDC track was biased.

The true and measured $z$ position of the $cog$ of the track are related
  linearly by
 \begin{equation}
    z_{\rm true} \,=\, \alpha_{\rm CDC} \, z_{\rm meas}
                 \,+\, \beta_{\rm CDC}.
\label{cdc_alpha}
 \end{equation}
With the muon data sample, the scale factor $\alpha_{\rm CDC}$ was
determined by defining a track using the $cog$ of the muon track, as
reconstructed in the first layer of muon spectrometer before the toroidal
magnet, and the vertex
$z$ position.  By comparing the expected and measured $z$ positions the
scale factor was determined to be
$\alpha_{\rm CDC} = 0.988 \pm 0.002$, where the error is the combination of
a small statistical component ($\pm 0.0003$) and the following systematic
components: the observed variations in  scale factor for different azimuthal
regions of the detector (0.001), observed variations for different polar
angles of the muon tracks (0.001), muon chamber alignment (0.0003), and
different methods to extract $\alpha_{\rm CDC}$ (0.0004). The offset
$\beta_{\rm CDC}$ is consistent with zero. In this analysis $\alpha_{CDC}$ =
0.988 and $\beta_{CDC}=0$ were used. The scale factor has been confirmed using
cosmic ray muon data.  The trajectory of cosmic ray muons traversing the
full detector was reconstructed using the non-magnetic inner volume of the
spectrometer. As before, the expected CDC track positions were compared to
their measured positions and the scaling behavior of the $cog$ of the CDC
track was, within its uncertainty, confirmed.

To verify the consistency of the electron angle determination, 
$Z\rightarrow ee$ events were studied.  Given the electron cluster 
position, their track intersections with the beamline were 
reconstructed from the $Z\rightarrow ee$ decay.  These intersections 
in general do not coincide.  After applying the correction to the CDC 
$cog$, the width of the distribution in the difference in $z$ positions 
of the two intersections was tracked by varying the calorimeter 
electron $z$ position.  The resultant width of the distribution was 
at a minimum without applying any correction to the electron calorimeter 
position, showing the internal consistency of the procedure.

\subsubsection{Measurement of 
$\mbox{${\hbox{${\vec E}$\kern-0.6em\lower-.1ex\hbox{/}}}_T^\prime$ }$}

The total missing transverse energy in the event is calculated by
summing over all calorimeter and ICD cells
\begin{equation}
\mbox{${\hbox{${\vec E}$\kern-0.6em\lower-.1ex\hbox{/}}}_T$ } = 
- \sum_i E_i \, \sin\theta_i {\hat{u}_i}
        = - \sum_i {\vec E}_T^i.
\label{eq:etmis}
\end{equation} Here $\hat{u}_i$ is a unit vector from the event
vertex to the $cog$ of the $i$-th calorimeter cell,  
$E_i$ the energy in the $i$-th 
calorimeter cell,
and $\theta_i$ is the
polar angle given by the event vertex and the $cog$ of the $i$-th calorimeter
cell.

The nominal event vertex was determined using all tracks in the CDC.
The transverse momentum of the electron was calculated using the 
total energy and direction of the cluster in the calorimeter. Since, the
electron direction, as computed above, may not intersect  
the nominal event vertex, a
recalculation of the 
$\mbox{${\hbox{$E$\kern-0.6em\lower-.1ex\hbox{/}}}_T$ }$ 
was done by using the vertex obtained from the electron alone.  
For $Z\rightarrow ee$ events, the event vertex and the electron polar
angles were determined using a constrained fit of the measured variables
of the two electron directions.
The missing transverse energy used in this analysis was based on 
calorimetric information alone and was not corrected for possible muons 
in the event.

\subsection{Final $W\rightarrow e \nu$ and $Z\rightarrow ee$ Data Sample}

After electron identification and   calculation of the missing transverse
energy, the final 
$W$ boson and $Z$ boson candidate samples were subjected 
to the following selection
criteria:

Fiducial requirements were placed on the electrons in the $W$ boson 
candidate events to select central electrons: $ | i_{\eta}^{e} | \le 
12.  $ Here, $ i_{\eta}^{e}$ is an integer index for the calorimeter 
tower containing the most energetic cell of the electron cluster in 
the third EM layer.  It is equal to $10\times \eta$ for particles 
which originate at $z=0$.  In order to ensure no energy leakage 
into the uninstrumented regions within modules, electrons were 
restricted from the readout edges in $\phi$ by requiring that their 
impact to be within the central 80\% of each module.  An additional 
requirement was imposed that no event have a jet in which the fraction 
of energy in the CH section of the calorimeter exceeds 0.4.  This 
eliminated events with spurious energy depositions from the Main Ring.

The kinematic and fiducial 
requirements that defined the $W$ boson candidate sample are: 
\begin{itemize}
        \item $|i_{\eta}^e| \le 12$
	\item $E_T^{e} > 25 $ GeV 
        \item $\mbox{${\hbox{$E$\kern-0.6em\lower-.1ex\hbox{/}}}_T$ }  
> 25 $ GeV 
        \item $p_T^W  < 30 $ GeV/$c$ 
\end{itemize}
This resulted in a sample of 7262 events. Additionally, including a 
transverse mass cut of 
$M_T < 110 $~GeV/$c^2$ left 7234 candidates.

$Z$ boson candidate events were accepted with the requirements:
\begin{itemize}
        \item $|i_{\eta}^e| \le 12$ or $|i_{\eta}^e| \ge 15$
	\item $E_T^{e_{1,2}} > 25 $ GeV \ . 
\end{itemize}
As in the $W$ boson sample, the
module boundary edge cut was made for CC electrons only.
For the final $Z$ data sample, only events with both electrons
  in the CC ($|i_{\eta}^e| \le 12$) were used.
This resulted in a sample of 395 candidate events with both 
electrons in the CC. 
A $Z$ boson mass cut which eliminated events outside
a window of $70< m_{ee} <110$~GeV/$c^2$ left 366 candidates.
For some studies, events with one electron in the
  forward region ($|i_{\eta}^e| \ge 15$) were included.
For resolution studies,  $Z\rightarrow ee$ events had the
  additional requirement that events were not accepted 
  when taken during the Main Ring cycle. 
  
\begin{figure}[t]
 \begin{center}
  \begin{tabular}{cc}
    \epsfxsize = 8.cm   \epsffile{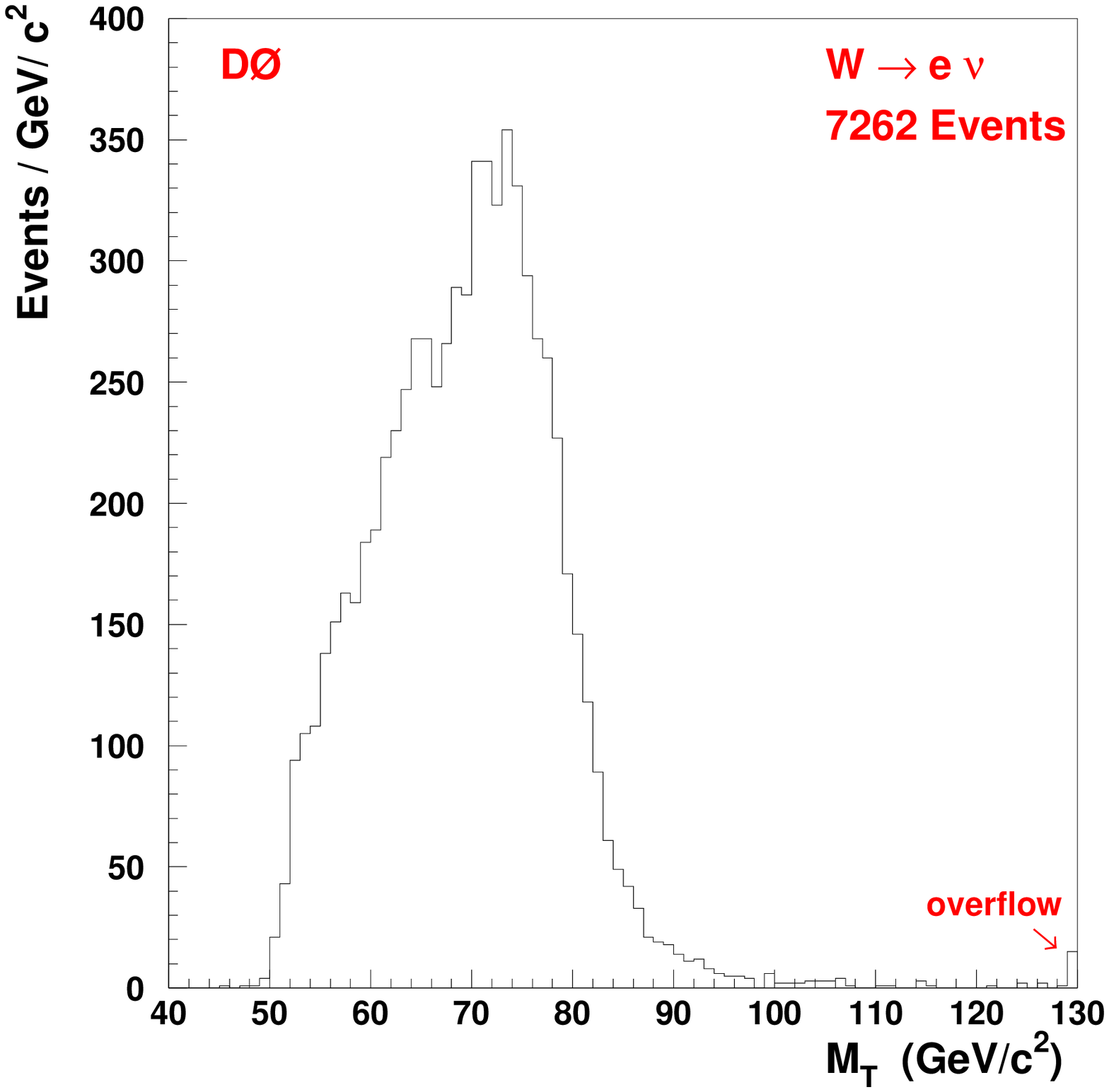}
    \epsfxsize = 8.cm   \epsffile{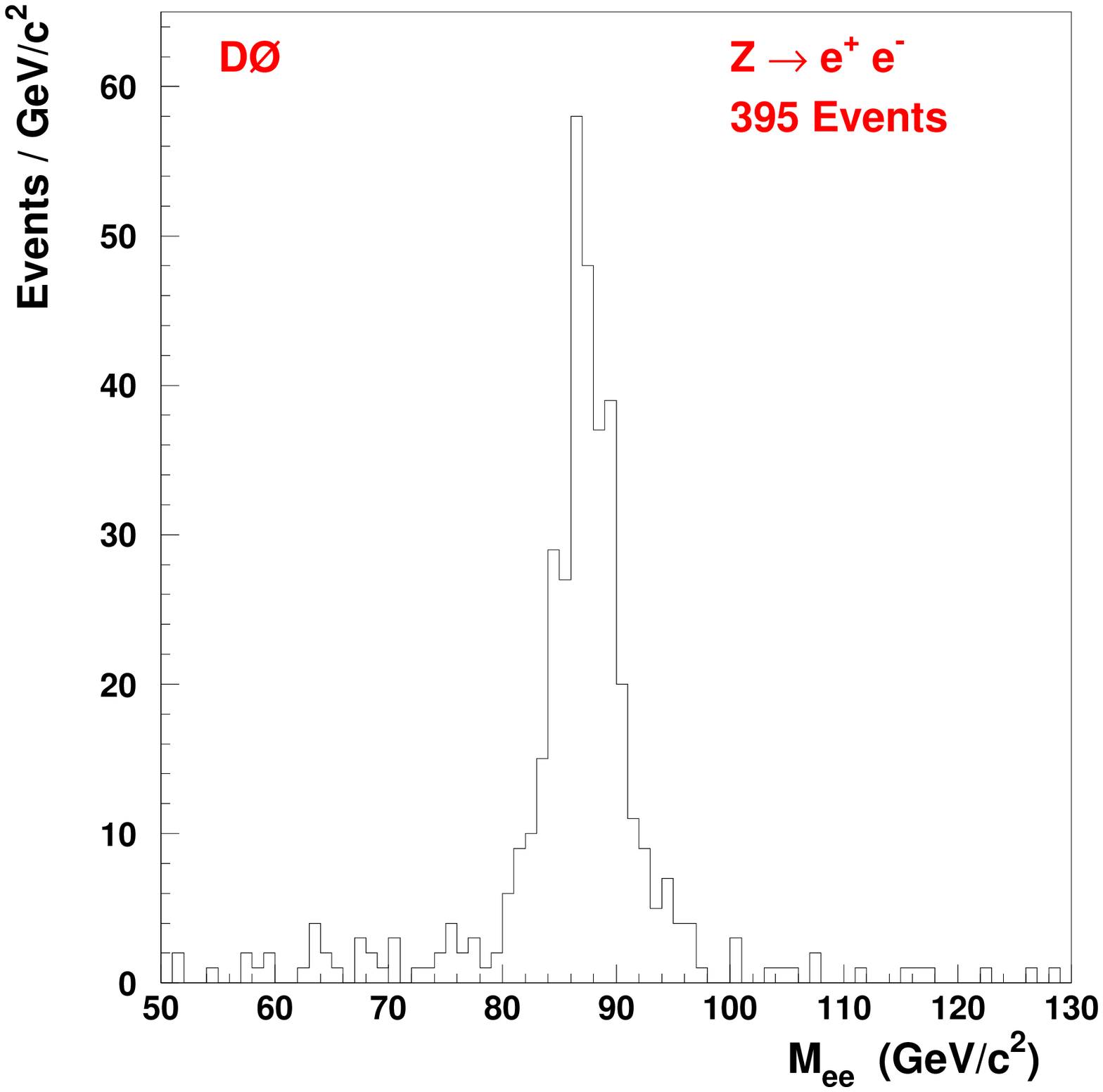}
  \end{tabular}
 \end{center}
\caption{(a) Transverse mass distribution of $W$ events and 
         (b) dielectron invariant mass distribution from $Z$ events.
           Both distributions are shown before the application of 
           fitting window cuts or energy scaling (see Section IV). }
\label{fig:data}
\end{figure}

Table~\ref{tab:num_wzevts}
lists the final number of events in the  samples. Figure \ref{fig:data}(a)
shows the transverse mass distribution  of the central $W$ 
boson candidate events,
before the transverse mass cut  and \ref{fig:data}(b) shows the invariant
mass distribution of the  central $Z$ boson candidate events. 
Neither distribution
is corrected for the electron energy scale determined {\it in situ} (see 
Section IV).

\subsection{Data Samples Used in the Analysis}

There were five primary data samples which are used in this analysis.
\begin{description}
\item{ $W\rightarrow e \nu$ sample:} 
    A sample of 7262 $W\rightarrow e \nu$ 
candidates (prior to fitting and transverse
mass cut) provided the main 
    data sample used to measure the $W$ boson mass .  
\item{$Z\rightarrow e^+e^-$ sample:} 
    A sample of 395 central $Z\rightarrow e^+e^-$ candidates 
    (prior to the $Z$ boson mass
cut) was used along 
    with the $W\rightarrow e \nu$ sample to measure the $W$ boson mass. 
    A slightly enlarged sample was used in the determination of detector 
    response parameters. 
\item{Minimum Bias sample:} A sample of approximately 
    50,000 triggers, taken at various luminosities,
    was used for modeling the underlying event.
\item{$J/\psi \rightarrow e^+e^-$ sample:} A data set of 
approximately 50 observed
$J/\psi$ candidates was used in 
    the electron energy scale determination.
\item{$\pi^0 \rightarrow \gamma \gamma \rightarrow e^+e^-e^+e^-$ sample:} A 
data set of approximately 2500 observed $\pi^0$
candidates was used 
  in the electron energy scale   determination.
\end{description}
\section{Determination of Parameters in Monte Carlo Simulation}

The extraction of $M_W$ relied on an accurate and fast  Monte Carlo
simulation.  The details of the physics model used in the  simulation
will be discussed in the next section.  However,  many parameters such
as calorimeter  response, efficiencies, and resolutions, were input to
the simulation  and were derived from the data. The focus of this section 
is a detailed description of how these parameters were determined.  
The use of these parameters
in the simulation appears in Section V and the systematic errors on
$M_W$ due to uncertainties inherent in these parameters is described in
Section VII.

\subsection{Electron Energy Scale}
\label{sec:escale}

All calorimetric measurements rely on the determination of the overall 
energy scale using particles of known momentum and/or on the reconstruction 
of the mass of well known particles. 
Both techniques have been used to calibrate the D\O\ calorimeter.   
Since the absolute energy scale
of the EM calorimeter was not known to the required precision,
the ratio of the measured $W$ boson and $Z$ boson masses and the world average
$Z$ boson mass were used to determine the $W$ boson mass.
A number of systematic effects, common to both measurements, cancel
in the ratio. Most notably, as will be discussed in more detail below,
the ratio was, to first order, insensitive to the absolute energy scale.
                                                                               
The initial calibration of the calorimeter was provided by transferring 
the calibration from a test beam to the collider detector, 
as discussed in Section~\ref{sec:evtreco}~\cite{escale}.
An important result of these test beam measurements was the demonstration
that the EM calorimeter is linear to better than 0.5\% for
electron energies exceeding 10~GeV. To complete the establishment of the
energy scale with the desired precision, it was necessary to determine to
what extent a possible offset in the energy response, as opposed to only
a scale factor, was responsible for the deviation of the ratio
$M_Z^{{\text D\O\ }} / M_Z^{\rm LEP}$ from unity.

A strategy for establishing the final energy scale and possible offset in the
response was implemented. Inherent to this program is the assumption
that the measured energy
$E^{\rm meas}$ is related to  the true energy, $E^{\rm true}$, by a
scale $\alpha$ and offset $\delta$:
\begin{eqnarray}
    E^{\rm meas} &=& \alpha E^{\rm true}+\delta.
\label{emeas}
\end{eqnarray}
Then, for a two body decay when $\delta \ll (E_1+E_2)$, 
the measured invariant mass of the decay products
$m^{\rm meas}$ is related to 
the true mass $m^{\rm true}$ by
\begin{equation}
    m^{\rm meas} \approx  \alpha m^{\rm true} \,+\, \delta \times f \ . 
\label{eq:m_meas}
\end{equation}
Here, $f$ is a parameter that depends on the kinematics of the decay 
and is given by
\begin{equation}
    f = {{(E_{1}^{\rm meas}+E_{2}^{\rm meas}) \, 
              (1-\cos\gamma)}   \over 
              {m^{\rm meas}}}
\end{equation}
where $E_{1,2}^{\rm meas}$ are the measured energies of the two 
decay products and $\gamma$ the opening angle between them.   
When $\delta$ is small, $f$ is nearly equal to
$\partial m^{\rm meas}/\partial \delta$.  
Hence, sensitivities to $\delta$ can be different, depending on $f$.  

Consequently, the dependence of the measured ratio of the $W$ boson to 
$Z$ boson masses on $\alpha,\delta$ can be estimated from the relation
\begin{eqnarray}
\lefteqn{\left. 
\frac{M_W (\alpha,\delta)}{M_Z (\alpha,\delta)}\right|_{\rm meas}
           = } \nonumber \\
   & & \left. \frac{M_W}{M_Z}\right|_{\rm true}
           \left[ 1 + \frac{\delta}{\alpha} \cdot
                      \frac{f_W \, M_Z - f_Z \, M_W}{M_Z \cdot M_W}
                      \right] \ .
\label{ratio_delta}
\end{eqnarray}
Here, $f_W$ and $f_Z$ correspond to average values of $f$ for the $W$
and $Z$  bosons,  respectively. Note that the 
determination of $M_W$ from this ratio
is insensitive to $\alpha$ if $\delta=0$, and that the correction due to
a non-vanishing  value  for $\delta$ is strongly suppressed due to 
the fact that  the $W$ and $Z$ boson masses are nearly equal. 

The values of $\alpha$ and 
$\delta$ were determined from the analysis of collider events containing 
two-body decays 
for which $m^{\rm true}$ is known from other measurements.  
The three decays used are the 
$Z \rightarrow ee$ decays, measurements of
$\pi^0 \rightarrow \gamma\gamma \rightarrow 4e$ states, 
and $J/\psi \rightarrow ee$ states. These three decays probe a 
useful  
range in $f$.
The reference mass values used as benchmarks are:
$M_Z = 91.1884 \pm 0.0022$~GeV/$c^2$\cite{zmass}, 
$m_{J/\psi} =  3.09688 \pm 0.00004$~GeV/$c^2$\cite{pdg}, 
and $m_{\pi^0} = 134.976\pm 0.0006$~MeV/$c^2$~\cite{pdg}.

\paragraph*{$Z\rightarrow e e$ Analysis:}

\begin{figure}[p]
   \epsfxsize=10.0cm
   \centerline{\epsffile[0 100 550 550]{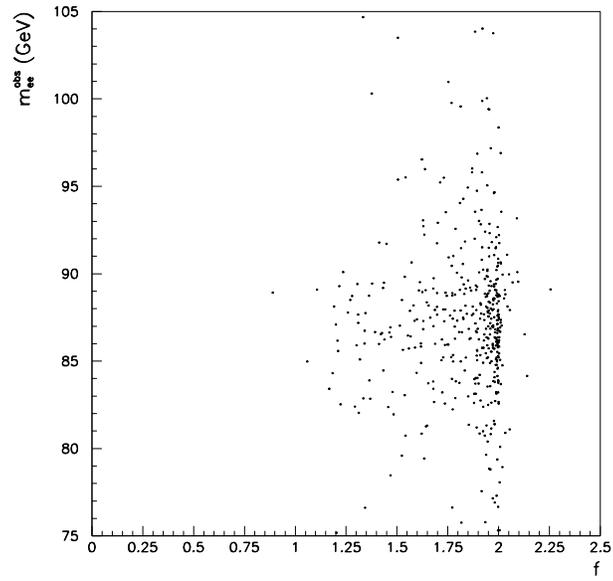}}
\caption{Observed invariant mass from $Z\rightarrow ee$ 
         decays, $m_{ee}^{\rm meas}$, versus $f$.}
\label{fig:mee_vs_f}
\end{figure}

\begin{figure}[p]
   \epsfxsize=10.0cm
   \centerline{\epsffile[0 100 550 550]{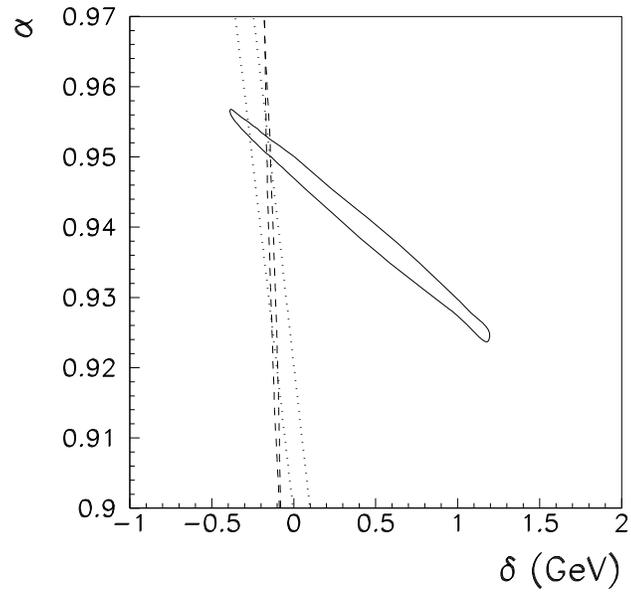}}
\caption
 {Constraints on $\alpha$ and $\delta$ from 
   a) $Z \rightarrow ee$  decays (solid contour), 
   b) $J/\psi \rightarrow ee$ decays (dotted lines), and 
   c) $\pi^0 \rightarrow \gamma\gamma \rightarrow 4e$ decays, 
      (dashed lines).}
\label{fig:Econtours}
\end{figure}

The strongest constraint on the energy scale uncertainty comes from
the $Z$ boson data. The fact that electrons from $Z$ boson decays are not
monochromatic is exploited by studying the invariant mass distribution as a
function of the variable $f_Z$. Small values of $f_Z$ correspond to the decay
of highly boosted $Z$ bosons with, on average, higher energies. The
dependence of the observed $Z$ boson mass as function of $f_Z$, shown in Fig.
~\ref{fig:mee_vs_f}, thus
directly translates into a constraint on the energy scale and offset. 
This analysis was based on $Z$ boson events with both electrons in the CC
 which were 
required to pass the same  selection criteria  as the final $Z$ boson sample, 
except that $E_T > 10$ GeV was required for both of the  electrons.   
The data were binned in $f_Z$ and the distribution in $m_{ee}^{\rm meas}$ 
 was fit using a convolution of the $Z$ boson Breit-Wigner resonance 
with a Gaussian resolution function. 
Using the standard Monte Carlo generator (described in the next section),  
sample distributions in $m_{ee}$ were generated under different assumptions 
for $\alpha$ and $\delta$. A $\chi^2$ comparison was performed between the 
data and the Monte Carlo and the $1\sigma$ constraint on $\alpha$
and $\delta$ from the $Z$ boson data, shown as the solid line in 
Fig.~\ref{fig:Econtours}, was determined.

\paragraph*{$\pi^0$ Analysis:}

\begin{figure}[p]
   \epsfxsize=10.0cm
   \centerline{\epsffile[0 100 550 550]{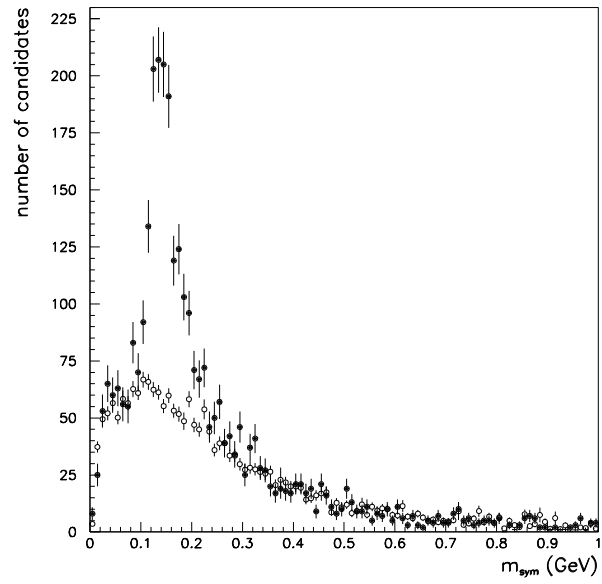}}
\caption
  {The invariant mass from  
   $\pi^0 \rightarrow \gamma\gamma \rightarrow  e^+e^-e^+e^-$ decay events
   (points). Also shown is the background contribution (open circles).}
\label{fig:pizero}
\end{figure}

\begin{figure}[p]
   \epsfxsize=10.0cm
   \centerline{\epsffile[0 100 550 550]{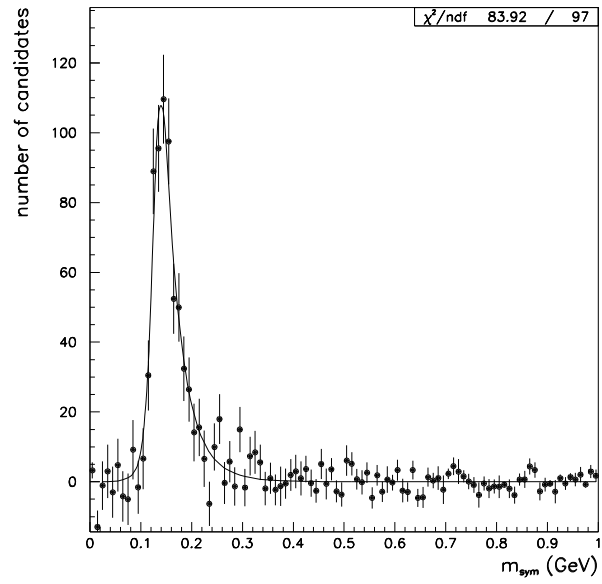}}
\caption{Background subtracted invariant mass from the $\pi^0$ event
    sample  (points) compared to the Monte Carlo simulation (line).}
\label{fig:pizero-model}
\end{figure}

$\pi^0$ mesons were observed through their two photon decay and subsequent
conversion to unresolved $e^+e^-$ pairs. There is a 10\% probability 
for each photon to convert in front of the CDC, so that when both 
photons convert the $dE/dx$ can be measured in the drift chamber 
and a strategy for the identification of $\pi^0$ decays is possible.
The identification requirement was that one electromagnetic cluster be observed
with two doubly ionizing tracks pointing to it.  
The diphoton opening angle, $\theta$, was calculated from the center of 
gravity of those two tracks and the measured vertex of the event. 
In this way, an approximation of the mass  was calculated as
\begin{equation}
    m_{\gamma \gamma}^{\rm meas} = 
    (E_{\rm clus})\sin{\theta \over 2},
\end{equation}
where $E_{\rm clus}$ is the cluster energy
which is equal to the sum of the photon
energies since the photons are not resolved in the calorimeter. 
This strategy  assumes a symmetric decay. 
Figure~\ref{fig:pizero} shows the signal and background, the latter 
determined from a single-conversion control sample. 
The invariant mass spectrum of the background-subtracted signal 
compares well with a Monte Carlo simulation shown as the
solid line in Fig.~\ref{fig:pizero-model}. 
The measured mass is $m_{\pi^0} = 135.4 \pm 10.0$~MeV/$c^2$.
The sensitivity to the energy scale and offset is  determined by 
varying both parameters in the Monte Carlo simulation and performing a
$\chi^2$ fit to the data.
Since the response given in Eq.~(\ref{emeas}) is the response per 
electron, the offset in response for the $\pi^0$ is  
$\delta_{\pi^0} = 4\delta$. 
This procedure maps out an allowed region in the
($\alpha,\delta$)-plane shown as the dashed line in
Fig.~\ref{fig:Econtours}.

\paragraph*{$J/\psi$ Analysis:}

\begin{figure}[p]
   \epsfxsize=10.0cm
   \centerline{\epsffile[0 100 550 550]{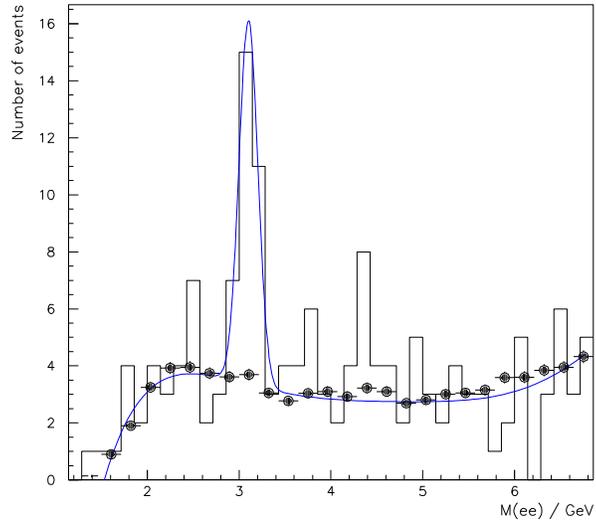}}
\caption{Mass distribution of the observed $J/\psi \rightarrow ee$ 
  decays (histogram) is shown above the background (points).
  The line is a fit to the signal plus background.}
\label{fig:Jpsi}
\end{figure}

A sample of $J/\psi \rightarrow ee$ events was also used in the EM 
energy scale determination. The data were collected in a set of special 
runs and had an integrated luminosity of 
$\approx  100~{\rm nb^{-1}}$.  
The L1 trigger required two EM triggers towers above a 2.5 GeV 
threshold with less than 1 GeV in the  corresponding hadronic towers.   
At L2, two EM energy clusters above 3 GeV were required and the 
isolation fraction was required to be $f_{\rm iso}< 0.4$. 
Since the major background is due to $\pi^0 \rightarrow \gamma\gamma$ 
and $\eta \rightarrow \gamma\gamma$ decays in which one of the photons converts 
before it reaches the tracking chamber, the track associated with the 
electron cluster was required to have an energy deposition in the 
 tracking chamber of less than 1.5 times the energy deposition of a 
minimum ionizing particle (MIP). In addition a cut was placed on the width 
of the cluster, defined as the average weighted distance of each cell 
of the cluster from its center. The weights are the 
same as those used in the position finding algorithm. 
The opening angle between the two electrons was determined from the event 
vertex and the cluster positions in the calorimeter. 
Figure~\ref{fig:Jpsi} shows a clear 
$J/\psi \rightarrow ee$ signal above  background.   
The background mass distribution for the $J/\psi$ signal 
was obtained independently by pairing EM energy clusters in
the calorimeter in which at least one of the EM clusters had no associated 
track. If there was an associated track, it was required to have an energy 
deposition greater than 1.5~MIP. 
The remaining requirements imposed on the EM energy clusters were 
the same as in the analysis of the $ee$ events.   The mass value fit for these
data is $m_{J/\psi} = 3.032\pm 0.035\pm 0.190$~GeV/$c^2$, where 
the first error is
statistical and the second is systematic.

The model used for comparison to the $m_{J/\psi}$ distribution was an 
{\footnotesize ISAJET}~\cite{isajet} based 
simulation for $b\bar b$ production and subsequent decay to $J/\psi$
followed by a {\footnotesize GEANT} detector simulation. 
Since there is additional jet activity close to the electrons
in $J/\psi$ decays from $b$ quarks, the 
contribution from the underlying event energy
was evaluated using the simulation. Two classes of events 
were generated consisting of only the $e^+e^-$ pair of the $J/\psi$ decay 
and events corresponding to the full $p\bar{p}$ collision. The difference 
between the fully simulated and 
the reconstructed mass was 80~MeV/$c^2$. This difference
was applied as a corection and an uncertainty of 100\% was 
assigned to this correction. 
Figure~\ref{fig:Econtours} shows the constraint on $\alpha$, $\delta$ 
from the $J/\psi$ analysis indicated by the dotted line.

\paragraph*{Underlying Event Contribution:}

Background energy and noise can contribute to the measurements of 
electron energies.  The different environments for $M_Z$, $m_{\pi 
^0}$, and $m_{J/\psi}$ final states led to different corrections for 
each.  Monte Carlo studies specialized to the scale analysis plus the 
understanding of instrumental effects lead to corrections for $M_Z$, 
$m_{\pi ^0}$, and $m_{J/\psi}$ of $0.17\pm 0.05$, $0.30\pm 0.10$, and 
$0.08\pm 0.08$~GeV/$c^2$, respectively.  The uncertainties on these 
measurements form the dominant uncertainties in the determination of 
$\alpha$ and $\delta$ for the $\pi^0$ and $J/\psi$ analyses.

\paragraph*{Combined Analysis:}

\begin{figure}[p]
   \epsfxsize=10.0cm
   \centerline{\epsffile[0 100 550 550]{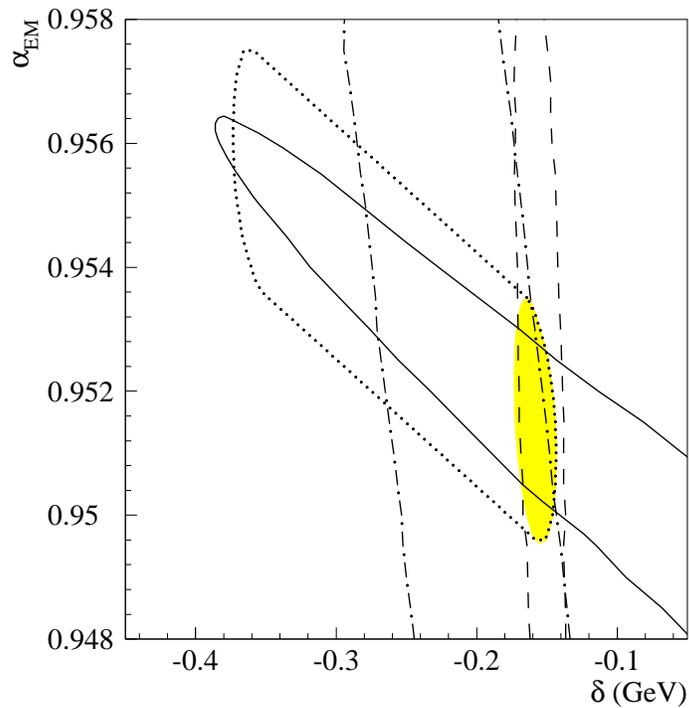}}
\vspace{2.0cm}
\caption{Expanded view of Fig.~\ref{fig:Econtours} showing $\alpha$ versus
         $\delta$ with the   combined best fit (shaded region). 
         The expanded lobe (dotted contour) to lower values of $\delta$ 
         is due to uncertainties in the low energy non-linear response 
         of the calorimeter.    
         The  contributions are from: 
            $Z \rightarrow ee$  decays (solid contour), 
            $J/\psi \rightarrow ee$ decays (dashed--dotted lines), 
            and $\pi^0 \rightarrow \gamma\gamma  \rightarrow 
               e^+e^-e^+e^-$ decays (dashed lines).}
\label{fig:Econt_blowup}
\end{figure}

The data from the three samples are combined by adding the $\chi^2$  
distributions. For the combined $\chi^2$ the minimum value is 
$\chi^2=53.8$ for 58 degrees of freedom, with a best fit of 
$\delta= -0.158\pm 0.016$ GeV  and 
$\alpha=0.9514 \pm 0.0018$.   
This is consistent with the result obtained from the $Z$ boson data only,  but
with substantially reduced errors.   
Figure~\ref{fig:Econt_blowup} shows an enlargement of the region where the 
contours from the three data samples overlap. 
The shaded area is the contour obtained from the combined analysis 
  for a unit change in the combined $\chi^2$.

The main contributions to the systematic uncertainties were the underlying 
event correction and possible non-linearities in the energy response. 
Varying the underlying event by the errors quoted above changes the value of
$\delta$ for which the combined 
$\chi^2$ is minimized by $\pm 2$~MeV  when varying the underlying
event contribution  to the
$J/\psi$ and   by $\pm 30$~MeV when varying the contribution  to the
$\pi^0$.  In addition the calorimeter $\pi^0$ response was  varied taking
$\delta_{\pi^0} = 3\delta$, rather than $4\delta$ as discussed above. This
 decreased the best  value for $\delta$ by 52~MeV. The dominant uncertainty
comes from a possible nonlinearity of the  calorimeter and has been
addressed by studying test beam data.  The test beam data permitted a small
nonlinear response of the  EM calorimeter and was parametrized by including 
a quadratic term in the energy response of Eq.~(\ref{emeas}),  which 
was
constrained by the test beam data to not  exceed 1 part in $10^4$.  
Allowing for a nonlinear response characterized by
such a quadratic term and repeating the above analyses results in 
an allowed region in $\alpha, \delta$
indicated by the  dotted line in Fig.~\ref{fig:Econt_blowup}. The result is to
decrease $\delta$ by 200 MeV. The energy scale parameters and their
uncertainties are thus 
\begin{eqnarray}
    \delta &=&  -0.158  \pm 0.016 {}^{+0.03}_{-0.21}  \quad {\rm GeV}, \\
    \alpha &=&    0.9514 \pm 0.0018 {}^{+0.0061}_{-0.0017} 
\end{eqnarray}
where the first error is statistical and the second systematic. The effect of a
possible quadratic response term was included  as the asymmetric 
contribution to the overall uncertainty shown on $\delta$.  

The result described in this section constitutes the calibration of the
central EM  calorimeter {\em after} the initial calibration based on a
transfer of the  calibration from the test beam given by Eq.~(\ref{SW}). In
practice, inserting the offset as defined in Eq.~(\ref{SW}) into
Eq.~(\ref{emeas}) leads to
\begin{equation}
E^{\rm meas}=\delta^{TB}
+ \delta + \alpha E^{\rm true}
\label{new16}
\end{equation}
demonstrating that the {\it in situ} determination of $\delta$ amounts to a
redetermination of $\delta^{TB}$.  Combining Eq.~(\ref{new16}) 
with Eq.~(\ref{SW})
leads to
\begin{equation}
    E^{\rm true}  = - {\delta  \over \alpha} \,+\,  
                 {\frac{C}{\alpha}}\sum_{i}\beta_{i} \, S_{i}\,.
\end{equation}
Using the redetermined values of $\alpha$ and $\delta$, an overall offset of
$ -\delta / \alpha = 158/0.9514= 166$~MeV was observed
consistent with the average energy loss by electrons in  the
material before the calorimeter which was predicted by  {\footnotesize GEANT}
Monte Carlo simulation studies.


\subsection{Hadronic Energy Scale}
\label{sec:ptwscale}

\begin{figure}[h]
   \epsfxsize=10.0cm
   \centerline{\epsffile[0 100 550 550]{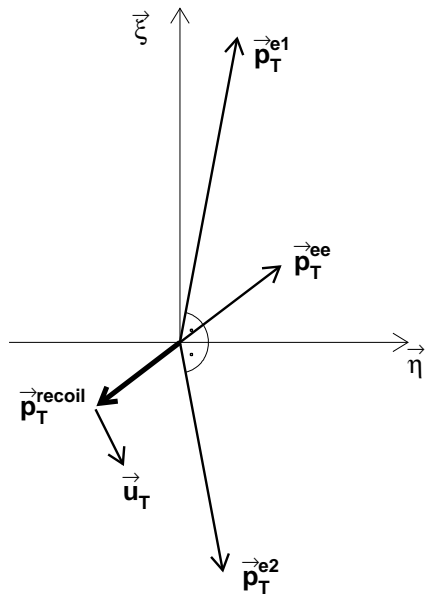}}
\caption{Definition of the $\eta$-$\xi$ coordinate system for $Z$ events. }
\label{fig:eta_xi}
\end{figure}

The scale of the measured recoil momentum differs from the electron  energy
scale because the recoil measurement also includes energy from hadronic
showers and  suffers from the loss of energy in uninstrumented regions of 
the calorimeter.   The response of the hadronic calorimeter relative to the
response of the  electromagnetic calorimeter was determined from 
$Z\rightarrow e e$ events.  In $Z\rightarrow e e$ events the
transverse momentum of the $Z$ boson $p_{T}^Z$  can be obtained from
either the  measurement of the transverse momenta of the two electrons 
${\vec p}_{T}^{\,ee}$ or from the recoil activity in the event 
$-{\vec p}_{T}^{\,{\text rec}}$.   The latter was the way in which 
$p_{T}^{W}$ was
measured.   To minimize the effects of the energy resolution in the 
determination of the hadronic energy scale relative to the electromagnetic 
energy scale, the  momentum imbalance was measured with  respect to the
($\eta,\xi$)-coordinate system\cite{UA2-92}.  The $\eta$ axis is  defined as
the bisector of the two electron transverse directions.   In the plane of the
electrons, the axis orthogonal to the $\eta$ axis  is the $\xi$ axis (see
Fig.~\ref{fig:eta_xi}). The $\eta$ imbalance is then  defined as
\begin{equation}
    \eta_{\rm imb}={\vec p}_{T}^{\,ee}  \cdot {\hat{\eta}}  \ + \ 
    {\vec p}_{T}^{\,\,{\text rec}} \cdot {\hat{\eta}} 
\end{equation}
with ${\hat{\eta}}$ a unit vector along the $\eta$ axis.  If the 
electromagnetic and hadronic responses are equal, 
$\eta_{\rm imb}$ is zero.  Since the positive $\eta$ axis is always in the 
direction of ${\vec p}_{T}^{\,ee}$, any systematic bias in the  measurement
of ${\vec p}_{T}^{\,{\text rec}}$ will manifest itself as 
a  bias in $\eta_{\rm imb}$. 
If the difference is due only to a  scale, then the relationship between the
two quantities can be  characterized by a proportionality constant $\kappa$.

\begin{figure}[p]
 \begin{center}
  \begin{tabular}{cc}
     \epsfxsize=8.0cm   \epsffile{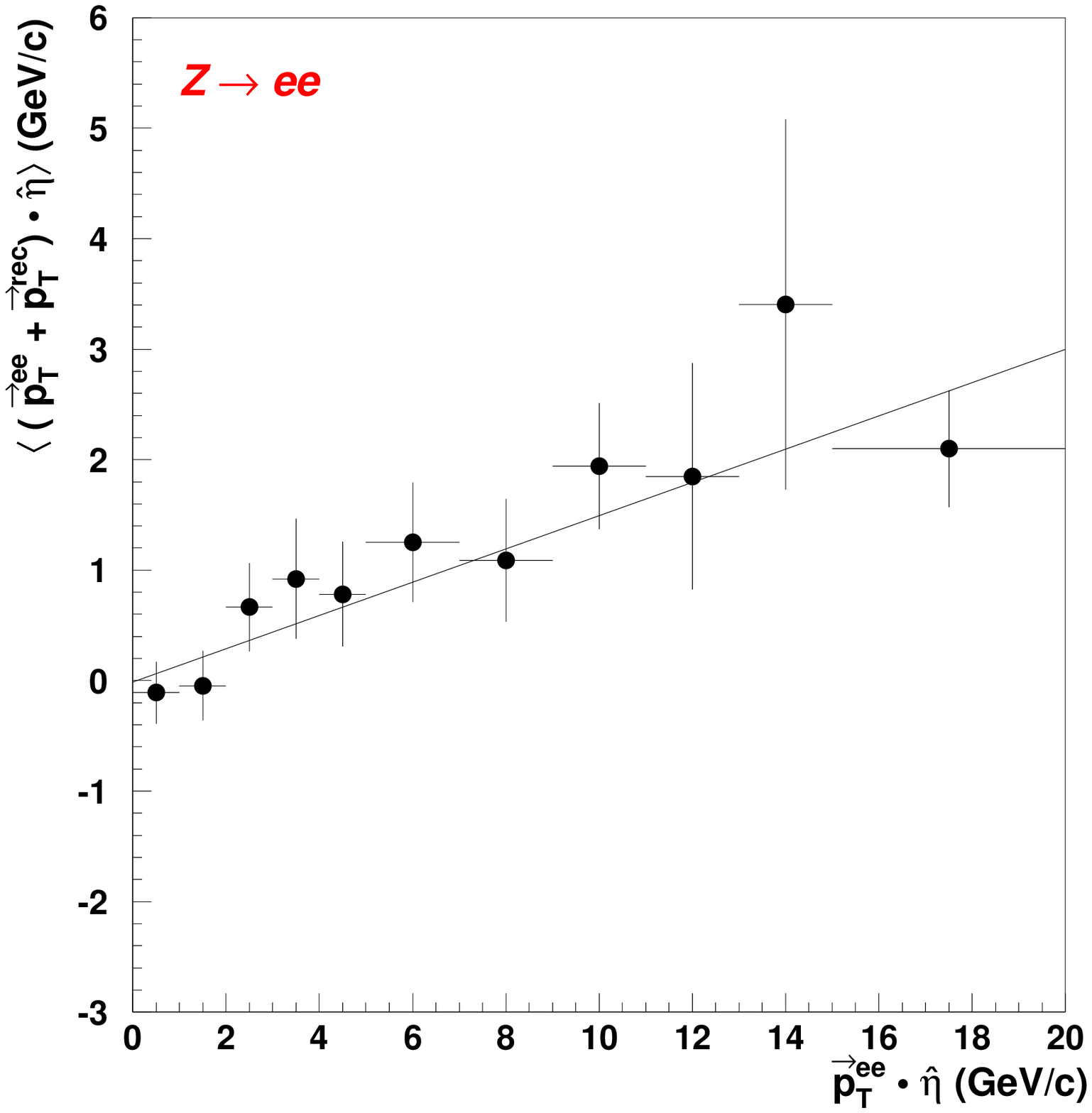} &
     \epsfxsize=8.0cm   \epsffile{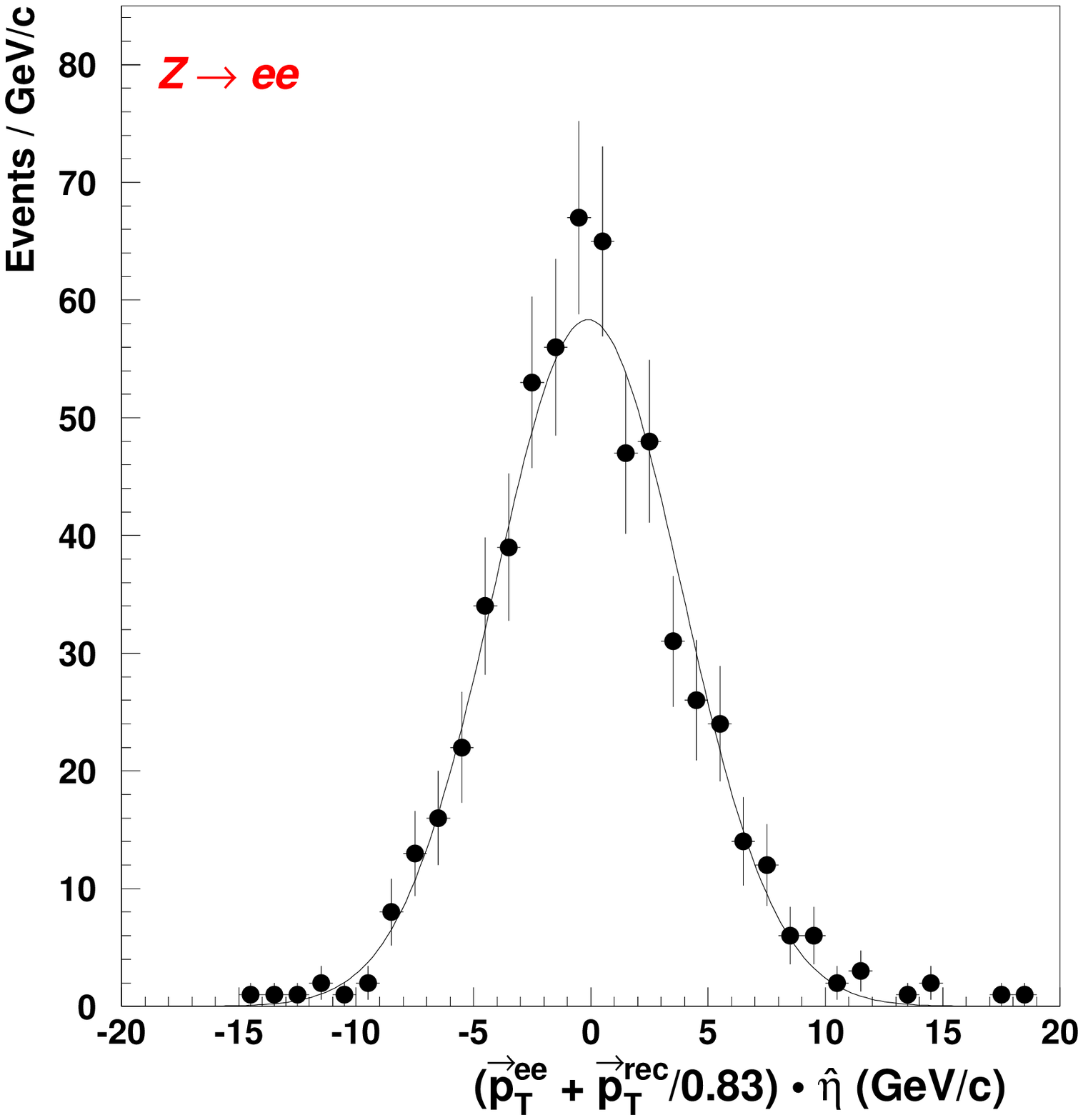}     \\
\end{tabular}
\end{center}
\caption{For $Z\rightarrow ee$ events (points) 
        (a) the average $\eta$ imbalance versus 
             $  {\vec p}_{T}^{\,ee} \cdot \hat \eta $ is shown
            along with the line obtained from a linear least-squared fit
             to the data and 
         (b) the $\eta$ imbalance with a hadronic energy scale factor 
                 $\kappa$ = 0.83 applied is shown with a Gaussian fit (curve).}
         
\label{fig:eta_balance}
\end{figure}

\begin{figure}[h]
    \epsfxsize=8.0cm
    \centerline{\epsffile{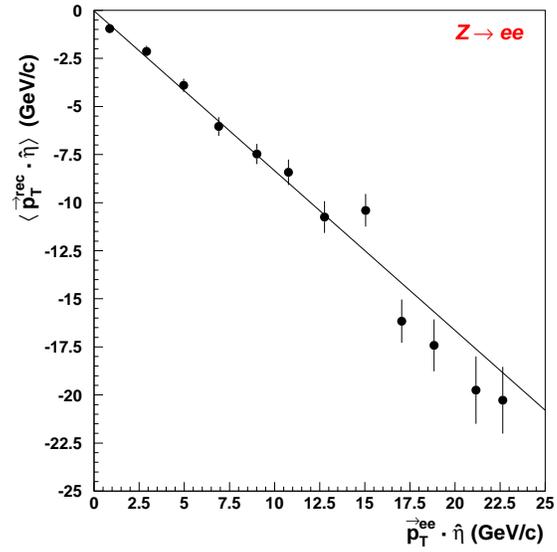}}
\caption[]{For $Z\rightarrow ee$ events (points) 
          with the same event topology   as $W$ events, the average value of
           ${\vec p}_{T}^{\,{\text rec}} \cdot {\hat{\eta}}$ is shown versus
           ${\vec p}_{T}^{\,ee}  \cdot {\hat{\eta}}$. 
           The line shown is obtained from a linear least-squares fit
           to the data.}
\label{fig:ptz_recoil}
\end{figure}

\begin{figure}[p]
 \begin{center}
  \begin{tabular}{c}
    \epsfxsize = 8.0cm    \epsffile{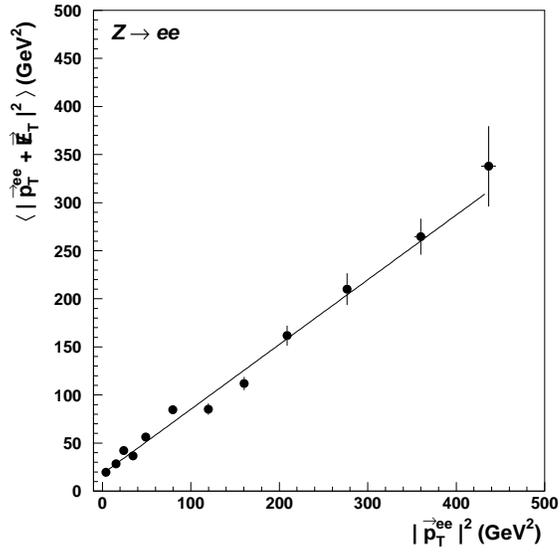}
  \end{tabular}
 \end{center}
\caption[]{Distribution of the average 
$ |{\vec p}_T^{\,e_1} + {\vec p}_T^{\,e_2} + \mbox{$\not\!\!E_T$})|^2 $ 
           versus $ | {\vec p}_T^{\,ee} |^2 $ for 
           $Z\rightarrow ee$ events.
           The line shown is obtained from a linear least-squares fit
           to the data.}
\label{fig:uvec}
\end{figure}

The determination of the hadronic energy scale factor requires  selection of 
$Z\rightarrow e e$ events with the same event topology as $W\rightarrow
e \nu$  events.  
$Z\rightarrow e e$ events were selected with at least one  electron in the
central calorimeter.
An additional cut to eliminate events which occur during the Main Ring
cycle was imposed to ensure that  no spurious calorimetric 
depositions affecting the
measurement of the  hadronic recoil were present.  As a consistency check, 
$Z$ boson events with both electrons in the central 
calorimeter have been used and
a  consistent result for the hadronic energy scale was obtained.   Three
related determinations of the hadronic energy response  relative to the
electromagnetic response have been carried out: 

\begin{enumerate}
\item The primary method of obtaining the calorimeter response used  
    was the measurement of the $\eta$ imbalance as function of 
    $|{\vec p}_{T}^{\,ee} \cdot \hat \eta | $, as shown in 
Fig.~\ref{fig:eta_balance}(a).  
     A least squares fit yields 
    $| {\vec p}_{T}^{\,{\text rec}} \cdot \hat \eta | = 
    \kappa \, | {\vec p}_{T}^{\,ee} \cdot \hat \eta| $, 
    with $\kappa = 0.83 \pm 0.03$.  The 
    offset in response, obtained from the intercept of the 
fit with the ordinate in 
    Fig.~\ref{fig:eta_balance}(a), was measured to be $-0.17
\pm 0.24$~GeV. This result is  
    consistent with zero. Figure~\ref{fig:eta_balance}(b) shows the 
    $\eta$ imbalance for $\kappa$ = 0.83.  The distribution is well 
    described by a Gaussian distribution, centered at zero, 
    with a width of 4.2~GeV.

\item A second, very similar approach 
    to fixing the scale $\kappa$ of the recoil system 
    with respect to the dielectron system was the measurement of 
    $|{\vec p}_{T}^{\,{\text rec}} \cdot {\hat{\eta}}|$ as function of 
    $|{\vec p}_{T}^{\,ee} \cdot {\hat{\eta}} | $, as shown in 
    Fig.~\ref{fig:ptz_recoil}. The linear dependence shows that, over the 
    $p_T$ range of interest to the $W$ boson mass measurement, the hadronic 
    recoil is related to the electromagnetic energy by a simple scale 
    factor.  The scale $\kappa$ was determined by a least squares fit to 
    the data, where the errors on 
    ${\vec p}_{T}^{\,{\text rec}} \cdot {\hat{\eta}}$ and 
    ${\vec p}_{T}^{\,ee}  \cdot {\hat{\eta}}$ 
    have been determined using the 
    known detector resolutions.  
    This method gives $\kappa = 0.84 \pm 0.03$.  
    The offset in response is $0.06 \pm 0.25$~GeV, consistent with 
    zero.  It should be noted that the contribution from the underlying 
    event ${\vec u_T}$ does not affect the determination of $\kappa$ since 
    it is distributed randomly with respect to the 
    ${\hat{\eta}}$ direction.

\item The hadronic energy scale $\kappa$ was also  determined using 
    a third method which yielded both the hadronic energy scale and 
    the magnitude of the underlying event vector.  The transverse
    momentum balance 
    in $Z\rightarrow e e$ events is given by
    \begin{equation}
        {\vec p}_T^{\,e_1} \ + \ {\vec p}_T^{\,e_2} \ + \ 
\mbox{${\hbox{${\vec E}$\kern-0.6em\lower-.1ex\hbox{/}}}_T^\prime$ } \ =
        -{\vec p}_T^{\,{\text rec}} \ - \ {\vec u_T}.
    \end{equation}
    Squaring both sides, one finds for the 
    average 
    \begin{eqnarray}
        | {\vec p}_T^{\,e_1} \ + \ {\vec p}_T^{\,e_2} \ + 
\ 
\mbox{${\hbox{${\vec E}$\kern-0.6em\lower-.1ex\hbox{/}}}_T^\prime$ } |^2 
\ & = & \ 
        | {\vec p}_T^{\,{\text rec}} \ + \ {\vec u_T} |^2    \nonumber \\ 
        \ & = & \ 
        \kappa^2 \, | {\vec p}_T^{\,ee} |^2 \ + \ | {\vec u_T} |^2 
    \end{eqnarray} 
    assuming again that 
    $ | {\vec p}_{T}^{\,{\text rec}} | = \kappa \, | {\vec p}_{T}^{\,ee} | $.  
    The cross term on the right-hand side averages to zero since the 
    underlying event vector is randomly distributed with respect to the 
    $Z$ boson recoil system. 
    Figure~\ref{fig:uvec} shows the distribution of 
    $ | {\vec p}_T^{\,e_1} \ + \ {\vec p}_T^{\,e_2} \ + 
\ \mbox{${\hbox{${\vec E}$\kern-0.6em\lower-.1ex\hbox{/}}}_T^\prime$ } |^2 $ 
    versus $ | {\vec p}_T^{\,ee} |^2 $ for  $Z\rightarrow e e$ events.  Again, 
    the data demonstrate that there is a linear relation between the 
    electromagnetic and hadronic energy scales.  The straight line is a fit 
    to the data and yields $\kappa = 0.83 \pm 0.03$.  This result is 
    consistent with the value determined using the other two methods.  The
    intercept of the straight line fit yields the 
    magnitude of the underlying event vector  $|{\vec u_T}| $ 
is $4.3 \pm 0.3$~GeV.
\end{enumerate}

Because there was no indication of a non-linear response of the 
hadronic calorimeter with respect to the electromagnetic calorimeter, 
nor a sign of a measurable offset, the  energy 
scale for the hadronic recoil was taken to be strictly proportional to
the electromagnetic scale  with a scale of $\kappa=0.83 \pm 0.04$, the
uncertainty of which was derived from the spread in the results among the
three different methods. 
No offset of the hadronic energy scale was included in the Monte Carlo model.
The effect of a possible non-linearity of the hadronic response was 
included when evaluating the systematic uncertainty on the $W$ 
boson mass. The only
use of the EC calorimeter in this analysis was in the 
determination of the missing
transverse energy. The  hadronic energy scale was determined from the 
CC-CC (both electrons in the CC calorimeter) and CC-EC (one electron 
in each calorimeter) $Z\rightarrow
ee$ events. The  hadronic response was 
the same as using the CC-derived scale alone
within errors which are negligible for this measurement.

\subsection{Resolutions}

\subsubsection{Electron Energy Resolution}

The electron energy resolution was parameterized according to 
the relation 
\begin{equation}
    \frac{\sigma_{e}}{E} = 
    \sqrt{ C^{2} + \left( \frac{S}{\sqrt{E_T}} \right)^{2} + 
                   \left( \frac{N}{E} \right)^{2}}
\label{eq:e_res}
\end{equation}
where $C$, $S$,  and $N$ are the coefficients of the constant, sampling, 
and noise terms, respectively.  The values of the sampling and noise 
terms were those as derived from test beam data.  Smearing in $E_{T}$ 
rather than in $E$ is used in the sampling term because the resolution 
should become poorer with increasing thickness of the absorber plates at 
large angles.  Replacing the usual $E$ with $E_{T}$ compensates for 
this and allows the coefficient $S$ to remain constant over all of the 
central calorimeter.  This observation was confirmed by test beam data 
\cite{D0nim,zhu}. The central values utilized in this analysis were
obtained from the test beam 
for the central calorimeter and are $C = 0.015$, $S = 0.13 \,\sqrt{\rm GeV}$, 
 and $N = 0.4$~GeV. For the EC, 
$S = 0.16 \,\sqrt{\rm GeV}$.

The value for the constant term was determined {\it in situ} by 
fitting the electron energy resolution to the observed width of the 
dielectron invariant mass distribution of the $Z\rightarrow e e$ events,
fixing  the width of the $Z$ boson to its measured value of 
$2.490 \pm 0.007$~GeV\cite{zmass}.   There was little 
sensitivity  for small values of
the constant term, since for relevant values of $E_T$  the energy  resolution
is dominated by the sampling term.   A constant term of $ C =
0.015^{+0.005}_{-0.015}$ was obtained, where the error is statistical only. 
The uncertainty on the shape of the background (discussed below) increased
the upper limit on the  error to $+ 0.6$\%.

\subsubsection{Electron Angular Resolution}
\label{sec:elec_angle}

The polar angle of the electron was determined using the 
$z$ position of the electromagnetic energy cluster in the calorimeter and
the $z$  position of the center of gravity of the CDC track.   The angular
resolution used in the Monte Carlo simulation was therefore determined by the
resolutions on these two quantities.

The resolution of the calorimeter hit position was determined  using
electrons from $W\rightarrow e \nu$ decays processed through a detailed
{\footnotesize GEANT} Monte Carlo. Because of the read-out  geometry of the
detector, it depended both on the angle of incidence of  the electron and its
cluster $z$ position,
$z_{\rm clus}$.  It was  parametrized as a Gaussian distribution having a width
\begin{equation}
    \sigma(z_{\rm clus}) \,=\, (p_1 + p_2\times |\varpi | )  \,+\, 
                           (p_3 + p_4\times |\varpi | ) \  |z_{clus}|
\label{eq_calres}
\end{equation}
where 
    $p_1 = 0.33183$~cm, 
    $p_2 = 0.52281\cdot 10^{-2}$~cm/degree,
    $p_3 = 0.41968\cdot 10^{-3}$, and 
    $p_4 = 0.75496\cdot 10^{-4}$~cm/degree.
The angle $\varpi$ is the polar angle of the electron (in degrees) as 
measured with respect to the $\eta$=0 axis of the detector. 

\begin{figure}[ht]
\begin{center}
\begin{tabular}{cc}
    \epsfxsize = 8.0cm  \epsffile{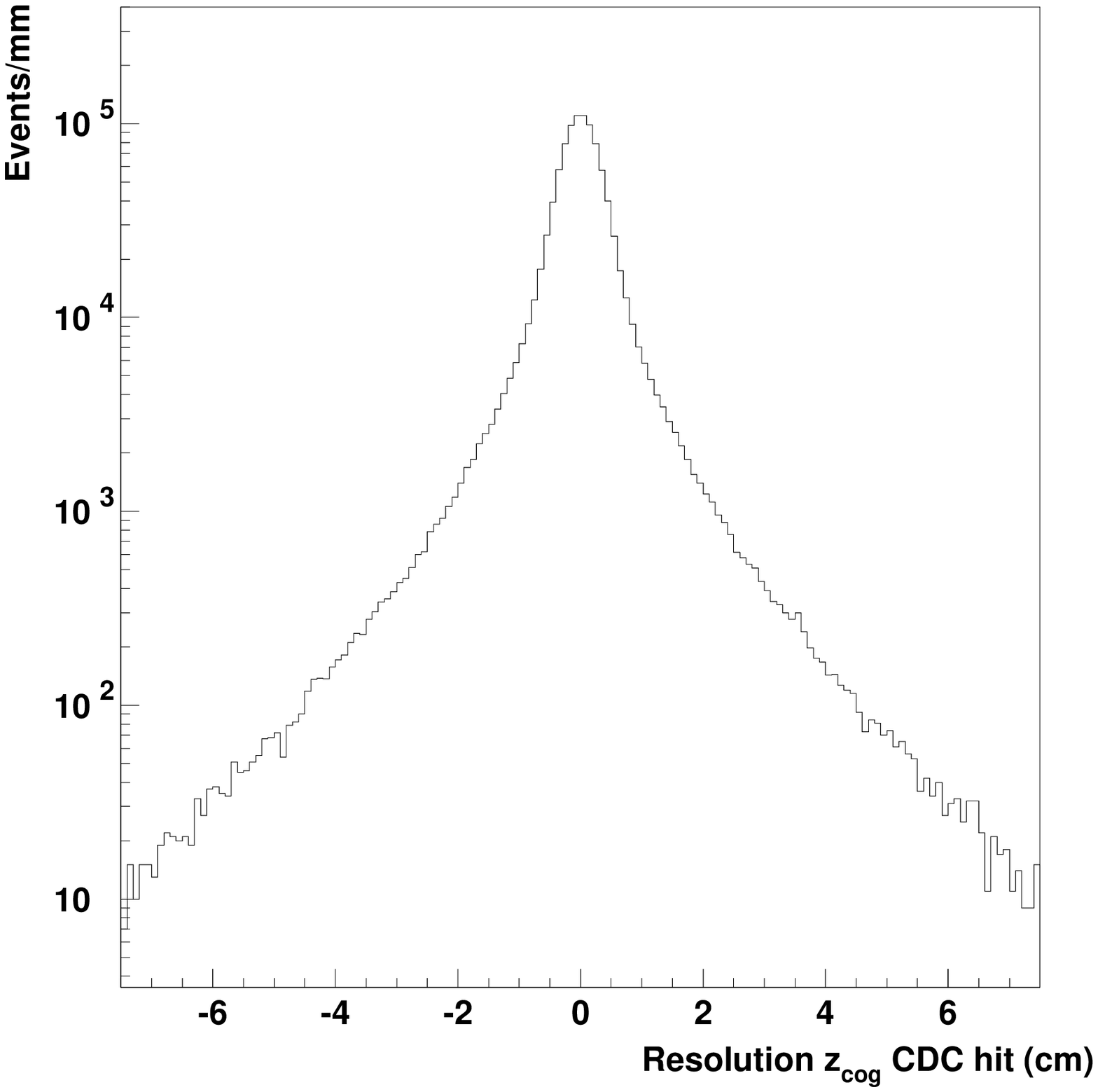} &
    \epsfxsize = 8.0cm  \epsffile{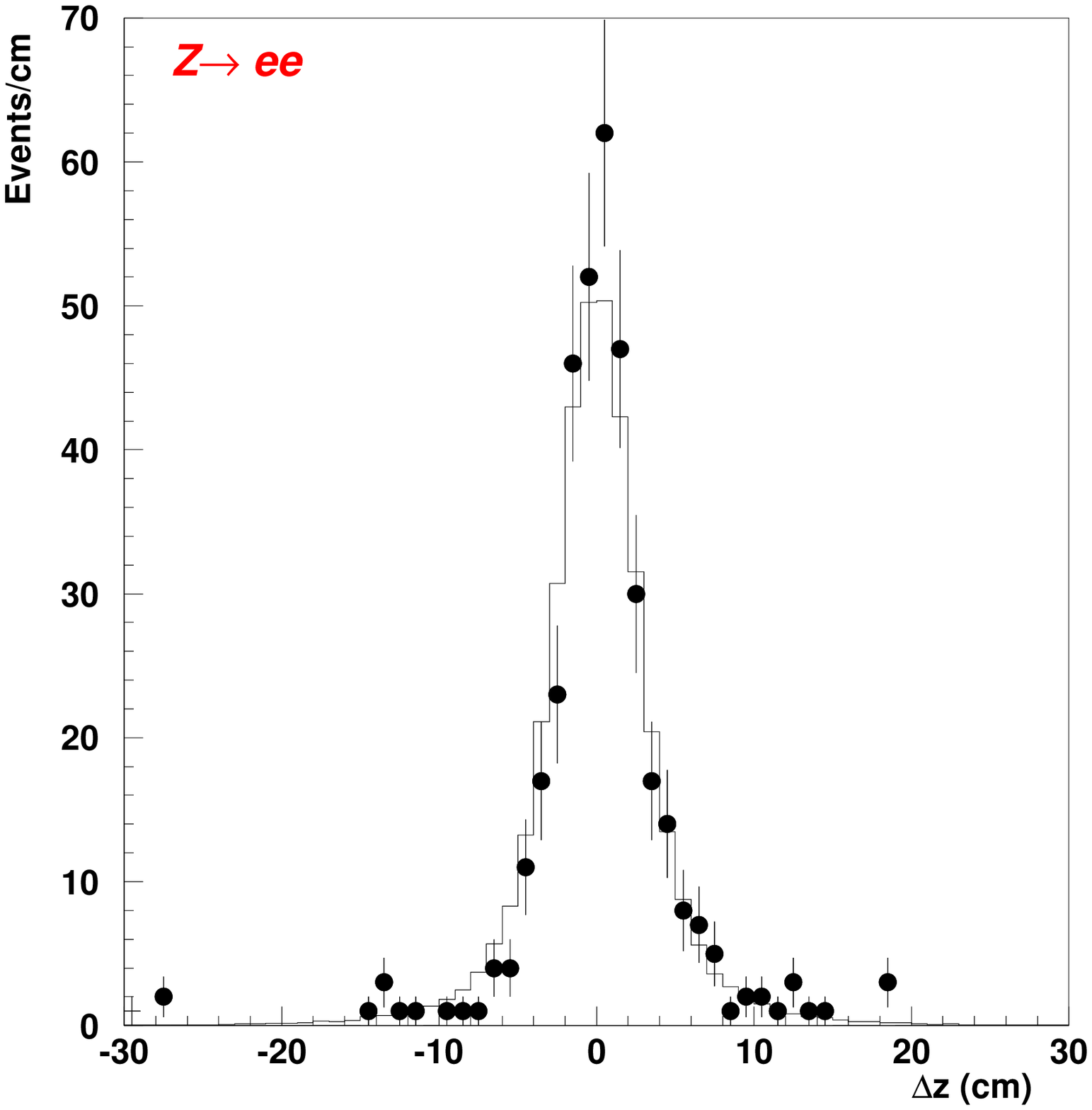} \\
\end{tabular}
\end{center}
\caption{(a) The modeled resolution of the $z$ position of the 
             center of gravity of  CDC tracks and 
         (b) the distribution in the difference  
             of the intersections of the $z$ axis  of the two
             electron tracks from $Z$ decays (points) compared with 
             the distribution from the Monte Carlo simulation.}
\label{fig:angres}
\end{figure}

The resolution of the $z$ position of the center of gravity~ of the track was 
measured from $Z\rightarrow e e$ events using the intersections of the
two  electron tracks with the beamline.  The distribution of the difference 
in $z$ position of the two intersections shows non-Gaussian tails  which 
were
represented in the Monte Carlo.   The simulation generates a resolution on
$z^{CDC}_{cog}$ as shown in  Fig.~\ref{fig:angres}(a). The model was verified by
comparing a Monte  Carlo generated distribution of the difference in the
intersections of  the two electrons from $Z$ decays with that obtained from
the data,  and is shown in Fig.~\ref{fig:angres}(b).

In the data analysis, the azimuthal angle of the electron was given by  the
$\varphi$ angle as measured by the CDC. The resolution was taken to  be the
CDC $\varphi$ resolution and is modeled as a Gaussian distribution with 
width $\sigma(\varphi) = 0.005$~radians. For some $Z\rightarrow e e$
studies electrons in the end calorimeters were also used.  The angular
resolutions of these electrons were modeled in the  Monte Carlo as Gaussian
distributions with resolution $\sigma(\varphi) =  0.015$~radians and
$\sigma(\theta) = 0.015$~radians.


\subsubsection{Hadron Energy Resolution}

The recoil against the vector boson was modeled by assuming it to be a single
jet.  The transverse momentum of the vector boson was smeared using the 
hadronic energy resolution determined both in the test beam and 
from analysis of jets {\it in
situ}. It was parameterized as
\begin{equation}
    \frac{\sigma_{had}}{E_T} = 
    \sqrt{ C^{2} + \left( \frac{S}{\sqrt{E_T}}\right)^{2} + 
                   \left( \frac{N}{E}\right)^{2}}
\label{eq:e_res2}
\end{equation}
with resolution parameters 
$C$ = 0.04, $S$ = 0.80 $\sqrt{\rm GeV}$ and $N$ = 1.5~GeV\cite{D0nim}.

\subsection{Efficiencies}

There were two main inefficiencies which affected this measurement: those 
related to the hardware trigger, and those related to electron 
identification criteria.  Both effects can potentially bias the measurement 
as these particular inefficiencies depend on the kinematics.  These 
efficiencies are determined from data as discussed below.

\subsubsection{Trigger Efficiencies}

The main data sample was recorded with an on-line  filter, which required an
electromagnetic cluster with 
$E_T^e > 
20$~GeV and $\mbox{${\hbox{$E$\kern-0.6em\lower-.1ex\hbox{/}}}_T$ } > 
20$~GeV.  The trigger efficiency as
function of the offline  electron and missing transverse energies was
determined using a single  electron trigger as well as triggers with lower
requirements.
After 27\% of the running was completed, 
the missing transverse  energy
calculation in the L2 trigger was changed to use the event vertex as
measured by  the L0 system,  rather than the nominal $z=0$ value. 
Therefore, two different threshold curves have been used in  this data
analysis.  
Both the $E_T^e$ and 
$\mbox{${\hbox{$E$\kern-0.6em\lower-.1ex\hbox{/}}}_T$ }$ 
requirements in the trigger
were  more than 99\% efficient for transverse energies greater than 30~GeV.

\subsubsection{Electron Identification Efficiency}

The recoil of the $W$ boson may affect the electron 
identification, especially if the recoil system is close to the 
electron.  A measure of the event selection biases can be obtained by studying 
identification efficiencies as a function of 
the quantity $ u_\parallel$, which is the
projection  of the momentum recoiling against the $W$ boson along the
electron\cite{CDF-90}:
  \begin{equation}
     u_\parallel \,\equiv\, {\vec p}_T^{\,{\text rec}} \cdot \hat{e},
  \end{equation}
where $\hat{e}$ is a unit vector in the electron direction.  A bias in 
the electron identification as function of $u_\parallel$ would distort 
the lepton $p_T$ spectra.  For example, an inefficiency of the 
electron identification at high positive values of $u_\parallel$, when 
the recoil is close to the electron, would result in a softer 
$p_T^\nu$ spectrum.

\begin{figure}[p]
    \epsfxsize = 8.0cm 
    \centerline{\epsffile{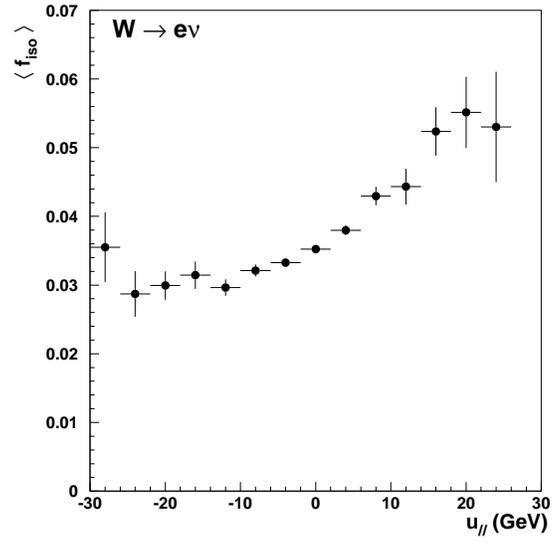} }
\caption[]{Average value of the isolation versus $u_\parallel$ for 
           electrons from       $W\rightarrow e\nu$ decays. }
\label{fig:upar_iso}
\end{figure}

\begin{figure}[p]
    \epsfxsize = 10.0cm
    \centerline{\epsffile{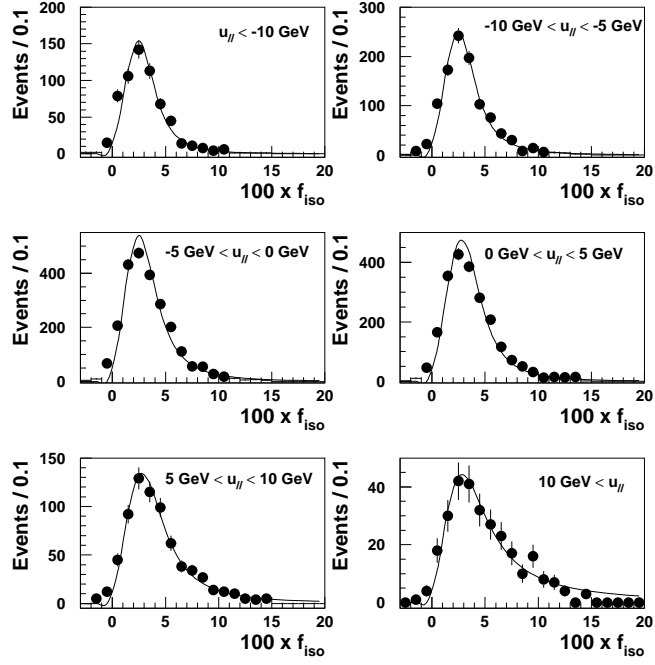} }
\caption[]{Distribution of the isolation value, $f_{iso}$,
           for electrons from $W\rightarrow e\nu$ decays for different 
           $u_\parallel$ ranges (points). 
           The curves are fits to the data. }
\label{fig:upar_isofit}
\end{figure}

\begin{figure}[p]
    \epsfxsize = 10.0cm
    \centerline{\epsffile{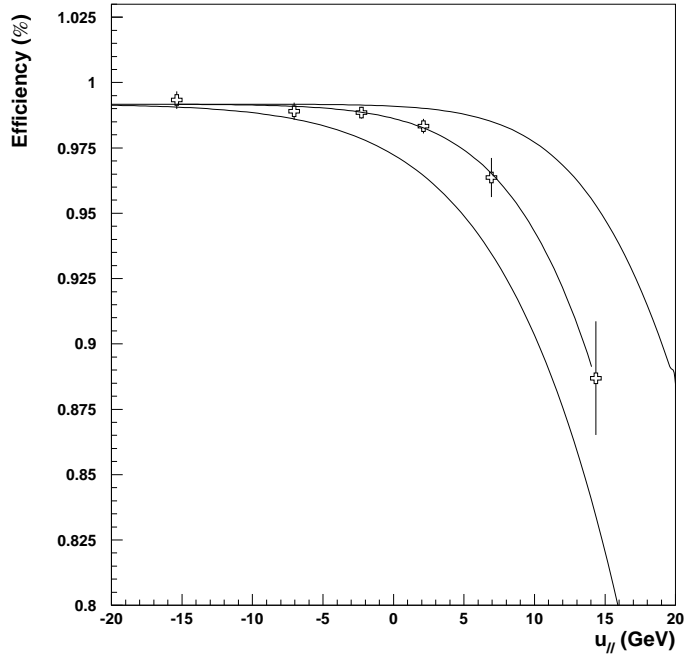} }
\caption[]
 {Electron identification efficiency as a function of 
  $u_\parallel$ (open crosses) from data.  The central curve is a fit to
   the data. The outer curves show the allowed 
  ranges for determining the systematic errors.}
\label{fig:upar_eff}
\end{figure}

The event selection efficiency as a function of $u_\parallel$ was 
determined by studying the behavior of the energy isolation fraction, 
$f_{\rm iso}$, of the electrons in the signal sample.  
Figure~\ref{fig:upar_iso} shows the average isolation versus 
$u_\parallel$ for the electrons in the $W$ boson data sample.  For 
negative values of $u_\parallel$, when the recoil jet is opposite the 
electron, the isolation is constant.  This indicates that for these 
event topologies the recoil system did not affect the electron, as 
expected.  For positive values of $u_\parallel$ the isolation 
increases linearly with $u_\parallel$, indicating that there was a 
\lq\lq halo\rq\rq\ of constant energy flow surrounding the direction 
of the recoil jet.  The electron identification efficiency was 
determined by modeling the distribution of the isolation variable for 
different ranges of $u_\parallel$ as shown in 
Fig.~\ref{fig:upar_isofit}.  The curves are the result of a fit to the 
data using a five parameter functional form.  To determine the 
electron identification efficiency as a function of $u_\parallel$, 
fits to the isolation distribution were integrated over $f_{\rm iso}$.  
The fraction of events with $f_{\rm iso}>0.15$ constituted a 
determination of the inefficiency due to the recoil jet spoiling the 
electron signature.  The efficiency as function of $u_\parallel$ is 
shown in Fig.~\ref{fig:upar_eff} where the curve is a parameterized 
fit.

The dominant systematic uncertainty stems from the shape of the 
isolation distribution for values of $f_{\rm iso} > 0.15$, above the 
trigger restriction.  This was addressed by studying $W$ boson events 
in which the electron cluster was rotated in azimuth, re-analyzed, and 
the isolation re-evaluated.  The tail of the isolation distribution 
obtained in this way was well described by the fitting function.  In 
addition, when fitting for the isolation distribution of the rotated 
sample, a maximum variation in the efficiency of 1.5\% was noted.  To 
be conservative, the efficiencies were shifted coherently by two 
standard deviations of their total uncertainties and refit.  The band 
in Fig.~\ref{fig:upar_eff} shows the resulting uncertainty on the 
efficiency.

\subsection{Backgrounds}

There was a dual approach to the treatment of backgrounds in this  analysis. 
The process $W \rightarrow \tau \nu \rightarrow e \nu\nu\nu$  is
indistinguishable from $W  \rightarrow e \nu$ and was therefore  
explicitly included in the Monte Carlo event  generation.
These decays were generated with a 17.9\% branching fraction for the decay 
$\tau \rightarrow e \nu\nu$, where the 
$\tau$ polarization was taken into account.  Other backgrounds are
characterized by data and were added to 
the final distributions of the fitted variables. 
The determination of these  background contributions is
discussed in this section.

\subsubsection {Backgrounds to $W\rightarrow e \nu$ Events } 

The dominant source of background to $W\rightarrow e \nu$ production 
was standard 
QCD multi-jet production, where one of the jets was misidentified 
as an electron and 
there was substantial $\mbox{$\not\!\!E_T$}$ from jet energy fluctuations or 
non-uniform energy response. 
This background has been estimated using data from an inclusive electron 
trigger that did not impose an isolation requirement at the trigger level. 
The background sample is selected by requiring the same kinematic and
fiducial cuts as in the $W$ boson event sample but imposing anti-electron 
identification cuts on the EM energy cluster.
These anti-electron selections are the combination of:
\begin{itemize}
  \item $f_{\rm iso} > 0.20$
  \item $\chi^2 > 250$
  \item $\sigma_{\rm trk} > 10.$
\end{itemize}
For multi-jet background events, it was assumed that the  shape of the
$\mbox{$\not\!\!E_T$}$ spectrum at low $\mbox{$\not\!\!E_T$}$ was 
the same independent of the electron 
quality cuts. The distribution in $\mbox{$\not\!\!E_T$}$ 
of the background sample was
then  normalized to the signal sample in the region 
$0 < \mbox{$\not\!\!E_T$} < 15$~GeV. 
The signal sample was selected from the same trigger by imposing  the
standard $W$ boson selection criteria.  The ratio of the number of events with
$\mbox{$\not\!\!E_T$} > 25$ GeV for the  signal and  
normalized background distributions
was then taken as  the amount of background in the sample. There was a $0.3\%$
variation in the amount of background due to how  the sample is normalized
and how the background sample was selected.

As a consistency check, the above procedure was repeated with data  taken
with an inclusive electron trigger that required the EM cluster  to be
isolated at the trigger level.  The signal sample was again taken as the
events that pass the $W$ boson event  selection cuts.  The background sample
consisted of those events which pass  the $\chi^2 > 250$ and $\sigma_{\rm trk} >
10$ cuts. Since there was an isolation requirement in the trigger, the
background sample does not have the anti-isolation cut applied as before.
The two methods yielded consistent results.  The overall background fraction
was taken to be 
$(1.6 \pm 0.8)$\%, the average of the two analyses. The uncertainty is the
total statistical and systematic  uncertainty and encompasses the error on
the two separate measurements. 

\begin{figure}[h]
    \epsfxsize = 10.0cm
    \centerline{\epsffile{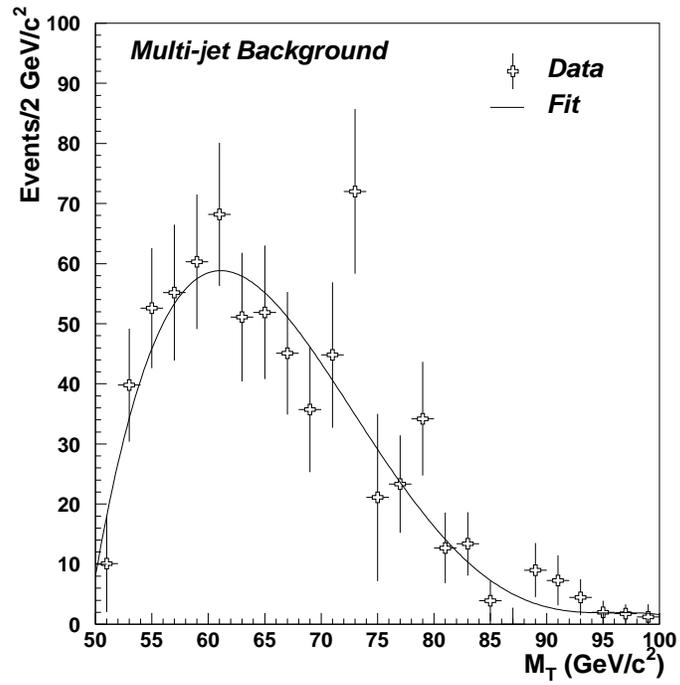}}
\caption[]{Transverse mass spectrum of the multi-jet background obtained
           from the data (open crosses). The solid line is a fourth-order 
           polynomial fit. }
\label{fig:qcd_bkg}
\end{figure}

Since very few background events survived the kinematic cuts, this method
yielded only  the overall background contribution leaving  the shape of the
background as a function of the transverse mass largely  undetermined. 
Employing the capability of the TRD to distinguish electrons, 
converted photons, and pions 
a likelihood function was constructed employing the energy 
deposition in the TRD, the track $dE/dx$ in the CDC, and the track  cluster
match.  
Using an anti-electron criterion based on this likelihood, slightly 
more background events survived the kinematic and acceptance
cuts,  allowing a determination of the dependence of the background as
function  of the relevant quantities. The data points in
Fig.~\ref{fig:qcd_bkg} show the calculated transverse  mass distribution of
the background obtained this way.  The line is a fourth order polynomial
fit.  The shape of the background in lepton transverse momentum  can be
described by an exponentially falling spectrum with  slope  
$-0.086 \pm 0.059$ GeV$^{-1}$ and 
$-0.129 \pm 0.055$ GeV$^{-1}$ for the $E_T^e$ and $\mbox{$\not\!\!E_T$}$  
spectra,
respectively. 

\begin{figure}[t]
\begin{center}
\begin{tabular}{ccc}
    \epsfxsize=6.0cm  \epsffile{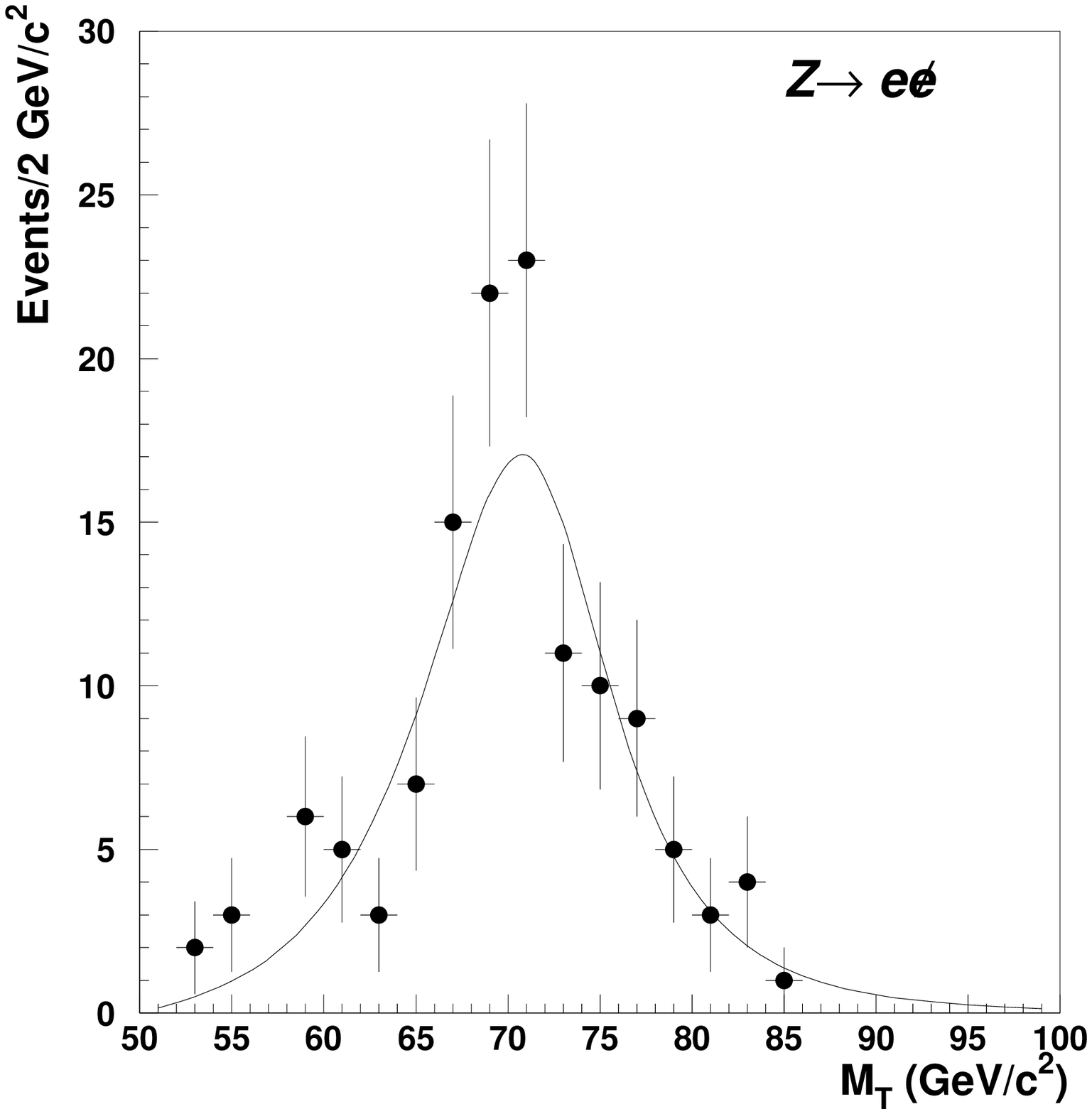}  &
    \epsfxsize=6.0cm  \epsffile{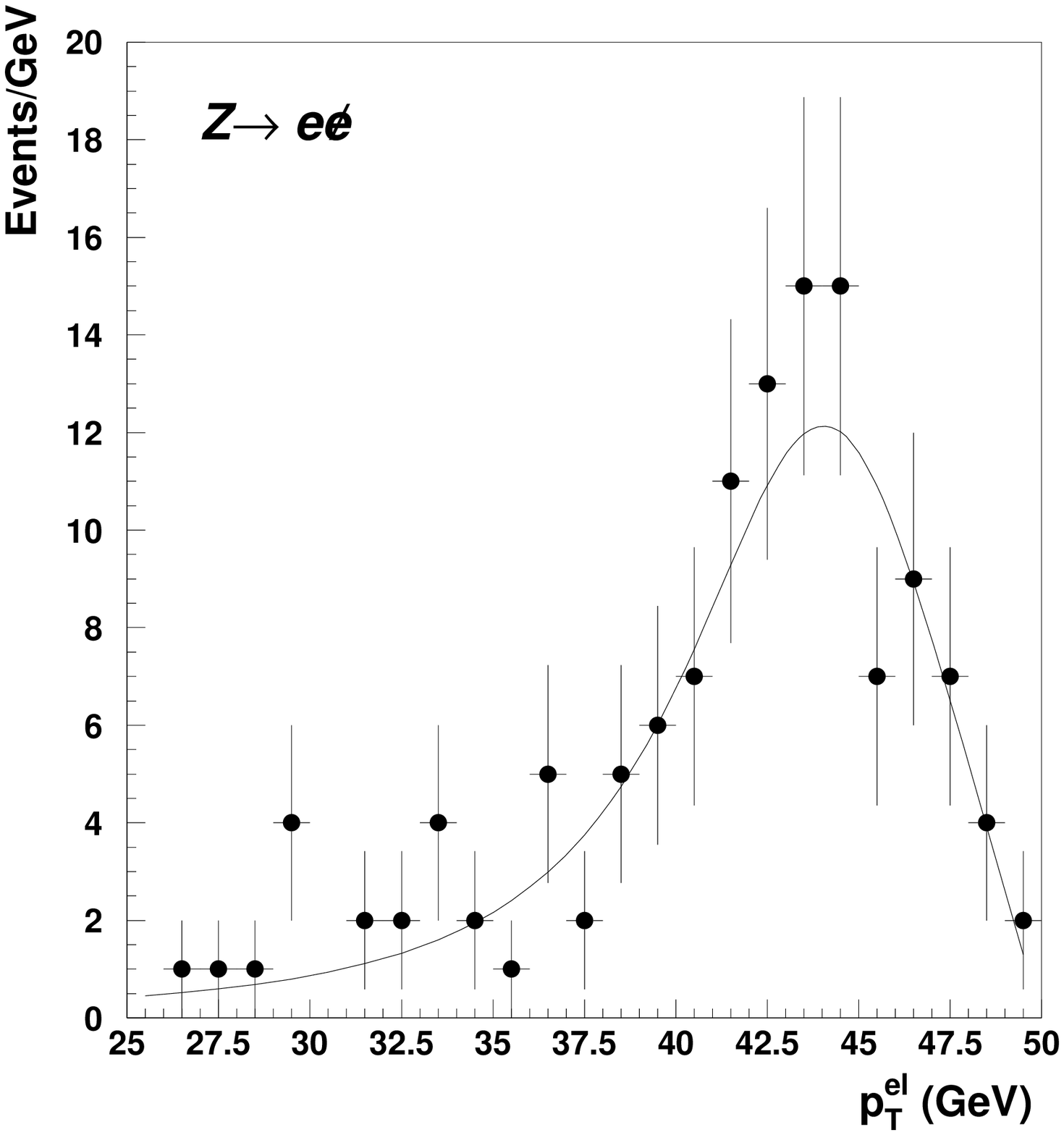}  &
    \epsfxsize=6.0cm  \epsffile{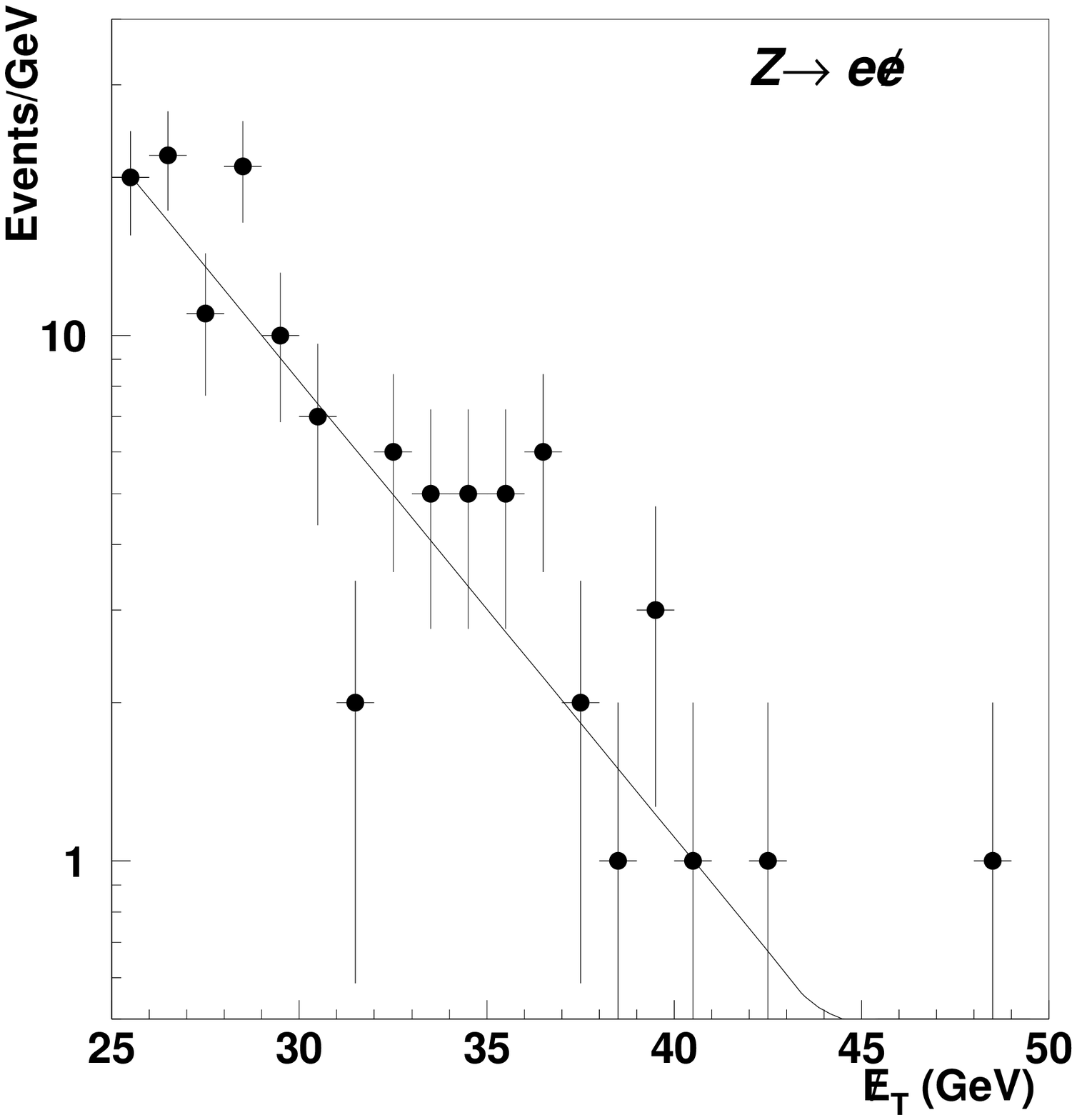} \\
\end{tabular}
\end{center}
\caption{Spectra in (a) $M_T$, (b) $p_T^e$ and (c) \mbox{$\not\!\!E_T$}
         for the $Z$ boson background in the $W$ boson sample. 
         The lines are fits to the data.}
\label{fig:zeebkg}
\end{figure}

The other background that has been considered is the process 
$Z\rightarrow e e$, where one electron escapes detection and is not 
measured (denoted by \mbox{${\hbox{$Z\rightarrow 
ee$\kern-0.4em\lower-.1ex\hbox{/}}}$ }) giving rise to a transverse 
momentum imbalance.  This background has been estimated using 
{\footnotesize ISAJET}~\cite{isajet}.  To appropriately model the 
underlying event in the {\footnotesize ISAJET} simulation, one minimum 
bias event was vectorially added to the $\mbox{$\not\!\!E_T$}$ for the 
Monte Carlo data.  The overall background contribution has been 
estimated to be $(0.43 \pm 0.05)$\%.  The $M_T$, $p_T^e$ and
$\mbox{$\not\!\!E_T$}$ spectra for this source are shown in 
Fig.~\ref{fig:zeebkg}.  The $\mbox{$\not\!\!E_T$}$ spectrum does not 
show a Jacobian edge because the detector is hermetic and the energy 
of the unidentified electron is typically well measured.  The solid 
lines for the $M_T$ and $p_T^e$ spectra are from a polynomial fit.
The $\mbox{$\not\!\!E_T$}$ spectrum was parametrized 
using an exponentially falling spectrum with slope $-0.20 \pm 
0.03$~GeV$^{-1}$.  The average $u_\parallel$ for this background is 
$-12.5 \pm 0.6$~GeV.

\begin{figure}[h]
\begin{center}
\begin{tabular}{c}
    \epsfxsize=13.0cm
    \epsffile{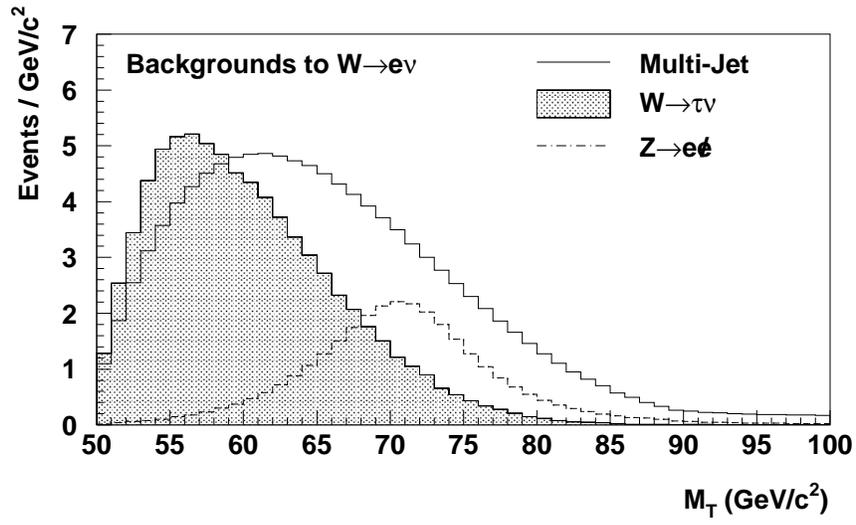}
\end{tabular}
\end{center}
\caption{Normalized distributions in transverse mass of the dominant background
         contributions to the $W$ boson event sample. }
\label{fig:mtw_bkgr}
\end{figure}

Figure~\ref{fig:mtw_bkgr} shows the distribution in transverse mass of 
the dominant background sources to the $W$ boson event sample.  The 
background has been normalized to the expected number of background 
events in the data sample.


\subsubsection {Backgrounds to $Z\rightarrow e e $ Events } 

The primary background to $Z\rightarrow e e$ events  came from multi-jet
production,  with the jets fragmenting into a leading $\pi^0$.  Since the
mass is determined  from the resonant cross section only,  a correction also
must be made  for Drell-Yan and $Z\gamma^*$ interference processes.  These
backgrounds were determined as a function of  invariant mass and were included
at the fitting stage.

The total background contribution was evaluated by fitting the 
$m_{ee}$ spectrum to a relativistic Breit-Wigner convoluted  with a Gaussian
resolution function plus a background falling  exponentially in $m_{ee}$.
For the mass range of interest, there is no distinction between a linear or
exponential model of the background. 
This method yielded a total QCD and Drell-Yan background under the 
$Z^0$ peak of 7.4\%, with a slope of $-0.0447 \pm 0.018$~(GeV/$c^2$)$^{-1}$ for
an invariant mass window of $70$ to $110$~GeV/$c^2$.

The Drell-Yan and $Z\gamma^*$ contribution  to
the total $Z$ boson production cross  section was determined
using an {\footnotesize ISAJET} simulation. In the mass range
$70 < m_{ee} < 110$~GeV/$c^2$  the Drell-Yan and $Z\gamma^*$ interference
terms contributed 3\% to the  total cross section. The background 
has an exponentially
falling spectrum with slope   $-0.03$~(GeV/$c^2$)$^{-1}$.  The
contribution to the background from multi-jet sources is thus 4.4\%.  Both
the overall background contribution and its shape are  in good agreement
with the background determination for the cross section 
analysis~\cite{wwidth}.

\section{Monte Carlo Simulation}

\subsection{Introduction}
	
The $W$ boson and $Z$ boson masses were extracted by comparing measured
distributions with those generated by a Monte Carlo simulation.
To determine the $W$ boson mass
the relevant distributions are those in transverse mass plus
the
electron and neutrino transverse momenta.   For determining the mass
of $Z$ bosons the relevent distribution is  
in the dielectron invariant mass.
The simulation was accomplished with a  generator   which
produced all of the basic processes,  incorporated the main features of the D\O\
detector, and was capable of generating tens 
of millions of simulated events in a few
hours. This section describes the  physics and detector simulation.  
A comparison 
between the Monte Carlo simulation and the data is   presented at the end.  
Table~\ref{tab:params} lists the parameters used in the  Monte
Carlo.


\subsection{$W$ and $Z$ Boson Production and Decay}
	
The simulation of $W$ and $Z$ bosons relied on the choice of a model for
the physics processes involved.   This physics was
divided  into three  parts:  
$i)$    the production model for $W$ and $Z$ bosons;  
$ii)$   the decay of the vector bosons and 
$iii)$  backgrounds. 
For the $W$ boson the basic processes generated were 
$W  \rightarrow e \nu$, 
$W  \rightarrow \tau \nu \rightarrow e \nu\nu\nu$ and 
$W  \rightarrow \gamma e \nu$; 
for the $Z$ boson they were
$Z  \rightarrow  e e $ and  
$Z  \rightarrow  \gamma  e e $.
As discussed in the previous section, all backgrounds, except for the 
$W  \rightarrow \tau \nu$ decay,  were
not a part of the $W$ or $Z$ simulation and were dealt with separately.

\subsubsection{Production of $W$ and $Z$ Bosons}

The triple differential cross section for vector boson production 
was assumed to factorize as 
\begin{equation}
    {d^{3}\sigma \over dp_{T}dydm}  
        \,=\, C \, {d\sigma \over dm} \times {d^{2}\sigma \over dp_{T}dy}.
\label{triple}
\end{equation}
Here, $C$ denotes the appropriate constants, $y$ is the rapidity of 
the vector boson, and the products on the right hand side refer to 
shapes rather than absolutely normalized quantities.  The double 
differential cross section was generated on a grid of $p_T$ and $y$ 
points over the region $-3.2 < y < 3.2$ and $0 < p_T < 50$~GeV/c, in 
steps of 0.2 in $y$ and 0.5~GeV/c in $p_T$.  Two choices of the 
production model, both based on the fully resummed theory of Collins 
and Soper~\cite{CSS}, were considered.  The double differential 
spectrum as given by Arnold and Kauffman\cite{ak_ptw} (AK) uses a 
next-to-leading order calculation for the high $p_T$ 
region~\cite{arnold_reno} with a prescription to match the low and 
high $p_T$ regions.  They used fits to Drell--Yan 
data\cite{davies_sterling} which have since been updated.  
The double differential cross section by Ladinsky and 
Yuan\cite{ly_ptw} (LY) employs a different parametrization for the 
non-perturbative functions describing the $p_T$ spectrum based on a 
fit to more recent data.  The differential spectra were generated for 
both models using various parton distribution functions as input.  
Alternative grids within the LY model were used, distinguished by a 
different choice of the non--perturbative parameters, $g_i$ (see 
Appendix~A for more details).  In order to properly keep track of the 
helicity states for the weak decay, annihilations involving different 
combinations of valence quarks and sea quarks were dealt with 
separately.  The default double differential cross section used the LY 
production model with the MRSA\cite{mrsa} parton distribution 
functions.

After generation of the kinematics of the $W$ boson, the mass 
dependence of the production cross section was folded in. 
A relativistic Breit-Wigner line shape was used to model the $W$ boson resonance
\begin{equation}
    {d\sigma \over dm}(m^2) \,\propto\, 
           \frac {m^{2} }
                 { (m^{2}-M_{W}^{2})^{2} +
                   \frac{m^{4} \Gamma_{W}^{2}}{M_{W}^{2}} }
\label{eq:bw}
\end{equation}
where $M_{W}$ and $\Gamma_{W}$ are the mass and width of the $W$ boson.
In $p\bar p$ production, however, the mass spectrum 
differs from the strict Breit-Wigner resonant line shape of the  
partonic cross section due to the variation of parton flux with
parton momentum.  
This mass dependence has been calculated by that the differential 
cross section is given by 
\begin{eqnarray}
    { d{\cal N} \over dm }    & \propto &
    { 2m \over s } \, 
    \int_{{m^2 \over s}}^1  \, {dx_1 \over x_1} \, 
           f_{q/p}(x_1) \, 
           f_{q^{\prime}/\overline{p}}
\left( {m^2 \over s x_1} \right) \, 
           {d\sigma \over dm}(m^2)       \nonumber     \\ 
                            & = & 
           {1 \over m } \  {\cal F} \
                           {d{\cal \sigma }\over dm}(m^2)
\end{eqnarray}
with 
\begin{equation}
    {\cal F }  =  { 2m^2 \over s } \,
                      \int_{{m^2 \over s}}^1   \, {dx_1 \over x_1} \,
                      f_{q/p}(x_1) \, 
                      f_{q^{\prime}/\overline{p}}
\left( {m^2 \over s x_1} \right)
\end{equation}
Here $f_{q/p(\overline{p})}(x)$ is the probability that a quark or 
antiquark $q$ in the (anti)proton carries a fraction $x$ of the 
(anti)proton's momentum. 
In this equation a sum over all $qq^{\prime}$ pairs that lead to 
$W$ boson production is implicit. 
The factor ${\cal F}$ is a dimensionless quantity which will be 
referred to as the parton luminosity~\cite{pl}. 
It has been parameterized as having an exponential mass 
dependence, $e^{-\beta m}$. 
The slope parameter $\beta$ has been treated as a single number, calculated by
evaluating the  integral using the available parametrizations of the parton
distribution functions~\cite{pdflib} at 
a mass of 80 and 91~GeV/$c^2$ for $W$ boson and $Z$ boson production, 
respectively. The small mass dependence of $\beta$ was included in 
the systematic uncertainty. 
Table~\ref{tab:beta} lists the values of $\beta$ which are used for $W$ boson
and $Z$ boson production for different  sets  of parton  distribution functions.
The most recent sets which use nearly identical modern input data are  
the MRSA and CTEQ3M\cite{cteq3m} sets.  
The relative contributions for vector boson 
production are listed separately for valence quarks 
and sea-sea quarks. 
In the  event  generation the  widths of
the intermediate  vector  bosons  were  fixed  to  their  measured  values,
$\Gamma(W) = 2.12$~GeV~\cite{wwidth}  and  
$\Gamma(Z) = 2.487$~GeV~\cite{zmass}.  

\subsubsection{Decay of $W$ and $Z$ Bosons}

The $W$ boson decay products were  generated in the $W$ boson 
rest frame  with an angular
distribution depending on which process,  
valence-valence/sea or sea-sea, is involved. 
$W^+$ bosons follow the angular distribution 
\begin{eqnarray}
    {d^2\sigma \over dy\,d\cos\theta^*}
        & \sim & (1 - \cos\theta^*)^2  \,\cdot\,
                 \left(\frac{1}{2} {d\sigma^{W^+}_s \over dy }        \,+\,
        {d\sigma^{W^+}_v \over dy } \right)\ + \, \nonumber \\  
              & &   (1 + \cos\theta^*)^2  \,\cdot\,
                  \frac{1}{2} {d\sigma^{W^+}_s \over dy }  
\end{eqnarray}
\noindent 
where the subscripts $v$ and $s$ refer to valence and sea contributions, 
respectively, and the $+z$ direction is chosen along the proton direction. 
Here $\theta^{\ast}$ is the center of mass angle 
between the electron direction and the $q\bar{q}$ axis.  

The $q\bar{q} \rightarrow  \ell \ell$
production cross section at the $Z$ boson resonance is proportional to 
\begin{eqnarray}
    ( {g_V^q}^2    \,+\, {g_A^q}^2 ) \, 
    ( {g_V^\ell}^2 \,+\, {g_A^\ell}^2 ) \,
    (1 + \cos^{2}\theta^{\ast}) \,+\, &  \nonumber \\
 \quad \quad \quad
     4 \, g_V^q \, g_A^q \, g_V^\ell \, g_A^\ell \, 
     \cos \theta^{\ast}. &
\end{eqnarray}
Because the lepton charge is unmeasured, the $\cos \theta^{\ast}$
term averages to zero.  
The leptons were therefore generated with a 
$(1 + \cos^{2}\theta^{\ast})$ angular distribution and the 
$u{\overline u}$ and $d{\overline d}$ contributions to the production 
weighted with their respective coupling strengths, 
$ {g_V^q}^2 \,+\, {g_A^q}^2 $. 
Here $ g_V^q $ and $ g_A^q $ are the vector and axial-vector 
coupling strengths of quark $q$ to the $Z$  boson 
\begin{eqnarray}
        g_V^q & = & I_3^q \,-\, 2\,Q_q\, \sin^2\vartheta_W    \\
        g_A^q & = & I_3^q
\end{eqnarray}
with $I_3$ the third component of the weak isospin and $Q_q$ the charge
of the quark.
$I_3^q$ is +$1\over 2$ for the charge $2\over 3$ quarks and 
-$1\over 2$ for the charge -$1\over 3$ quarks. 
The value $\sin^2 \vartheta_W = 0.2317$\cite{pdg} was used.

\subsubsection{Radiative Processes}

Radiative $W$ boson and $Z$ boson decays,
$q' \bar{q} \rightarrow W \rightarrow e \nu  (\gamma)$and 
$q  \bar{q} \rightarrow Z \rightarrow e e (\gamma)$,
must be properly simulated in the Monte Carlo program to extract correct 
values for $M_W$ and $M_Z$  (see
Appendix B).  Because the $ee$ and
$e \nu$ invariant masses are smaller in these decays than 
the corresponding vector boson masses,
the experimentally measured mass distributions were 
shifted toward lower values. 

The rates and distributions in lepton and photon momenta were generated 
to ${\cal O}(\alpha)$ following reference~\cite{Berends}.  Using this 
calculation, in the decay of the $Z$ boson either of the electrons 
(but not both) may radiate.  In $W$ boson decays, the electron or $W$ 
boson may radiate the photon.  Approximately 30\% of $W \rightarrow 
e\nu $ decays and 60\% of
$Z \rightarrow e e$ decays had a photon of 50 MeV or more
in the final state.
The calculation does not include processes that in the limit of a zero 
width boson would be considered $W\gamma$ or $Z\gamma$ production. 
Therefore, initial state radiation was not included in the  
calculation, nor was the production of a virtual high mass
$W$ boson which decays to an on-shell $W$ boson and a photon.  In $W\gamma$ and
$Z\gamma$ production,  $M_W$ and $M_Z$ were correctly obtained from the
dilepton invariant masses ($e\nu$ or $ee$) and the $\gamma$ direction 
was
not strongly correlated with that of either lepton.  Its presence
produced a  background not fundamentally different from that of other
processes.

In implementing radiative decays in the Monte Carlo simulation, three
experimental scenarios were considered:
$i)$ When the $\gamma$ was produced inside the electron  cone, 
taken to be a radius
of
$R=0.2$ in $\eta-\phi$ space,  the $\gamma$ was measured as part of the
electron.   The neutrino momentum, obtained from the missing transverse
energy  in the event, was calculated correctly.   Therefore, the invariant
mass of the $ e \nu $ system is the 
$W$ boson mass, and the transverse mass and transverse 
momentum of the $W$ boson  was
properly calculated.
$ii)$ If the $\gamma$ was far from the electron, that is outside a cone of 
radius $R=0.4$, the photon retains its identity.  The electron energy 
was
measured correctly, and $p^\gamma_{T}$ becomes part of the recoil against
the $W$ boson, 
$\vec{p}_{T}^{\,W}({\rm meas}) = \vec{p}_{T}^{\, W}({\rm generated}) - 
 \vec{p}_{T}^{\, \gamma}$.   The transverse mass of the $e\nu$ system 
 was
calculated correctly, but was shifted downward because the $e\nu$ invariant
mass is smaller than the $W$ boson mass.   Therefore, $M_{W}$ was mismeasured.
$iii)$ If the $\gamma$ was produced in the region between
$R=0.2 $ and $R=0.4$, it alters the shape of the electron shower.  Isolation
and electron identification cuts then resulted in inefficiencies  that can
affect the $W$ boson mass if not properly simulated in the Monte Carlo. 

In the Monte Carlo simulation, the fraction of the electron's energy 
in the region between $R=0.2$ and $R=0.4$ was generated according to 
the experimental distribution measured in $W$ boson events.
The photon energy was added to the electron energy and the event was discarded 
if it failed the isolation cut.
If the event survived the isolation cut and the radial distance $R_{e\gamma}$
between the $\gamma$ and $e$, was less
than 0.3, the $\gamma$ momentum was added to the electron's and the 
$W$ boson mass was correctly calculated, as in the first case above.  
If the radial distance was greater than $R_{e\gamma}=0.3$, the $\gamma$ 
momentum was not added to the electron's and the reconstructed $W$ boson mass 
and transverse momentum were shifted downward.


\subsection{Detector Simulation}
\label{sec:ue} 

The production of likelihood templates in
$M_T$ required large Monte Carlo samples. Twenty million generated  events
were required  to sufficiently eliminate  effects of statistics in
the likelihood function.   To  study the  effects of systematic 
uncertainties  many  complete analyses  were needed. The combination of these
requirements made a fast detector simulation essential. 

After production and decay
products were boosted into the laboratory frame, 
the parameters whose measurements
were described in the previous  section were 
utilized in this simulation as follows.  
\begin{itemize}
\item The energies of the generated
electrons and radiative photons, if they  were present and retained their
identity, were scaled by the  measured EM energy scale.  The generated
transverse momenta were then smeared according to the  measured resolution,
as was the generated electron angle.    
\item The transverse momentum of the
recoil system was taken to be the  negative of the generated transverse
momentum of the $W$  boson, 
${\vec p}_T^{\,{\text rec}} = - {\vec p}_T^{\,W}$. 
Its magnitude was scaled by  the
product of the measured EM energy scale and the relative response  of the
hadronic and EM calorimeters. Smearing was added according  to the jet
energy resolution.  The hadronic content of $Z\rightarrow e e$ events and
the electrons from the $Z$ boson  decay were modeled in the same fashion as
$W\rightarrow e \nu$ events.
\item The underlying event, denoted by ${\vec u}_T({\cal L})$, was modeled using
collider minimum bias events, which mimic the 
debris in the event due to spectator
parton interactions and the pile-up associated with multiple interactions.
The use of  minimum bias events  properly includes any residual energy
which might be present from previous crossings 
as well as detector effects. A library
of minimum bias triggers was created in bins of luminosity in order to correctly
simulate overlapping event and noise characteristics 
of the data. Events were chosen
according to the distribution of instantaneous 
luminosities observed during the run
as shown in  Fig.~\ref{fig:w_lum}. (See the discussion in Appendix~C.)
Figure~\ref{fig:minb} shows the
$\mbox{$\not\!\!E_T$}$ and total scalar $E_T$ distributions of the minimum
bias events used.
The average $\mbox{$\not\!\!E_T$}$ is 3.93~GeV with an rms of 2.69~GeV.
The mean total scalar $E_T$ is 67.1~GeV with an rms of 39.8~GeV.
(The total scalar $E_T$ distribution is shown for completeness
only, as this quantity is not used in the event modeling.)
\item The generated and smeared recoil hadronic 
energy vector and the underlying event
hadronic energy vector were superimposed on one 
another to form a simulation of the
total hadronic deposition.
\item The vertex for each
generated event was taken to be that of the minimum bias event.
\item The efficiencies and cuts were applied to the smeared quantities.
\end{itemize}

\subsubsection{Underlying Event Discussion}

In the data, the contribution from the underlying event cannot be
separated from the measured recoil energy.
In the simulation of the $W$ events the recoil and the underlying event
were treated separately.
The superposition in the Monte Carlo of the underlying event
and the production of the $W$ boson and its decay
is laced with intricate details.
Although the average energy deposition per read-out
tower in minimum bias events was very small, its effect on the $W$ boson mass
measurement is of crucial importance mainly because the corrections 
were correlated
with the electron direction. Its presence affects not only the
measurement of the electron energy but 
also the measurement of the missing transverse
energy. Equation~\ref{mtmom} shows that the 
neutrino transverse momentum differs from
the measured missing transverse energy because of the
presence of ${\vec u}_T$.

In addition to incorporating the effects of the energy flow of the 
underlying event in the event model, detector effects needed to be 
taken into account, in particular the effect of the zero-suppression.  
Calorimeter depositions were only read out if the absolute value of 
the magnitude of the energy fell outside the zero-suppression limits.  
Low energy tails of the electron shower were thus suppressed.  A 
convolution of the shaping electronics and the natural radioactivity 
of uranium caused the pedestal distribution to be asymmetric with a 
long tail towards positive energies.  Therefore, even when no particle 
struck a read-out tower, the energy registered for that tower, when 
read out zero-suppressed, would on average not be zero.

In the following subsection, first the effect of the energy flow of the
underlying event on the measured energy of the electron will be discussed.
The corrections to the measured electron energy introduced by the
zero-suppression will then be detailed.

\subsubsection{Underlying Event Energy and Electron Simulation}

 Because the clustering algorithm for electron
identification  used in this analysis was dynamic, the cluster
size can vary from event to event and a description of the underlying
event contribution to the electron would get rather involved.
However, the  clustering approach was found to be 
numerically equivalent to a window
in $(\eta, \varphi)$ space having a constant size of $0.5\times 0.5$.
As this will facilitate the discussion, this window analog
consisting of a fixed set of 25 towers with
$(\Delta\eta \times \Delta\varphi) = 0.1\times 0.1$
will be used to illustrate the size of the effects of the underlying
event on the electron energy measurement.

Within the $0.5\times 0.5$ window in $(\eta, \varphi)$ centered
around the electron was contained
not only the energy of the electron but also the energy
from the underlying event. The measured electron transverse energy,
 ${}_{{}_m}{\vec p}_{T}^{\,e}$,
is thus given by:
\begin{eqnarray}
    {}_{{}_m}{\vec p}_{T}^{\,e} &\simeq& {\vec p}_{T}^{\,e} \,+\,
                                         {\vec u}_T^{\rm \, 25}
\label{eq:pte_meas}
\end{eqnarray}
where ${\vec p}_{T}^{\,e}$ refers to the true electron transverse energy,
folded with the appropriate resolution, and ${\vec u}_T^{\rm \, 25}$ is
the underlying event contribution inside the 25 towers defining the
electron cluster.
The latter term has been estimated from $W$ events by
rotating the electron cluster in azimuth and measuring the average
energy flow per tower. Care was taken to ensure that the rotated
cluster was isolated and was not in  proximity to any jet activity.
The energy flow per tower was found to be 16.8~MeV.
The average energy flow under the electron is therefore
${\vec u}_T^{\rm \, 25} = 25 \times 16.8\, {\hat e} = 420\, {\hat e}$~MeV,
with ${\hat e}$ a unit vector in the electron direction.

This contribution has also been determined from minimum bias events,
spanning an appropriate range in luminosity.
An average energy flow of 15.3~MeV per tower was found.
The difference of 1.5~MeV between the two methods is attributed to the
presence of the $W$ recoil.
A value of 16.8~MeV per tower has been used in the simulation.
The uncertainty on the average energy flow is reflected in the
systematic uncertainty due to this source.

To each of the two terms in Eq.~(\ref{eq:pte_meas}), a correction
needed to be applied due to zero-suppression.
Under normal running conditions, calorimeter cells were not read
out if the signal was within 2$\sigma$ of the mean pedestal for that
channel; that is, the read-out was zero-suppressed. As a consequence,
the tails of the electron shower which fell within the zero-suppression
limits ${\vec u}_{T_{\rm zs}}^{\,e}$ were lost.
The average energy that was lost below the $2\sigma$ zero-suppression threshold
was estimated to be
${\vec u}_{T_{\rm zs}}^{\,e} = - 152 \, {\hat e}$~MeV
for electrons from $W$ decays,
using a detailed {\footnotesize GEANT} simulation.

Because the absorber medium in the calorimeter is uranium, which is
a natural $\beta$-emitter, the pedestal distributions were asymmetric.
Additionally, some
asymmetry in the pedestal distributions was introduced due to
the shaping electronics\cite{D0nim}.
Therefore, even when no particle strikes a read-out tower,
the zero--supressed energy read out for that tower was on average not  zero.
This zero-suppression contribution has been studied by analyzing
non zero-suppressed minimum bias events.
By comparing the energy per $0.1\times 0.1$ read-out tower measured in
these events to the energy that results after applying the zero suppression
offline, the energy per read-out tower of EM and the first FH layers,
was 7.55~MeV higher than in non zero-suppressed events.
It should thus be realized that the average energy flow of 16.8~MeV
per tower, derived above, has two contributions. The first contribution
is from the true energy flow in the event, determined to be 9.23~MeV per
tower. The second contribution is an artifact of the zero-suppression,
due to the asymmetric pedestal distributions, which adds an energy
of 7.55~MeV per tower to the read-out.

As mentioned before, a minimum bias event was used to model
the event underlying the $W$ boson. The presence of the electron from the $W$
decay affected the energy flow in the underlying event. Notably, the
read-out towers occupied by the electron had a very large energy
deposition and therefore were not affected by the zero-suppression
correction. For the $W$ data, the electron occupied on average
$8\pm 3$ towers.
Therefore, applying the zero-suppression correction to all 25
read-out towers of the electron cluster, which has been assumed above,
is incorrect. This was corrected by applying the correction to only the
17 channels within the cluster that on average were zero-suppressed or,
equivalently, by subtracting out the zero-suppressed pedestal energy from
the 8~cells that on average were read-out with the electron.
Thus, a correction
${\vec u}_{T_{\rm zs}}^{{\rm ue}} = - 8 \times 7.55 \, {\hat e}$~MeV
needed to be applied to the energy flow under the electron,
${\vec u}_T^{\rm \, 25}$.

To summarize, the measured electron transverse energy, in MeV,
is given by
\begin{eqnarray}
    {}_{{}_m}{\vec p}_{T}^{\,e}
            &=& {\vec p}_{T}^{\,e} \,+\,
                {\vec u}_{T_{\rm zs}}^{e} \,+\,
                {\vec u}_T^{\rm \, 25} \,+\,
                {\vec u}_{T_{\rm zs}}^{{\rm ue}} \nonumber \\
            &=& {\vec p}_{T}^{\,e} \,-\,
                152 \, {\hat e} \,+\,
                25 \times (9.23 + 7.55) \, {\hat e}  \,-\, \nonumber \\
            & &  \quad   8 \times 7.55 \, {\hat e}
\end{eqnarray}
with
\begin{itemize}
\item   ${\vec p}_{T}^{\,e}$ the true electron transverse energy folded
        with its resolution;
\item   ${\vec u}_{T_{\rm zs}}^{e}$ the energy of the tails of the
        electron shower lost due to the zero-suppression, determined
        to be $-152 \, {\hat e}$~MeV;
\item   ${\vec u}_T^{\rm \, 25}$ the energy flow from the underlying
        event within the $0.5\times 0.5$ window in $(\eta, \varphi)$
        defining the electron cluster, given by
        $25 \times (9.23 + 7.55) \, {\hat e}$~MeV; and
\item   ${\vec u}_{T_{\rm zs}}^{{\rm ue}}$ the correction to the
        energy flow of the underlying event due to the presence of the
        electron which corrects for the zero-suppression effect of the
        underlying event for the towers occupied by the electron,
        $- 8 \times 7.55 \, {\hat e}$.
\end{itemize}
When all of these effects were taken into account,
an addition of an average of 207 MeV  to the generated electron along
the electron direction was required in order to correctly simulate the measured
electron $p_T$.

\subsubsection{Underlying Event Energy and Recoil Energy }

The measured recoil energy in the detector is a combination of the
true recoil of the $W$ boson and the contribution of the underlying
event. In the simulation the true recoil of the $W$ boson was taken
to be ${\vec p}_T^{\,{\text rec}} $ and the underlying
event was simulated using a minimum bias event.
Therefore the measured recoil was given by
\begin{equation}
    {}_{{}_m}{\vec p}_{T}^{\,{\text rec}} \simeq - {\vec p}_{T}^{\, W} \,+\,
                                             {\vec u}_T.
\end{equation}
The underlying event vector ${\vec u}_T$ was taken to be 
the sum of the $E_T$ of all calorimeter cells in the minimum bias event.
However, a correction needed to be applied to the underlying
event energy vector due to the presence of the electron in $W$ events.
Recall that in the data analysis the recoil momentum was determined
by subtracting the electron transverse energy from the total measured
transverse energy in the event. Therefore the energy flow under the
electron from the underlying event, pointing along the electron,
should be subtracted from ${\vec u}_T$. 
In the simulation, the recoil was thus calculated as
\begin{eqnarray}
    {}_{{}_m}{\vec p}_{T}^{\,{\text rec}}
                &=& - {\vec p}_{T}^{\, W} \,+\,
                      {\vec u}_T  \,-\,
                      {\vec u}_T^{\rm \, 25}   \nonumber \\
                &=& - {\vec p}_{T}^{\, W} \,+\,
                      {\vec u}_T  \,-\,
                      25 \times (9.23 + 7.55) \, {\hat e}  \ .
\end{eqnarray}
Note the absence of the ${\vec u}_{T_{\rm zs}}^{{\rm ue}}$ term,
which does not need to be applied here since ${\vec u}_T$ is from
a minimum bias event in which no high $p_T$ electrons are present.

\begin{figure}[t]
    \epsfxsize = 10.cm
    \centerline{\epsffile{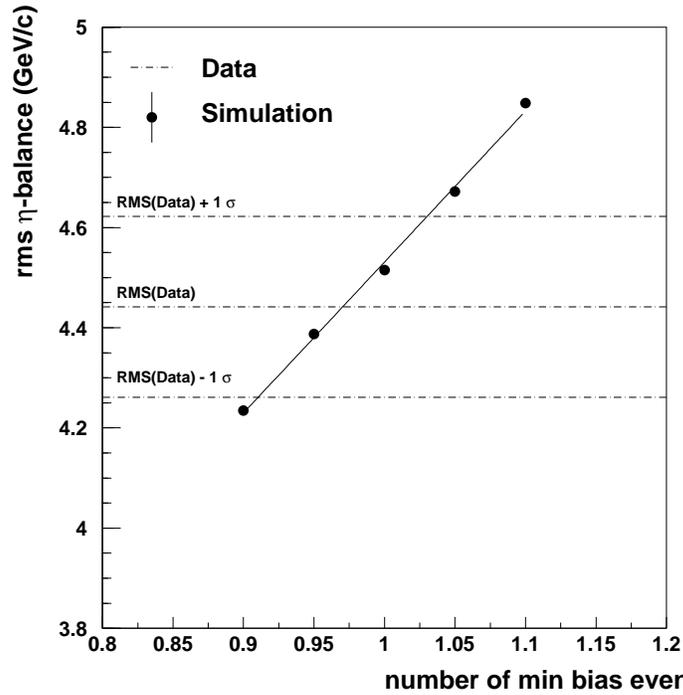}}
\caption[]{Sensitivity of the width of the $\eta$ imbalance distribution
           to the number of minimum bias events used to simulate the
           underlying event in the Monte Carlo simulation (points).
           The line is the result of a linear least-squares fit.
           The bands (dotted-dashed) correspond to the nominal and
           $\pm\,1\sigma$ measurements of the width in $Z$ boson events. }
\label{fig:num_minb}
\end{figure}

The underlying event model and the resolution in $p_T^{\,{\text rec}}$ has been
verified using the $\eta$ imbalance in $Z$ boson events, defined
previously. 
Since the magnitude of the $E_T$ in minimum bias events was of the same order as
that of the $p_T$ of the vector boson, the width of the distribution
of the $\eta$ imbalance (see Fig.~\ref{fig:eta_balance}(b)) was very sensitive
to the underlying event contribution.
The rms of the $\eta$ imbalance distribution in
Fig.~\ref{fig:eta_balance}(b), after the correction 
for the hadronic energy scale has been
applied, is $\sigma = 4.44 \pm 0.18$~GeV.  
This is the band shown in Fig.~\ref{fig:num_minb}.
By varying the number of minimum bias events in the Monte Carlo that mimic the 
underlying event, the width of the
$\eta$ imbalance determined the number of minimum bias events to be added
in the simulation.
The points in Fig.~\ref{fig:num_minb} show the Monte Carlo predicted
widths as function of the number of minimum bias events.
The number of minimum bias events preferred by the
data was $N_{\rm min. bias} = 0.98 \pm 0.06 $ events.
Since this number is consistent with 1.0, one minimum bias event was used to 
model the underlying event in $W$ and $Z$ boson production.

\subsubsection{Underlying Event and the Neutrino Momentum}

The neutrino momentum is a derived quantity which follows
directly from the electron and recoil measurements:
\begin{eqnarray}
\mbox{${\hbox{${\vec E}$\kern-0.6em\lower-.1ex\hbox{/}}}_T$ } 
&=& - {}_{{}_m}{\vec p}_{T}^{\,{\text rec}} \,
                - {}_{{}_m}{\vec p}_{T}^{\,e}          \nonumber     \\
            &=& - {\vec p}_{T}^{\,{\text rec}} \,-\,
                  {}_{{}_m}{\vec p}_{T}^{\,e} \,-\,
                  [ {\vec u}_T  \,-\, {\vec u}_T^{\rm \, 25} ] \nonumber \\
            &=& - {\vec p}_{T}^{\,{\text rec}} \,-\,
                  {\vec p}_{T}^{\,e} \,-\,
                  {\vec u}_T  \,-\,
                  {\vec u}_{T_{\rm zs}}
\end{eqnarray}
where
$ {\vec u}_{T_{\rm zs}} =
  {\vec u}_{T_{\rm zs}}^{e} +
  {\vec u}_{T_{\rm zs}}^{{\rm ue}} $.
Note that $ {\vec u}_T  \,-\, {\vec u}_T^{\rm \, 25} $ represents the
energy vector of the underlying event with the region that the electron
occupies excised.

There are two equivalent ways to view the effect of the underlying event.
If one uses for the neutrino momentum the second line above,
then the measured electron energy, including the contribution from zero
suppression and the energy from the underlying event,
appears in the neutrino and the electron in $W$ decays
and in both electrons in $Z$ decays.
This correction then cancels in the ratio of the two masses. Then what is
important is the amount of the underlying event energy which
should be excluded from the determination of the $W$ boson recoil energy
because it is inside the electron cluster.
Alternatively, if one examines the expression for the neutrino momentum
given in the third line above, only the total recoil momentum and the
total underlying energy enter.
The zero suppression correction is still irrelevant,
appearing in the neutrino, the $W$ electron, and the two electrons
from the $Z$ boson decay.
Now the correction to the electron energy from the energy flow from the
underlying event that appears inside the electron cluster
does not cancel completely in the $M_W/M_Z$ ratio.

The missing transverse momentum
differs from the neutrino momentum because of the presence of
${\vec u}_T$. This effect has no counterpart in $Z$ boson decays and
it changes the measured transverse mass and must be properly modeled.
As described above, in the Monte Carlo simulation
${\vec u}_T$ was obtained from minimum bias events.
If there were a biased region of the calorimeter which made 
${\vec u}_T$ directional, this effect would be accounted for in the 
Monte Carlo events.
Although the above is dependent on properly extracting
small energies in the calorimeter, many of the effects cancel in the
ratio $M_W/M_Z$.


\subsection{Application of efficiencies }

After simulating the vector boson event kinematics,  the efficiencies of the
trigger as well as the electron identification  efficiency as a function of
$u_{||}$ were applied, using the measured  kinematic quantities.  Fiducial
cuts in $\eta$ and $\phi$ were made as in the data.  Using the measured
quantities, the transverse mass was calculated and  the same selection
criteria as in the data were applied: 
$m_{T} > 50$~GeV/$c^{2}$; $E_T^e > 25$~GeV/$c$;  
$\mbox{$\not\!\!E_T$} > 25$ GeV;  and $p_T^{\,W} < 30$~GeV/$c$.   


\subsection{ Comparison of Data with Monte Carlo}
 
Comparisons of various distributions of the simulated quantities with 
data are shown in this section. 
The distributions comparing the data and the results of the simulation 
are area normalized. The Monte Carlo was generated at the final $W$ boson mass 
value of this analysis obtained from the transverse mass fit. 
In the comparisons the data are generally shown as points with 
statistical errors; the simulation is shown as the histogram.


\subsubsection {Characterization of the $W\rightarrow e \nu$ Candidates }

\begin{figure}[t]
\begin{center}
\begin{tabular}{c}
    \epsfxsize=10.cm   \epsffile{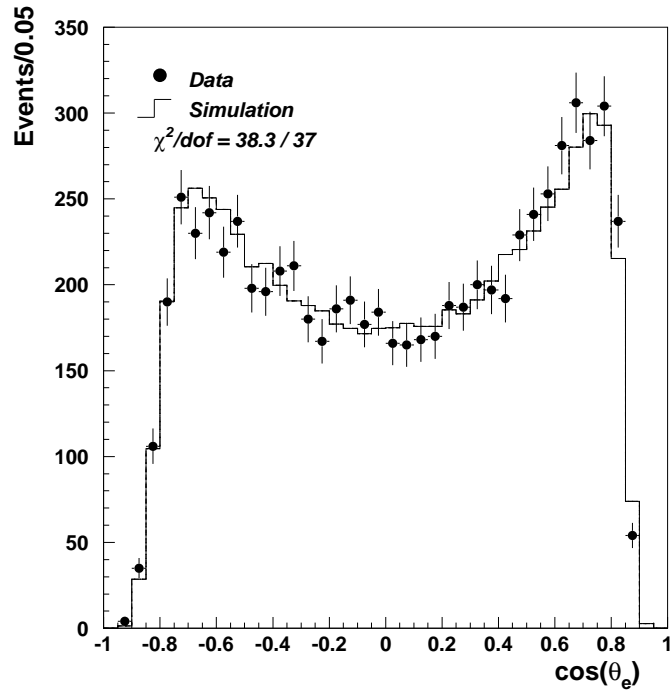} 
\end{tabular}
\end{center}
\caption[]{Angular distribution of electrons 
           from $W\rightarrow e\nu$ decays (points)
           compared to the simulation (histogram). The asymmetry is
due to the fact that the luminous region was not located at $z=0$~cm
in the D\O\ detector, but was rather centered at $z= -7.98$~cm.} 
\label{fig:mw_datamc_costh}
\end{figure}

\begin{figure}[h]
\begin{center}
\begin{tabular}{c}
    \epsfxsize=10.cm  \epsffile{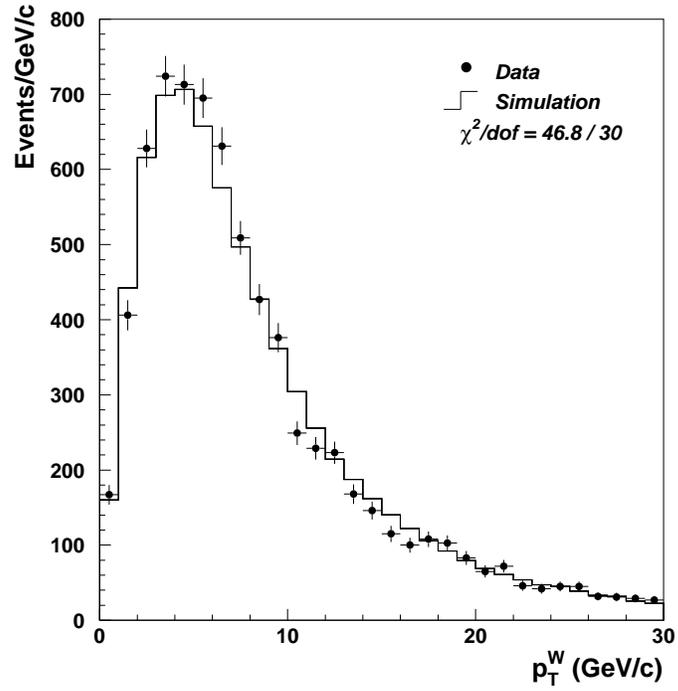}    
\end{tabular}
\end{center}
\caption[]{Distribution of $p_T^W$ 
           from $W\rightarrow e\nu$ decays (points)
           compared to the simulation (histogram). }
\label{fig:mw_datamc_ptw}
\end{figure}

The primary measurables in $W\rightarrow e \nu$ events are the energy
and direction  of the electron $\vec E$ and the transverse momentum of the
recoil
$\vec p_T^{\, {\text rec}}$. In addition, there are a 
variety of derived quantities
which are  especially sensitive to the presence of inefficiencies or  biases
which serve as important checks. The comparison between the data and the 
Monte Carlo simulation for $W\rightarrow e \nu$ events in  the electron
polar angle $\cos(\theta_e)$ and  the transverse momentum of the $W$ boson
$p_T^W$  are shown in Figs.~\ref{fig:mw_datamc_costh} and 
\ref{fig:mw_datamc_ptw}.  There is reasonable agreement 
between the simulation and
the data in both  distributions. 

\begin{figure}[h]
\begin{center}
\begin{tabular}{c}
    \epsfxsize=10.cm   \epsffile{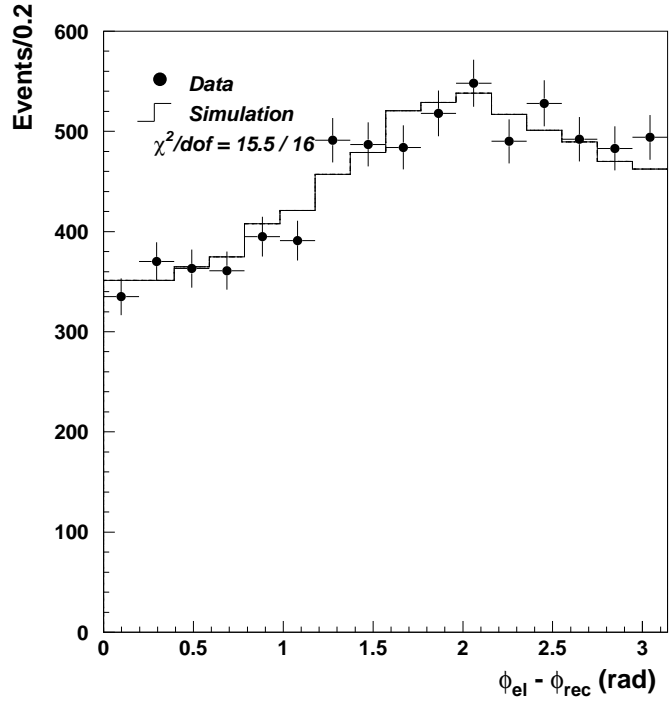}  
\end{tabular}
\end{center}
\caption[]{Distribution of the angle between the recoil 
           jet and the electron in the transverse plane 
           from $W\rightarrow e\nu$ decays (points)
           compared to the simulation (histogram). } 
\label{fig:upar_phi}
\end{figure}

Because of its strong correlation with the lepton transverse momenta, 
$u_\parallel$, defined previously, is an  important quantity. As was 
noted in Section
IV~D2, a bias in
$u_\parallel$ distorts the available momentum phase  space of the  leptons
and results in a softer or harder lepton 
$p_T$ spectrum, depending on that bias.  Since $u_\parallel$ involves both
the electron identification efficiency  and the hadronic energy scale, it is
advantageous to study the  distribution in the angle between the recoil
system and the electron, as 
well as a distribution in $u_\parallel$ itself. Figure~\ref{fig:upar_phi} shows
the distribution  in $\varphi_{el} - \varphi_{{\text rec}}$.

Note that for small $p_T^W$, assuming perfect electron 
identification, the $W$ boson recoil would be distributed uniformly in 
$\varphi$ around the electron direction.  However, the distribution in
$\varphi_{el} - \varphi_{{\text rec}}$ is asymmetric.  There are two sources for
this asymmetry.  The dominant effect is simply the kinematics of
$W\rightarrow e \nu$ decays.  For transversely boosted $W$ bosons, on
average the electron carries away 
$p_T^e \approx p_{T}^W/2$ along the $\vec{p}_T^{\, W}$ direction, having a 
magnitude of $\approx \,M_{W}/2$ for small values of $p_{T}^W$.   This implies
that 
$\langle u_\parallel \rangle \approx - \langle {p_T^{W}}^{2} \rangle / M_{W}$.  
Since the mean value of $p_{T}^W$ is approximately 9~GeV/c (see
Fig.~\ref{fig:mw_datamc_ptw}),
$\langle u_\parallel \rangle $ is about $-1$~GeV and  the distribution in
the difference in azimuthal angle of the  electron and the recoil tends to
favor negative values of $u_\parallel$.   The second effect which enhances
the asymmetry is due to a decrease in  electron identification efficiency as
function of $u_\parallel$. The value of $u_\parallel$ is an indication of
the proximity of the recoil jet to the electron.  For high positive values
of $u_\parallel$ the recoil jet is  close to the electron and can spoil its
signature.  The observed excellent agreement between the simulation and the 
data indicates that the event kinematics and the electron identification 
efficiency are modeled adequately.

\begin{figure}[t]
\begin{center}
\begin{tabular}{c}
    \epsfxsize=10.cm   \epsffile{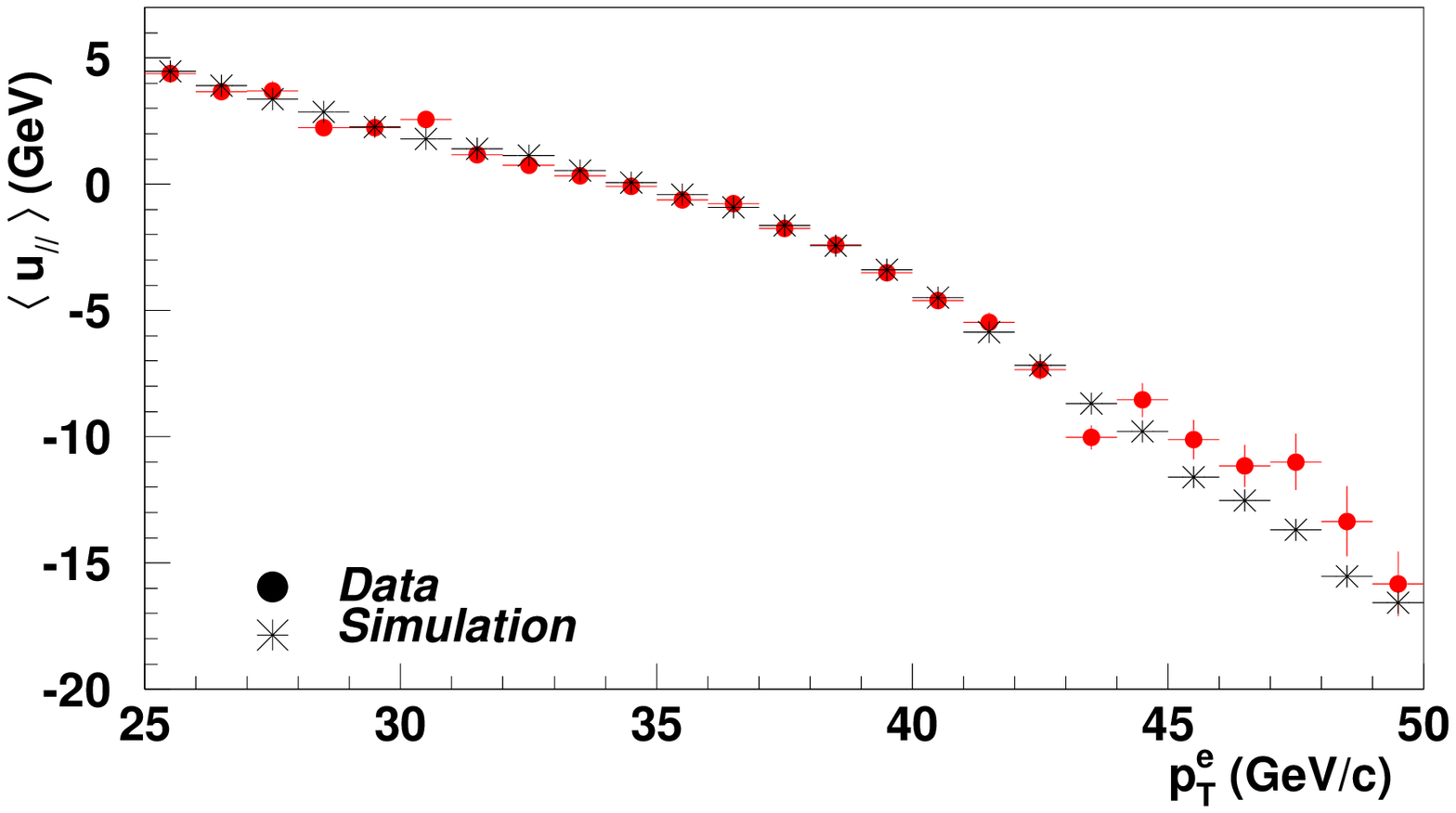}   
\end{tabular}
\end{center}
\caption[]{Distribution of the mean $u_\parallel$ versus
         $p_T^e$   from $W\rightarrow e\nu$ decays (points) compared to
         the simulation ($\ast$). }
\label{fig:upar_pte}
\end{figure}

\begin{figure}[t]
\begin{center}
\begin{tabular}{c}
    \epsfxsize=10.cm    \epsffile{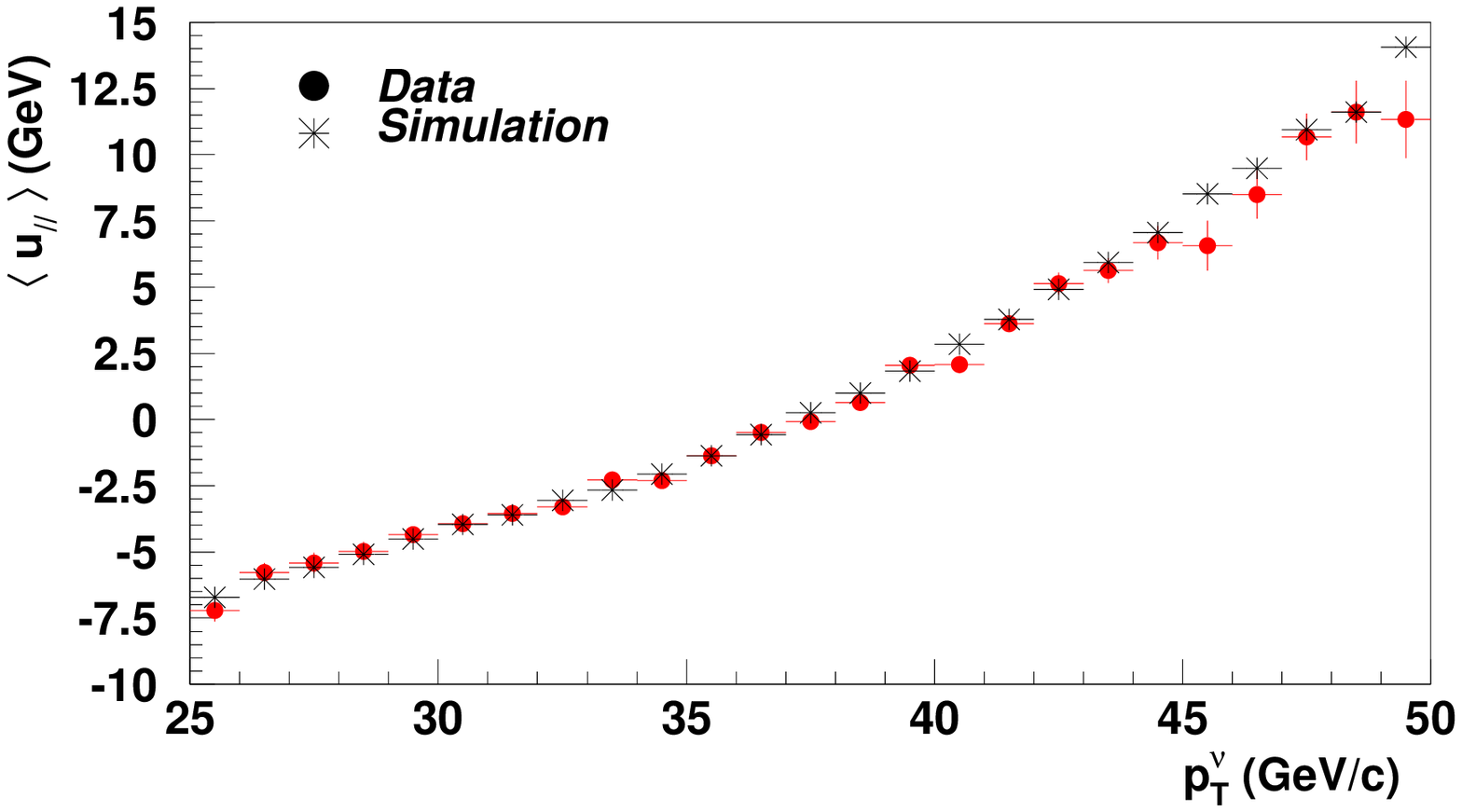}  
\end{tabular}
\end{center}
\caption[]{Distribution of the mean $u_\parallel$ versus
         $p_T^{\nu}$   from $W\rightarrow e\nu$ decays (points) compared to
         the simulation ($\ast$). }
\label{fig:upar_ptnu}
\end{figure}

\begin{figure}[t]
\begin{center}
\begin{tabular}{c}
    \epsfxsize=10.cm     \epsffile{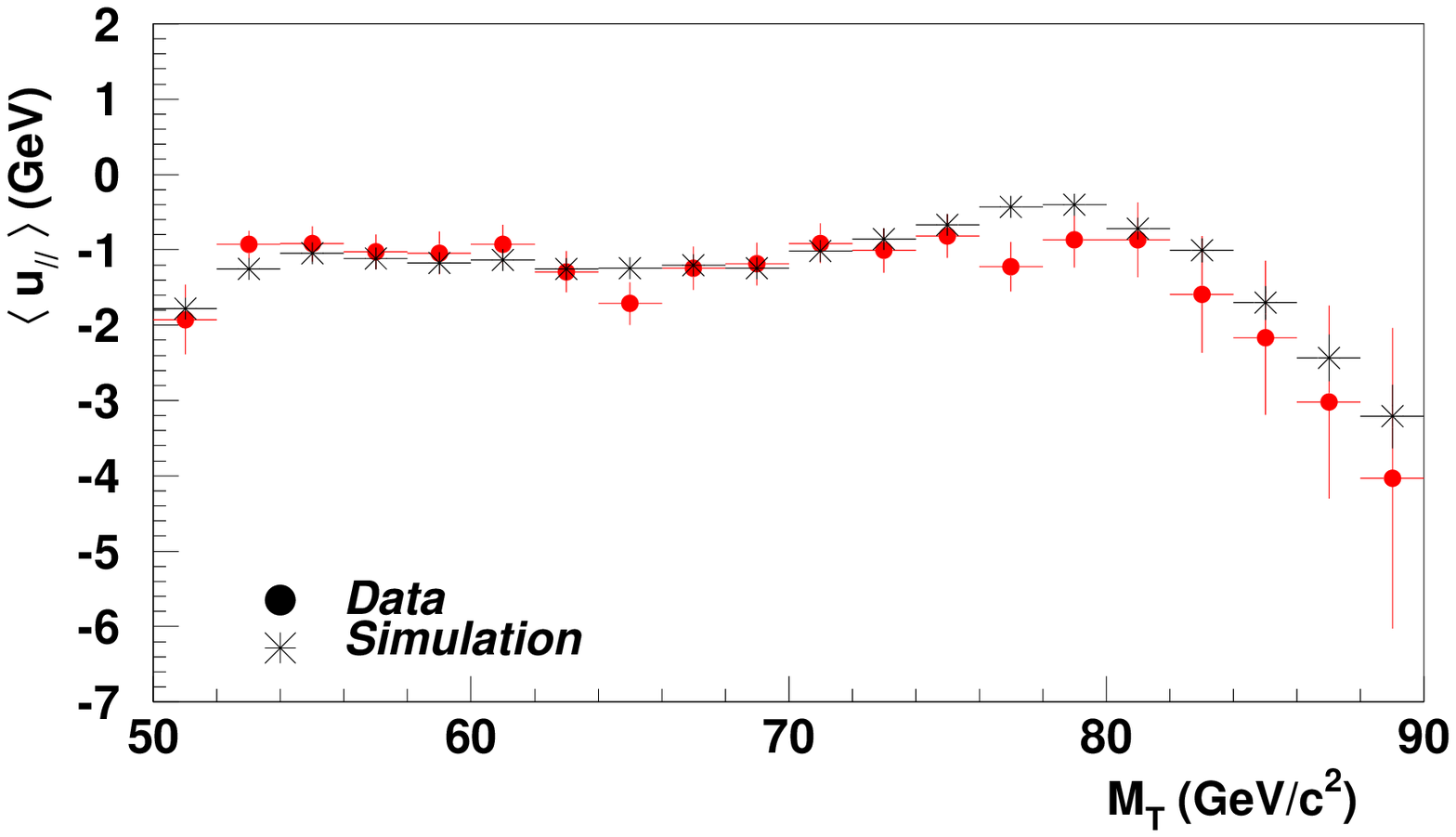}  
\end{tabular}
\end{center}
\caption[]{Distribution of the mean $u_\parallel$ versus
         $M_T$   from $W\rightarrow e\nu$ decays (points) compared to
         the simulation ($\ast$). }
\label{fig:upar_mt}
\end{figure}

\begin{figure}[t]
\begin{center}
\begin{tabular}{c}
    \epsfxsize=10.cm    \epsffile{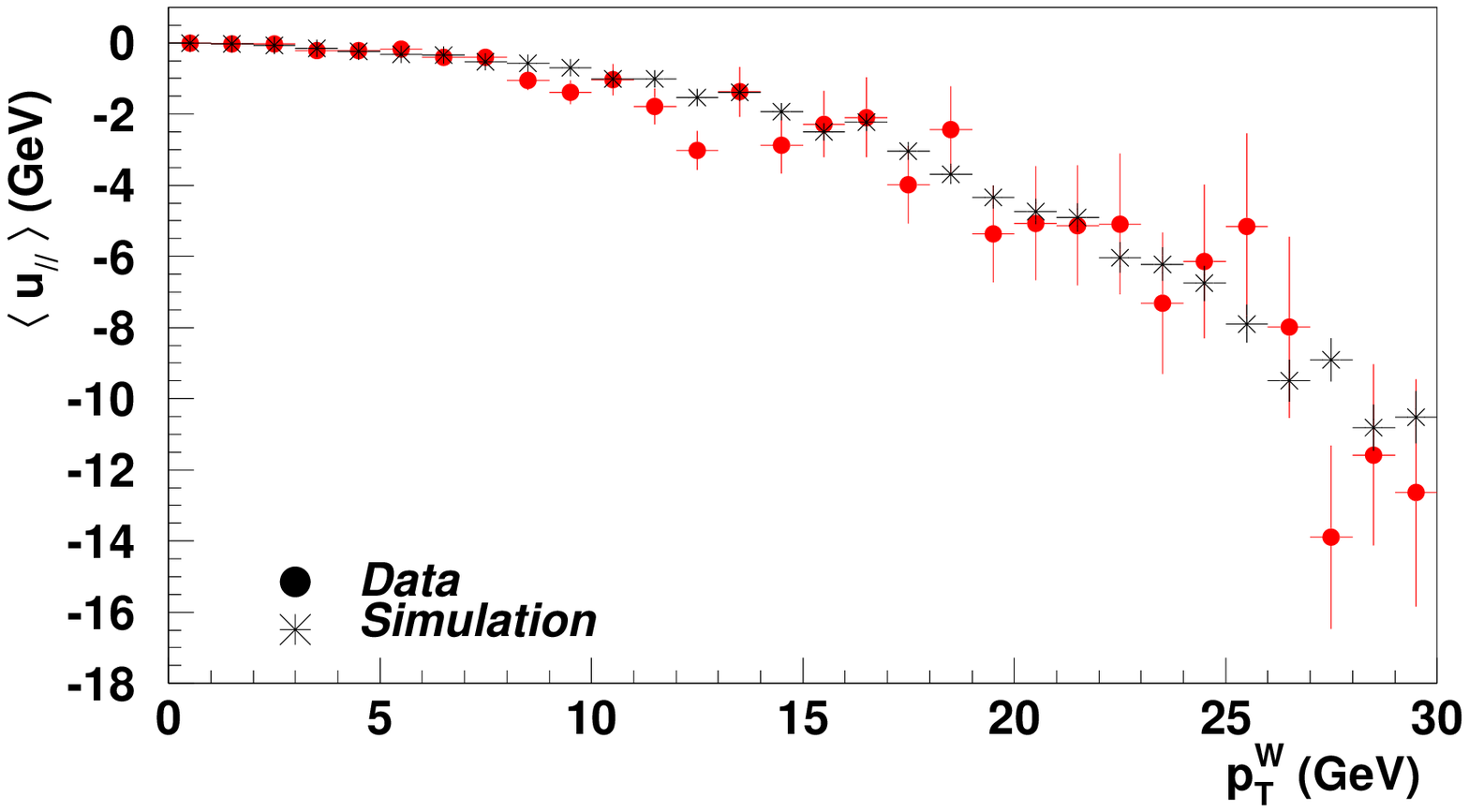} 
\end{tabular}
\end{center}
\caption[]{Distribution of the mean $u_\parallel$ versus
         $p_T^W$   from $W\rightarrow e\nu$ decays (points) compared to
         the simulation ($\ast$). }
\label{fig:upar_ptw}
\end{figure}

Figures~\ref{fig:upar_pte} and \ref{fig:upar_ptnu} show the correlation 
between $\langle u_\parallel \rangle$ and $p_T^e$ and $p_T^\nu$.   An
important feature of the transverse mass is that, unlike 
$p_T^e$ and $p_T^\nu$, $M_T$ is relatively uncorrelated with
$u_\parallel$  as shown in Fig.~\ref{fig:upar_mt}.   This shows clearly one
of the advantages of using the transverse  mass 
to obtain the $W$ boson mass.  The
correlation between $u_\parallel$ and $p_T^W$ is shown in 
Fig.~\ref{fig:upar_ptw}.

\begin{figure}[t]
\begin{center}
\begin{tabular}{c}
    \epsfxsize=10.cm   \epsffile{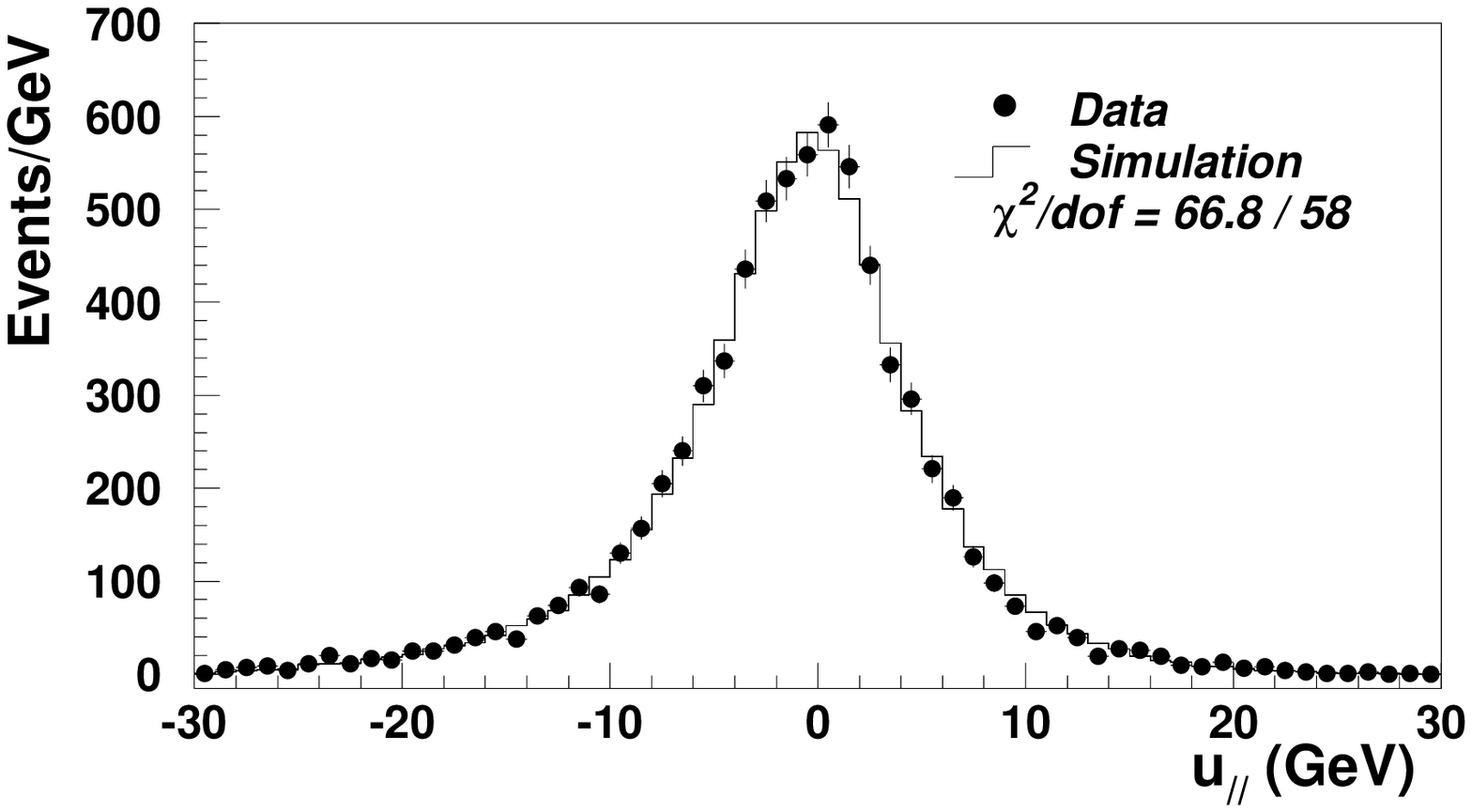}
\end{tabular}
\end{center}
\caption{Comparison of the $u_\parallel$ distribution from
         $W\rightarrow e\nu$ events (points) and the Monte
         Carlo simulation (histogram). }
\label{fig:upar}
\end{figure}

Figure~\ref{fig:upar} shows the distribution in $u_\parallel$ itself. 
Note that there has not been a subtraction for background.
The mean value of $u_\parallel$ for the data is 
$\langle u_\parallel \rangle = -1.19 \pm 0.08$~GeV whereas the 
simulation gives $\langle u_\parallel \rangle = -1.13 \pm 0.02$~GeV. 
An average correction for the QCD and $Z\rightarrow ee$ background 
has been applied to the value just quoted for
$\langle u_\parallel \rangle $ for the Monte Carlo. 

\begin{figure}[t]
\begin{center}
\begin{tabular}{c}
    \epsfxsize=10.cm   \epsffile{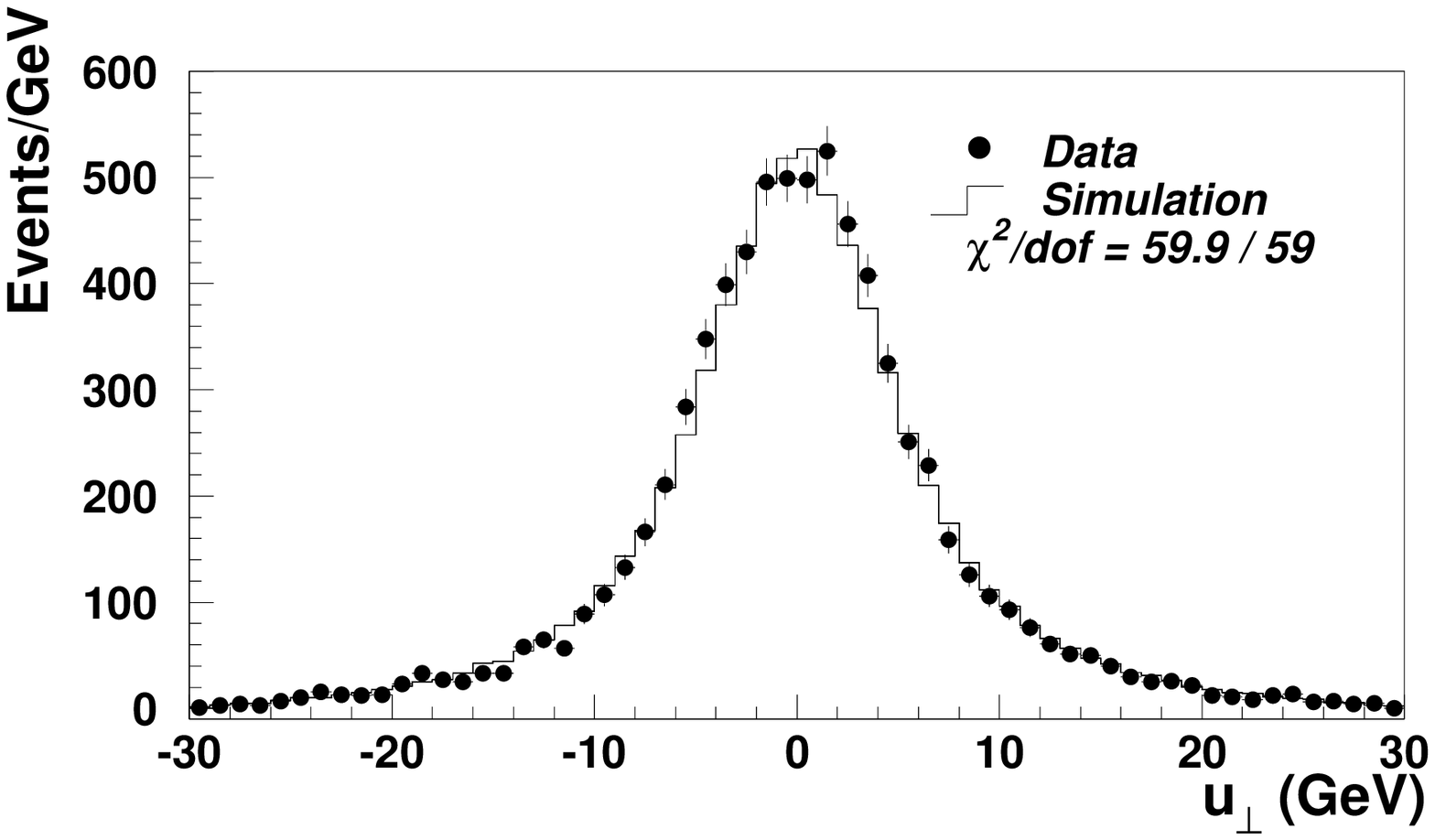} 
\end{tabular}
\end{center}
\caption{Comparison of the $u_\perp$ distribution from
         $W\rightarrow e\nu$ events (points) and the Monte
         Carlo simulation (histogram). }
\label{fig:uperp}
\end{figure}

The distribution of $u_\perp$, Fig.~\ref{fig:uperp},  defined as the
projection of the recoil jet onto the axis perpendicular to the electron
direction,  is a measure of the resolution of the recoil system.  Its mean
value is close to zero, as expected.   For the data   $\langle u_\perp
\rangle = 0.025 \pm 0.087$~GeV with  an rms of 7.4~GeV;  the simulation
gives $\langle u_\perp \rangle = 0.024 $~GeV   with an rms of 7.5~GeV.


\subsubsection {Characterization of the $Z\rightarrow e e$ Candidates } 

\begin{figure}[tp]
\begin{center}
\begin{tabular}{c}
    \epsfxsize=10.cm \epsffile{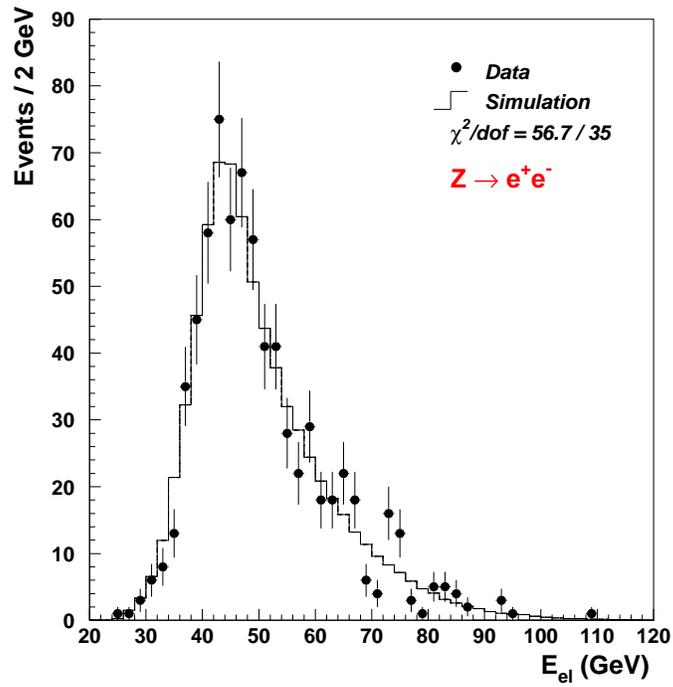}    
\end{tabular}
\end{center}
\caption[]{Comparison of the electron energy distribution 
           from $Z\rightarrow ee$ events (points) and the
           simulation (histogram). }
\label{fig:mz_datamc_e}
\end{figure}

\begin{figure}[tp]
\begin{center}
\begin{tabular}{c}
    \epsfxsize=10.cm   \epsffile{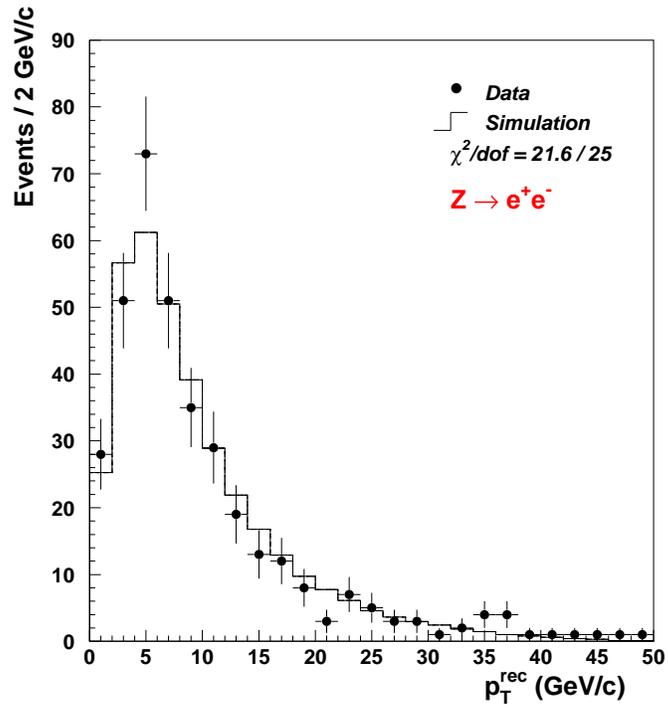}   
\end{tabular}
\end{center}
\caption[]{Comparison between the $Z$ boson transverse momentum distribution as 
           measured from the recoil system in $Z\rightarrow ee$ events (points)
and the 
           simulation (histogram). }
\label{fig:mz_datamc_ptz_rec}
\end{figure}

The measured quantities in $Z\rightarrow e e$ events are the energy and
direction  of both electrons and the transverse momentum of the recoil
system.  Equally important are the determination of derived quantities  of
the $Z$ boson kinematics.  Figures~\ref{fig:mz_datamc_e}
--~\ref{fig:mz_datamc_ptz_rec} show  the comparison in  electron energy
$E_{el}$ and the transverse momentum distribution from the 
recoil system, $p_T^{{\text rec}}$.

\section {Fitting Procedure } 

The Monte Carlo event generation was performed for 21 equidistant mass values
binned at intervals of 100~MeV/$c^2$ in the transverse mass for $W$ boson
spectra, 200~MeV/$c^2$ in invariant mass for $Z$ boson spectra, and 100~MeV/c
 for  the transverse momentum spectra. 

An unbinned maximum likelihood fit was used to determine the vector boson 
mass using the normalized Monte Carlo spectra as templates. 
The log-likelihood   was calculated
for the data for the 21 different  generated masses.  Since the templates 
were
binned whereas the data were unbinned, a  quadratic interpolation
between adjacent bins in the templates was  performed.  The log-likelihood
values for the 21 different vector boson masses were  fit to a parabola and
the minimum was taken to be the fitted mass value.  A decrease of half a unit
in the log-likelihood is the quoted single standard  deviation statistical
uncertainty. 

The likelihood distribution need not be Gaussian, depending on the range  of
the parameter fit, the intrinsic shape of the spectrum and the  resolution
function.   This is particularly true for spectra with a sharp edge like the
Jacobian  peak in the distributions considered here.  Both quadratic and
cubic polynomial  fits were performed to the log-likelihood. The differences
were small and for all results presented here, the values from the quadratic
fit are quoted.  

\begin{figure}[tp]
\begin{center}
\begin{tabular}{cc}
    \epsfxsize=6.5cm     \epsffile{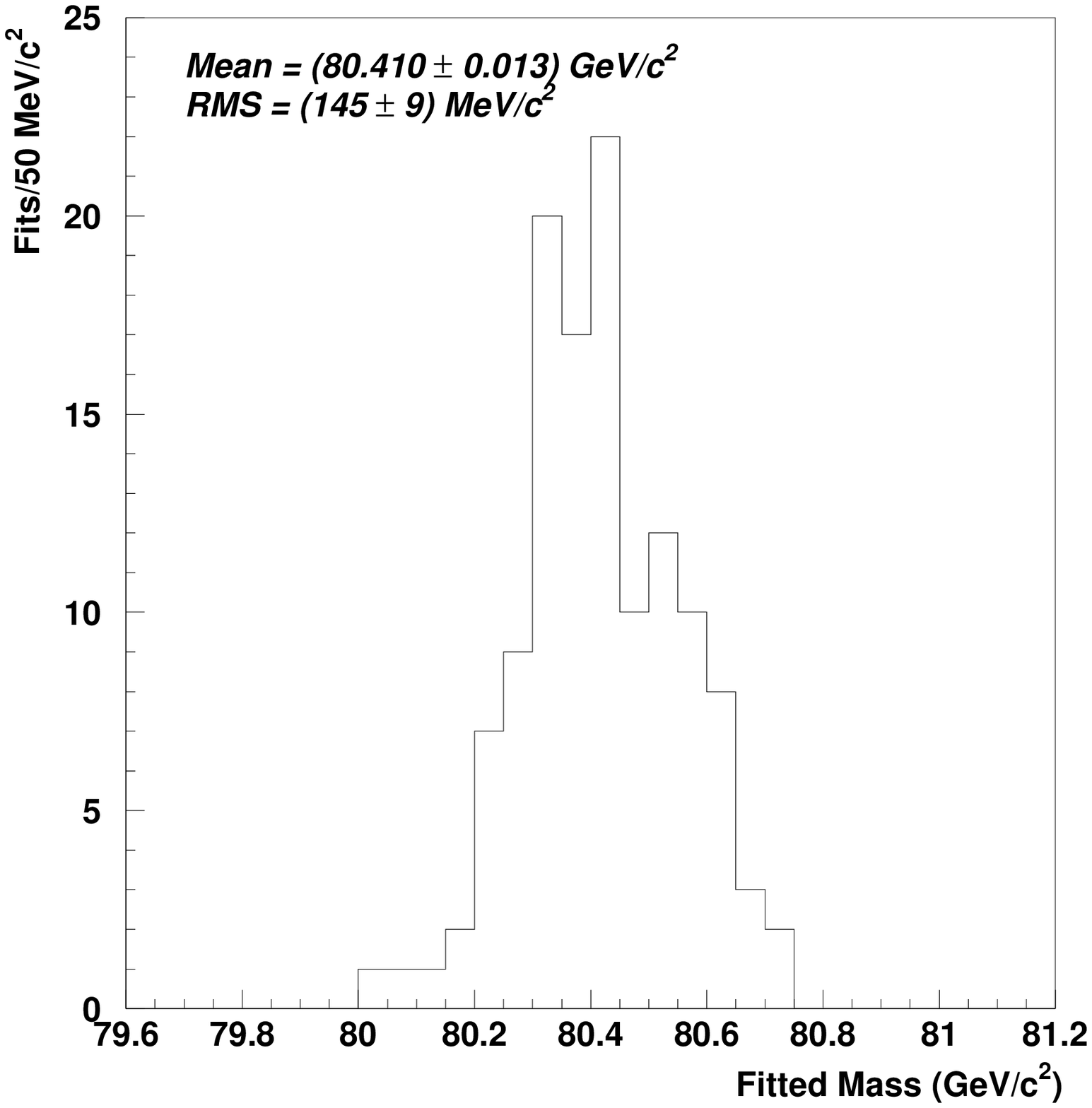}  &
    \epsfxsize=6.5cm     \epsffile{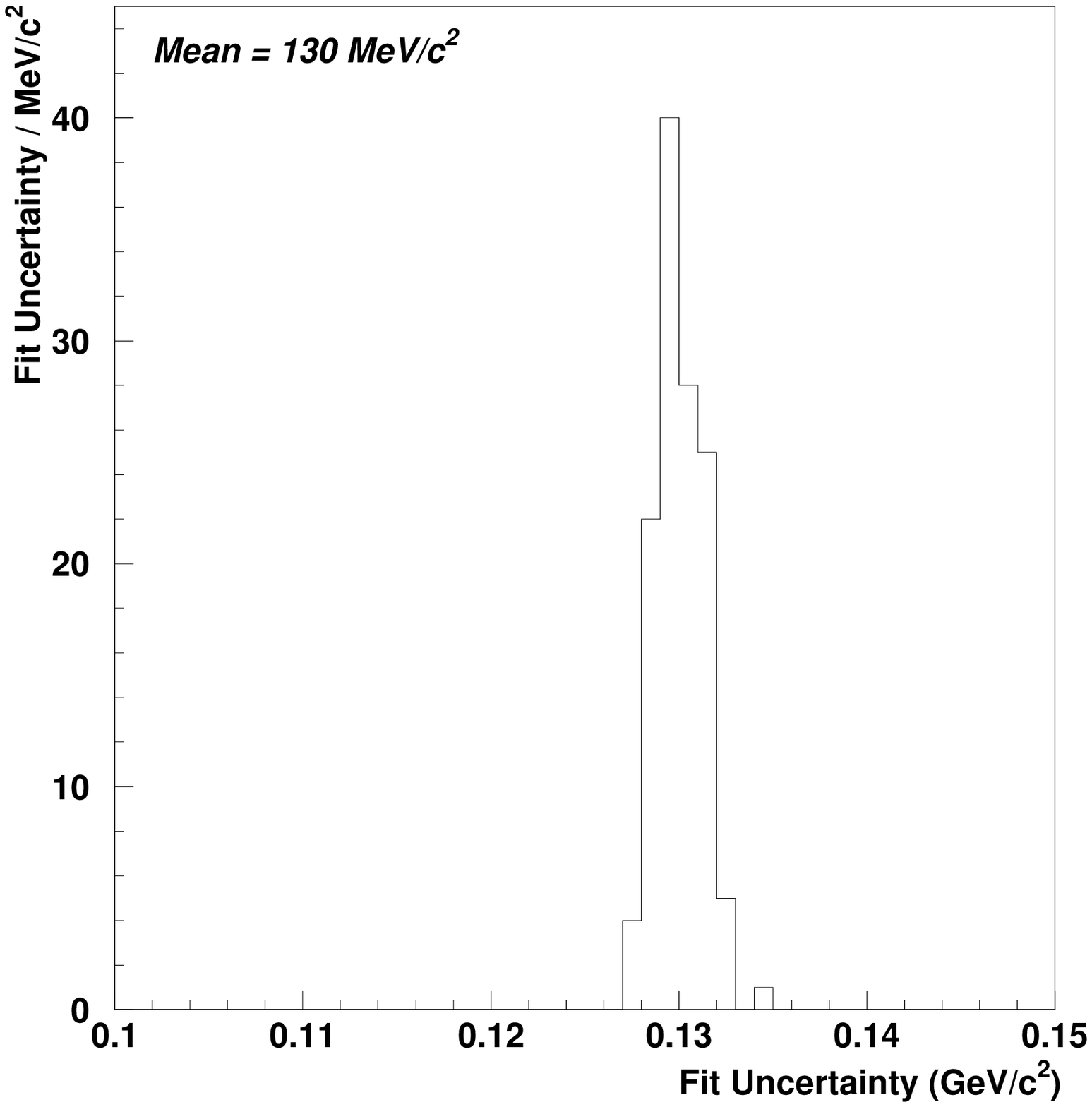}     \\
\end{tabular}
\end{center}
\caption{Distribution of (a) the fitted masses and (b) the fit uncertainties 
         from fits to the transverse mass distributions 
         for an ensemble of 125 Monte Carlo generated data samples 
         of 8000 $W\rightarrow e\nu$ decays. } 
\label{fig:mc_consistent}
\end{figure}

Any Monte Carlo--based fitting procedure should satisfy the requirements
that, if the procedure is applied to an ensemble of Monte Carlo generated 
data samples, it returns the input values with which the events were
generated and, secondly, that the rms spread of the values for the fitted 
parameter be consistent with the mean statistical uncertainty of the fit to
each individual data sample. This was done for an ensemble of
125 generated data samples of 8000 events each.
The average statistical error for each of the three different $W$ 
boson mass fits is: $\delta(M_T) = 130 $, 
$\delta(p_T^e) = 183 $ and 
$\delta(p_T^\nu) = 248 $~MeV/c$^{2}$, respectively.
The average fitted mass values are 
$M_W(M_T) = 80.410 \pm 0.013 $, 
$M_W(p_T^e) = 80.398 \pm 0.017 $ and 
$M_W(p_T^\nu) = 80.420 \pm 0.021$~GeV/$c^2$, 
in good agreement with the input value of 80.400~GeV/$c^2$ 
within the statistical accuracy of the generated templates. 
They are consistent with the rms spread of the distribution of the fitted 
masses, 
${\rm rms}(M_T) = 145 \pm 9 $, 
${\rm rms}(p_T^e) = 188 \pm 12 $ and 
${\rm rms}(p_T^\nu) = 237 \pm 15 $~MeV/$c^2$, respectively.
Figure~\ref{fig:mc_consistent} shows the
distribution of fitted mass values and fit uncertainty for $W$ bosons as
obtained from a fit to the transverse mass for this ensemble. 

As discussed in the previous section, backgrounds were not included in
the  event simulation.  Their effect on the mass determination was taken
into  account through inclusion of the shape of the background spectrum 
in the likelihood distributions.  The background was properly  normalized
to the expected background fraction in the relevant fitting  range. All results
were corrected for backgrounds.

\subsection {Results of $Z$ Boson Mass Fits } 

\begin{figure}[tp]
\begin{center}
\begin{tabular}{cc}
    \multicolumn{2}{c}{
    \epsfxsize=10.cm \epsffile{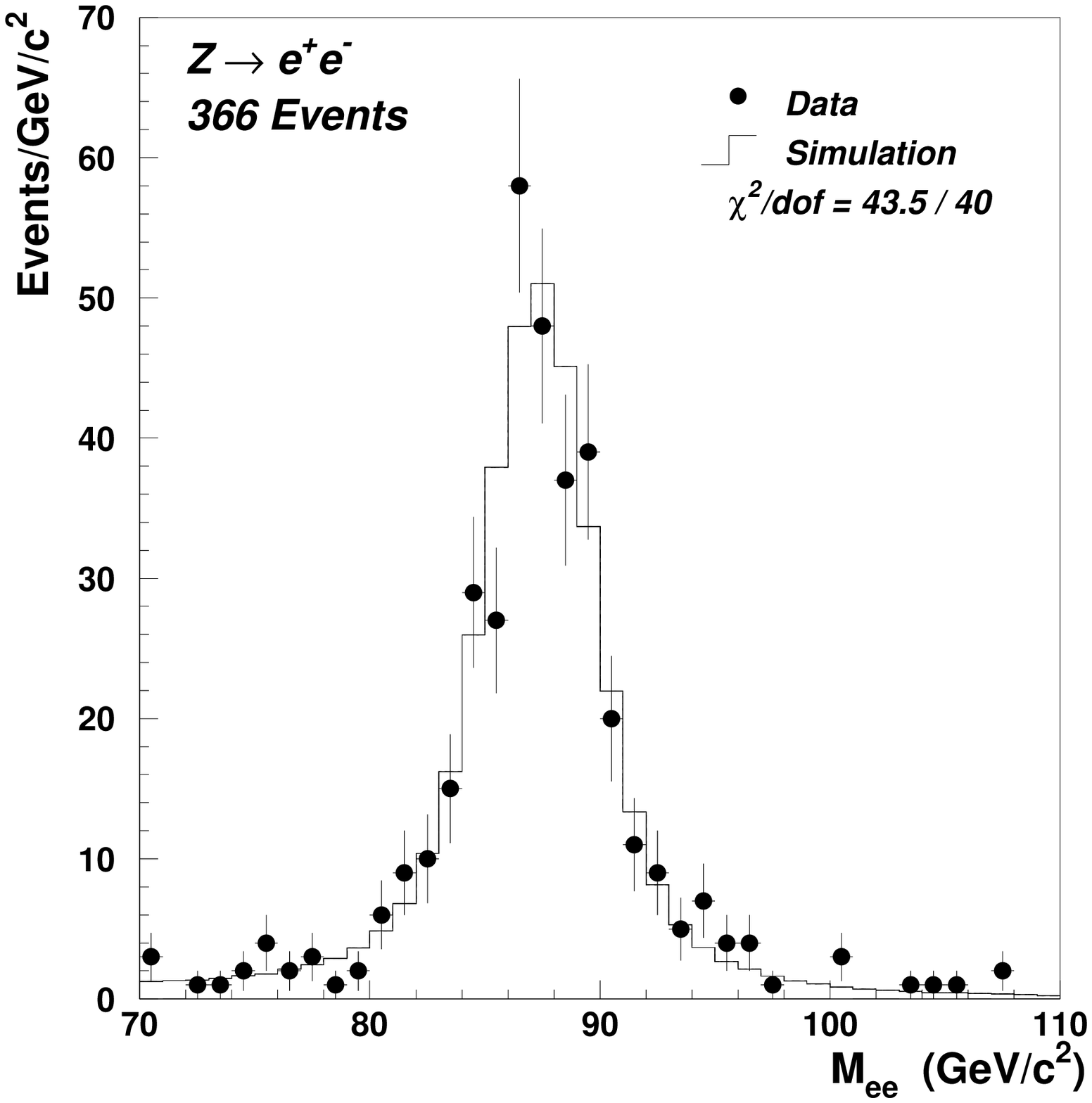} } \\
    \epsfxsize=7.cm  \epsffile{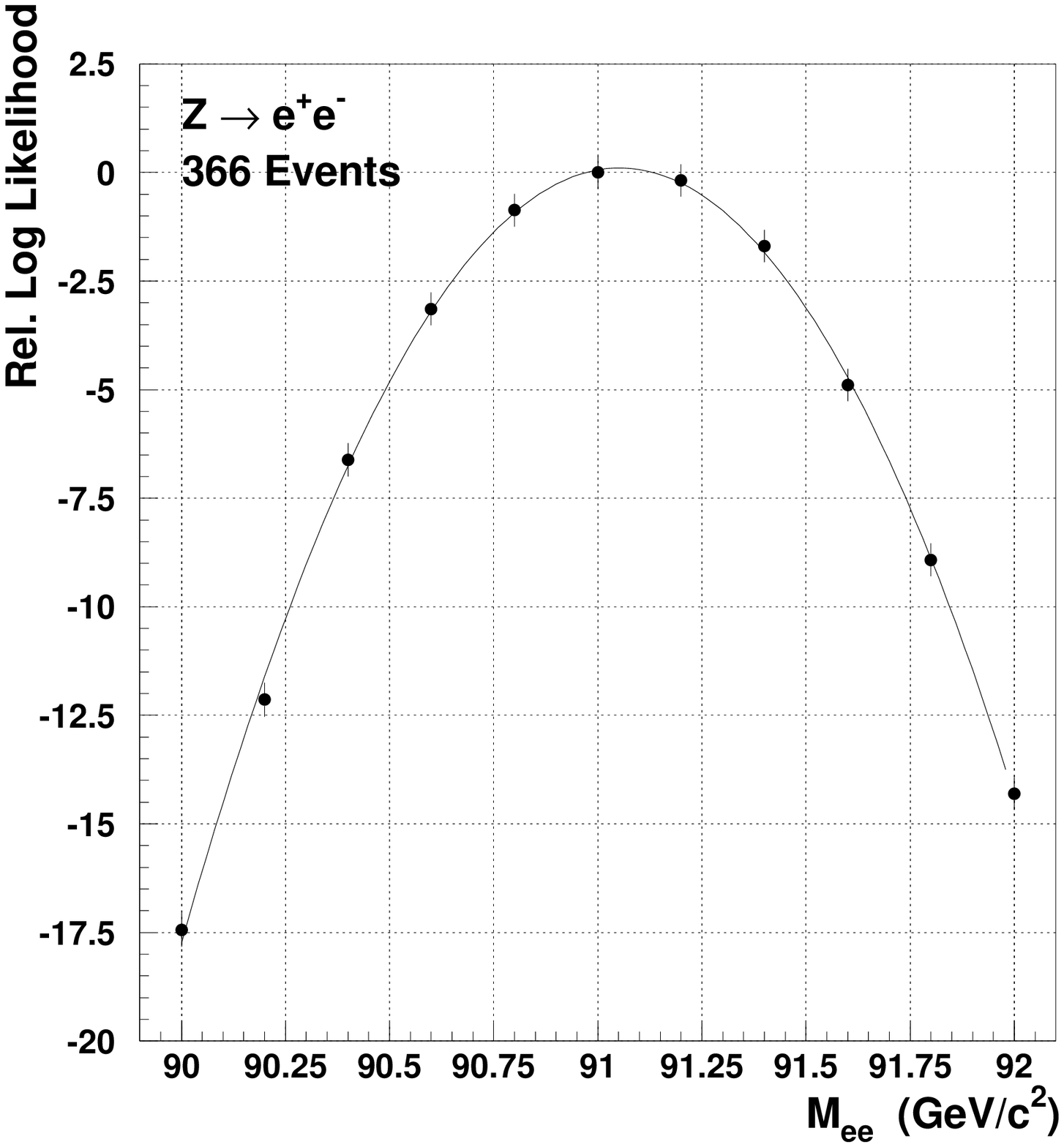}     & 
    \epsfxsize=7.cm   \epsffile{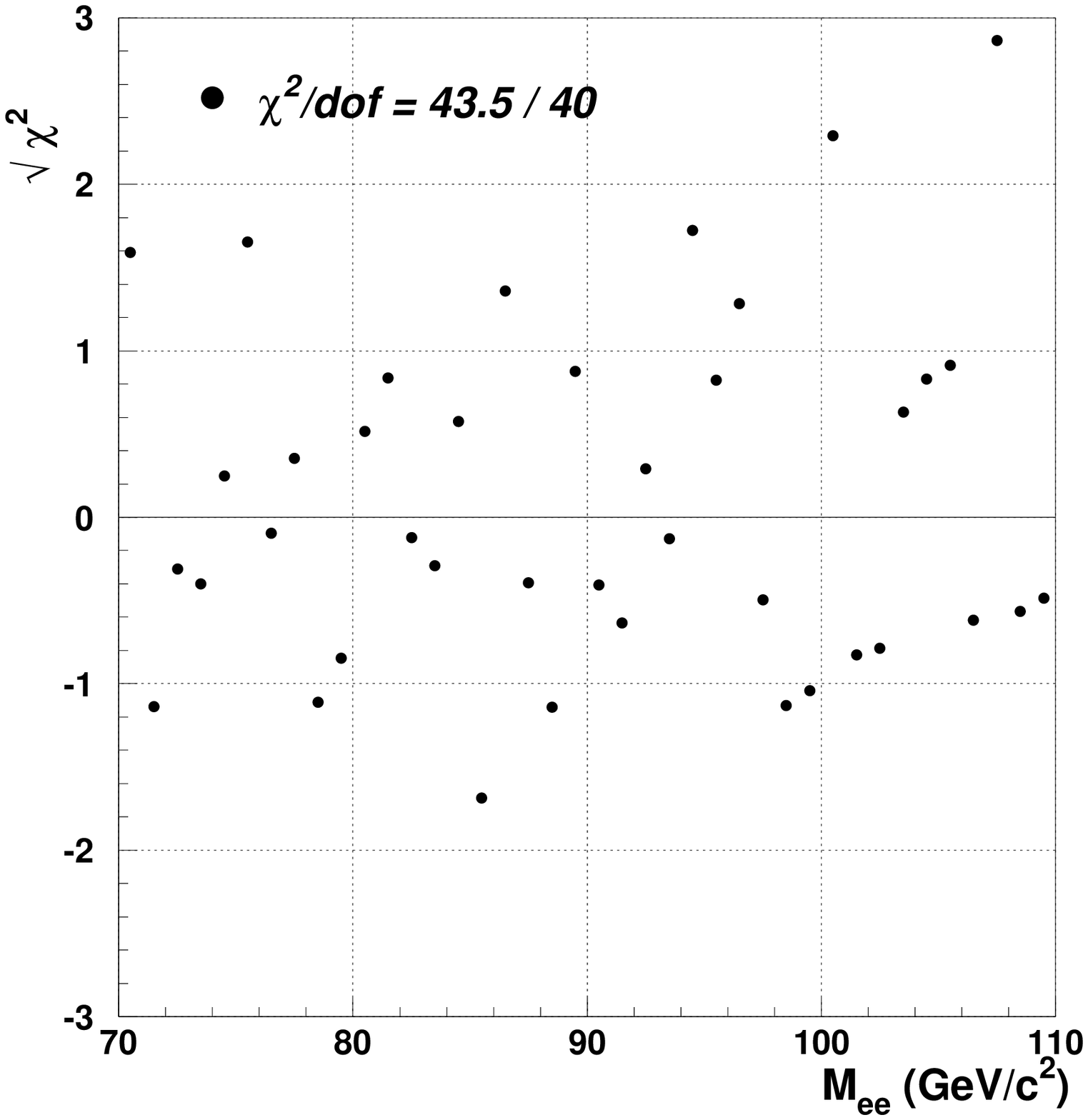}     \\
\end{tabular}
\end{center}
\caption[]{(a) The central dielectron invariant mass distribution 
           for $Z$ events (points) and the best fit of the simulation
           (histogram), (b) the corresponding relative log-likelihood 
           distribution and (c)   signed $\sqrt{\chi^2}$  distribution. }
\label{fig:mz_cc}
\end{figure}

The dielectron invariant mass spectrum for the  central-central (CC-CC) event
topology, with the corresponding best fit of the templates to the data,  
is shown in
Fig.~\ref{fig:mz_cc}.  The events in the mass range $ 70 < m_{ee} < 
110$~GeV/$c^2$  were
used to extract the $Z$ boson mass.  The final measured
$Z$ boson mass for events which require both electrons in the central 
calorimeter is:
\begin{equation}
  M_Z  = \ 91.070  \pm 0.170 \, {\rm GeV/}c^2.
\end{equation}
The error is statistical only.  Figure~\ref{fig:mz_cc} also
shows the relative likelihood distribution and signed $\sqrt {\chi^2}$ of 
the fit
for   central-central electrons.

\subsection {Results of $W$ Boson Mass Fits } 

\begin{figure}[tp]
\begin{center}
\begin{tabular}{cc}
    \multicolumn{2}{c}{
    \epsfxsize=10.cm   \epsffile{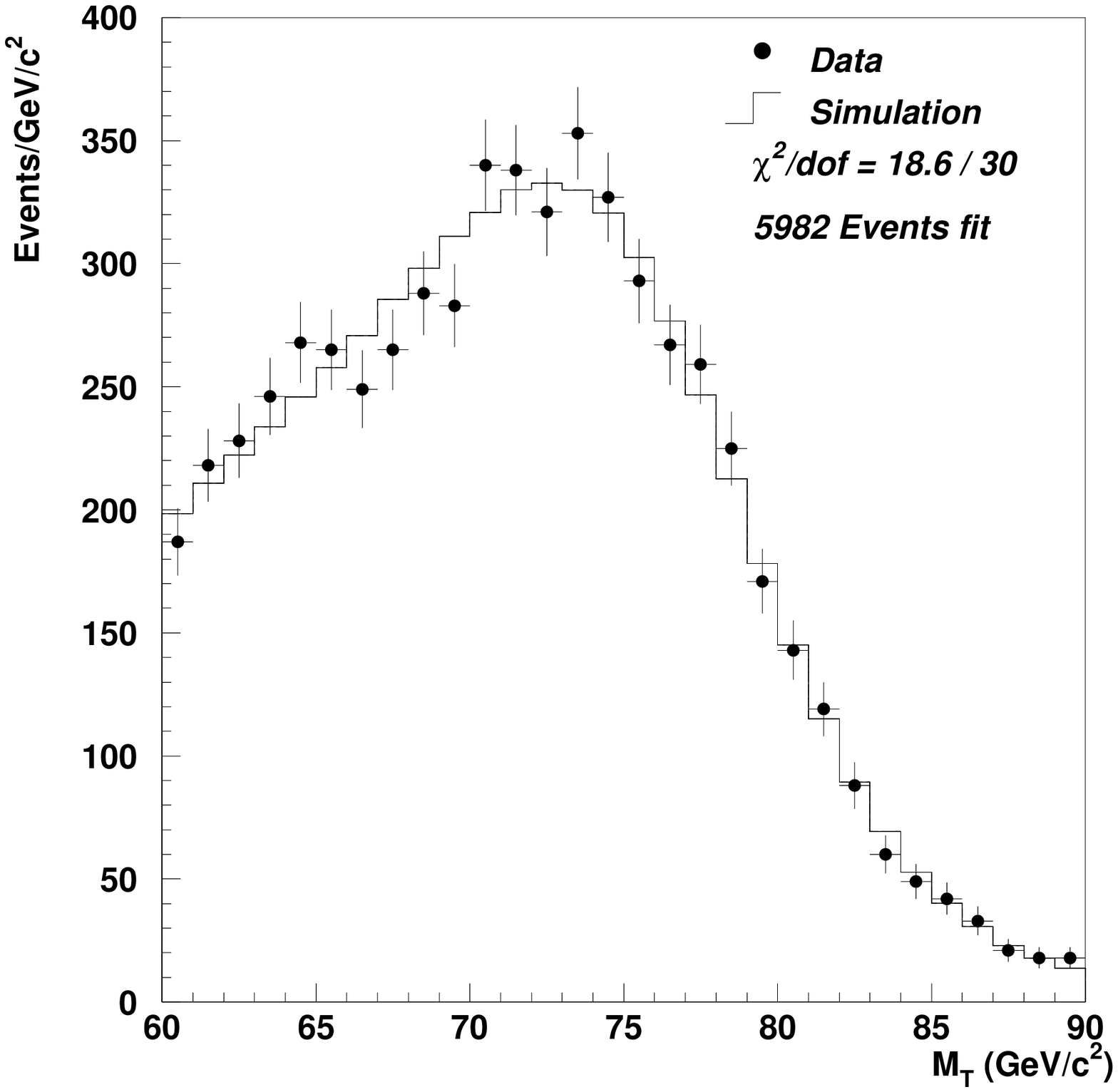} }  \\
    \epsfxsize=7.cm     \epsffile{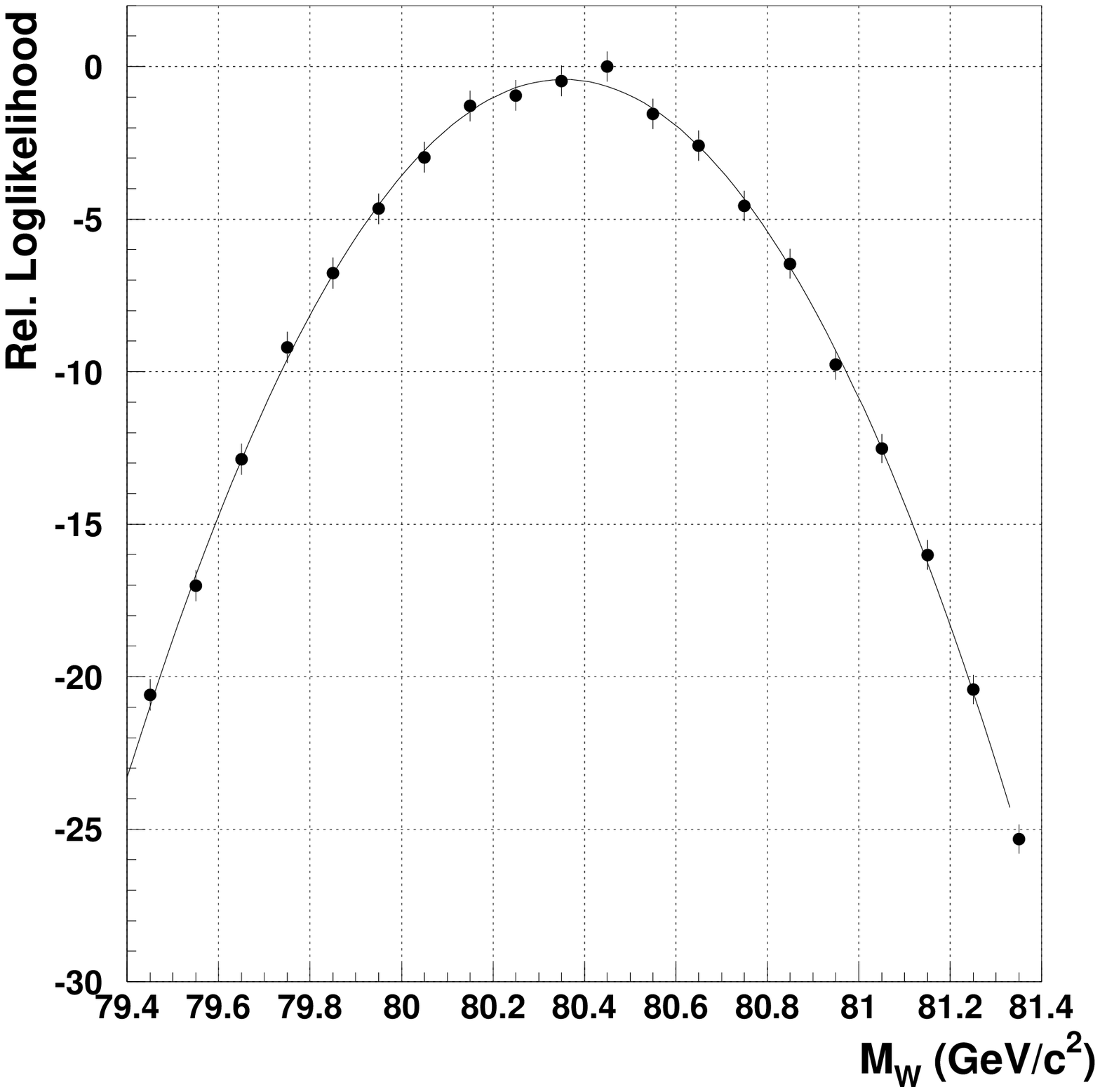}     & 
    \epsfxsize=7.cm     \epsffile{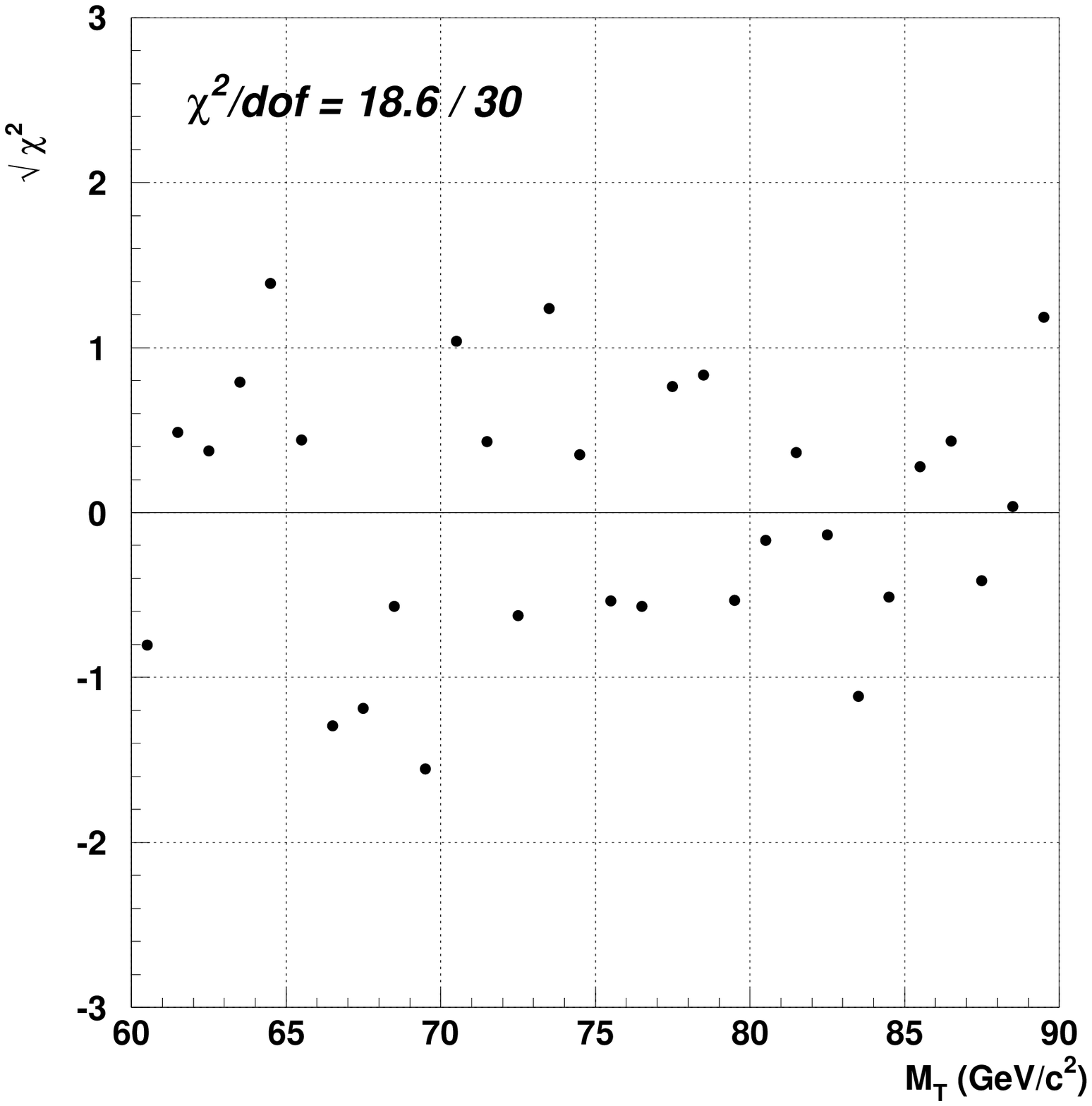}     \\
\end{tabular}
\end{center}
\caption[]{(a) The transverse mass distribution 
           for $W$ events (points) and the best fit of the simulation
           (histogram), (b) the corresponding relative log-likelihood 
           distribution and (c)   signed $\sqrt{\chi^2}$  distribution. }
\label{fig:mt_fit}
\end{figure}

\begin{figure}[tp]
\begin{center}
\begin{tabular}{cc}
    \multicolumn{2}{c}{
    \epsfxsize=10.cm     \epsffile{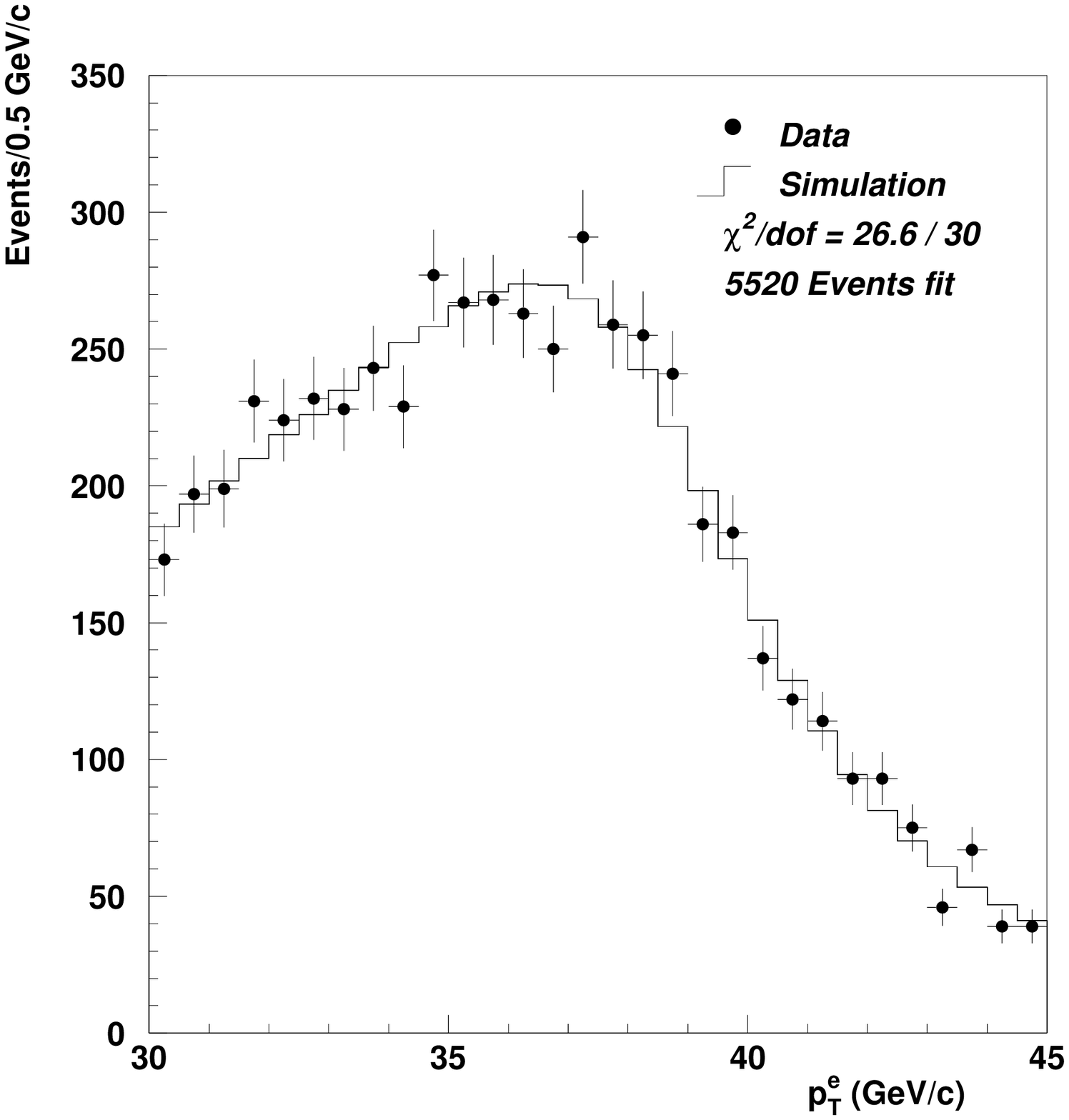} }    \\
    \epsfxsize=7.cm    \epsffile{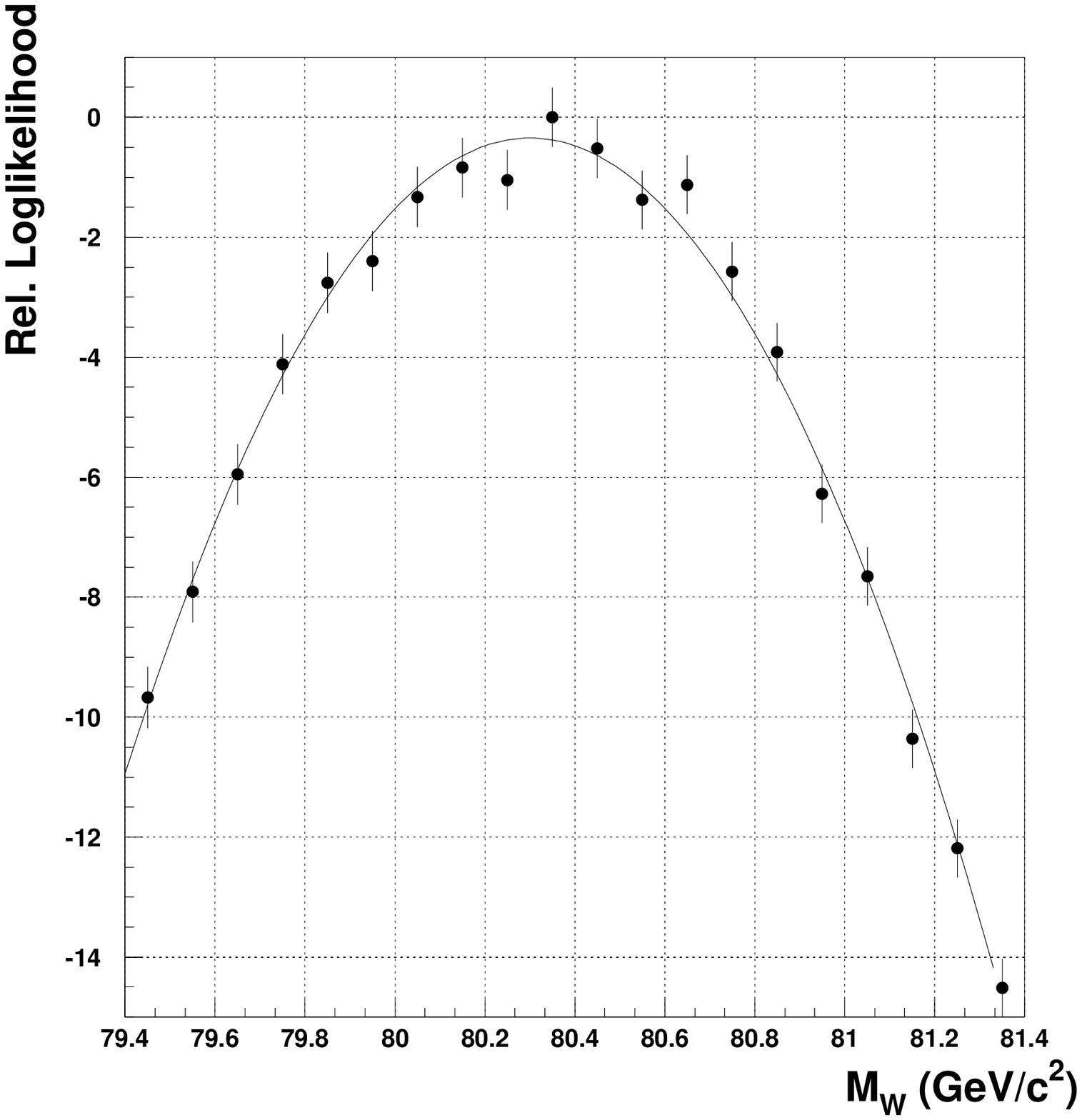}   &
    \epsfxsize=7.cm    \epsffile{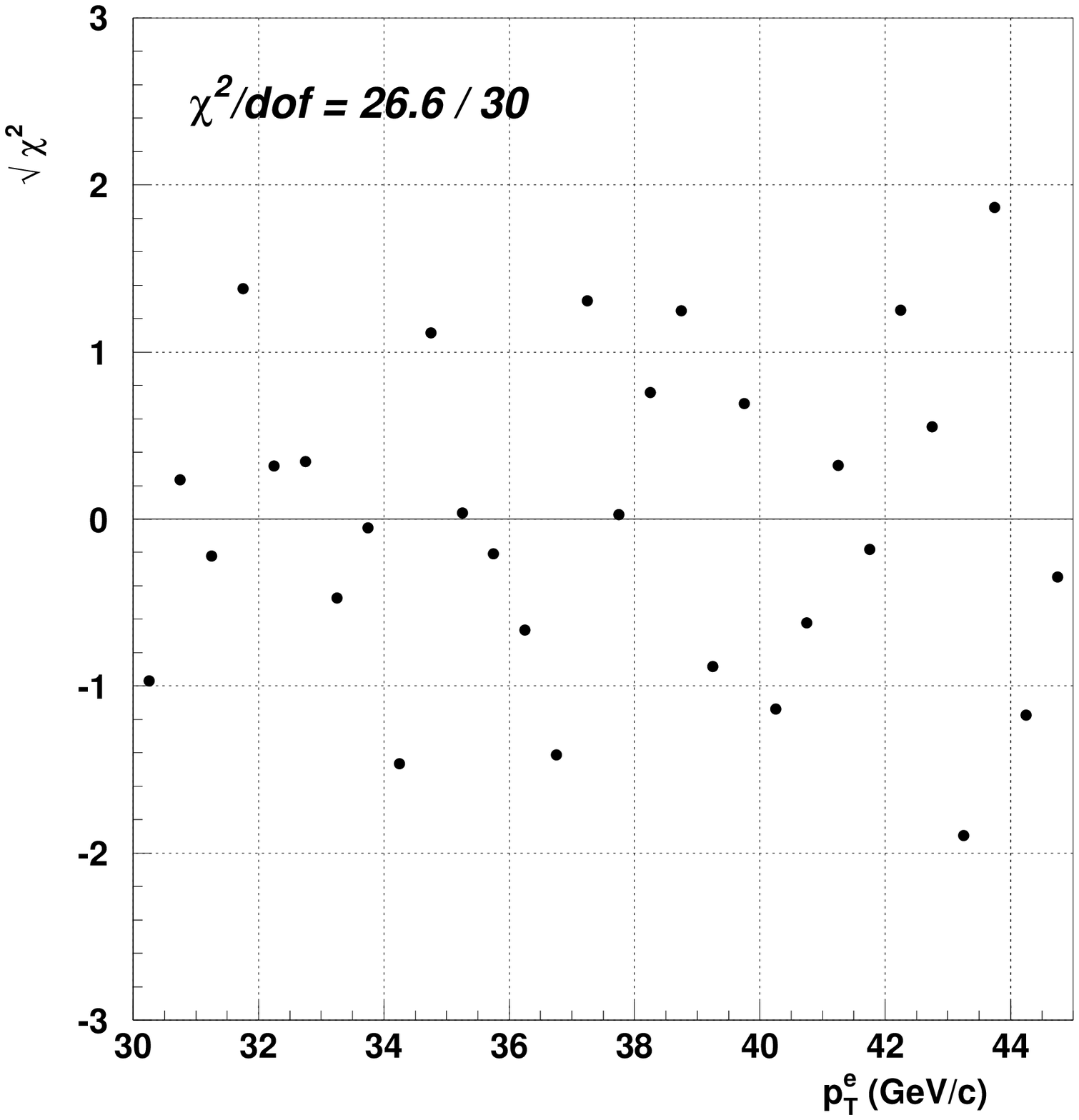}   \\
\end{tabular}
\end{center}
\caption[]{(a) The electron transverse momentum distribution 
           for $W$ events (points) and the best fit of the simulation
           (histogram), (b) the corresponding relative log-likelihood 
           distribution and (c)   signed $\sqrt{\chi^2}$  distribution. }
\label{fig:pte_fit}
\end{figure}

\begin{figure}[tp]
\begin{center}
\begin{tabular}{cc}
    \multicolumn{2}{c}{
    \epsfxsize=10.cm     \epsffile{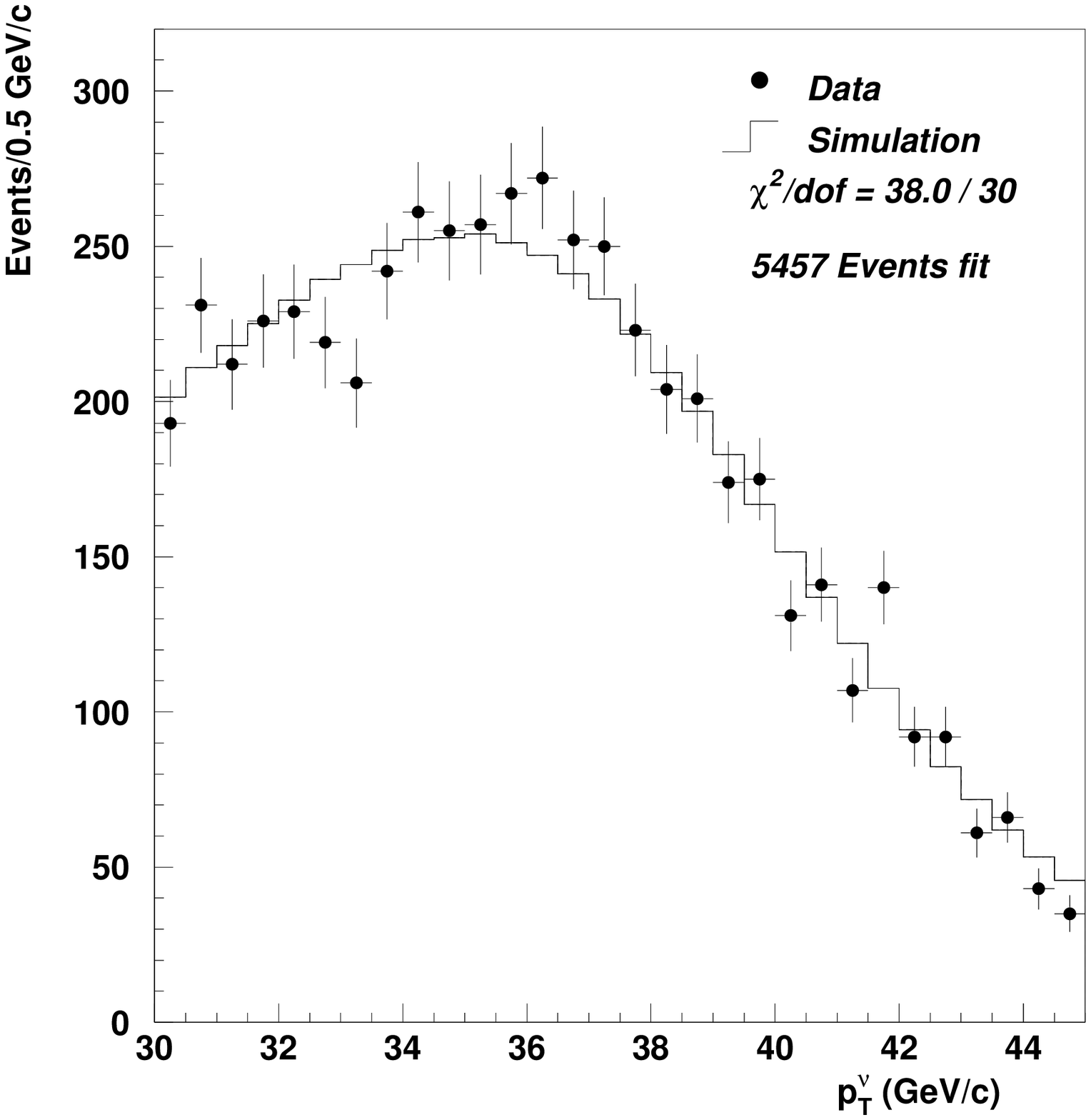} }    \\
    \epsfxsize=7.cm   \epsffile{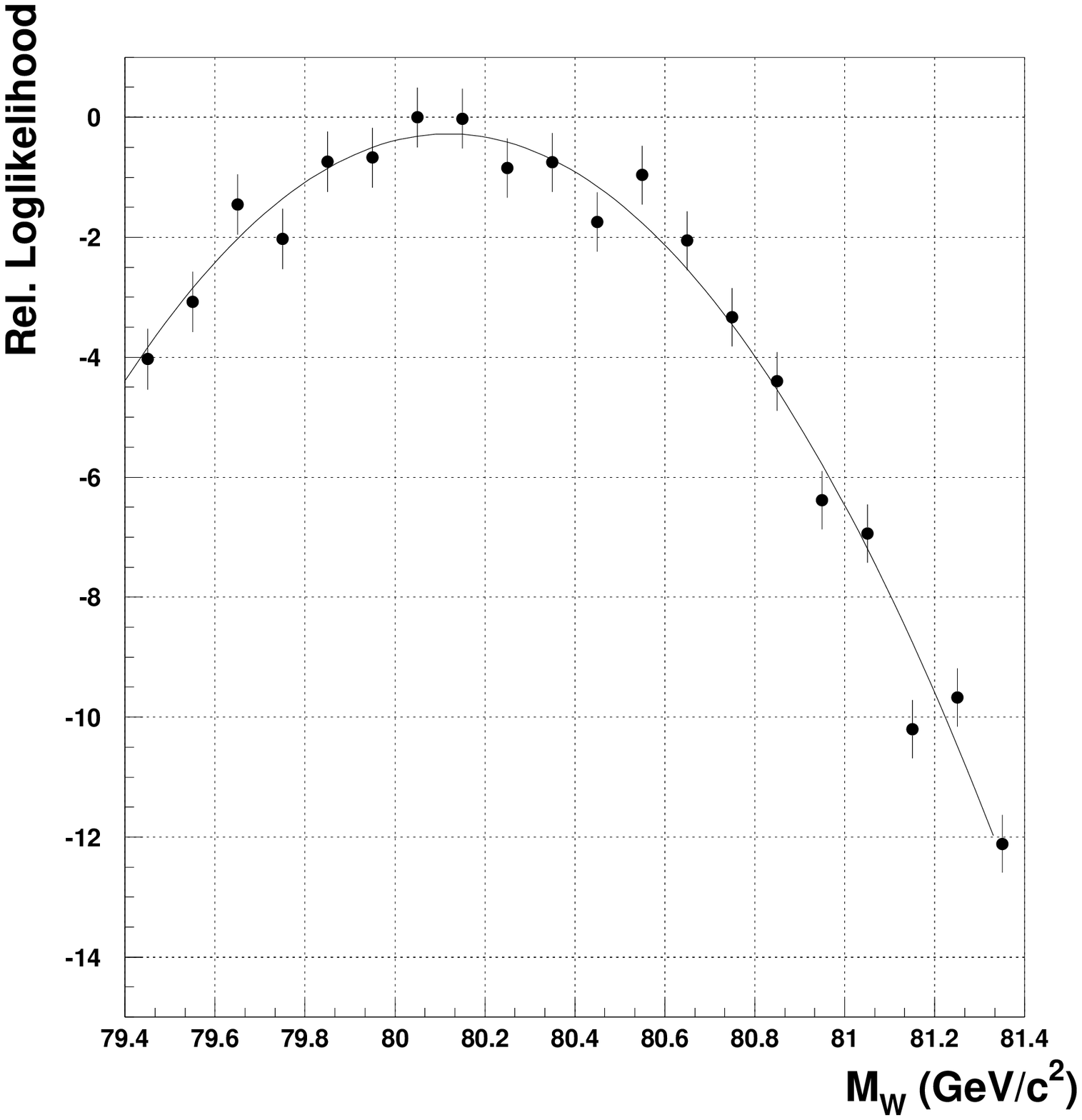}  &
    \epsfxsize=7.cm   \epsffile{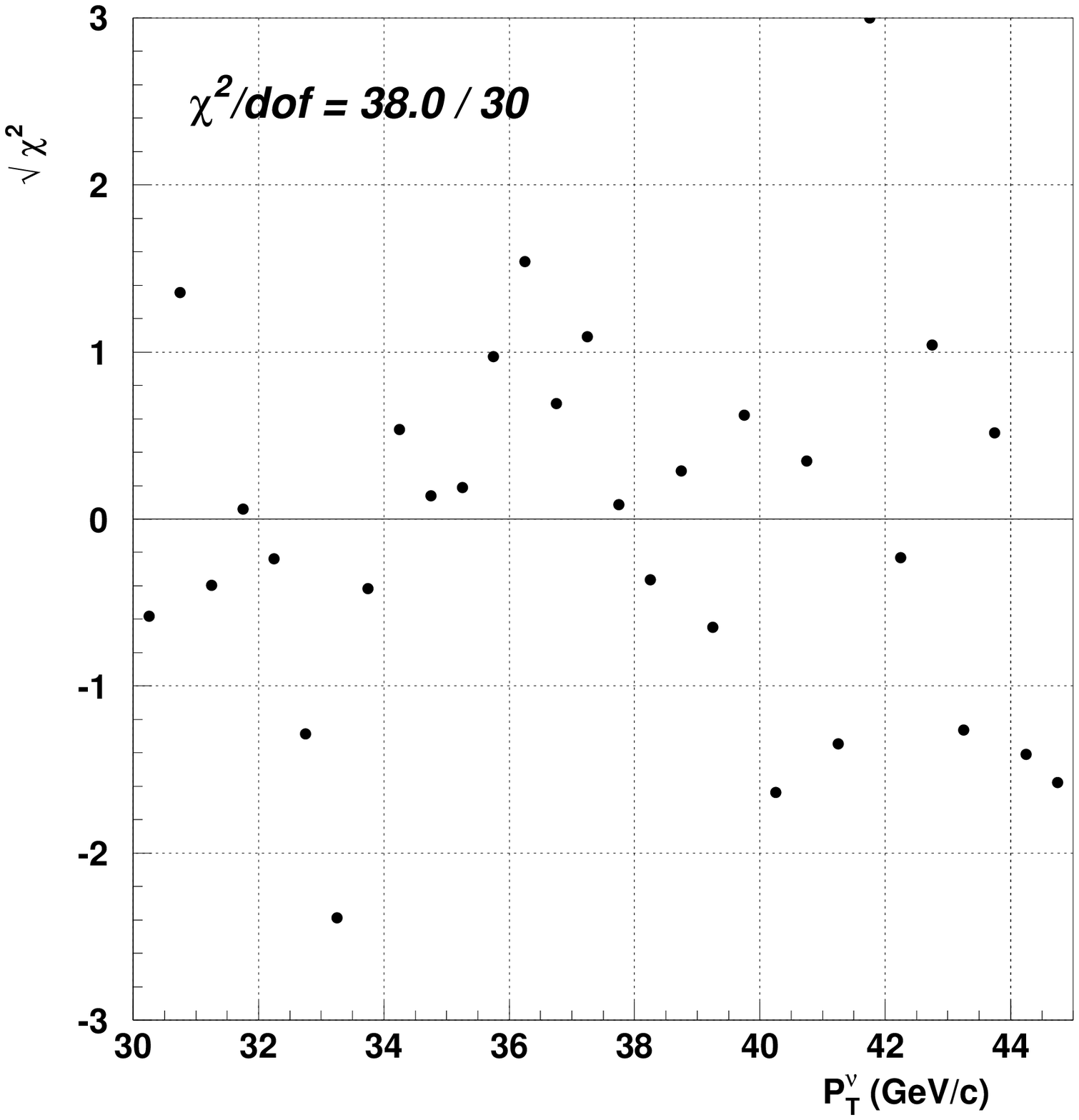}  \\
\end{tabular}
\end{center}
\caption[]{(a) The neutrino transverse momentum distribution 
           for $W$ events (points) and the best fit of the simulation
           (histogram), (b) the corresponding relative log-likelihood 
           distribution and (c)   signed $\sqrt{\chi^2}$  distribution. }
\label{fig:ptnu_fit}
\end{figure}

The $W$ boson mass was obtained from fits to the transverse mass of  
the $W$ boson,
$M_T$  (Fig.~\ref{fig:mt_fit}),  the electron $p_T$ 
(Fig.~\ref{fig:pte_fit})  and the neutrino $p_T$ spectrum 
(Fig.~\ref{fig:ptnu_fit}).  The transverse mass fit was performed over
the range 
$60 < M_T < 90$~GeV/$c^2$, which contains 5982~events.  Placing the lower
edge at 60~GeV/$c^2$ removed most of the  QCD background. Since the
probability for finding events in the  very high transverse mass
tail was small, relatively small  fluctuations in the number of observed
high transverse mass  events can significantly affect the fitted mass. 
Given that the high transverse mass tail of the QCD  background  was
rather poorly known, a high $M_T$ cut of 90 GeV/$c^2$ was also  
imposed. A transverse momentum range of 30 to 45~GeV/c was used for fits
to  the transverse momentum spectra. There were 5520~events in the
fitting range for the electron transverse  momentum spectrum and
5457~events for the neutrino transverse momentum  spectrum.  It should
be noted that  the fitting windows were placed on \lq\lq
uncorrected\rq\rq\ energies,  that is, electron energies 
which had not been scaled 
as described in
Section IV.  

The final fitted masses from the three spectra are
\begin{eqnarray}
  M_W (M_T)  &=& \ 80.350 \pm 0.140  \, {\rm GeV/}c^2    \\
  M_W (p_T^e)  &=& \ 80.300 \pm 0.190  \, {\rm GeV/}c^2  \\
  M_W (P_T^{\nu})  &=& \ 80.045 \pm 0.260  \, {\rm GeV/}c^2  
\end{eqnarray}
The errors are again statistical only.  Note that the $W$ boson mass
determination using the transverse mass  is the most precise. 
After  taking into account the small offset, which resulted in a
change of the $W$ boson mass of 5~MeV/$c^2$ as described in
Section IV~A, the measured mass ratio is
\begin{equation}
M_W / M_Z = 0.88114\pm 0.00154
\end{equation}
where the error is statistical only.

\section {Systematic Shifts and Uncertainties} 
\label{sec:sys}

In this analysis, the $W$ boson mass was obtained from a fit to the spectrum in
transverse mass defined in Eq.~(\ref{mtdef}). The $Z$ boson mass was obtained
from a fit to the spectrum in invariant mass of the two electrons,
defined in Eq.~(\ref{meedef}). In this section the uncertainties in the
measured masses that could arise  from mismeasurements  of the terms in
these equations are described. Note that the errors quoted will be those for
the measured $W$ boson mass which is extracted from the ratio of the fitted
$W$ boson and $Z$ boson masses; correlations between the 
two masses have been taken
into account.

Unless otherwise noted, the determinations of the shifts
in mass due to the various uncertainties have been obtained through 
Monte Carlo studies and are labeled \lq\lq Monte Carlo\rq\rq\ in 
the tables.
In these studies, high statistics Monte Carlo event samples 
were generated with the parameter in question varied within 
its allowed range.
These samples were then fit to the templates with the nominal settings
to determine the systematic error.
The errors on these shifts reflect the statistical error on the simulation. 
The sensitivity, ${\partial M_W \over \partial P}$, where $P$ is the
parameter that has been varied, was determined from a linear fit 
to the shifts in mass over a representative range around the 
nominal value of the parameter.
Values in the tables labeled \lq\lq Data\rq\rq\ are the shifts in mass 
when the data are fit to a template in which one of the
parameters deviates from its preferred value, with the others unchanged.
No error is quoted for these data shifts, since it would be meaningless.

%
\subsection {Electron Energy Scale Uncertainty} 

As discussed in Section~\ref{sec:escale}, 
many systematic effects due to the calorimeter scale 
which are common to the measurement of both the $W$ 
and $Z$ bosons cancel in the ratio of their masses. 
However, there are small effects that can  bias the measured 
$Z$ boson mass in ways which do not cancel in the ratio, $M_W/M_Z$, 
and they are discussed in the next section. 

\subsubsection{Uncertainties in $M_Z$}
The first source of a possible bias in the $Z$ boson mass measurement is the 
background under the $Z$ boson resonance. 
The nominal multi-jet background in the $Z\rightarrow ee$ sample and the 
Drell-Yan contribution caused a shift in the reconstructed $Z$ boson mass of 
$+39 \pm 12$~MeV/$c^2$. 
The uncertainty on this correction has been estimated by varying the 
slope of the background which resulted in a change in the 
overall background level from  3.2\% to 8.2\%. 
Such a variation in the background results in a variation of 
20~MeV/$c^2$ in $M_Z$, 
which was taken to be the systematic uncertainty on the $Z$ boson mass from the
background contribution. Other uncertainties arose due to  parton
distribution functions, radiative corrections, and a small fitting error.
Among these, the change in parton luminosity for the different 
parton distribution functions was most significant. 
Varying the parton luminosity slope $\beta$ within the range 
given by the various parton distribution functions considered in this 
analysis, $ 1.030\times 10^{-2}< \beta < 1.113\times 10^{-2} $, 
along with the other effects results in an overall 35~MeV/$c^2$ uncertainty in
the $Z$ boson mass. 

\subsubsection{Total $M_W$~Uncertainty Due to Electron Scale}
As was noted in Section~\ref{sec:escale}, the largest contribution to 
the overall scale uncertainty was due to the number of $Z$ boson 
events. This statistical component was  $150$~MeV/$c^2$. In addition, 
the uncertainty due to the possible nonlinearity in the calorimeter 
response as determined by the combined $m_{J/\psi}$, $m_{\pi^{0}}$, 
and $M_{Z}$ analysis (related to the uncertainty in $\delta$) was 
assigned as 25~MeV/c$^{2}$. 
Combining these in quadrature with the  systematic uncertainties just 
discussed resulted in the overall scale uncertainty assignment which 
is rounded up to 160~MeV/$c^2$.


\subsection{Uniformity of Electron Energy Response Uncertainty} 

The data were   corrected for the observed azimuthal variations in energy 
response of the different calorimeter modules, reducing the error from 
this source to a negligible level. 
Any residual non-uniformity in response was taken into account
through the constant term in the energy resolution. 

A non-uniform response in $\eta$, however,  can
introduce a bias in the mass determination, arising 
from the fact that the kinematic distribution of 
electrons from $Z$ boson decays differs from that in $W$ boson decays. 
The electrons from $Z$ boson decays have a different average $\eta$
than the electrons from $W$ boson decays, even when event 
samples are very large.
Moreover, a non-uniformity can distort the differential distributions. 
A possible $\eta$ dependence of the calorimeter response will thus not 
cancel in the ratio of the two masses.

To address this,  
the response of the different $\eta$ regions of the detector were scaled 
in the Monte Carlo with respect to the nominal uniform response.
Two sets of scale factors were used, corresponding to 
the response of two EM modules measured in the 1991 test beam. 
These scale factors were applied in discrete steps in $\eta$, following the
read-out geometry of the calorimeter, 
and varied from $0.985$ to $1.013$ over the central pseudorapidity range. 
The observed shifts in fitted mass are listed in 
Table~\ref{tab:eta_resp}. 
Assuming the $\eta$ response of the test beam modules typified the 
variation in uniformity, a systematic uncertainty on the 
$W$ boson mass from the transverse mass fit of 10~MeV/$c^2$ 
was assigned due to this uncertainty.


\subsection{Electron Energy Resolution Uncertainty}
\label{sec:eres}

The electron energy resolution in the central calorimeter was
parameterized as discussed in Section IV~C3. 
Most effects which degrade the resolution affected the
resolution function constant term.
For example, spatial non-uniformities in the detector response
and electronics gain variations 
contributed to the constant. 
The sampling term varies very little, 
from 1.9--2.4\%, as the electron $p_{T}$ is varied over the 
range 30--45 GeV/$c$.  
Therefore changing only the constant term and noting
the change in the $W$ boson mass was sufficient to accommodate most 
sources of uncertainty in the energy resolution.

To study the dependence of the $W$ boson mass on the  resolution, the 
constant term was varied in the Monte Carlo simulation.
The $W$ boson mass increases if a resolution 
smaller than actually exists in the data is used in the Monte Carlo.
Better resolution in the Monte Carlo results in a sharper Jacobian edge 
and the fitted mass shifts upward to accommodate the larger resolution 
tail in the data.  
The transverse mass distribution was most sensitive, since 
the Jacobian edge was  best preserved. 
For the $p_T$ spectra the edge is smeared, due in part to the 
transverse boost of the $W$ boson. 
Table~\ref{tab:constant_mw} lists the changes in $W$ boson mass for 
all three fits when varying the 
constant term by 0.5\% from its nominal value of $1.5\%$.
An uncertainty in the measured $W$ boson mass for the transverse mass fit of 
70~MeV/$c^2$ was assigned according to this variation.


\subsection{Electron Angle Uncertainty} 

The electron polar angle is defined by the 
position of the electromagnetic cluster in the calorimeter 
and the position of the {\it cog} of the CDC track.
Recall from Eq.~(\ref{cdc_alpha}) in Section III that a 
scale factor, $\alpha_{\rm CDC}$, 
was applied during the data analysis to correct
the bias in the $z$ position of the {\it cog} of the CDC track.
The uncertainty in the $W$ boson mass due to the uncertainty in 
$\alpha_{\rm CDC}$ has been determined by applying 
varying scale factors to the $z$ position of the 
CDC {\it cog} in the $W$ boson and $Z$ boson data and 
fitting to the standard templates.
By varying the CDC scale factor around the nominal value within its 
tolerance of 0.002 for the $W$ boson and $Z$ boson data sample simultaneously, 
the uncertainty on the $W$ boson mass was determined to be 50~MeV/$c^2$.


\subsection {Hadronic Energy Scale Uncertainty} 
\label{sec:ptwscale_sys}

The energy scale of the vectors 
${\vec u_T}$ and ${\vec p}_{T}^{\,{\text rec}}$, 
which both include hadronic energy, was not the same as the scale of 
${\vec p}_{T}^{\,e}$, which contains only electromagnetic energy
and was calibrated by the $Z$ boson mass.
The relative hadronic to electromagnetic energy scale is set 
using $Z$ boson events
and the scale obtained is $ \kappa = 0.83 \pm 0.04\,$.
The sensitivity of the measured $W$ boson mass was obtained by varying
the value of $\kappa$ within its uncertainty in the Monte
Carlo generation of the templates.
The 0.04 variation in hadronic energy scale produced a 
50~MeV/$c^2$ uncertainty on the $W$ boson mass from the transverse 
mass fit, where an increase in the scale factor resulted in 
an increase of the measured $W$ boson mass. 
Table~\ref{tab:ptw_mw} lists the change in $W$ boson mass when varying the 
hadronic energy scale factor by 0.04 from its nominal value for all 
three fits. 
The mass obtained from the $p_T^e$ fit was affected by a change in the 
hadron energy scale through the electron identification
efficiency as function of $u_\parallel$.


\subsection {Hadron Energy Resolution Uncertainty} 

The resolution in $p_T^W$ had two components:  the energy  resolution of the
recoil jet which is aligned with the recoil 
direction\cite{onejet}, and the underlying
event vector $\vec{u}_T$ which was randomly oriented  with respect to the
recoil. In the Monte Carlo the recoil momentum ${\vec p}_T^{\,{\text rec}}$ 
was 
simulated by assuming it is a  jet with  resolution $ \sigma_{had} /E = 80
\%/\sqrt{E}$ as discussed above. All of the uncertainty due to this quantity 
was
presumed to be accounted for through variations in the sampling term alone.   
The second contribution, that from ${\vec u_T}$, dominated the
overall resolution in $p_T^W$.  It was obtained directly from the experiment
using minimum bias events  chosen at the proper luminosity to simulate the
underlying event. 

The data constrained the number of minimum bias events to 
$N_{\rm min. bias} = 0.98 \pm 0.06$.  The nominal value used in the
simulation was $1.0$. The change in $W$ boson 
mass for various values of the number
of underlying events is listed in Table~\ref{tab:syserr_minb}.  This 
includes the effect of resolution broadening and  the neutrino scale shift
which results from changing ${\vec u_T}$.  The application of the randomly
oriented underlying event has the effect of adding an azimuthally symmetric
component to
the overall  resolution for the total hadron energy vector.
The systematic uncertainty on the measured 
$W$ boson mass due to the uncertainty on
the number of minimum bias events is  
$60~\rm{MeV/}c^2$ for the transverse mass fit. 

The mass determined from the $p_T^e$
spectrum was, within errors, not  affected by the hadron energy resolution.  
The $W$
boson mass determined from the other two spectra would increase if 
a smaller average  number of minimum bias events underlying the $W$ boson
were used in the Monte Carlo
since the resolution  improves. 

The jet energy resolution also contributed to the  uncertainty attributed to
the overall hadronic energy resolution.  Varying the sampling term in the jet
energy resolution  from 0.6 to 1.0 changes the $W$ boson  mass by 65~MeV/$c^2$,
which was taken to be the  systematic error due to this source. 
Table~\ref{tab:syserr_hadres} lists the change in the mass from the 
different fits when varying the sampling term of the hadronic energy 
resolution.


\subsection{Energy Under the Electron Uncertainty}

The measured electron energy not only consisted of the electron energy itself,
smeared by the detector  resolution, but also included a contribution from the
underlying event.  In addition, there was a bias in the  electron energy due to
zero-suppression in the readout electronics. Following the discussion in
Section~\ref{sec:ue}  the measured electron $p_T$ 
was modeled as a combination of
four terms,
${\vec p}_{T}^{\,e}$, ${\vec u}_{T_{\rm zs}}^{e}$, ${\vec u}_T^{\rm \, 25}$,
and ${\vec u}_{T_{\rm zs}}^{{\rm ue}}$. 
The additional contributions to the electron energy point, to a good
approximation, along the electron direction with the magnitude of 207~MeV. 
The uncertainty
on this has been estimated to be approximately 50~MeV.  The measured neutrino
momentum can be written in two equivalent ways:
\begin{eqnarray}
    \mbox{${\hbox{${\vec E}$\kern-0.6em\lower-.1ex\hbox{/}}}_T$ } 
&=& - {\vec p}_{T}^{\,{\text rec}} \,-\,
                  {}_{{}_m}{\vec p}_{T}^{\,e} \,-\,         
                  [ {\vec u}_T  \,-\, {\vec u}_T^{\rm \, 25} ]  \nonumber \\
            &=& - {\vec p}_{T}^{\,{\text rec}} \,-\,
                  {\vec p}_{T}^{\,e} \,-\,
                  {\vec u}_T  \,-\, 
                  {\vec u}_{T_{\rm zs}}             
\end{eqnarray}
Using the second equation,  the total recoil momentum and the total
underlying energy enter in the calculation of the neutrino momentum. Both
were well determined by the $W$ boson and $Z$ boson data. Using this 
approach, the
overall uncertainty
 derived from  the measured electron energy  in a manner which 
did not completely cancel  in the ratio $M_W / M_Z$. The zero suppression
correction here was quite small, since it contributed to  the neutrino 
and the
$W$ boson electron, as well as the two $Z$ boson electrons.

Using the first equation the measured electron energy appeared in both the
measured neutrino momentum  and the measured electron momentum for 
$W$ boson decays
and in both electrons for $Z$ boson decays. The correction to the 
electron energy
then canceled completely in  the ratio of the $W$ boson and $Z$ boson 
masses.  What is
important is how much of the underlying energy and $W$ boson 
recoil energy should
be excluded from the event for the 
${\vec p}_{T}^{\,{\text rec}}$ determination, because it was inside the electron
cone. The method used
to determine the uncertainty on the $W$ boson mass from the  contribution due to
energy under the electron followed this approach.

\begin{figure}[ht]
    \epsfxsize=10.cm
    \centerline{\epsffile{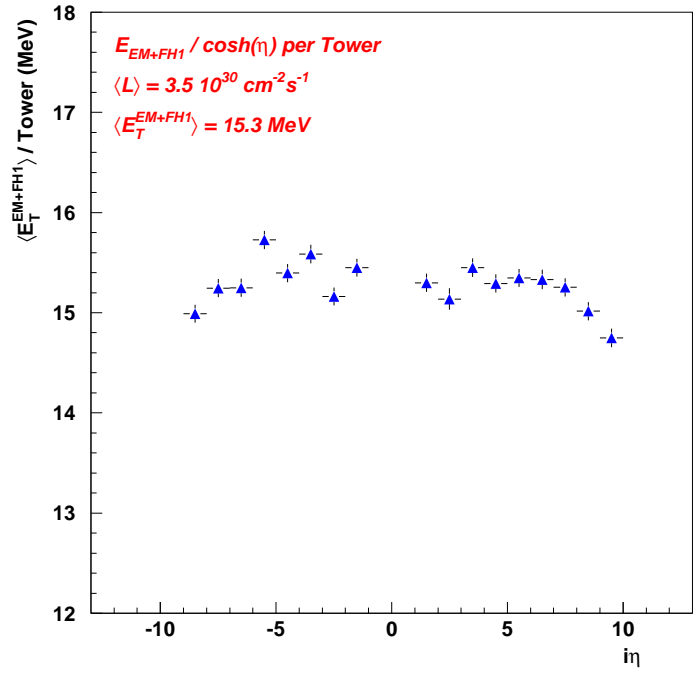}}
\caption{Average transverse energy flow per electron cluster tower 
         as a function of $\eta$ measured from minimum bias events.}
\label{fig:minb-eflow}
\end{figure}

Three effects were identified that contribute to this uncertainty.
Figure~\ref{fig:minb-eflow} shows the average transverse energy flow in an
EM tower plus the first FH  layer versus tower index  ($i_{\eta}^{e}$).  
It is seen that the energy flow was constant in $\eta$ within 0.5~MeV
for the central calorimeter.  In the Monte Carlo a uniform $E_T$
distribution was assumed and the deviation of a flat 
distribution from that shaped
like the data contributed an uncertainty of  
approximately~20~MeV/$c^2$ on the $W$
boson mass.

The second source of uncertainty stems from the fact that the underlying
energy in $W$ boson events was measured to be 16.8~MeV per tower in the EM
plus FH1 layers, whereas  minimum bias events yielded 15.3~MeV. In the Monte
Carlo an energy flow of 16.8~MeV was assumed.  This difference of 1.5~MeV, most
likely due to the presence of the $W$ boson 
recoil,  was treated as an uncertainty
on the mass which is equal to
$(25\times 1.5)/2 \simeq 20$~MeV/$c^2$. 

The third source is due to the uncertainty on the number of  towers to be
excluded from the $E_T$ of the underlying event.  In the Monte Carlo, a
region of $5\times 5 = 25$ towers was  excluded. In the data, the number of
towers used by the electron  in the clustering algorithm varied event by
event.  This uncertainty on the $W$ boson mass was evaluated by repeating the
analysis using another electron clustering algorithm that removed  this
error completely  (see Section \ref{sec:window}).  The difference in $W$ boson
mass between the two electron clustering approaches  led to a 20~MeV/$c^2$
uncertainty due to this effect. These three uncertainties were summed in
quadrature to obtain the total uncertainty on the $W$ boson mass of 35~MeV/$c^2$
due to the uncertainty in the energy flow underlying the electron.


\subsection{Production Model Uncertainty}

In the generation of the $W$ boson and $Z$ boson 
events a theoretical model  for the
vector boson transverse momentum and rapidity spectrum was used.  This
production model had an uncertainty associated  with it which led to an
uncertainty in the measured $W$ boson mass.  Since parton distributions and the
spectrum in $p_T^W$ are correlated,  this correlation was addressed in
the determination  of its uncertainty on the $W$ boson mass.  To constrain the
production model, both the measured $p_T^Z$ spectrum as  well as the
published CDF $W$ boson charge asymmetry data~\cite{cdf_asym} were used. 

\begin{figure}[t]
    \epsfxsize = 10.0cm
    \centerline{\epsffile{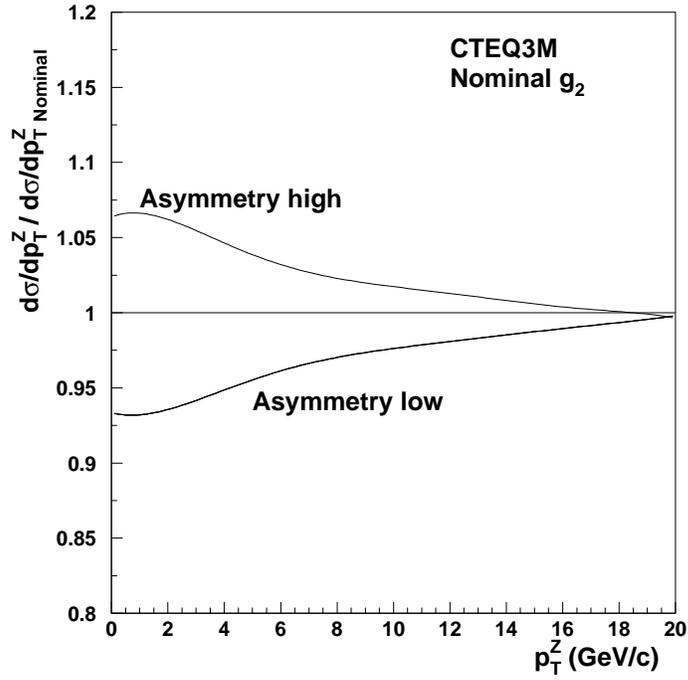}}    
\caption{Ratio of predicted differential cross section in $p_T^Z$ 
         and the nominal cross section for new 
         parameterizations of the CTEQ3M parton distribution function. } 
\label{fig:ptz_asym}
\end{figure}

The parton distribution functions were 
constrained by the CDF measured $W$ boson charge asymmetry  data. 
To accommodate the variation allowed by the asymmetry data  while at the
same time utilizing the available data from all other  experiments, new 
parametrizations of the CTEQ3M parton distribution function were 
obtained\cite{private-wuki}.  The fit used to 
obtain these parametrizations included the CDF
$W$ boson  asymmetry data with all data points 
moved coherently up or down  by one
standard deviation.  These parametrizations will be referred to in the
following as 
\lq\lq asymmetry high\rq\rq\ and \lq\lq asymmetry low\rq\rq ,  respectively. 
Figure~\ref{fig:ptz_asym} shows the relative change in the theoretical
$p_T^Z$ spectrum for these new parametrizations of the  
CTEQ3M parton distribution function with
respect to the nominal spectrum. 

The $p_T$ spectra of the vector bosons were most sensitive  to variations in
the parameter $g_2$, which describes the 
$Q^2$ dependence of the parametrization of the non-perturbative  functions
(see Appendix~\ref{apx1}).  Figure~\ref{fig:ptz_g}a shows the change in the 
$p_T^Z$ spectrum when the parameter $g_2$ is varied significantly 
from its nominal value.  Note that for low $p_T$, the cross
section varies by approximately a factor of two.  
Figure~\ref{fig:ptz_g}b shows the
constraint on $g_2$ by the $Z$ boson data as given 
by a simple $\chi^2$ test.  For
the estimate of the uncertainty on the $W$ boson mass, the range for 
$g_2$ was limited to $-2\sigma < g_2 < 4\sigma$, which are conservative 
bounds in
agreement with the $Z$ boson data. 

To assess the uncertainty due to parton distribution functions and 
$p_T^W$ input spectrum, the change in $W$ 
boson mass was noted when varying  both
the parton distribution function, as determined 
by varying  the measured $W$ boson
charge asymmetry, and the $g_2$ parameter  simultaneously.  The results of
the change in $W$ boson mass are listed in  
Table~\ref{tab:mw_sys_theor}. A total
error on the
$W$ boson mass of 65~MeV/$c^2$  has been assigned due  to the uncertainty on the
parton distribution functions and the  input $p_T^W$ spectrum. 

The change in $W$ boson mass obtained  from  high statistics
Monte Carlo studies for different  parton distribution functions, compared
to  the nominal MRSA parton distribution function is shown in 
Table~\ref{tab:pdf}. An uncertainty of 50~MeV/$c^2$ in the measured $W$ boson
mass  could be attributed to the choice of parton distribution function.  Note
that this uncertainty is only listed for completeness. The more conservative
estimate, varying both the parton distribution functions and  the $p_T^W$
spectrum simultaneously, was taken as the final uncertainty due to these
sources. 

Finally, the finite width of the $W$ boson was 
taken as $2.1\pm 0.1$~GeV and the effect
on the $W$ boson mass due to its uncertainty was found to be 20 MeV/$c^2$.


\subsection{Background Uncertainty}

The presence of background caused a bias in the determination of 
the mass. 
The shift in mass has been determined by  including the 
nominal background spectra in the likelihood templates. 
Systematic uncertainties arose due to the uncertainty on the 
overall background contribution and the shape of the background spectrum. 

\begin{figure}[ht]
    \epsfxsize = 10.0cm
    \centerline{\epsffile{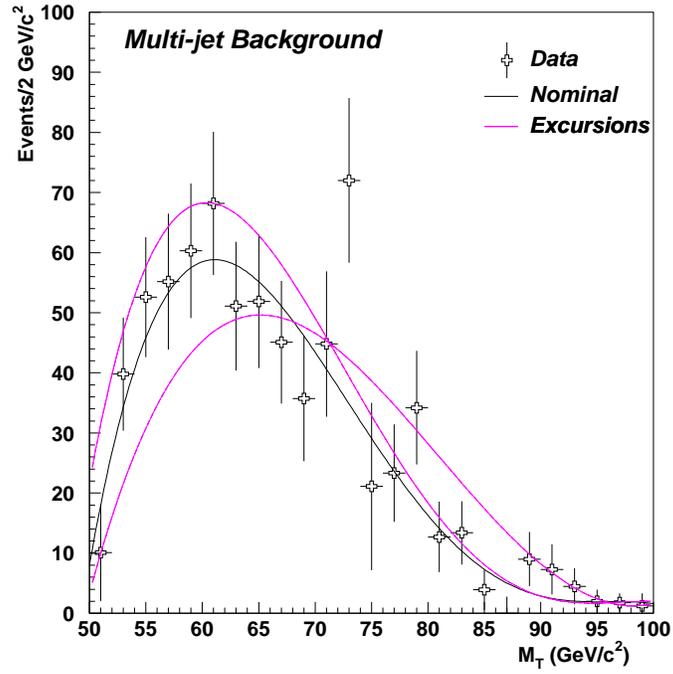}}
\caption[]{The measured multi-jet background distribution versus $M_T$
           from the data (open crosses).
           The allowed variations in the shape of the 
           transverse mass spectrum (dotted lines) are shown.
           The solid line indicates the nominal background distribution. } 
\label{fig:bkg_mt_err}
\end{figure}

The QCD multi-jet background contribution to the signal sample is 
$(1.6 \pm 0.8)$\%. The contribution of $Z\rightarrow e e$, in which one
electron is  not identified, is  $(0.43 \pm 0.05)$\%. 
The presence of these backgrounds 
introduced a shift in measured mass of +33~${\rm MeV/}c^2$ and +4~${\rm
MeV/}c^2$, respectively, for the transverse mass fit.  The background levels
have been varied within the quoted uncertainties.  The shape of the QCD
multi-jet background for the transverse mass  distribution was  varied as
shown by the curves labeled 
\lq\lq excursions\rq\rq\ in Fig.~\ref{fig:bkg_mt_err}. 
Similarly, the shape of the $Z\rightarrow e e$ background was varied.  The total
systematic uncertainty on
$M_W$ due to the variations in the QCD and $Z\rightarrow e e$ background  is
$30~{\rm MeV/}c^2$ and $15~{\rm MeV/}c^2$, respectively. 
An overall uncertainty of
$35~{\rm MeV/}c^2$ has been assigned to the uncertainty in the
background. 
  

\subsection{Radiative Decay Uncertainty} 

The parameters used in the modeling of radiative decays were the minimum
separation between the electron and photon for the photon to retain its
identity $R_{e\gamma}$  and the minimum energy of the radiated photon
$E_{\gamma}^{\rm min}$. The uncertainty in the value of these parameters
led to an  uncertainty in the measured $W$ boson mass. Uncertainties can also
arise from inefficiencies caused by the photon affecting the electron shower
shape,  the effect of upstream material on the energy measurement of photons
and from theoretical uncertainties. 

The electron photon separation parameter $R_{e\gamma}$ was varied by $\pm 0.1$
from its nominal value of 0.3 and  the effect on the $W$ boson mass  was noted. 
From this, an uncertainty of 10~MeV on  $M_W$ was determined.
In a second independent analysis the correlation 
between the effect of a photon on the
isolation as well as the topological requirements was taken into account
through a full detector simulation. The four-vectors of the decay products
from radiative decay events  were input to the {\footnotesize GEANT} 
simulation. The events, processed using the standard reconstruction 
algorithms, were then subjected to the same selection criteria as the  data
and electron identification 
efficiencies were determined  as a function of
$E_T^\gamma$ and $R_{e\gamma}$.  Modeling the resulting variation of 
the
efficiencies determined 
in this fashion
in the Monte Carlo led, again, to an uncertainty on  
the $W$ boson mass of 10~${\rm MeV/}c^2$ which is the same as that found in 
the other method.

The dependence of $M_W$ on $E_{\gamma}^{\rm min}$  was negligible. The
choice of $E_{\gamma}^{\rm min} = 50$ MeV was sufficiently low that within
the accuracy of the measurement it was insensitive to this parameter.

In the modeling of radiative decays, only order $\alpha_{\mbox{\tiny EM}}$
corrections  to the
lowest order diagrams have been considered and processes  in which two or
more photons are radiated have been ignored.   Also, initial state radiation
and finite lepton masses were not included  in the calculation. This effect
has been estimated to be  10~${\rm MeV/}c^2$
 and confirmed by a recent  theoretical calculation\cite{baur_rad}.

Since the effect of radiative decays was large and changed the 
$W$ boson and $Z$ boson masses in a way that did not cancel in the ratio,  
it was
important to also evaluate the effect when the photon is  produced by
bremsstrahlung in the central detector.  For the photon to have an effect on
the measured $W$ boson mass, it  must be 
separated from the electron in  ($\eta ,
\varphi$) space by at least $R = 0.2$. 

The probability for radiating a photon is very strongly peaked at  small
angles (see Appendix B), with very little 
dependence  on the fraction of the electron's
energy carried by  the photon~\cite{bremsstrahlung}.  The photon never
separates
from the electron  
beyond a cone of 0.2
by radiation alone  and therefore external
bremsstrahlung has no effect on the $W$ boson mass. 

As shown in Appendix B, the  electron and photon can also separate if the
electron undergoes multiple  scattering through a large angle.  
The angles resulting
from multiple scattering are generally larger than those produced in the
radiation itself, particularly when the electron is low in energy.  In spite
of the possibly large angles between the electron and the photon, the 
probability for this to occur is negligible and it  can  safely be 
concluded that  bremsstrahlung and
multiple scattering have no effect on the measured 
$W$ boson mass.

A last issue regarding radiation is the  energy loss by ionization and by
radiative processes where,  for example, the electron radiates a photon that
does not reach the  calorimeter but produces an $e^+e^-$  pair 
that loses energy by
${dE\over dx}$.  These processes  affect the $W$ boson 
and $Z$ boson mass and produce an
offset  in the energy scale, which was included in the energy scale
determination. Small offsets produced in  this way cancel to first order 
in the mean of the ratio of the 
$W$ boson to $Z$ boson masses, since the energy is lost  
to both the neutrino and the
electron in each $W$ boson event in which it occurs.   
In $Z$ boson events only one
electron loses the energy but the  probability of such loss is twice
as large. Using a {\footnotesize GEANT} simulation, a study of the effect  of
upstream material on the photon energy response was carried out. The photon
response observed in the {\footnotesize GEANT} simulation  was consistent
with the response measured {\it in situ}, as described in 
Section~\ref{sec:escale}. Notably, the offset in  response was found to be
consistent with the {\it in situ} measurement.  Combining all effects an overall
systematic uncertainty of 
$20~\rm{MeV/}c^2$ was assigned to $M_W$ due to radiative effects.


\subsection{Efficiency and Bias Uncertainties} 
\subsubsection{Trigger Efficiencies } 

The effect of the uncertainty in the trigger efficiency 
has been studied by varying the nominal trigger efficiency distributions in the 
Monte Carlo within the range determined by the data.
This resulted in an uncertainty on the $W$ boson mass of 
20, 20 and 60~MeV/$c^2$ from the 
$M_T$, $p_T^e$ and $p_T^\nu$ fits, respectively. 
In addition, the $W$ boson mass was determined from a data sample that 
did not have the $\mbox{$\not\!\!E_T$}$ requirement 
imposed at the trigger level. 
The fitted mass from this sample was consistent with the nominal fit result 
within the statistical uncertainty, taking into account the large overlap 
between the two data samples.

\subsubsection{Efficiency as a function of $u_\parallel$} 

The transverse mass is relatively uncorrelated with the uncertainty in 
$u_\parallel$, unlike the fits to the lepton transverse momentum spectra,
which are  very sensitive to this efficiency. The nominal variation in  the
electron identification efficiency encompasses the band shown in
Fig.~\ref{fig:upar_eff}.  The results of large 
statistics Monte Carlo data samples
generated  with the nominal variations of the efficiency  are given in
Table~\ref{tab:upar_mw}.  Also listed are the results of the change in mass when
fitting the  data to templates generated with 
the different efficiencies. It is seen
that the Monte Carlo studies and the data exhibit  the same behavior.  The
corresponding electron identification uncertainty on the
$W$ boson mass is  20, 70 and 115~MeV/$c^2$  from the 
$M_T$, $p_T^e$ and $p_T^\nu$ fits, respectively.

\subsection{Error in the Fitting Procedure}  

The $W$ boson mass was obtained from an unbinned maximum likelihood fit  in
which the data were fit to transverse mass spectra which were generated  for 21
different values of the $W$ boson mass. The 
log-likelihood values for the different
vector boson masses were fit to a parabola and the minimum was taken to be the
fitted mass value. A decrease of half a unit in the log-likelihood was the one
standard deviation  statistical error. The likelihood distribution need not
be Gaussian, depending on the range of the parameter fit, the intrinsic
shape of the spectrum and the resolution function. The resulting
log-likelihood curve was then non-quadratic. In addition, there will be
fluctuations in the  log-likelihood reflecting the Monte Carlo statistics. 
In order to determine the uncertainty, the
fitting was redone with a cubic polynomial 
parameterization and the mass spacing was
altered. This led to the assignment of 5~MeV/$c^2$ 
for the uncertainty due to the
$M_T$ fitting procedure.

\subsection{Results of Systematic Errors}  

The systematic errors on the $W$ boson mass as obtained from the transverse 
mass, electron transverse momentum and neutrino transverse momentum 
are summarized in Table~\ref{tab:mw_sys_err_sum}. The measured 
mass results
from this analysis are:
\begin{eqnarray*}
M_W (M_T) 
& = & \ 80.350\pm 0.140 \pm 0.165 \pm 0.160 \quad {\rm GeV/}c^2  ; \nonumber \\
M_W (p_T^e) 
& = & \ 80.300\pm 0.190 \pm 0.180 \pm 0.160 \quad {\rm GeV/}c^2  ; \nonumber \\
M_W (p_T^{\nu}) 
& = & \ 80.045\pm 0.260 \pm 0.305 \pm 0.160 \quad {\rm GeV/}c^2; \nonumber \\
\end{eqnarray*}
\noindent and from the transverse mass analysis,
\begin{equation}
M_W / M_Z =  0.88114 \pm 0.00154 \pm 0.00181 \pm 0.00175. \nonumber
\end{equation}
In each result, the first uncertainty is due to statistics, the second is
due to  systematic  effects, and the third is due to the electron
energy scale determination.


\section{Consistency Checks} 

To verify the stability of the $W$ boson mass result,  consistency checks have
been performed in which the $W$ boson mass is determined from various 
modified
data samples. These samples include those in which the fitting window 
was 
varied,  additional selection criteria were applied, and a
different electron clustering algorithm was used.  Fully overlapping data
samples were used to check the consistency of the results obtained from fits
to the $p_T^e$ and $p_T^{\nu}$ spectra. Also, two-dimensional
fits were done to check the consistency of parameters used in the Monte Carlo
simulation.

In general, the data sample was reduced or enlarged in these consistency
studies. There was a large overlap between the original data sample and the
samples used to verify the result. In order to quantify 
this verification, define
the mass from the original data sample
$M_W^{\rm nom}$, and that from the  sample used in the verification
$M_W^{\rm con}$. Then the estimator of the  independent statistical error on the
difference in the two results that were used is 
$\sigma( M_W^{\rm nom} - M_W^{\rm con}) = \sigma \, \sqrt{{N_2 \over N_1}}$. 
Here $\sigma$ is the statistical error on the original data sample, 
consisting of $N_1 + N_2$ events.  The sample used for the consistency
check contained $N_1$ events.  This is the error that is quoted for
the difference in mass for the  consistency checks.

\subsection{Additional Selection and Fitting Criteria } 

To investigate the effect of multiple interactions, events were selected with 
low hit multiplicity and a narrow time distribution in the small--angle
scintillation counters (see Section II~D). This yielded a sample in which
approximately 77\% were single interaction events.  Also events  with one and
only one reconstructed event vertex and with  only one track from the central
detector in the  electron road were selected.  The latter cut removed mainly
events with a random track from the  underlying event. The change in fitted
$W$ boson mass from the transverse mass spectrum,  with respect to the nominal
mass value for each of these cross checks,  is listed in
Table~\ref{tab:cross_check}. Note that the errors are statistical only.  Any
systematic error on the shifts is not included. 

To test the event modeling, the $p_T^W$ cut was tightened to  10~GeV/c and
the result is listed in Table~\ref{tab:cross_check}. When requiring
$p_T^W$ to be less than 10~GeV/c, there  is an additional uncertainty due to
the error on the hadronic  energy scale factor and change in background
contribution, which have  not been included in the error estimate.  

Another 
important check of the event modeling is  testing  both the sensitivity and 
consistency of the result by tracking the change in mass during the  process
of applying different cuts.  As an example, the first column in
Table~\ref{tab:upar_consist} lists  the change 
in $W$ boson mass from the nominal
fit when $u_\parallel$ in the  data is required to be less than 10~GeV
without modifying the templates.  The change was rather dramatic for the mass
from the $p_T^\nu$ spectrum.  The second column lists the change in mass
when the templates are made  consistent with 
the data.  Even though the $W$ boson
mass is rather sensitive to the cut on 
$u_\parallel$, the fitted masses agreed well with the nominal  values when
data and Monte Carlo were treated consistently, indicating  that both the
$p_T^W$ scale and $u_\parallel$ efficiency were modeled correctly. 

To check for any systematic bias in detector response, event samples were
selected with different fiducial requirements.  For $W$ boson events the $\eta$
range of the electron was restricted to  electrons produced in the
central region.  For $Z$ boson events the restriction was placed on only one of
the two electrons, ensuring that a variation in detector response to the
electron in
$W$ boson events was tracked in an identical manner in $Z$ boson events.
Table~\ref{tab:eta_cut} lists the resultant change in the ratio of masses. 
The errors on the change were again  statistical errors
only.  The ratio,  tabulated in Table~\ref{tab:eta_cut}, was with respect to the
normalized ratio for  the nominal
$\eta$ range.  The ratio
of masses did not change within errors.  When the restriction was placed on
both electrons in $Z$ boson events, the  ratio also did not change but 
had a large
statistical uncertainty  due to the significant loss of events. 

\begin{figure}[tp]
\begin{center}
\begin{tabular}{cc}
    \epsfxsize=8.cm \epsffile{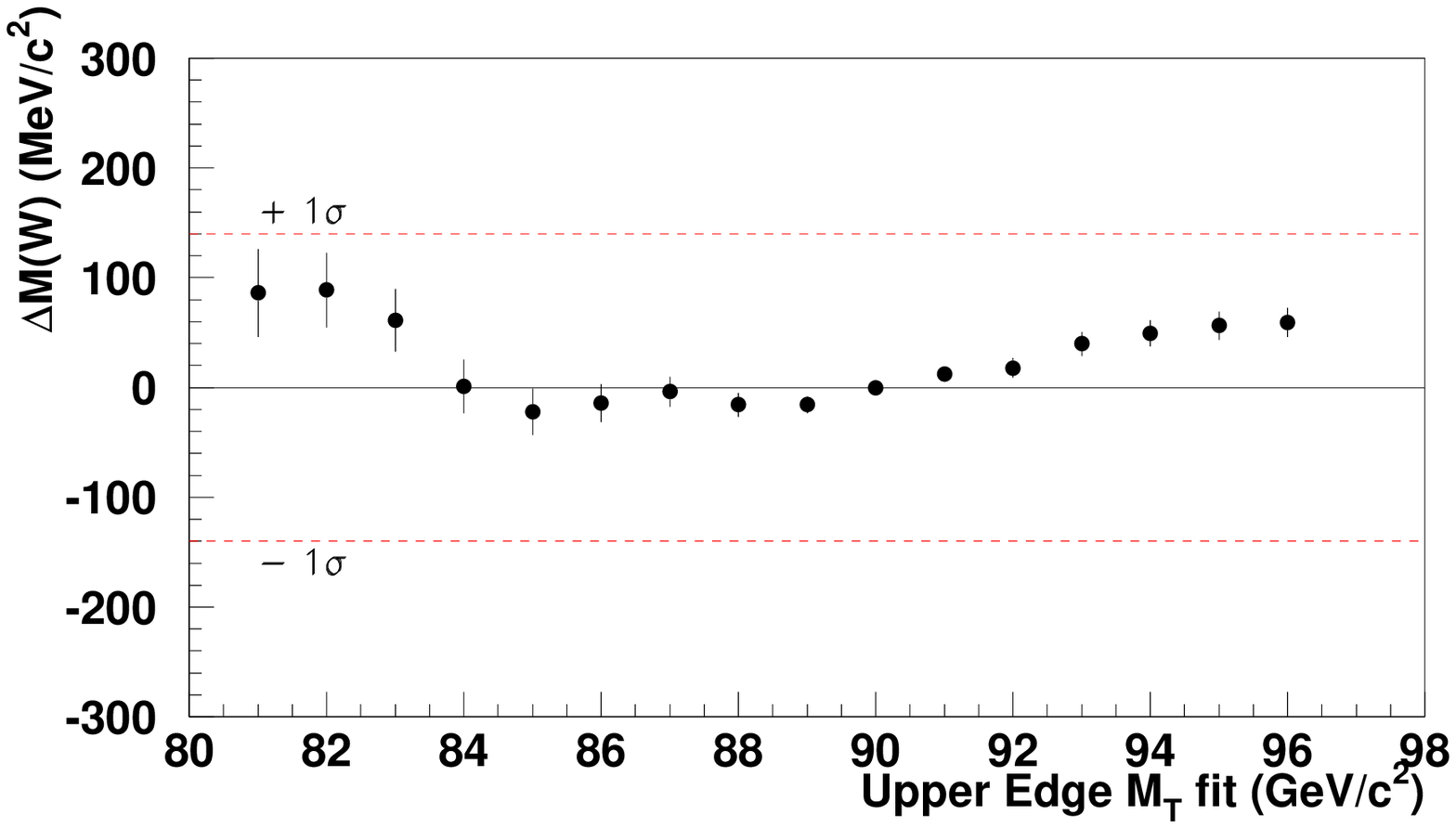} &
    \epsfxsize=8.cm \epsffile{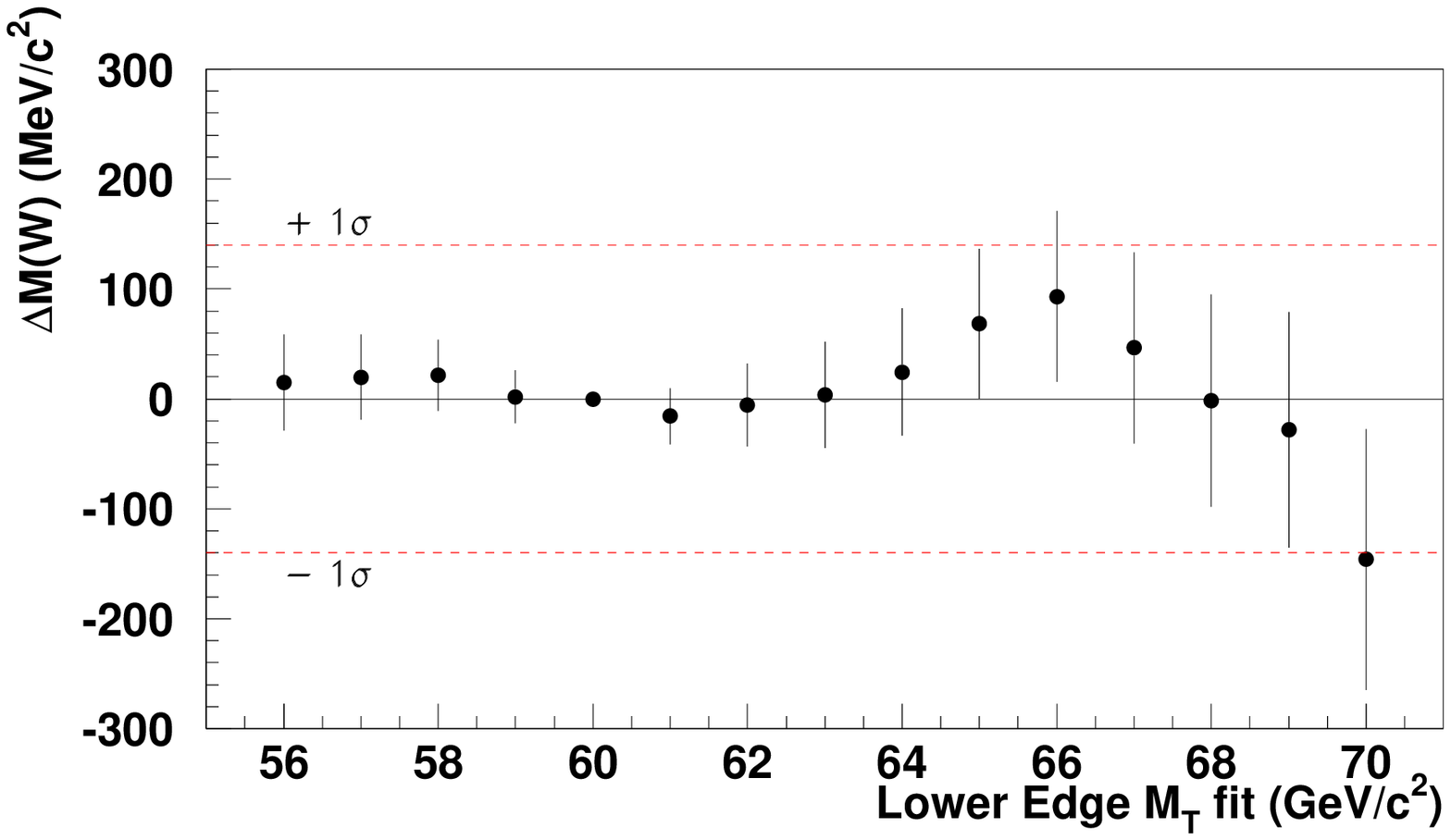} \\
\end{tabular}
\end{center}
\caption{Change in fitted $W$ boson mass when varying the (a) upper and 
         (b) lower edge of the fitting window from the fit to the 
          transverse mass spectrum (points).
          The horizontal bands indicate the $1\sigma$ statistical error
          on the nominal fit.}
\label{fig:fit_window}
\end{figure}

The variation in mass was also tracked when the nominal fitting range  in
transverse mass was varied.  Figure~\ref{fig:fit_window} shows the change in
$W$ boson mass when varying the  lower and 
upper edge of the fitting window for the
fit  to the $M_T$  distribution.  Changing 
the fitting window led to a negligible
systematic trend.

\subsection{Modified Electron Clustering }
\label{sec:window}

Electron clusters were found by the reconstruction program using a nearest
neighbor clustering algorithm~\cite{youssef}. The number of calorimeter
towers included in the cluster was dynamic and  depended on the
environment  of the electron.  This algorithm thus introduced an uncertainty
on the amount of underlying event energy included in the electron energy
cluster,  and therefore an uncertainty on how much energy was excluded from
the  underlying event for the calculation of the $p_T^W$. In the discussion
in Section~\ref{sec:ue} the energy assignments and the modeling of the
underlying event are described using a window algorithm for the
reconstruction of the electron energy.  The corrections necessary to
translate these results to the  cluster algorithm then had to be dealt with
properly.  These ambiguities can be completely circumvented if a fixed
electron definition is used. To verify the internal consistency,  the $W$ boson
mass was also determined using a fixed size electron cluster. 

The definition employed for the fixed size cluster is the 
\lq\lq 5$\times$5 window algorithm\rq\rq .  In this procedure, the electron
energy was defined as the energy  in the 25~towers in the region $\pm$0.2~in
$\eta$ and $\varphi$ from  the most energetic tower of the electron cluster
as found by the  nearest neighbor algorithm.  Using the original $W
\rightarrow e\nu$ data sample, the electron
energies were recalculated using the  5$\times$5 window algorithm and the
$p_T^W$ was calculated with  respect to the electron vertex, excluding the
5$\times$5 window occupied  by the electron.  The region excluded from the
underlying event for the calculation of  
the $\mbox{${\hbox{$E$\kern-0.6em\lower-.1ex\hbox{/}}}_T$ }$ was thus exactly
known for each event.  Subjecting these events to the standard event
selection criteria  yielded 7167~events,  7131~of which were in the
nominal data sample. The fitted $W$ boson mass obtained from this data sample,
using the window  algorithm to define the electron, was $12~\rm{MeV/}c^2$
lower than when using the nearest neighbor algorithm. 
As noted above, a systematic uncertainty on the 
$W$ boson mass of 20~MeV/$c^2$ has been attributed 
due to the difference in  these
two approaches and  has been included in the underlying event uncertainty in
Table~\ref{tab:mw_sys_err_sum}.

\subsection{Fully Overlapping Data Samples } 

\begin{figure}[h]
\begin{center}
\begin{tabular}{cc}
    \epsfxsize=8.0cm     \epsffile{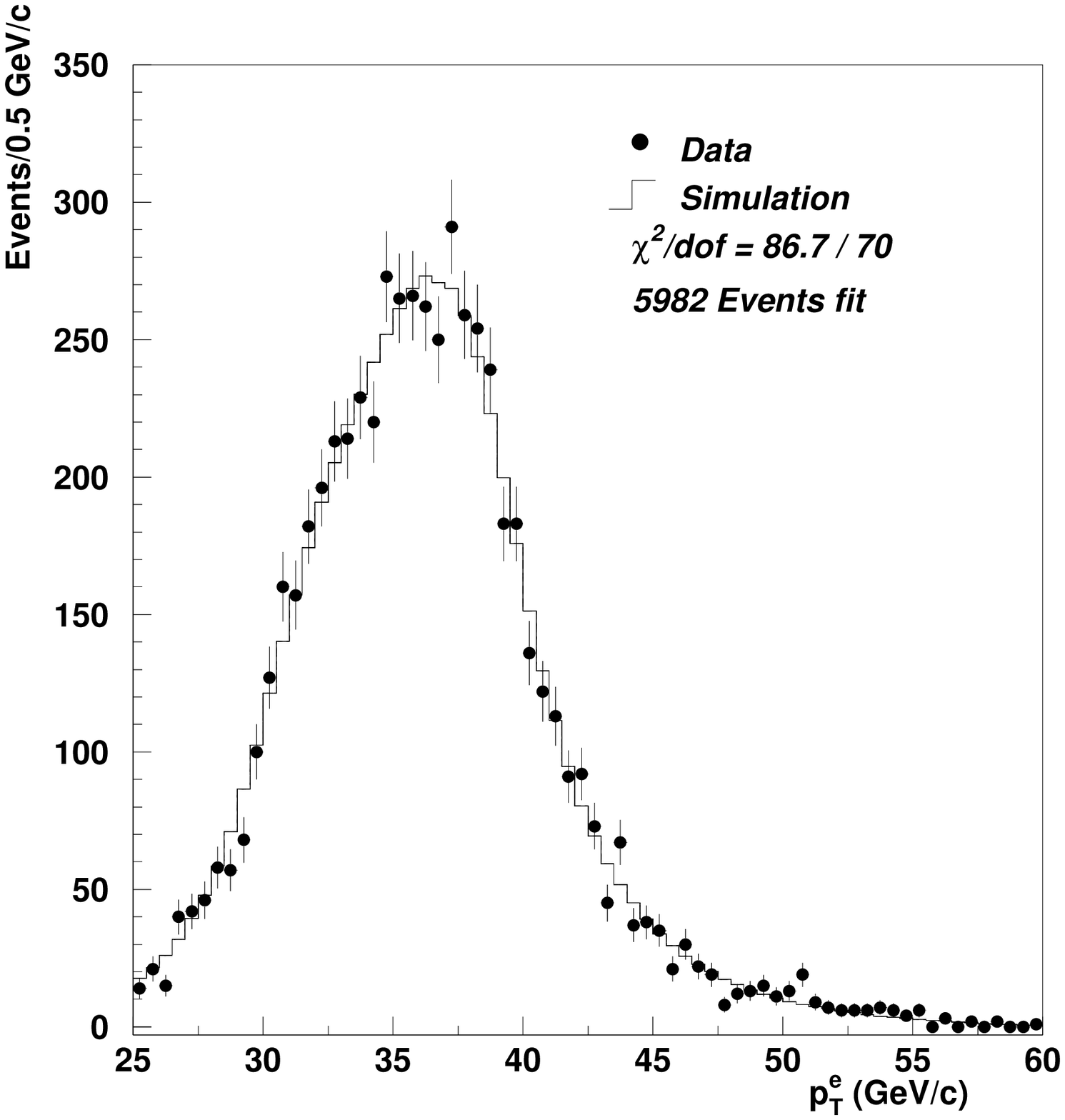} & 
    \epsfxsize=8.0cm     \epsffile{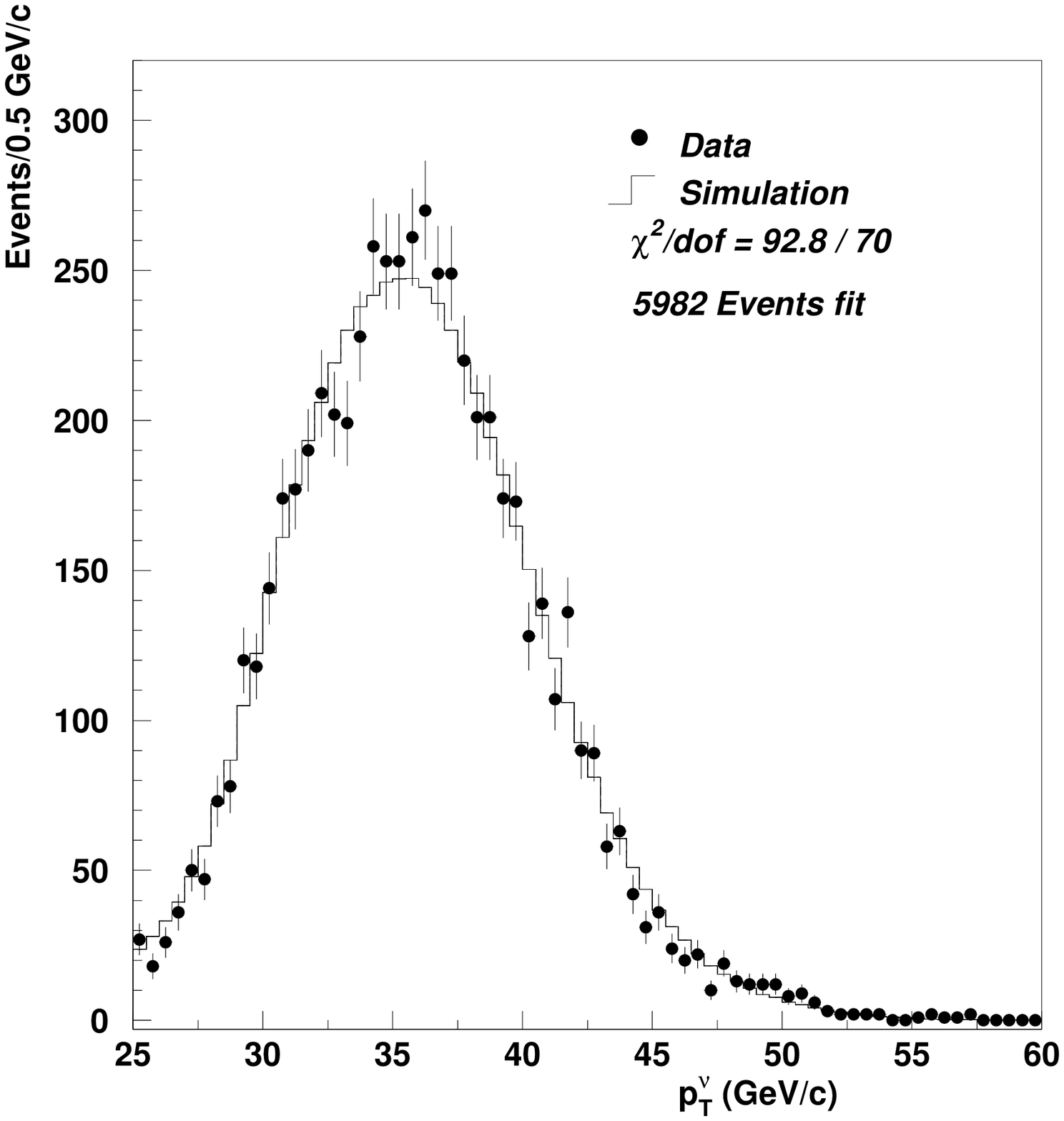} \\
\end{tabular}
\end{center}
\caption{ The (a) electron and (b) neutrino transverse momentum distribution 
         for the events in the transverse mass window 
         $60 < M_T < 90$~GeV/$c^2$ (points).
         The histograms are the best fits of the simulation. }
\label{fig:pt_overlap}
\end{figure}

The nominal fits to obtain the $W$ boson mass 
were performed using events  within a
certain range either in transverse mass or in transverse momentum.  These
event samples did not fully overlap.  Fully overlapping event samples are
obtained  when applying a fitting window in one variable and then utilizing
the full  unrestricted spectra in the other two variables, using all events
in  this window.  Figure~\ref{fig:pt_overlap} shows the $p_T^e$ and
$p_T^\nu$ spectra with only the requirement
that $60 < M_T < 90$~GeV/$c^2$.  The change in 
$W$ boson mass obtained from a fit
to these spectra is
$+84 \pm 55$~MeV/$c^2$ from the fit to the $p_T^e$ spectrum and 
$+54 \pm 81$~MeV/$c^2$ from fitting the $p_T^\nu$ spectrum.  The errors on
the shift in mass are the statistical errors due to  the different number of
events fit. Again, the results are   consistent with the nominal
results for the fits to the transverse  momentum spectra.

\subsection{Two-Dimensional Fits}

\begin{figure}[thp]
\begin{center}
\begin{tabular}{c}
    \epsfxsize=11.0cm \epsffile{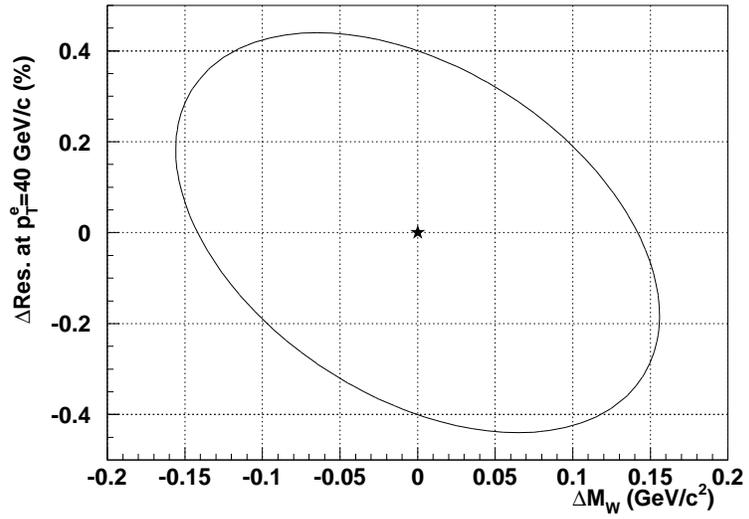} \\
\end{tabular}
\end{center}
\caption{The 1$\sigma$ contour in the change in $M_W$ and 
  the electron energy resolution  at $p_T^e$ = 40 GeV/c 
  from fits of the simulation, in which the constant term 
  is allowed to vary, to $W$ boson events.}
\label{fig:2d_mw_const}
\end{figure}

\begin{figure}[thp]
\begin{center}
\begin{tabular}{c}
    \epsfxsize=12.0cm   \epsffile{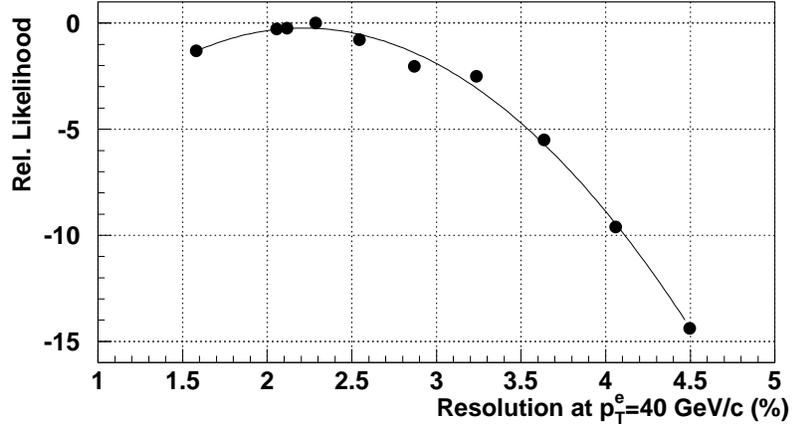} \\
    \epsfxsize=12.0cm   \epsffile{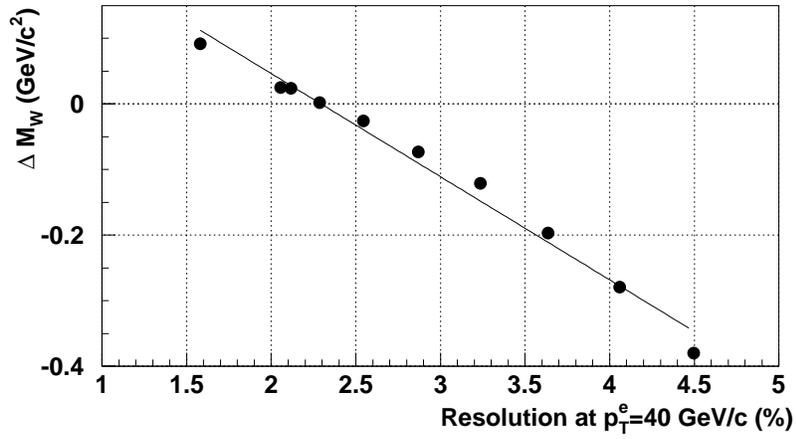} \\
\end{tabular}
\end{center}
\caption{From fits of the simulation, in which the constant term is allowed to 
  vary, to $W$ events:
  (a) 
    The relative likelihood 
    versus the electron energy resolution  at $p_T^e$ = 40 GeV/c 
  (b)
    The change in the fitted $W$ mass 
    versus the electron energy resolution  at $p_T^e$ = 40 GeV/c. }
\label{fig:mw_sys_mw_const}
\end{figure}

Two-dimensional fits were carried out to check the stability and 
correctness of parameters used in the Monte Carlo simulation.
The first two-dimensional fit was performed on the $W$ boson mass and the 
constant term in the electron energy resolution. 
Rather than expressing the likelihood in terms of the constant term, 
which resulted in a very asymmetric likelihood distribution, it was 
expressed in terms of the energy resolution at an electron $p_T$ of 40~GeV/$c$
\begin{equation}
    {\cal R}_{40} \,\equiv\, 
    \sqrt{ C^{2} + \frac{S^2}{40} }
\end{equation}
where $S$ and $C$ are the coefficients of the sampling and constant 
term, respectively. 
The sampling term was taken to be 0.13 and the constant term is varied. 
The error matrix for the fit in $M_W$ and ${\cal R}_{40}$ is: 
\begin{equation}
\begin{array}{c} 
    \left(  
    \begin{array}{cc} 
         0.0243  & -0.0286 \\ 
        -0.0286  &  0.1933 
    \end{array} 
    \right) 
\end{array} 
\end{equation}
with a correlation coefficient of $\rho = -0.4155 \,$.
Figure~\ref{fig:2d_mw_const} shows the contour in 
$M_W$ and ${\cal R}_{40}$ for a change of 0.5 units in the log-likelihood. 
The values on the axes are with respect to the central value of the fit. 
The fitted $W$ boson mass was higher by 26~MeV/$c^2$ compared to the value 
obtained when the constant term in the energy resolution was fixed at 
1.5\%, in agreement with the nominal fit. 
The error on the mass from the two-dimensional fit was 156~MeV/$c^2$. 
For a fixed value of the constant term the error would be 
$0.156 \, \sqrt{1-\rho^2} = 142$~MeV/$c^2$, 
consistent with the nominal fit result. 
The fitted value for the resolution (see Fig.~\ref{fig:mw_sys_mw_const}a) 
was ${\cal R}_{40} = (2.34 \pm 0.440)$\% which, assuming a sampling term of 
$S = 0.13$, corresponded to a constant term of $C = (1.1^{+0.8}_{-1.1})$\%. 
This is again consistent with the result obtained from fitting the width 
of the $Z$ boson resonance from which the constraint on the 
resolution is actually slightly tighter.
The correlation between the $W$ boson mass and ${\cal R}_{40}$ was given by 
$\rho \, { \sigma(M_W) \over \sigma({\cal R}_{40}) } = -147$~MeV/$c^2$
  per 1\% change in ${\cal R}_{40}$, 
which was also consistent within errors with the result obtained 
in Section~\ref{sec:eres} and shown in Fig.~\ref{fig:mw_sys_mw_const}(b). 

\begin{figure}[tp]
\begin{center}
\begin{tabular}{c}
    \epsfxsize=11.0cm
       \epsffile{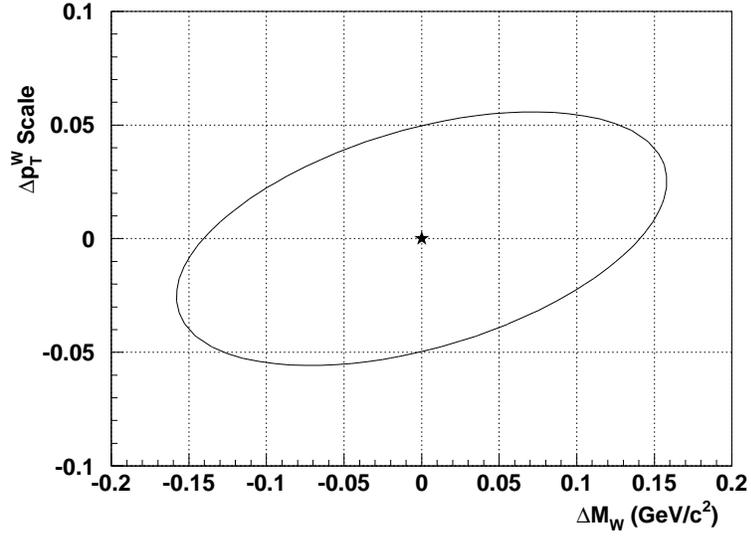} \\
\end{tabular}
\end{center}
\caption{The 1$\sigma$ contour in the change in $M_W$ and $p_T^W$ scale 
  from fits of the simulation, in which the $p_T^W$ scale factor is
  allowed to vary, to $W$ boson events.}
\label{fig:2d_mw_ptwscale}
\end{figure}

A two-dimensional fit was also performed in mass and hadronic 
energy scale factor $\kappa$. 
The error matrix for this fit in $M_W$ and $\kappa$ is: 
\begin{equation}
\begin{array}{c} 
    \left(  
    \begin{array}{cc} 
         0.0250  &  0.0043 \\ 
         0.0043  &  0.003121 
    \end{array} 
    \right) 
\end{array} 
\end{equation}
with a correlation coefficient of $\rho = 0.457 \,$.
Figure~\ref{fig:2d_mw_ptwscale} shows the one $\sigma$ contour in 
$M_W$ and $\kappa $. 
The values on the axes are with respect to the central value of the fit. 
The fitted $W$ boson mass was lower by 7~MeV/$c^2$ compared to the value 
obtained when the $p_T^W$ scale was fixed at 0.83, in
agreement with the nominal fit. 
The error on the mass from the two-dimensional fit is consistent with 
the nominal fit keeping the hadronic energy scale factor fixed. 
The fitted value for the hadronic energy scale factor was $\kappa = 0.834 \pm
0.056$,  consistent with the result obtained from the $Z$ boson data. 
The error is large because the $W$ boson mass was not very sensitive 
to the hadronic energy scale. 
The correlation between the $W$ boson mass and $\kappa $ is given by 
$\rho \, { \sigma(M_W) \over \sigma(\kappa )} = 12.8$
MeV/$c^2$ per 1\% change in scale factor. 
This is to be compared to the sensitivity of 12.1~MeV/$c^2$
per 1\% change in scale factor obtained in Section~\ref{sec:ptwscale}. 

In conclusion, the mass values obtained for different subsamples of the
nominal data sample were all consistent within the quoted statistical
uncertainty. Moreover, when leaving crucial
parameters in the event modeling as free parameters in the fit, 
the $W$ boson data preferred values for these parameters which were completely 
consistent with those obtained from external constraints, a strong
indication of the stability of the result. 

\section{Conclusion }

A measurement of the $W$ boson mass determined from the transverse mass 
distribution using electrons
in the central region of the  D\O\ detector from the 1992--1993 
Fermilab Tevatron running 
in 12.8~pb$^{-1}$
has been described. The determination of $M_W$ was based on a ratio of the
measured $W$ boson and $Z$ boson masses, normalized to 
the world average $Z$ boson mass
as determined by the LEP experiments. This measurement yielded a
$W$ boson mass value of 
\begin{equation}
M_W =  80.350 \pm 0.140 \pm 0.165 \pm 0.160 \quad {\rm GeV/}c^2. \nonumber
\end{equation}
and has an uncertainty comparable to that of other recent measurements in a
single channel. The first uncertainty is due to statistics, the second is
due to  systematic  effects, and the third is due to the electron
energy scale determination.
The 160~MeV/$c^2$ uncertainty due to the uncertainty on the absolute 
energy scale has a contribution of 150~MeV/$c^2$ due to the limited 
$Z$ boson statistics. The measured ratio of the 
$W$ boson and $Z$ boson masses is 
\begin{equation}
M_W / M_Z = 0.88114 \pm  0.00154 \pm 0.00181 \pm 0.00175. \nonumber
\end{equation} 
Here, the first uncertainty is due to statistics, the second is
due to  systematic  effects, and the third is due to the electron
energy scale determination.

Based on this measurement alone, the values for $\Delta r$ and
$\Delta r_{\rm res}$, as defined in Eqs.~(\ref{delr},\ref{delr-res}),
were determined to be:
\begin{eqnarray*}
    \Delta r            & = &  0.0372  \pm 0.0160 \quad {\rm and}   \nonumber \\
    \Delta r_{\rm res}  & = & -0.0236  \pm 0.0170. \nonumber
\end{eqnarray*}
This measurement alone is thus sensitive to quantum corrections 
in the Standard Model at the $2.3 \sigma$ level with evidence for 
bosonic radiative corrections with a significance of $1.4 \sigma$. 

An average $W$ boson mass can be  determined by 
combining the current result with
recent previous measurements.  
The  measurements are weighted with their uncorrelated   uncertainties.  
The  correlated  uncertainty for the most recent   measurements is that  
due to   proton   structure as parameterized in
global parton distribution function fits.  
For each measurement,   the 
uncertainty due to  the common  effect is removed to determine the  uncorrelated
error.  
Based on  the UA2~\cite{UA2-92}  and most recent CDF  
publication~\cite{CDF-95},  the  common uncertainty is taken to be 
85~MeV/c$^2$,  the largest of the  individual uncertainties due to the
uncertainty  on the structure of  the proton.  
This procedure then    yields  a 
world    average  $W$ boson   mass  of  
$M_W  =   80.34 \pm 0.15$~GeV/c$^2$.        

\begin{figure}[t]
 \begin{center}
  \begin{tabular}{cc}
    \epsfxsize = 10.cm   \epsffile{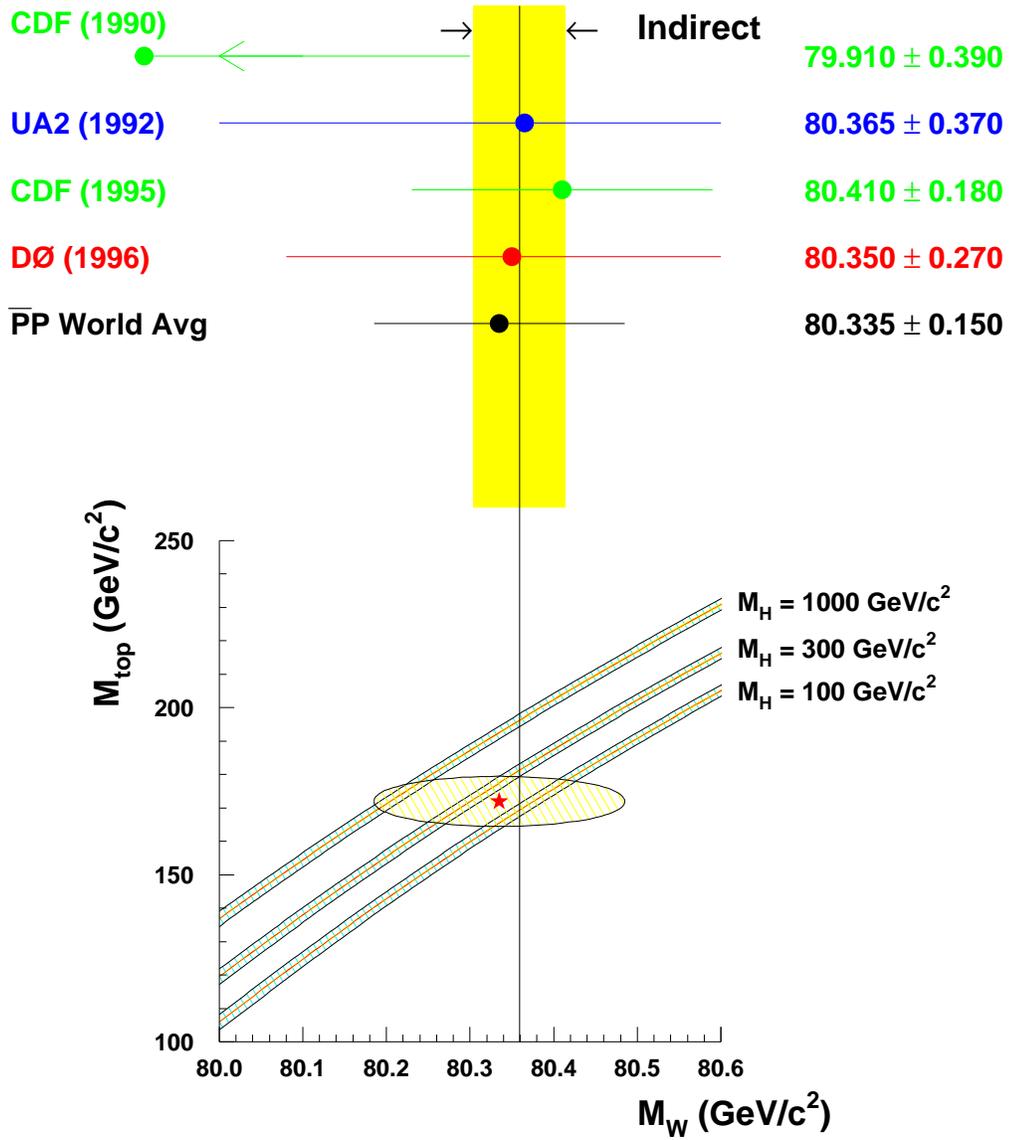}
  \end{tabular}
 \end{center}
\caption{The  upper half of the figure shows the D\O\ determination of $M_W$ 
along with recent results from 
  other hadron collider experiments and the $p\bar p$ world average 
  (see the text for a discussion of the world average calculation).  
  The band is the Standard Model prediction from the combined LEP
   results. The lower half of the figure shows the D\O\ determination of the 
mass of the top quark
   versus the world average determination of $M_W$ ($\star$).  
   The contour shows the allowed range in each value.
   The Standard Model prediction (see the text)
    for various assumptions of the Higgs boson  mass is indicated
   by the bands.}
\label{fig:mw_mt}
\end{figure}

Figure~\ref{fig:mw_mt}~(top)  presents a  comparison of the
world's direct determinations of $M_W$ 
including  this measurement and the overall
$p\bar{p}$ world average. 
Also shown (band) is the Standard Model 
prediction using the LEP data as calculated by the
LEP Electroweak Working Group\cite{zmass}. 
Figure~\ref{fig:mw_mt}~(bottom) shows 
the  recently measured  top quark  
mass\cite{d0_top} from the D\O\ collaboration 
versus 
the world average $W$ mass. The top quark mass value used is 
\begin{equation}
m_t(D\O\ ) = 172.0\pm 5.1 \,({\rm stat})\pm 5.5 \, ({\rm sys} )~{\rm Gev/}c^2
\end{equation}
which is from the combined measurement of the 
dilepton and lepton plus jets channels. 
The Standard Model prediction for different  values of the 
Higgs mass\cite{mw-higgs-band} is also shown as the colored bands.

Using the world average $W$ boson mass, the derived values for the quantum
corrections in the SM are
\begin{eqnarray*}
    \Delta r & = & 0.03834  \pm 0.00885  \quad {\rm and}               \\
    \Delta r_{\rm res}  & = & -0.0224  \pm 0.00944  \ .
\end{eqnarray*}
The direct measurement of the $W$ boson mass at $p\bar{p}$ colliders indicates
the existence of radiative corrections in the
Standard Model at the $\sim 4.3 \sigma$ level  
and evidence of bosonic radiative corrections at  the $\sim 2.4
\sigma$  level.

We thank the staffs at Fermilab and collaborating institutions for their
contributions to this work, and acknowledge support from the 
Department of Energy and National Science Foundation (U.S.A.),  
Commissariat  \` a L'Energie Atomique (France), 
State Committee for Science and Technology and Ministry for Atomic 
   Energy (Russia),
CNPq (Brazil),
Departments of Atomic Energy and Science and Education (India),
Colciencias (Colombia),
CONACyT (Mexico),
Ministry of Education and KOSEF (Korea),
and CONICET and UBACyT (Argentina).

\appendix
\section{$W$ and $Z$ Boson Production Model}
\label{apx1} 

The theory and phenomenology of production of $W$ and $Z$
bosons can be divided into three regions of the 
$p_T$ of the vector boson. 
These regions are imprecisely ordered as follows:
\begin{enumerate}
	\item  The high--$p_t$ region in which perturbation theory is 
	expected to be valid. This region is roughly 50~GeV/$c$~and above.

	\item  The low--$p_T$ region where perturbation theory 
	is not helpful and soft gluons are freely emitted. There is a model 
	for this process, and the validity of this theory is roughly below  
	 15~GeV/$c$. By far, the bulk of the cross section 
	for $W$ boson and $Z$ boson production is in this region.

	\item  The intermediate region  for which there is no 
	theoretical description. Some analyses  attempt to  smoothly connect 
	the two regions, beyond that which occurs naturally by simply adding 
	the cross sections from region 1 to those of
	region 2.
\end{enumerate}

The Monte Carlo 
generation of the vector bosons relied on the resummation formalism 
of Collins, Soper, and Sterman (CSS)\cite{CSS} which treats the emission of 
soft gluons in region 2 by summing all contributions in impact parameter space. 
There are few free parameters in this model and it is shown below that it 
satisfactorily matches the D\O\ data. 
The triple differential cross section for production 
of a $W$ boson can be written
\begin{eqnarray}
\lefteqn{{{d\sigma (AB\to W)} \over {dp^2_T dydQ^2}}  = 
\frac{\pi}{s}\sigma_0 \; \delta(Q^2-M^2_W) {1 \over {(2\pi )^2}}
\int{d^2\vec{b}e^{i\vec{p}_{T}\cdot \vec{b}}}\cdot } \nonumber  \\
   &  &  \sum\limits_{ij}\widetilde 
W_{ij}(b^*,Q,x_A,x_B)e^{-S(b^{*},Q)}F_{ij}^{NP}(b,Q,Q_0,x_A,x_B) \nonumber \\ 
   &  & + Y(p_T,Q,x_A,x_B).
\label{css}
\end{eqnarray}
\noindent Here, $\widetilde W_{ij}(b^*,Q,x_A,x_B)$ includes the 
convolution of parton densities for partons $i,j$ and the splitting 
functions, the Cabibbo-Kobayashi-Maskawa elements, and the electroweak 
parameters. The quantity $Q$ is the invariant 
mass of the annihilating partons, while 
$x_{A,B}$ is the Bjorken $x$ variable representing the 
fraction of the colliding hadron's 
momenta carried by the annihilating partons. $Q_0$ 
is taken to be the lowest scale
where perturbation theory is presumed to be sensible. The quantity
$\sigma_0$ is for normalization. The  Sudakov 
form factor $S(b,Q)$ is fixed by the
order  in $\alpha_s$ and is an integral over a running scale. 
The combination  of these terms describe region 2. 
The quantity $Y(p_T,Q,x_A,x_B)$  contains terms which are less singular 
than ${p_T}^{-2}$ and is the term which dominates in the perturbative regime, 
region 1. 

The complication inherent in this formalism is the  
Fourier transformation of the cross section, which involves an integral over 
all values of the impact parameter $b$. 
This is dealt with by regulating  $b$ to behave well near the origin, forcing
it to  tend to a constant as $b\rightarrow 0$. 
In the CSS formalism, this amounts to a replacement of
$b \rightarrow b^{*}\equiv {{b}\over {\sqrt{1+b^{2}/b^{2}_{max}}}}$.
The price for making this modification is the  obligation to add 
a term to ``replace'' the missing contribution to the integral from 
this $b\rightarrow b^{*}$ substitution. 
This extra factor is the so-called non-perturbative 
function, represented in Eq.~(\ref{css}) as $F_{ij}^{NP}$. 
Theoretical arguments fix the form of $F_{ij}^{NP}$, up to 
phenomenological parameters. 

There have been two efforts to determine the non-perturbative  function. 
One such recent fit is by Ladinsky and Yuan\cite{ly_ptw}  (LY) who
parameterized the non--perturbative function as
\begin{eqnarray}
\lefteqn{F_{ij}^{NP} (b,Q,Q_0,x_A,x_B)  =   } \nonumber \\
& & \exp \left[  -b^2 g_1-g_2 b^2 \ln \left( {Q \over {2Q_0}} \right)
 - g_1 g_3 b \ln (100 x_A x_B)   \right] .
\end{eqnarray}
\noindent The $g$ parameters are not specified by theory, but are 
measurable. A much earlier effort by Davies and
Stirling\cite{davies_sterling}  (DS) used an identical parameterization, but 
essentially with $g_{3}=0$.  Recently, Arnold and Kaufman\cite{ak_ptw} (AK)
employed the  CSS formalism including the DS fits, a NLO calculation for the
$Y$  term\cite{arnold_reno} (region 1), and a strategy of dealing with 
region 3.  A computer program has been available for the AK approach. 
Likewise, the LY calculation was done with an independent  computer program
which is identical in its coding of the CSS theory,  but utilized a simple
${\cal O}(\alpha_{S})$ calculation for the $Y$  term.  LY made no attempt to
match the two regions.  Both computer codes have been used in this analysis.

This description of $W$ boson and $Z$ boson production is taken as the {\it 
ansatz} for the Monte Carlo production model.
The more recent LY fits to modern Drell-Yan and collider $Z$ boson data 
constrain the $g$ parameters 
and have been used here as representative of the best available 
information. 
In this sense the $g$ parameters  function operationally like the 
parton distribution functions.  
The LY fits result in  
$g_1=0.11^{+0.04}_{-0.03}$ GeV$^2$,
$g_2=0.58^{+0.10}_{-0.20}$ GeV$^2$, and 
$g_3=-1.5\pm 0.10$ GeV$^{-1}$.
These  central values have been 
used as the nominal production model for $W$ and $Z$ bosons, with the 
major sensitivity to $g_2$. 

\section{Bremsstrahlung}
             
\begin{figure}[h]
\begin{center}
\begin{tabular}{cc}
    \epsfxsize=7.0cm
    \epsffile{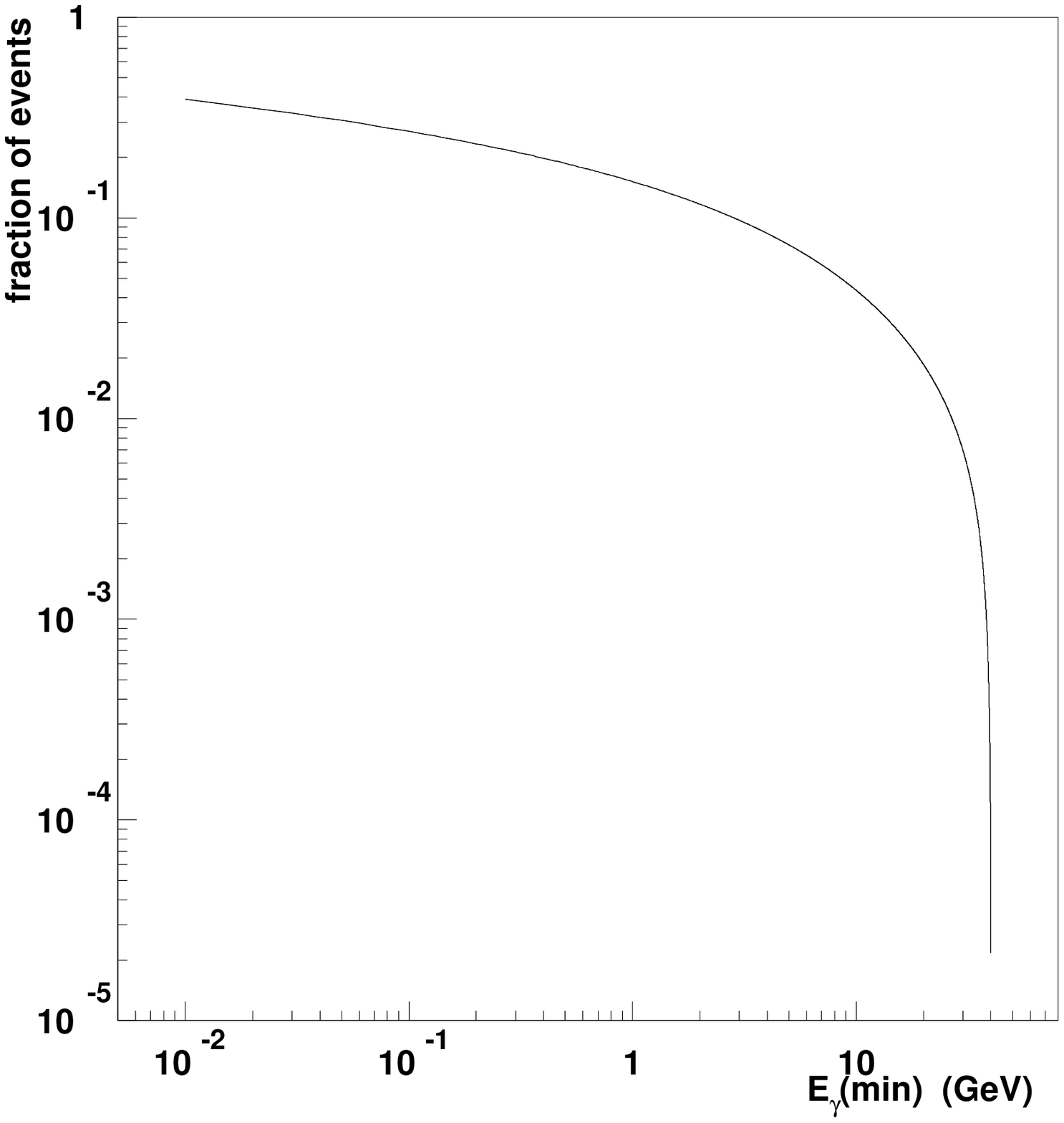}
    \epsfxsize=7.0cm
    \epsffile{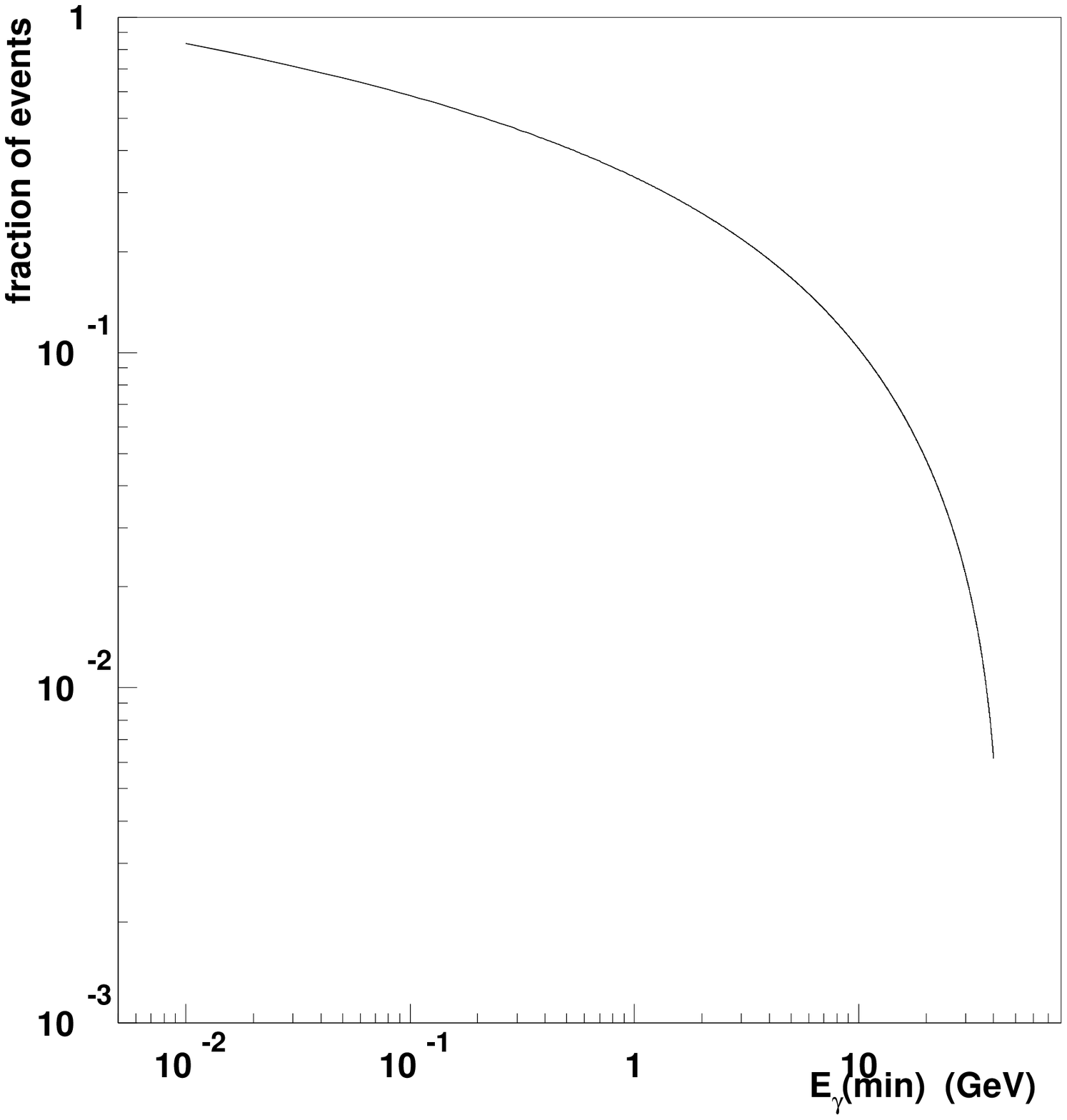}
\end{tabular}
\end{center}
\caption{Fraction of radiative (a) $W$ decays and (b) $Z$ decays 
         as function of the minimum photon energy. }
\label{fig:radfrac}
\end{figure}

The fraction of decays which involves radiation depends on the minimum photon 
energy, $E_{\gamma}^{min}$, which was taken to be 50~MeV. 
Figure~\ref{fig:radfrac} shows this fraction as function of 
$E_{\gamma}^{min}$ for (a) $W$ boson and (b) $Z$ boson decays.  
For $Z$ boson decays the
fraction of radiative decays is about a factor of two  higher than for $W$ boson
decays, as expected.  For the default $E_{\gamma}^{min}$, 31\% of the $W$ boson
decays and 66\% of the $Z$ boson decays were radiative.  Only order
$\alpha_{\mbox{\tiny EM}}$  corrections have been included and so  
processes in which two or more photons are radiated were not generated.  

For radiative $W$ boson decays, $W \rightarrow e\nu\gamma$,  it is important to
determine the minimum spatial separation between the photon  and electron
that would result in the photon energy not being included with  that of the
electron by the reconstruction program.   For events with $R =
\sqrt{\Delta\eta^2 + \Delta\varphi^2}$ above approximately 0.2  
the photon energy may not be added to that of the electron. Instead, it 
was
reconstructed as part of the $W$ boson recoil.  The neutrino energy 
was
unchanged, but the electron energy is too low.   The $W$ boson and $Z$ 
boson masses were
then too low in a manner which  does not cancel in the ratio. Since
this effect is large, it is important to evaluate the effect  when the photon is
produced by bremsstrahlung in the central detector. 

For the photon to have an effect on the measured $W$ boson mass, it 
must be separated from the electron in 
($\eta , \varphi$) space by at least $R_{e\gamma} = 0.2$, that is, 
\begin{equation}
\Delta\eta^2 + \Delta\varphi^2  \,>\, R^2_{e\gamma}.  
\end{equation}
With $\Delta\eta = {\Delta\vartheta \over \sin\vartheta } = 
\cosh\eta \, \Delta\vartheta $, this can be written as
\begin{equation}
   \left( {\Delta\vartheta \over \sin\vartheta } \right) ^2  \,+\, 
    \Delta\varphi^2  \,>\, R^2_{e\gamma}
\end{equation} 
Switching to coordinates measured with respect to the electron
\begin{eqnarray}
\Delta\vartheta  \,&=\,   \omega\cos\alpha        \\
\Delta\varphi    \,&=\, { \omega\sin\alpha \over \sin\vartheta } \ ,
\end{eqnarray}
where $\omega$ is the angle between the electron and the photon and 
$\alpha$ the azimuthal angle of the photon with respect to the electron, 
one can write:
\begin{equation}
  \left( \omega \cos\alpha \right)^2  \,+\, 
  \left( \omega \sin\alpha \right)^2  \,>\, 
  \left( { R_{e\gamma} \over \cosh\eta } \right)^2 
\end{equation}
or 
\begin{equation}
  \omega \,>\, { R_{e\gamma} \over \cosh\eta }. 
\end{equation}
The angle between the electron and photon must be greater than 0.2
rad for $\eta = 0$ and greater than 0.13 rad for $\eta = 1$.  In units of
$m_e \over E$, where $m_e$ is the electron mass and 
$E$ the electron energy, this corresponds to 
${ E \over m_e } \times \omega > 13,000$ for an electron energy $E$ of 
50~GeV and $\eta = 1$. 

\begin{figure}[t]
\begin{center}
\begin{tabular}{cc}
    \epsfxsize=8.0cm
    \epsffile{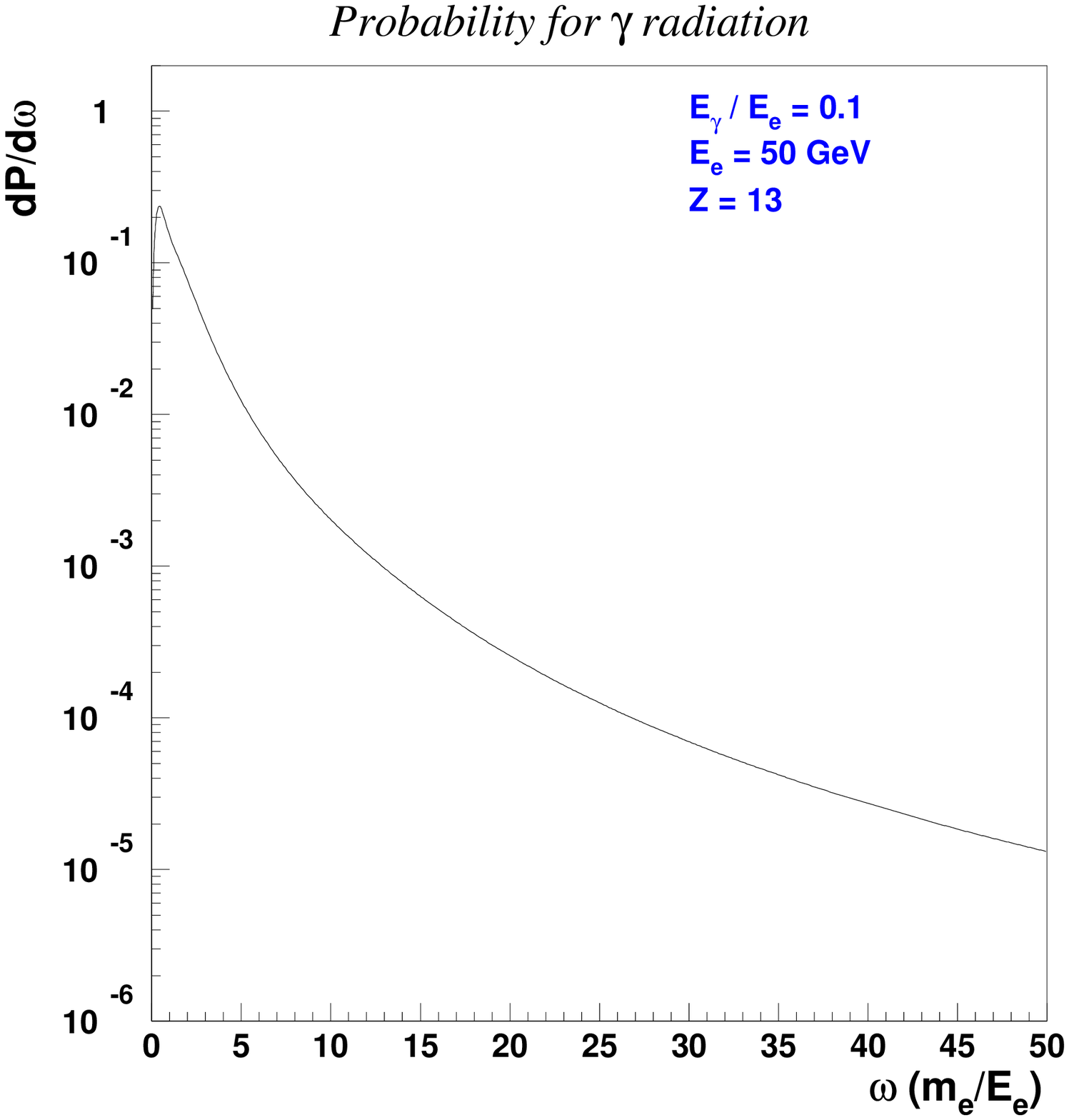}
    \epsfxsize=8.0cm
    \epsffile{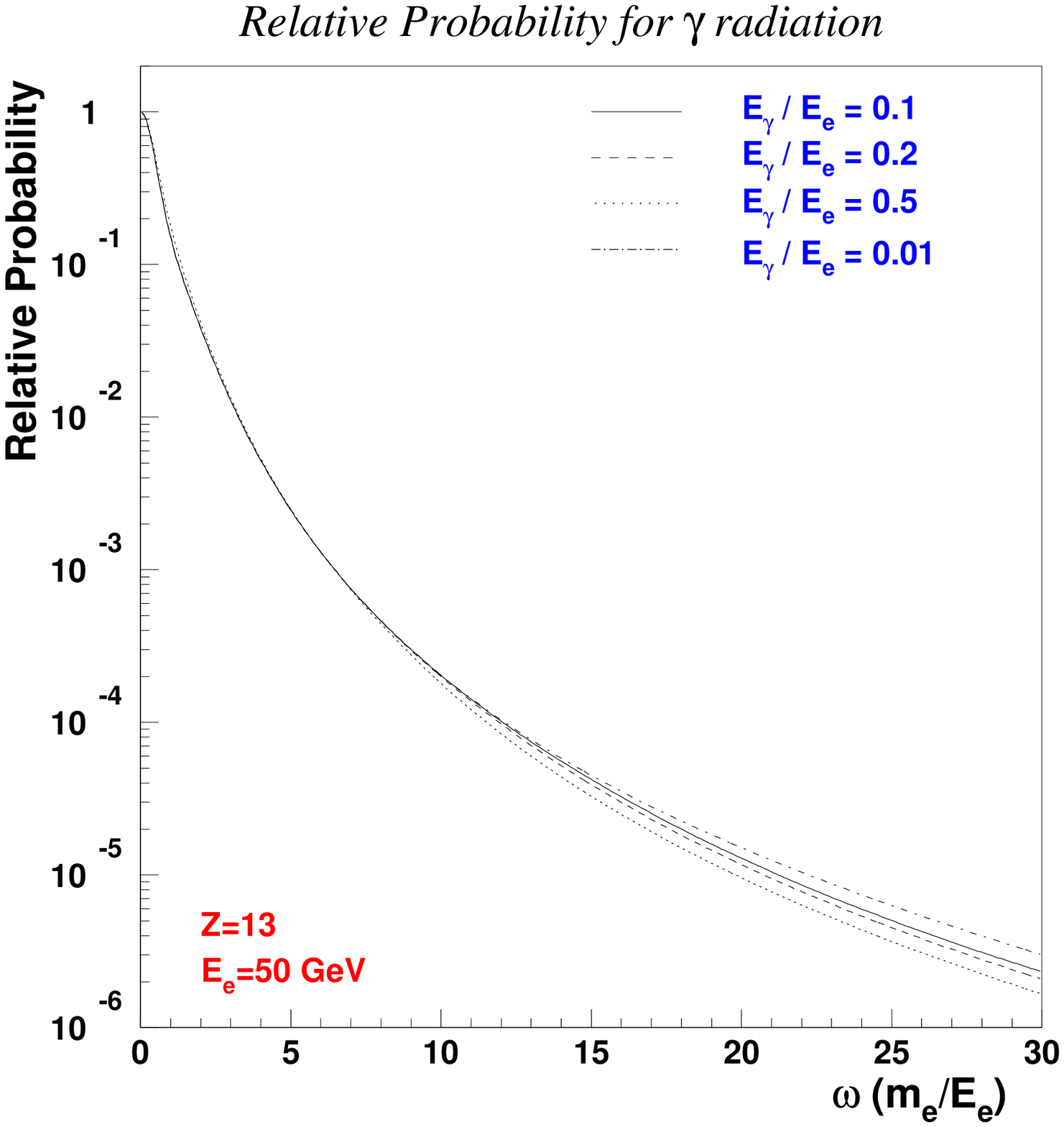}
\end{tabular}
\end{center}
\caption[]{(a) Probability for an electron to radiate a photon in aluminum 
               ($Z$=13) as function of the angle $\omega$ between the electron 
               and the photon in units of $m_e \over E$, where $m_e$ is the 
               electron mass and $E$ its energy. 
           (b) Relative probability for radiating a photon for different 
               values of $y= {k\over E}$, with $k$ the photon 
               energy. }
\label{fig:brem_prob}
\end{figure}

Figure~\ref{fig:brem_prob}(a) shows the probability $d P \over d\omega $ for 
radiating a photon at an angle $\omega$ for the case $y$=0.1,  where $y$ is
the fraction of the electron's energy carried by  the
photon~\cite{bremsstrahlung}.  The angle $\omega$ is expressed in units of
$m_e \over E$.  For all calculations  in this analysis, $Z$=13 (aluminum) has
been assumed  and the energy of the electron has been fixed to $E = 50$~GeV. 
The probability decreases by four orders of magnitude at 
$m_e \over E$-scaled angles
of 50.   Figure~\ref{fig:brem_prob}(b) shows the relative probability  for
radiating a photon at an angle $\omega$ and its dependence on $y$.  Although
the probability for radiating a photon is larger at small $y$, after
normalization, there is  little $y$ dependence of the angle at which the
photon is radiated.   Since scaled angles of 13,000 or more are needed,  the
photon never separates from the electron by radiation alone  and therefore
bremsstrahlung has no effect on the $W$ boson mass. 

The electron and photon can also separate if the electron  scatters through
a large angle.  The probability that an electron radiates a photon of momentum
between $k$ and $k+dk$ in
$dx$ in a medium with  radiation length $X_0$ is \cite{bremsstrahlung,Berends} 
\begin{equation}
  P(E,k) dk \, dx \, = \, {dx \over X_0} \,
                         {dk \over k} \, 
                  \left( {4\over 3} \,-\, {4\over 3} y \,+\, y^2 
                  \right).
\end{equation}
The quantity $y$ is  the fraction of the electron's energy carried by
the photon, $y= {k\over E}$, and $E$ and $k$ are the electron 
and photon energies. 
Integrating from $k = k_{\rm min}$ to $E$ one finds 
\begin{eqnarray}
  P(E,k > k_{\rm min})dx \,& = &\, {dx \over X_0} \,
          [      -{4\over 3} ( \ln y_{\rm min} \,+\, 1 \,-\,
                                        y_{\rm min} ) \, \nonumber \\
                                        &  & +\, 
                       {1\over 2} ( 1 \,-\, y_{\rm min}^2 )]dx.
\end{eqnarray}
For $y_{\rm min}$ close to 1 this gives 
\begin{equation}
  P(E,k > k_{\rm min})dx \, =\, {dx \over X_0} \, 
                              ( \zeta \,+\, {1 \over 6} \zeta^2 )dx 
\end{equation}
with $\zeta \equiv 1 - y_{\rm min}$. 
For example, the probability that a 50~GeV electron radiates a 49~GeV
photon in 0.15~$X_0$ is
\begin{equation}
      0.15 \times 0.02  \approx 3 \cdot 10^{-3}.
\end{equation}
The 1~GeV electron can then scatter through a large angle~\cite{pdg}
\begin{equation}
  \omega_{\rm rms} \,=\, \sqrt{2} \ \times 
                         {13.6 \  {\rm MeV} \over 1 \ {\rm GeV}} \ 
                         \times
                         \sqrt{0.15} \,=\, 7.4 \ {\rm mrad}.
\end{equation}
The angles resulting from multiple scattering are generally larger than
those produced in the radiation itself, particularly when the electron is
low in energy $(\omega_{\rm rms} \approx {1\over E})$.   Nevertheless, it is
still difficult to separate the photon  and electron sufficiently.   The
7.4~mrad angle calculated above translates, in units of $m_e \over E$,  into
${ E\over m } \times 7.4 \times 10^{-3}  \approx 800$,  still small compared to
13,000.   The falloff in scattering is rapid (Gaussian). If one considers a
50~GeV electron radiating 99.8\% of its energy, the probability becomes
small, $3 \times 10^{-4} $.  The resulting 100~MeV electron, however, can now
multiple scatter through a large angle, 80 mrad or 8000 in units of $m_e\over
E$.

\begin{figure}[t]
\begin{center}
\begin{tabular}{cc}
    \epsfxsize=8.0cm
    \epsffile{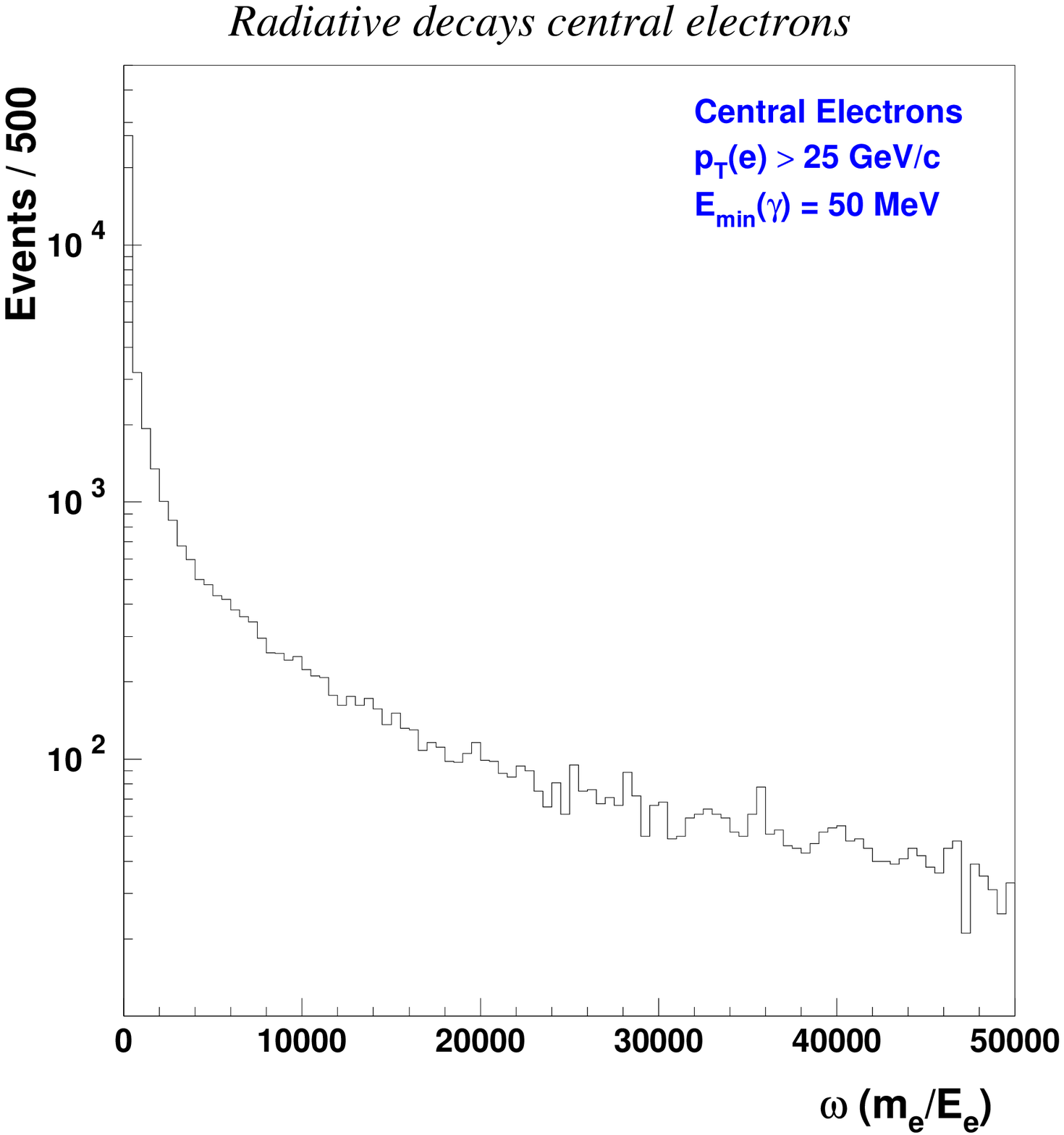}
    \epsfxsize=8.0cm
    \epsffile{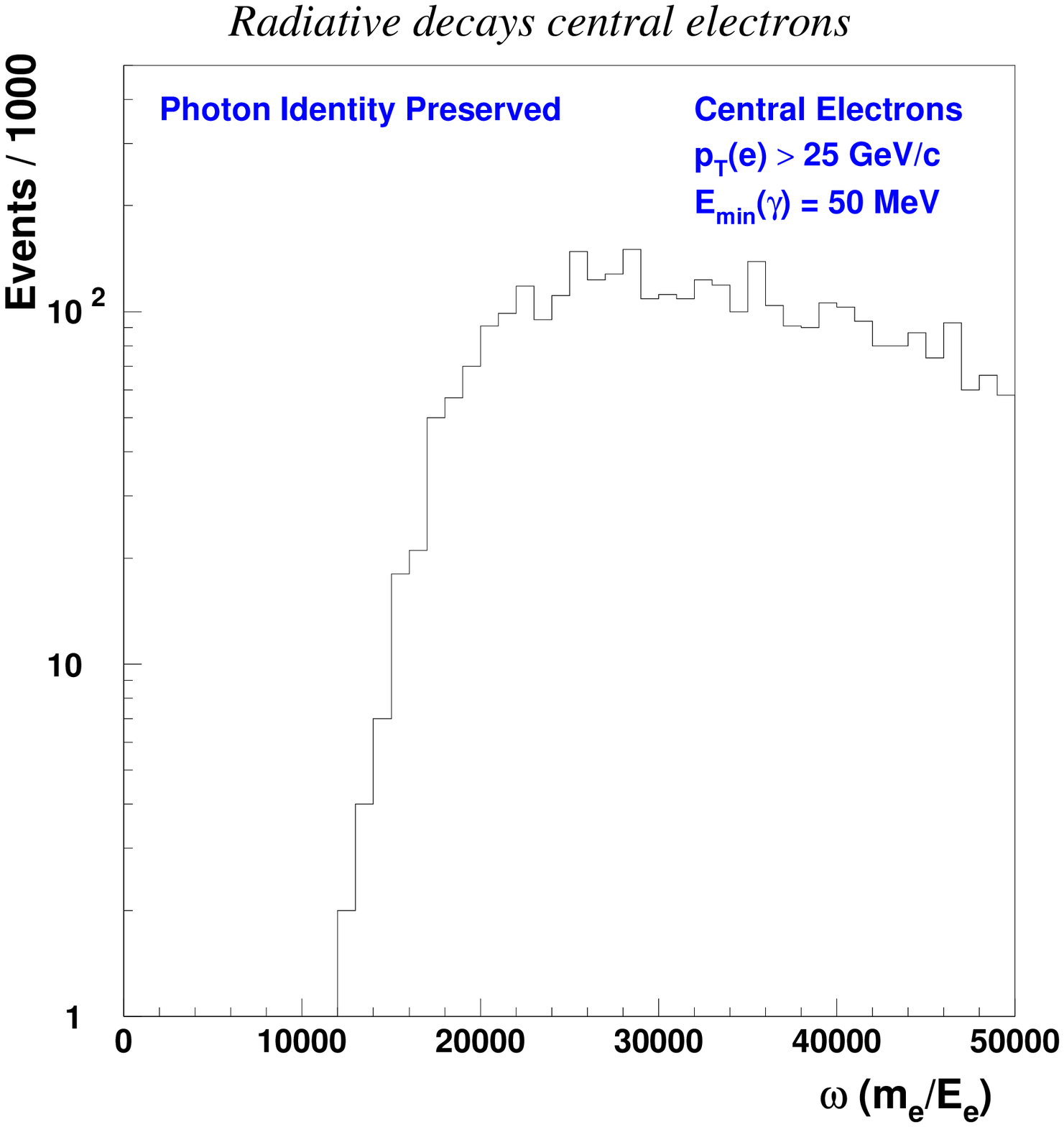}
\end{tabular}
\end{center}
\caption{Distribution in $\omega$, in units of $m_e \over E$, for radiative 
         $W$ boson events in the central calorimeter and (b) 
for events where the 
         the photon retains its identity. } 
\label{fig:rad_omega}
\end{figure}

This situation is compared with radiative $W$ boson decays in 
Fig.~\ref{fig:rad_omega}(a) which shows the distribution in $\omega$, in units 
of $m_e \over E$, for radiative $W$ boson 
events with the electron in the  central
calorimeter with $p_T^e > 25$~GeV/c.  The distribution has a very long tail
extending to values of 50,000  for $\omega$.  At small angles of ${E\over m_e}
\times \omega = 10,000$ the cross section 
is down by a factor of approximately 200. 
Nevertheless, 21\% of the events have angles greater than 5000. 
Figure~\ref{fig:rad_omega}(b) shows the event distribution in 
$\omega$ for events in which the photon and electron reconstruct as 
separate entities with the photon retaining 
its identity.  As was estimated above, the
threshold at
${E\over m_e} \times \omega$ is approximately 15,000. 

\section{Mean Number of Interactions}
\label{mean}
The library of minimum bias events was stored in bins of luminosity 
according to the following rule.  Given a $W$ boson event, recorded at a 
luminosity ${\cal L}$ with corresponding average number of  interactions per
crossing $\langle n \rangle$, the minimum bias event,  mimicking the
underlying event, was taken at a scaled value of the  instantaneous
luminosity, ${\cal L}^\prime$. ${\cal L}$ was chosen so that 
the mean  multiple interaction rate
in Monte Carlo generated $W$ boson events is the  
same as in the $W$ boson data sample.

The probability of getting a $W$ boson trigger in a crossing in which there  are
$n$ interactions is given by
\begin{equation}
  P(W,n) \,=\,  n \, P(n) \,  {\sigma_W \over \sigma_{\rm inel}} \ .
\end{equation}
Here $P(n)$ is the Poisson probability of $n$ interactions in the 
crossing, ${\sigma_W \over \sigma_{\rm inel}}$ the probability that the 
inelastic interaction is one in which a $W$ boson is produced. The factor 
$n$ represents the number of ways one can choose the $W$ boson interaction 
from the $n$ interactions in the crossing.  Note that the probability 
of getting a $W$ boson in a crossing is then
\begin{equation}
  P(W) \,=\, \sum_n \, n \, P(n) { \sigma_W \over \sigma_{\rm inel} }
       \,=\, \langle n \rangle  \, { \sigma_W \over \sigma_{\rm inel} },
\end{equation}
which is the expected rate when $\langle n \rangle $ is written in 
terms of the luminosity and the inelastic cross section, 
$\sigma_{inel}$.  The probability distribution of getting $n$ 
interactions in a crossing in which there is a $W$ boson is
\begin{equation}
  P(n|W) \,=\,  n \, P(n),
\end{equation}
and has a mean value of $\langle n \rangle + 1 $.  This shows that the mean 
number of interactions in a crossing in which there is a $W$ is 
$\langle n \rangle + 1 $.

For the minimum bias trigger, the average number of interactions per 
crossing $\langle n_{min} \rangle$ given that there is at least one, 
is
\begin{equation}
  \langle n_{min} \rangle 
      \,=\, 
              { \sum_{n^\prime =0}^\infty n^\prime \, P(n^\prime ) \over 
                \sum_{n^\prime =1}^\infty P(n^\prime ) } 
      \,=\, 
              { \langle n^\prime \rangle  \over 
                1 - e^{- \langle n^\prime \rangle } }.
\label{eq:nminb}
\end{equation}
The minimum bias events are chosen at a luminosity ${\cal L}^\prime$ 
such that the mean $\langle n_{min} \rangle$, as given by 
equation~\ref{eq:nminb}, is equal to $\langle n \rangle + 1$, where 
$\langle n \rangle$ is the mean number of interactions at luminosity 
${\cal L}$ at which the $W$ event was recorded.  This guaranteed that 
the mean number of interactions was correct.  
The distributions in the number of interactions per crossing are 
somewhat different, though. The minimum bias 
distribution is a Poisson distribution, cut off at $n=1$, while the number of 
interactions in $W$ events is a Poisson distribution, beginning at $n=1$.  The 
impact of this difference in this analysis is negligible.

%




\newpage

\newpage

\newpage

\newpage

\newpage

  \newpage

  \newpage

  \newpage

\vspace{2.0cm}
  \newpage

  \newpage

  \newpage

  \newpage

  \newpage

  \newpage

\newpage
  \newpage

  \newpage

\newpage
  \newpage

  \newpage

  \newpage

\begin{figure}[t]
\begin{center}
\begin{tabular}{c}
   \epsfxsize=10.cm  \centerline{\epsffile{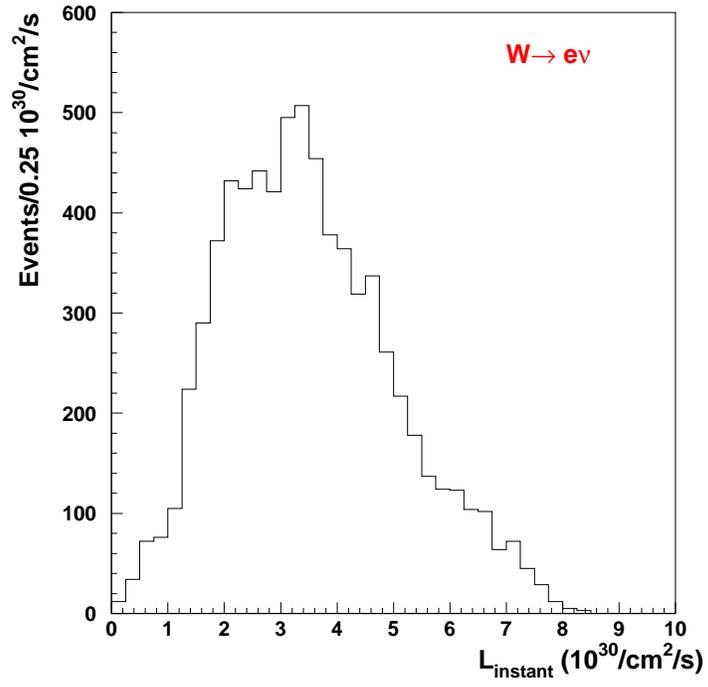}} 
\end{tabular}
\caption{Distribution in instantaneous luminosity of the 
         $W$ events used in the $W$ boson mass measurement.}
\label{fig:w_lum}
\end{center}
\end{figure}  \newpage

\begin{figure}[p]
 \begin{center}
  \begin{tabular}{cc}
    \epsfxsize = 8.cm \epsffile{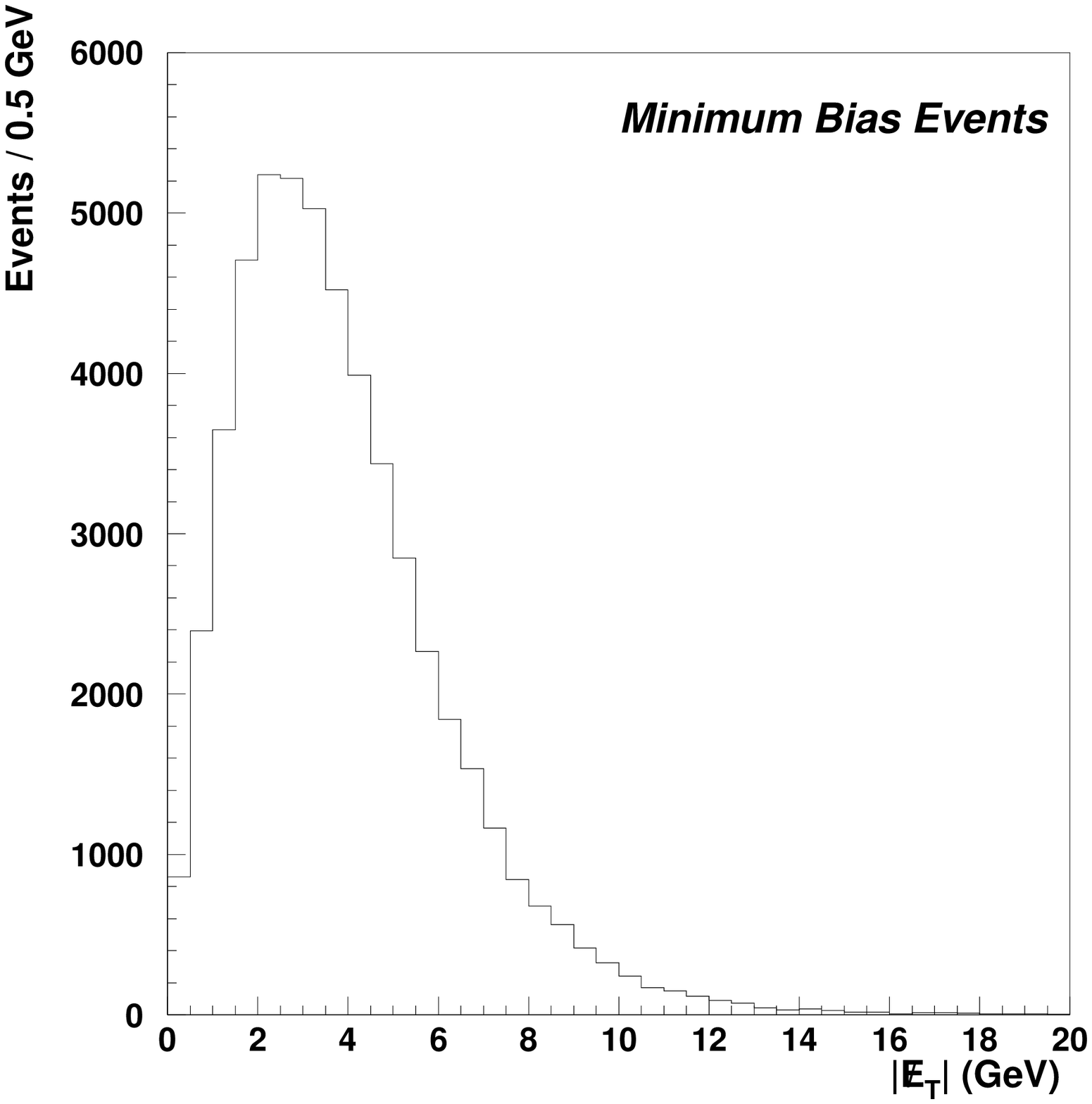}     &
    \epsfxsize = 8.cm \epsffile{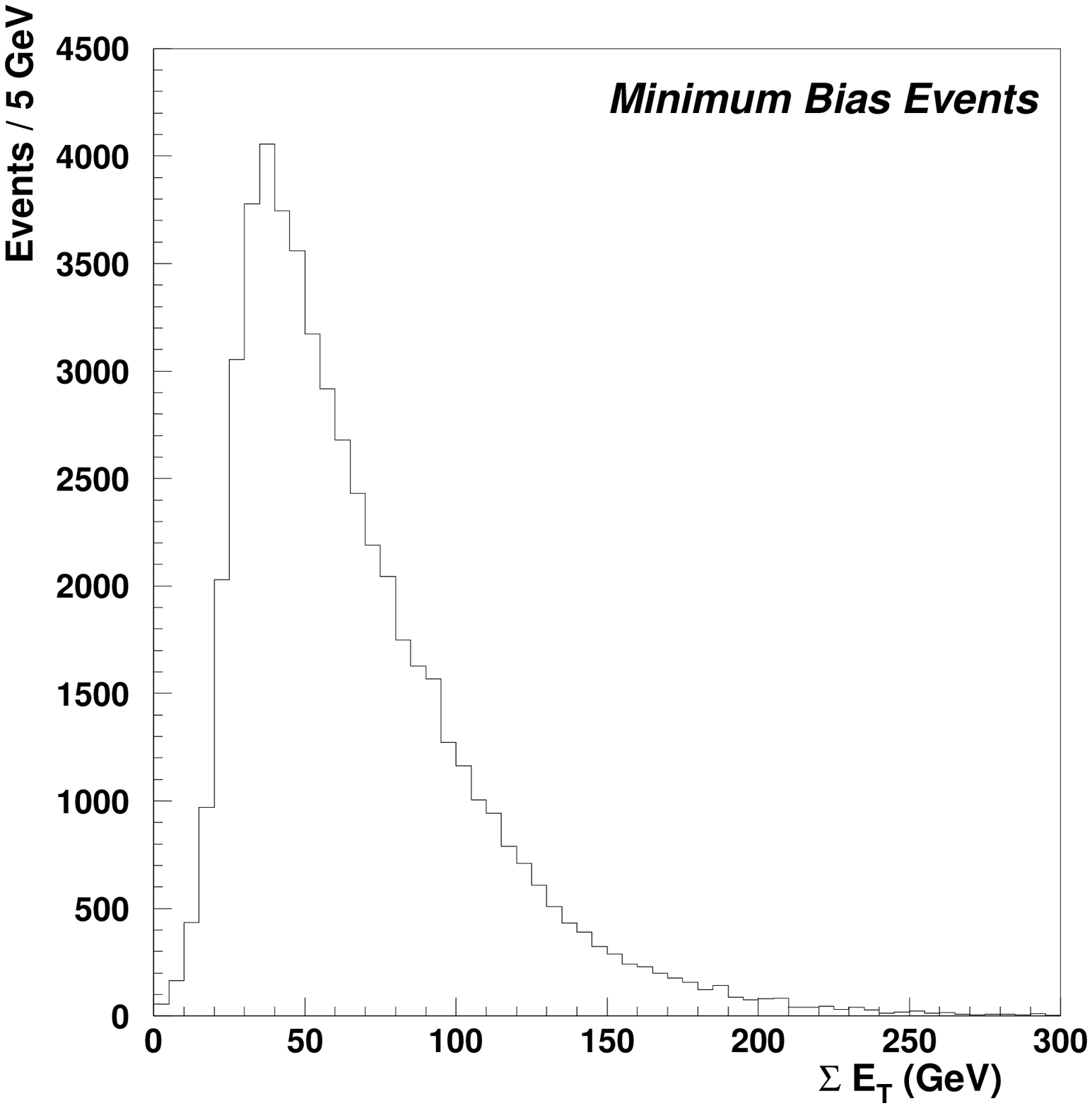} \\
\end{tabular}
\end{center}
\caption{(a) $\mbox{$\not\!\!E_T$}$ and (b) $\Sigma E_T$ 
distributions of the minimum bias 
         events used to model the $W\rightarrow e \nu$ 
and $Z\rightarrow e e$ underlying event. }
\label{fig:minb}
\end{figure}  \newpage

  \newpage

  \newpage

  \newpage

  \newpage

  \newpage

  \newpage

  \newpage

  \newpage

  \newpage

  \newpage

  \newpage



  \newpage

  \newpage

  \newpage

  \newpage

  \newpage

  \newpage

  \newpage

  \newpage

\begin{figure}[h]
\begin{center}
\begin{tabular}{cc}
    \epsfxsize=8.cm  \epsffile{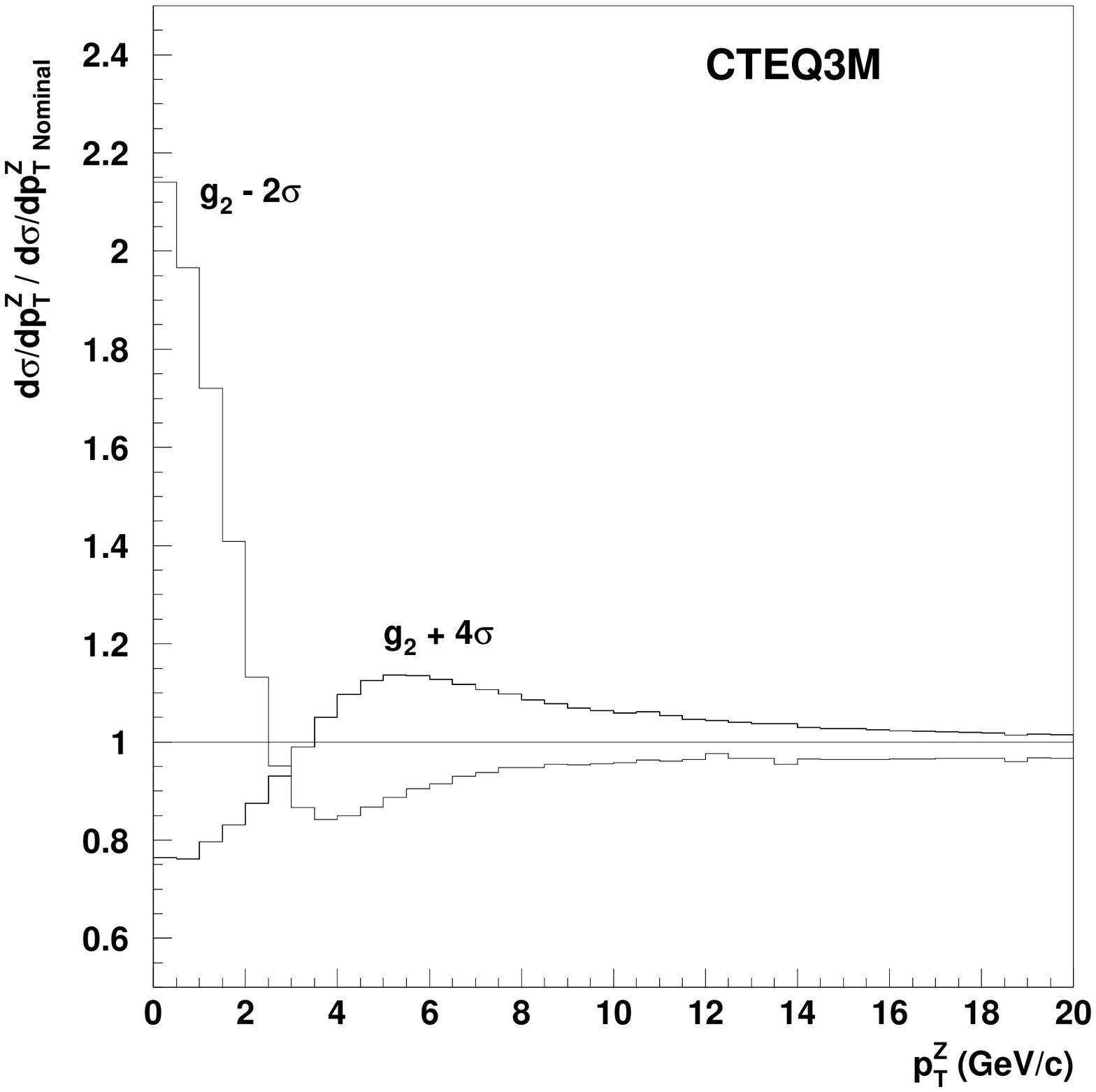}  &
    \epsfxsize=8.cm  \epsffile{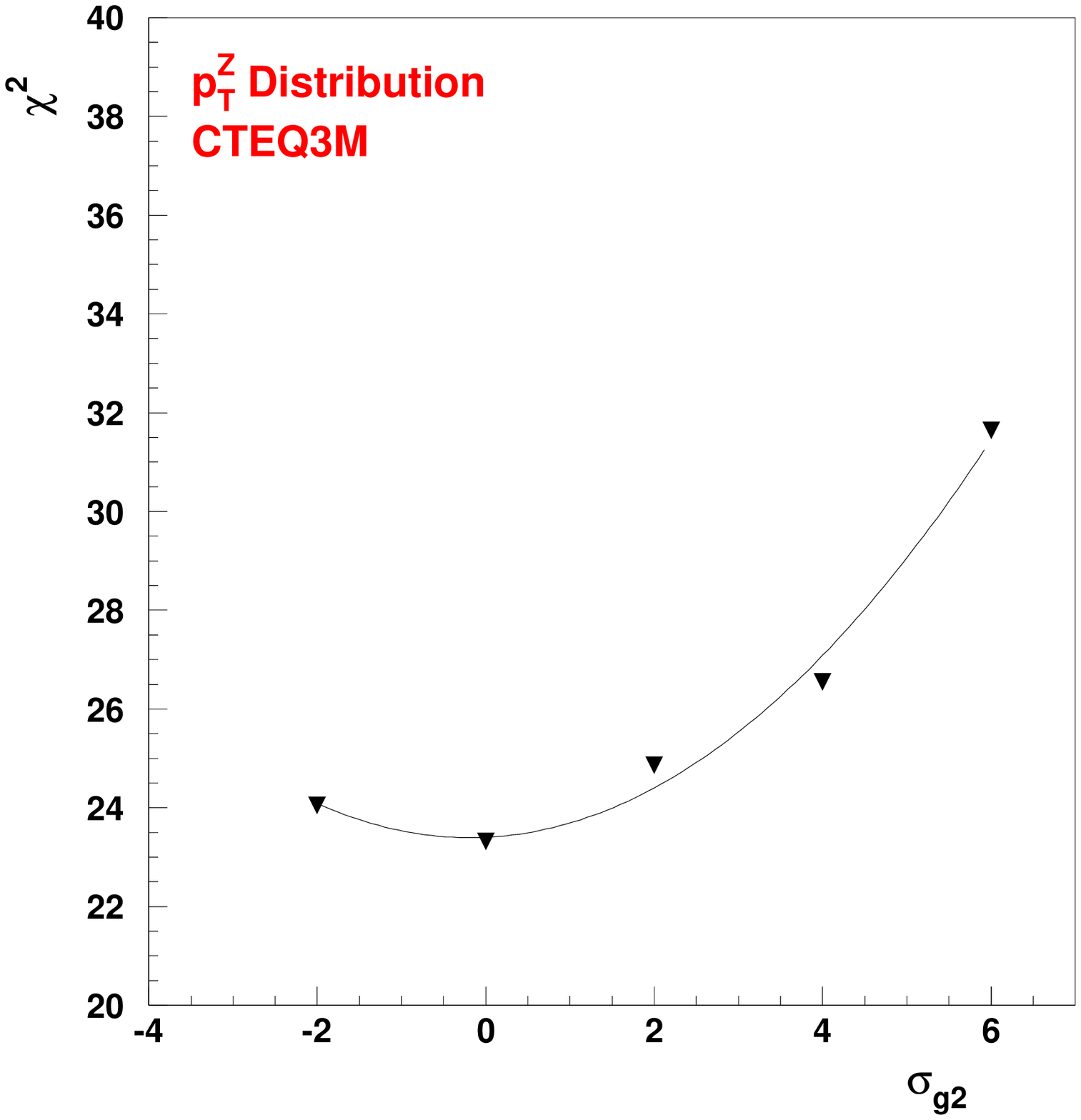}\\
\end{tabular}
\end{center}
\caption[]{(a) Ratio of predicted differential cross section in $p_T^Z$ 
                and the nominal cross section versus $p_T^Z$
               when the parameter $g_2$ is varied by multiple 
               standard deviations from its nominal value in the 
               Ladinsky-Yuan prediction and  
           (b) the distribution in $\chi^2$ for a comparison between data and 
               Monte Carlo of the $p_T^Z$ spectrum versus the variation of
                $g_2$ in units of its standard deviation. }
\label{fig:ptz_g}
\end{figure}  \newpage

\newpage
  \newpage

  \newpage

  \newpage

  \newpage

  \newpage

  \newpage

  \newpage

  \newpage

  \newpage

  \newpage


\begin{table}[t]
	\centering
	\begin{tabular}{ccc}
	Experiment & channel & $M_W$ (GeV/$c^2$) \\
	\hline
  UA1(1983)\cite{UA1-83} & $e\nu$ & $81\pm 5$  \\
  UA2(1983)\cite{UA2-83} &$e\nu$ &$80^{+10}_{-6}$ \\
  UA1(1986)\cite{UA1-86} &$e\nu$ &$83.5^{+1.1}_{-1.0}\pm 2.7$ \\
  UA2(1987)\cite{UA2-87} &$e\nu$ &$80.2\pm 0.6\pm 0.5\pm 1.3$ \\
  UA1(1989)\cite{UA1-89} &$e\nu$ &$82.7\pm 1.0 \pm 2.7$ \\
  UA2(1990)\cite{UA2-90} &$e\nu$ &$80.49\pm 0.43 \pm 0.24$ \\
  UA2(1992)\cite{UA2-92} &$e\nu$ &$80.35 \pm 0.33 \pm 0.17$\\
  CDF(1989)\cite{CDF-89} &$e\nu$ &$80.0\pm 3.3 \pm 2.4$ \\
  CDF(1990)\cite{CDF-90} &$e\nu$ &$79.91\pm 0.35 \pm 0.24\pm 0.19$ \\
  CDF(1995)\cite{CDF-95} &$e\nu$ &$80.490\pm 0.145 \pm 0.175$ \\
		\hline
  UA1(1984)\cite{UA1-84} &  $\mu \nu$ & $81^{+6}_{-7}$  \\
  UA1(1989)\cite{UA1-89} &  $\mu \nu$ & $81.8^{+6.0}_{-5.3}\pm 2.6$  \\
  CDF(1990)\cite{CDF-90}& $\mu \nu$ & $79.90\pm 0.53 \pm 0.32\pm 0.08$ \\
  CDF(1995)\cite{CDF-95}& $\mu \nu$ & $80.310\pm 0.205 \pm 0.130$ \\
                 \hline
  UA1(1989)\cite{UA1-89} &  $\tau \nu$ & $89\pm 3 \pm 6$  \\
		\hline
  CDF(1990)\cite{CDF-90} & $e\nu +\mu \nu$ & $79.91\pm 0.39$ \\
  CDF(1995)\cite{CDF-95} & $e\nu +\mu \nu$ & $80.410\pm 0.180$ \\
		\hline
	\end{tabular}
	\caption{Previously published hadron collider measurements of $M_W$. 
In each case
 the first uncertainty listed is statistical, the second is systematic, and 
 the third is due to energy scale. For the latest CDF values, the energy
 scales have been incorporated into the total systematic uncertainty.}
 \vspace{0.12 in}
	\label{whistory}
\end{table}


\begin{table}[t]
\begin{center}
\begin{tabular}{lcc} 
 ~ & $W\rightarrow e\nu$ candidates  & $Z\rightarrow ee$ candidates \\
 \hline
 L1 trigger requirements &  1 EM tower with $E_T> 10$ GeV & 2 EM 
 towers with $E_T> 7$ GeV \\
 L2 filter requirements  & 1 EM cluster with $E_T> 20$ GeV & 2 EM 
 clusters with $E_T> 10$ GeV      \\
 ~ &   $f_{iso}< 0.15$  & $f_{iso}< 0.15$               \\
~  &   $\mbox{${\hbox{$E$\kern-0.6em\lower-.1ex\hbox{/}}}_T$ } > 
20$ GeV & ~                           \\ 
\end{tabular}
\caption[]{L1 and L2 trigger requirements for $W$ and $Z$ event data
           samples. Here, $f_{iso}$ is defined by Eq.~\ref{fiso}.}
\label{trigger}
\end{center}
\end{table}


\begin{table}[t]
\begin{center}
\begin{tabular}{ccc}
\multicolumn{3}{c}{\em $W$ boson event sample } \\ 
\hline
          ECN         & CC        & ECS     \\ 
          1838        & 7234      & 1681    \\ 
\end{tabular}
\begin{tabular}{lccccc}
\multicolumn{6}{c}{\em $Z$ boson event sample } \\ 
        & ECN-ECN     & ECN-CC    & CC-CC     & CC-ECS    & ECS-ECS \\ 
\hline
Mass Measurement
        & 48          & 147       & 366       & 134       & 39      \\ 
Resolution Studies
        & 46          & 143       & 344       & 130       & 35      \\ 
\end{tabular}
\end{center}
\caption{Event samples for $W$ and $Z$ bosons. Here, ``N'' and ``S'' refer
to the end calorimeters on the north and the south. }
\label{tab:num_wzevts}
\end{table}


\begin{table}[p]
\begin{center}
\begin{tabular}{ll}
Descriptor  & Nominal value \\
\hline 
EM energy resolution, sampling (CC)   & $S$ = 13.0\%                   \\
EM energy resolution, constant (CC)   & $C$ = 1.5\%                    \\
EM energy resolution, noise (CC)      & $N$ = 0.4 GeV                  \\
HAD energy resolution, sampling (CC)  & $S$ = 80.0\%                   \\
HAD energy resolution, constant (CC)  & $C$ = 4.0\%                    \\
HAD energy resolution, noise (CC)     & $N$ = 1.5 GeV                  \\
HAD energy scale                      & $\kappa$ = 0.83                \\
Electron Underlying Event             & $E^{el}_{UE}$ = 205 MeV        \\
$W$ Width                             & $\Gamma_{W}$ = 2.1 GeV         \\
$Z$ Width                             & $\Gamma_{Z}$ = 2.5 GeV         \\
\# minimum bias events                & 1.0                            \\
minimum $E_\gamma$                    & $E_\gamma^{min}$ = 50 MeV      \\ 
$\Delta R (e\gamma)$                  & $\Delta R (e\gamma)$ = 0.3     \\ 
Calorimeter position resolution       & $\sigma(z) \approx 0.7$ cm     \\
CDC $z_{cog}$ resolution              & rms $z_{cog}$ = 2.0 cm       \\
$\varphi$ resolution                  & $\sigma(\varphi)$ = 0.005 rad   \\
\end{tabular}
\end{center}
\caption[]{Parameters used in the fast Monte Carlos. }
\label{tab:params}
\end{table}
%


\begin{table}[t]
\begin{center}
\begin{tabular}{lcccccc}
     &            \multicolumn{3}{c}{ $W^\pm$ production }
                 & \multicolumn{3}{c}{ $Z^0$ production   }  \\ \hline
pdf  &  $\beta \times 100 $ & $v-v$ and $v-s$ & $s-s$ 
                 & $\beta \times 100 $ & $v-v$ and $v-s$ & $s-s$ \\ 
     & {\small (GeV$^{-1}$) }  & (\%)  & (\%)  
                          & {\small (GeV$^{-1}$) }  & (\%)  & (\%)    \\ 
\hline  
MRS E$^\prime$   & 0.980  & 82.7  & 17.3  & 0.869  & 84.7  & 15.3   \\ 
MRS B            & 1.054  & 82.7  & 17.3  & 0.897  & 85.0  & 15.0   \\ 
HMRS B            & 1.048  & 75.5  & 24.5  & 0.932  & 77.7  & 22.3   \\ 
KMRS B\O         & 1.022  & 79.2  & 20.8  & 0.908  & 81.4  & 18.6   \\ 
MRS D0~$^\prime$  & 1.220  & 78.9  & 21.1  & 1.077  & 80.9  & 19.1   \\ 
MRS D$^\prime$-   & 1.277  & 79.9  & 20.1  & 1.097  & 81.7  & 18.3   \\ 
MRS H            & 1.264  & 79.0  & 21.0  & 1.104  & 81.0  & 19.0   \\ 
MRS A             & 1.282  & 79.6  & 20.4  & 1.101  & 81.0  & 19.0   \\ 
MRS G             & 1.297  & 80.3  & 19.7  & 1.107  & 81.6  & 18.4   \\ 
MT B1            & 1.076  & 83.1  & 16.9  & 0.925  & 85.4  & 14.6   \\ 
CTEQ 1M         & 1.204  & 79.6  & 20.4  & 1.038  & 81.3  & 18.7   \\ 
CTEQ 1MS         & 1.206  & 79.9  & 20.1  & 1.030  & 81.6  & 18.4   \\ 
CTEQ 2M         & 1.274  & 79.4  & 20.6  & 1.078  & 81.0  & 19.0   \\ 
CTEQ 2MS         & 1.231  & 79.7  & 20.3  & 1.043  & 81.2  & 18.8   \\ 
CTEQ 2MF         & 1.225  & 78.7  & 21.3  & 1.054  & 80.2  & 19.8   \\ 
CTEQ 2ML         & 1.310  & 79.7  & 20.3  & 1.113  & 81.4  & 18.6   \\ 
CTEQ 3M           & 1.224  & 79.7  & 20.3  & 1.051  & 81.1  & 18.9   \\ 
GRV H\O          & 1.237  & 82.0  & 18.0  & 1.095  & 80.5  & 19.5   \\  
\end{tabular}
\end{center}
\caption[]{Parton luminosity slope, valence-valence ($v-v$), 
valence-sea ($v-s$) and sea-sea ($s-s$)
           contributions to the $W$ and $Z$ boson production cross section at 
           $\sqrt s$= 1.8~TeV. }
\label{tab:beta}
\end{table}



\begin{table}[h]
\begin{center}
\begin{tabular}{lcccc} 
    $\eta$ response     & $\Delta M_W$  & $\Delta M_W$   
                        & $\Delta M_W$   & $\Delta M_Z$     \\  
                        & $M_T$ fit ~(MeV/$c^2$)     & $p_T^e$ fit ~(MeV/$c^2$)
                        & $p_T^\nu$ fit ~(MeV/$c^2$) & $m_{ee}$ 
fit ~(MeV/$c^2$)      \\ \hline
    module A            &  $-6$ $\pm$ 16   &  $-7$ $\pm$ 22
                        & $-49$ $\pm$ 30   &  $-2$ $\pm$  6      \\
    module B            &  $+5$ $\pm$ 16   & $-15$ $\pm$ 22 
                        & $-26$ $\pm$ 30   &  $-8$ $\pm$  6      \\
\end{tabular}
\caption[]{Change in $W$ and $Z$ boson masses in MeV/$c^2$ if a non-uniform 
           calorimeter $\eta$ response is assumed, bracketed by the 
           variations observed for two EM modules exposed in a test beam. }
\label{tab:eta_resp} 
\end{center}
\end{table}

\begin{table}[h]
\begin{center}
\begin{tabular}{cccc}   
    fitted spectrum   & Monte Carlo   & Sensitivity     & Data      \\
                      & $\Delta M_W$~(MeV/$c^2$)  
                      & ${\partial M_W \over \partial C}$ 
                        (${{\rm MeV/}c^2 \over \% }$)
                  & $\Delta M_W$~(MeV/$c^2$)                                \\
\hline
    & & & \\
    $M_T$     & ${ }^{+58}_{-44}  \pm 17 $ 
              & $-112$ $\pm$ 19 
              & ${ }^{+43}_{-44}         $ \\
    $p_T^e$   & ${ }^{+44}_{-8}   \pm 22 $ 
              & $-54$ $\pm$ 14 
              & ${ }^{+11}_{-27}        $ \\
    $p_T^\nu$ & ${ }^{+64}_{-20}  \pm 30 $ 
              & $-56$ $\pm$ 19 
              & ${ }^{+47}_{-5}         $ \\
\end{tabular}
\caption[]{Uncertainty on the $W$ boson mass in MeV/$c^2$ due to a change in 
           the constant term of the electromagnetic energy resolution 
           by 0.5\%. The upper numbers correspond to the lower constant term.} 
\label{tab:constant_mw} 
\end{center}
\end{table}

\begin{table}[h]
\begin{center}
\begin{tabular}{cccc}   
    fitted spectrum   & Monte Carlo   & Sensitivity     & Data      \\
                      & $\Delta M_W$~(MeV/$c^2$)  
                      & ${\partial M_W \over \partial \kappa}$ 
                        (${{\rm MeV/}c^2 \over 0.01 }$)
                      & $\Delta M_W$~(MeV/$c^2$)                       \\
\hline
    & & & \\
    $M_T$     & ${ }^{+55}_{-73}  \pm 17 $ 
              & $+12.1$ $\pm$ 1.3 
              & ${ }^{+42}_{-80}         $ \\
    $p_T^e$   & ${ }^{+38}_{-29}  \pm 23 $ 
              & $+6.7$ $\pm$ 1.7 
              & ${ }^{+4}_{-38}             $ \\
    $p_T^\nu$ & ${ }^{-161}_{+94} \pm 30 $ 
              & $-30.3$ $\pm$ 2.5 
              & ${ }^{-125}_{+100} $ \\
\end{tabular}
\caption[]{Uncertainty on the $W$ boson mass due to the change 
in $p_T^W$ scale by 
           0.04. 
           The upper numbers are the change in mass when the $p_T^W$ scale 
           factor increases and the hadronic response is closer to the 
           electromagnetic response. } 
\label{tab:ptw_mw} 
\end{center}
\end{table}

\begin{table}[h]
\begin{center}
\begin{tabular}{cccc}
    fitted spectrum   & Monte Carlo   & Sensitivity     & Data      \\
                      & $\Delta M_W$~(MeV/$c^2$)
                      & ${\partial M_W \over \partial \# min.bias }$
                        (${{\rm MeV/}c^2 \over 0.1 }$)
                      & $\Delta M_W$~(MeV/$c^2$)                        \\
\hline
    & & & \\
    $M_T$     & ${ }^{-105}_{+121}  \pm 17 $
              & $-117$ $\pm$ 5
              & ${ }^{-253}_{+201}          $ \\
    $p_T^e$   & ${ }^{-14}_{+29}    \pm 23 $
              & $-20.0$ $\pm$ 7.0
              & ${ }^{-55}_{+9}          $ \\
    $p_T^\nu$ & ${ }^{-245}_{+318}  \pm 30  $
              & $-286$ $\pm$ 14
              & ${ }^{-535}_{+554}         $ \\
\end{tabular}
\caption[]{Uncertainty on the $W$ boson mass due to a change by 0.1 
in the number
           of minimum bias events underlying the $W$ event.
           The upper numbers are the change in mass for a higher average
           number of minimum bias events. }
\label{tab:syserr_minb}
\end{center}
\end{table}

\begin{table}[h]
\begin{center}
\begin{tabular}{cccc}   
    fitted spectrum   & Monte Carlo   & Sensitivity     & Data      \\
                      & $\Delta M_W$~(MeV/$c^2$)  
                      & ${\partial M_W \over \partial S }$ 
                        (${{\rm MeV/}c^2 \over 10\% }$)
                      & $\Delta M_W$~(MeV/$c^2$)                         \\
\hline
    $M_T$     & ${ }^{-74}_{+52}   \pm 17 $ 
              & $-31.5$ $\pm$ 6.0 
              & ${ }^{-52}_{+31}         $ \\
    $p_T^e$   & ${ }^{+2}_{-8}     \pm 23 $ 
              & $-2.5$ $\pm$ 7.8 
              & ${ }^{-4}_{-26}         $ \\
    $p_T^\nu$ & ${ }^{+95}_{-58}   \pm 30  $ 
              & $-38.3$ $\pm$ 11.0 
              & ${ }^{-87}_{+32}         $ \\
\end{tabular}
\caption[]{Uncertainty on the $W$ boson mass due to the change 
in the sampling term 
           of the hadronic energy resolution by 0.2\,. 
           The upper numbers are the change in mass for a larger resolution. }
\label{tab:syserr_hadres} 
\end{center}
\end{table}

\begin{table}[t]
\begin{center}
\begin{tabular}{lccccc}   
    PDF (CTEQ3M)  & $g_2 - 2\sigma$  & $g_2$    & $g_2 + 2\sigma$  
                  & $g_2 + 4\sigma$  & fit      \\
                  & $\Delta M_W$~(MeV/$c^2$) & $\Delta 
                  M_W$~(MeV/$c^2$) & $\Delta M_W$~(MeV/$c^2$) & $\Delta 
                  M_W$~(MeV/$c^2$) \\
\hline
CDF Asym. high    & $+32$   &  $+14$  &   $+50$  &  $+11$   & $M_T$ \\
CDF Asym. nominal & $-14$   &  0    &   $-37$  &  $-30$   & $M_T$ \\
CDF Asym. low     & $-55$   &  $-67$  &   $-69 $ &  $-65$   & $M_T$ \\
\hline
CDF Asym. high    & $+125$  &  $+51$  &   $+36$  &  $-60$   & $p_T^e$ \\
CDF Asym. nominal & $+45$   &  0    &   $-93$  &  $-137$  & $p_T^e$ \\
CDF Asym. low     & $-48$   &  $-127$ &   $-169$ &  $-197$  & $p_T^e$ \\
\hline
CDF Asym. high    & $+64 $  &  $+80$  &   $+77$  &  $-17$   & $p_T^\nu$ \\
CDF Asym. nominal & $+40$   &  0    &   $-43$  &  $-78$   & $p_T^\nu$ \\
CDF Asym. low     & $-64$   &  $-69$  &   $-141$ &  $-121$  & $p_T^\nu$ \\
\end{tabular}
\caption[]{Shift in the $W$ boson mass in MeV/$c^2$ when using different 
           parametrizations of the parton distribution functions and 
           $p_T^W$ spectrum. 
           There is a statistical uncertainty of 17, 24 and 31~MeV/$c^2$ 
on each 
           value for the $M_T$, $p_T^e$ and $p_T^\nu$ fit, respectively.}
\label{tab:mw_sys_theor} 
\end{center}
\end{table}

\begin{table}[ht]
\begin{center}
\begin{tabular}{lrrr}
    PDF                  & \multicolumn{1}{c}{$\Delta M_W$ }     
                         & \multicolumn{1}{c}{$\Delta M_W$ }     
                         & \multicolumn{1}{c}{$\Delta M_W$ }   \\
                         & \multicolumn{1}{c}{$M_T$ fit ~(MeV/$c^2$)} 
                         & \multicolumn{1}{c}{$p_T^e$ fit ~(MeV/$c^2$)}
                         & \multicolumn{1}{c}{$p_T^\nu$ fit ~(MeV/$c^2$)}   \\
    \hline
    MRSA                    & \multicolumn{1}{c}{---}   
                            & \multicolumn{1}{c}{---}        
                            & \multicolumn{1}{c}{---}        \\
    MRSB$^{(\ast)}$  
                            & $-90$  $\pm$ 19  & $-196$ $\pm$ 24   
                            & $-86$  $\pm$ 34                    \\
    MRSE$^{(\ast)}$  
                            & $-136$ $\pm$ 19  & $-168$ $\pm$ 24    
                            & $-198$ $\pm$ 34                    \\
    HMRSB$^{(\ast)}$ 
                            & $-157$ $\pm$ 19  & $-280$ $\pm$ 24    
                            & $-204$ $\pm$ 34                    \\
    KMRSB\O$^{(\ast)}$ 
                            & $-175$ $\pm$ 19  & $-238$ $\pm$ 24    
                            & $-244$ $\pm$ 34                    \\
    MRSD0$^\prime$          & $-74$  $\pm$ 19  & $-109$ $\pm$ 24    
                            & $-26$  $\pm$ 34                    \\
    MRSD$^\prime$-          & $-31$  $\pm$ 19  & $-9$   $\pm$ 24    
                            & $+8$   $\pm$ 34                    \\
    MRSH                    & $-30$  $\pm$ 19  & $-47$  $\pm$ 24    
                            & $-70$  $\pm$ 34                    \\
    MTB1$^{(\ast)}$  
                            & $-135$ $\pm$ 19  & $-260$ $\pm$ 24    
                            & $-144$ $\pm$ 34                    \\
    CTEQ1MS$^{(\ast)}$ 
                            & $-29$  $\pm$ 19  & $-109$ $\pm$ 24    
                            & $-1$   $\pm$ 34                    \\
    CTEQ2M                  & $+20$  $\pm$ 19  & $+1$   $\pm$ 24    
                            & $+53$  $\pm$ 34                    \\
    CTEQ2MS                 &   0  $\pm$ 19  & $-26$  $\pm$ 24    
                            & $+62$  $\pm$ 34                    \\
    CTEQ2MF                 & $-59$  $\pm$ 19  & $-112$ $\pm$ 24    
                            & $-84$  $\pm$ 34                    \\
    CTEQ2ML                 & $+29$  $\pm$ 19  & $+19$  $\pm$ 24    
                            & $+57$  $\pm$ 34                    \\
    CTEQ3M                  & $-31$  $\pm$ 19  & $-75$ $\pm$ 24    
                            & $-102$  $\pm$ 34                    \\
    GRVH\O                  & $-47$  $\pm$ 19  & $-88$  $\pm$ 24    
                            & $-50$  $\pm$ 34                    \\
\end{tabular}
\caption[]{Change in the $W$ and $Z$ boson masses in MeV/$c^2$ with varying 
           parametrizations of the structure of the proton for 
           transverse momentum spectra. Amounts quoted are relative to the 
           MRSA fit.
           The asterisk indicates those parton distribution functions considered
obsolete for 
           this analysis. }
\label{tab:pdf}
\end{center}
\end{table}

\begin{table}[h]
\begin{center}
\begin{tabular}{ccc}   
                      & \multicolumn{1}{r}{Monte Carlo}  
                      & \multicolumn{1}{r}{Data}                  \\ \hline 
    fitted spectrum   & $\Delta M_W $~(MeV/$c^2$)
                      & $\Delta M_W $~(MeV/$c^2$) \\
\hline
    $M_T$     & ${ }^{+37}_{-9}    \pm 17 $ 
              & ${ }^{+2}_{-13}           $                         \\
    $p_T^e$   & ${ }^{-46}_{+52}   \pm 23 $ 
              & ${ }^{-63}_{+41}          $                         \\
    $p_T^\nu$ & ${ }^{+124}_{-143} \pm 30 $ 
              & ${ }^{+136}_{-95}         $                         \\
\end{tabular}
\caption[]{Uncertainty on the $W$ boson mass due to uncertainty on the electron 
identification efficiency as a function of the quantity $u_\parallel$.
           The upper numbers are the change in mass when the 
           overall efficiency decreases. } 
\label{tab:upar_mw} 
\end{center}
\end{table}

\begin{table}[ht]
\begin{center}
\begin{tabular}{llccc}
Source & 
  Variation Used& 
  $\sigma(M_W)$   & 
  $\sigma(M_W)$   &
  $\sigma(M_W)$    \\
~       &     ~       & $M_T$ Fit & $p_T^e$ Fit & $p_T^\nu$ fit  \\
~       &     ~       & ~(MeV/$c^2$)  & ~(MeV/$c^2$) & ~(MeV/$c^2$) \\
\hline
Statistical                 &  & 140  & 190   & 260   \\
\hline  
Energy Scale                &  & 160  & 160   & 160   \\
\hline  
Other Systematic Errors     &  & 165  & 180   & 305   \\
\hline
EM energy resolution     &
     $ C = (1.5^{+0.6}_{-1.5}) $  &
     70 &
     35 &
     35 \\
      CDC $z$ scale$^{(\ast)}$ 
    & $ \alpha = (0.988 \pm 0.002 )  $ 
    & 50  
    & 55  
    & 55  \\
      Hadronic energy resolution    
    & $ S_{had} = 0.8 \pm 0.2 $ 
    & 65 
    & 5 
    & 80 \\
      Underlying event$^{(\ast)}$   
    & $ E^{\rm Tower}_T = (16.8 \pm 1.5) $  MeV 
    & 35    
    & 35 
    & 35 \\
      $\Gamma_W$                   
    & $ \Gamma_{W} = (2.1 \pm 0.1) $ GeV
    & 20 
    & 20 
    & 20 \\
      Hadronic energy scale         
    & $ \alpha_{had} = (0.83 \pm 0.04) $ 
    & 50 
    & 30 
    & 120 \\
      Number of minimum bias events      
    & $ (1.0 \pm 0.06) $ 
    & 60 
    & 10 
    & 150 \\
      QCD background               
    & $ (1.6 \pm 0.8) \% $ 
    & 30 
    & 35 
    & 35 \\
      {\small $Z\rightarrow e e$ } background
    & $(0.43 \pm 0.05) $ \% 
    & 15  
    & 20  
    & 20  \\
     Electron ID  efficiency  
    & parametrization 
    & 20 
    & 70 
    & 115 \\
      Radiative decays  
    & $E_\gamma^{min},\ R_{e\gamma}, \ \chi^2 $ 
    & 20 
    & 40 
    & 40 \\
      $ p_T(W) $, pdf 
    & $ p_T(W) $ variation 
    & 65    
    & 130    
    & 130 \\
      Trigger efficiency 
    & efficiency spread
    & 20 
    & 20 
    & 60 \\
      Non-uniformity in $\eta$
    & test beam 
    & 10 
    & 10 
    & 25 \\
    Fitting error
    &     & 5 & 10 & 10 \\
Total  &~ & 275 & 315 & 435 \\
\end{tabular}
\caption[]{Summary of systematic errors on the $W$ boson mass from the 
           three mass fits. 
           Those errors that are strongly correlated with the measured 
           $Z$ boson mass are indicated by an asterisk.      }
\label{tab:mw_sys_err_sum}
\end{center}
\end{table}


\begin{table}[t]
\begin{center}
\begin{tabular}{cc}   
    Data subsample               &   $\Delta M_W~\rm{(MeV/c^2)}$    \\
\hline
    One track in electron road$^{(\ast)}$       &   -2 $\pm$ 54     \\
    One reconstructed event vertex$^{(\ast)}$   &  -76 $\pm$ 76     \\
    Single interaction events $^{(\ast)}$ & -107 $\pm$ 95     \\
    $p_T^W < 10 $ GeV/c                           & -166 $\pm$ 90     \\
\end{tabular}
\caption[]{Change in $W$ mass from nominal for different subsamples of 
           the data. 
           Those subsamples for which the Monte Carlo templates were not 
           modified are indicated by an asterisk $\ast$. 
           Errors are statistical only.}
\label{tab:cross_check} 
\end{center}
\end{table}

\begin{table}[h]
\begin{center}
\begin{tabular}{ccc}   
    fitted spectrum   & $\Delta M_W~\rm{(MeV/c^2)}$  & 
                        $\Delta M_W~\rm{(MeV/c^2)}$        \\
                      & Data, no $u_\parallel$ cut  
                      & Data, $u_\parallel < 10 $ GeV       \\
                      & MC, $u_\parallel < 10 $ GeV         
                      & MC, $u_\parallel < 10 $ GeV         \\
\hline
    & & \\
    $M_T$     & $+78$         & $-16$       \\
    $p_T^e$   & $-280$        & $+40$       \\
    $p_T^\nu$ & $+810$        & $-45$       \\

\end{tabular}
\caption[]{Change in $W$ mass from nominal when applying a cut on 
           $u_\parallel$ of 10~GeV.}
\label{tab:upar_consist} 
\end{center}
\end{table}

\begin{table}[h]
\begin{center}
\begin{tabular}{lc}   

    $\eta$ range \qquad \qquad      
                                  & $R = {M_W \over M_Z } $             \\
\hline
  $|\eta^e | < 1.0$  & 1.0003 $\pm$ 0.0005 \\
  $|\eta^e | < 0.8$  & 1.0010 $\pm$ 0.0012 \\
  $|\eta^e | < 0.6$  &  1.0011 $\pm$ 0.0019 \\
\end{tabular}
\caption{The ratio of 
         the $W$ and $Z$ boson masses when restricting the $\eta$ range of the 
         electron. 
         The errors are the independent statistical errors with respect to 
         the nominal fitted mass. } 
\label{tab:eta_cut} 
\end{center}
\end{table}


\begin{references}
\bibitem{hbarc} Here, natural units in which $\hbar=c=1$ are used, except for
units of momentum (GeV/c) and mass (GeV/$c^2$).
\bibitem{light} For the light leptons, momenta will be equated with energies.
Transverse energy will then sometimes appear as a vector, $\vec{E}_T$.
\bibitem{reduce} The  effect of the width is to reduce the  height  and
smear  the distribution  slightly to the high side of the peak.  The transverse 
motion of the $W$ boson reduces the height of the  peak and  smears the shape 
toward the  low $M_T$ side;
$p_{T}^{W}$   will not   significantly  affect the   high edge of the
spectrum.
\bibitem{def_eta}
           The coordinate system in D\O\ is right-handed, with the $+z$ axis
pointing in the direction of the proton beam (south); 
the $+x$ axis horizontal, with the 
positive direction away from the center of the accelerator; 
and the $+y$ axis, up. The
origin of this coordinate stystem is centered at the midpoint 
of the detector where the nominal
beam intersection would occur. Radial directions are measured 
from that point. Polar
anges $\theta$ are measured with respect to the $+z$ axis, 
and azimuthal angles, $\phi$,
are measured with respect to the $+x$ axis. 
Pseudorapidity is defined as $\eta = -\ln\tan{\theta\over 2}$  
            where $\theta$ is the polar angle with respect to the proton beam.
\bibitem{track-top} The top quark analysis published in reference\cite{topprd}
cut this quantity at $\sigma_{trk} < 5$ which was 94\% efficient.
\bibitem{onejet} In
approximately 40\% of all
$W$ boson events in the data sample there is at least one jet 
produced with the $W$
boson with
$p_T > 8$~GeV/c.
\end{references}
\end{document}